\documentclass[12pt,draft]{ucthesis}

\input psfig.sty

\def\dsp{\def\baselinestretch{2.0}\large\normalsize}
\dsp
\def\CC{{\rm\kern.24em \vrule width.04em height1.5ex depth-.07ex
\kern-.30em C}}
\def\RR{{\rm\kern.24em \vrule width.04em height1.5ex depth-.07ex
\kern-.30em R}}
\def\bmath#1{\mbox{\boldmath$#1$}}
\def\NN{{\rm\kern.24em \vrule width.04em height1.2ex depth-.07ex
\kern-.30em N}}

\begin{document}

\newtheorem{definition}{Definition}[section]
\newtheorem{conjecture}{Conjecture}[section]
\newtheorem{lemma}{Lemma}[section]
\newtheorem{theorem}{Theorem}[section]
\newtheorem{open}{Open Question}[section]

% Declarations for Front Matter

\title{Decoherence, Control, and Symmetry in Quantum Computers}
\author{Dave Morris Bacon}
\degreeyear{2001} \degree{Doctor of Philosophy} \chair{Professor K. Birgitta
Whaley} \othermembers{Professor Raymond Chiao\\ Professor Dan Stamper-Kurn \\
Professor Umesh Vazirani} \numberofmembers{4} \prevdegrees{Double B.S. with
Honors, (California Institute of Technology) 1997} \field{Physics}
\campus{Berkeley}

\maketitle \approvalpage \copyrightpage

\ssp

\begin{abstract}
Computers built on the physical principles of quantum theory offer the
possibility of tremendous computational advantages over conventional computers.
To actually realize such quantum computers will require technologies far beyond
present day capabilities.  One problem which particularly plagues quantum
computers is the coupling of the quantum computer to an environment and the
subsequent destruction of the quantum information in the computer through the
process known as decoherence.  In this thesis we describe methods for avoiding
the detrimental effects of decoherence while at the same time still allowing
for computation of the quantum information.  The philosophy of our method is to
use a symmetry of the decoherence mechanism to find robust encodings of the
quantum information.  The theory of such decoherence-free systems is developed
in this thesis with a particular emphasis on the manipulation of the
decoherence-free information.  Stability, control, and methods for using
decoherence-free information in a quantum computer are presented.  Specific
emphasis is put on decoherence due to a collective coupling between the system
and its environment.  Universal quantum computation on such collective
decoherence decoherence-free encodings is demonstrated.  Along the way,
rigorous definitions of control and the use of encoded universality in the
physical implementations of quantum computers are addressed.  Explicit gate
constructions for encoded universality on ion trap and exchange based quantum
computers are given.  The second part of the thesis is devoted to methods of
reducing the decoherence problem which rely on more classically motivated
reasoning for the robust storage of information.  We examine quantum systems
that can store information in their ground state such that decoherence
processes are prohibited via considerations of energetics.  We present the
theory of supercoherent systems whose ground states are quantum error detecting
codes and give examples of supercoherent systems which allow universal quantum
computation.  We also give examples of a spin ladder whose ground state has
both the error detecting properties of supercoherence as well as error
correcting properties.  We present the first example of a quantum error
correcting ground state which is a natural error correcting code under
reasonable physical assumptions. We conclude by discussing the radical
possibility of a naturally fault-tolerant quantum computer.
 \abstractsignature
\end{abstract}

\begin{frontmatter}

\begin{dedication}

\null \vspace{3cm}

\begin{center}
Dedicated to my parents  \\
 \vspace{1cm}
 Nancy and Larry Bacon \\
 \vspace{1cm}
\end{center}

\noindent``My home...I think I would be lost without it.  I spend so much time
in my head, and what I do there is so different...so bizarre to most people. If
I didn't have a center, a place to return to, I'd get lost.'' \flushright{
--Patricia Vasquez, {\em Eon}\cite{Bear:95a}}
\\

\begin{center}
 \vspace{2cm}
and to the memory of my grandfathers \\ \vspace{1cm}
 Lee Morris (1912-1996)\\
 Glen Bacon (1908-1994)
\end{center}
\end{dedication}

\tableofcontents \listoffigures \listoftables

\begin{acknowledgements}

I would like to thank my parents first and foremost, not that ``thanks'' can
express my deep gratitude for what they have contributed to me.  This thesis is
dedicated to their love and unwavering support.

It is with great pleasure that I also wish to thank my advisor, Professor K.
Birgitta Whaley, for her support and guidance throughout my years as a graduate
student.  Anyone who can put up with such a pig-headed advisee like me deserve
a medal.  I would especially like to thank Professor Whaley for allowing me the
freedom to explore my many interests.

Thanks also are long overdue to Professor (mere Doctor just seconds ago!)
Daniel Lidar who leverage me from the depths of astrophysics back into the
bright light of quantum computation.  Thanks, Daniel, for listening to my crazy
ideas and putting up with my scatter-brained inefficiency.

This thesis would not be possible were it not for the long list of
collaborators from whom I have learned so much: Ken Brown, Guido Burkard,
Andrew Childs, Professor Isaac Chuang, Professor Richard Cleve, Dr. David
DiVincenzo, Dr. Julia Kempe, Dr. Debbie Leung, Professor Daniel Lidar,
Professor K. Birgitta Whaley, and Xinlan Zhou all put up with the sporadic
behavior of my collaboration.

As the culmination of formal education, my doctorate would not exists were it
not for the lineage of teachers who dedicated their hard work towards my mere
education.  Mr. Alexander let me find my own path and knew that science should
be done for its own sake and not to earn a grade or get a degree (and, yes, I
can derive this thesis from first principles!)  The encouragement of my
earliest science teachers Mr. McGonigal and Mr. Taylor are gratefully
acknowledged.  My English teachers through the years, especially Mrs. Perry,
Mr. Cummings, and Ms. Barker slowly peeled my eyes away from solely scientific
concerns and showed me a world of beauty and grandeur beyond science's rigor.
Finally I owe a tremendous debt to Professor Thomas Tombrello of the California
Institute of Technology. A hick fresh from the sticks, not knowing whether my
passion for science could translate into actually becoming a scientist,
Professor Tombrello showed me what it really meant to be a scientist and gave
me the confidence to believe in my own scientific abilities.

The two final teachers to whom I owe so much are not even official teachers:
Jun Cai (a.k.a Mike Cai a.k.a. studbud) and Luis de la Fuente.  Thanks for
being good friends and teaching me so much about myself.

I would also like to express my thanks to the lineage of scientists from whom I
have learned not only scientific principles, but also what it means to be a
scientist.  These are my heroes.  They have been my candles in the dark, the
bearers of my soul's burden from which I take solace and from whom I seek
guidance towards my own betterment.  David Bohm, John Bell, Charlie Bennett,
Isaac Chuang, Michael Nielsen, Roger Penrose, John Preskill, Charles Townes,
Alan Turing, John Archibald Wheeler, and William Wooters all showed me the
world of science in broad beautiful strokes. I owe more, perhaps, to the late
Richard Feynman than even I would ever be able to acknowledge.  The courage to
be unafraid of being a scientist, the unbelievable beauty of the universe as
seen through a physicists eyes, and the intense desire to explore the universe
are among the few traits which I found expressed in Feynman and which I hope to
express in myself.

I would especially like to express my thanks to the quantum computists at
Berkeley: my fellow-graduate-students-in-arms--Ken Brown, Julia Kempe, and
Simon Myrgren.  They laugh and ridicule today--but the future is not written on
a stone tablets in some god's closet--we will build a quantum computer.

Many thanks go out to the members of my qualifying examination committee,
Professor Eugene Cummins, Professor Raymond Chiao, Professor Robert Littlejohn,
and  Professor Umesh Vazirani and to the members of thesis committee, Professor
Raymond Chiao, Professor Dan Stamper-Kurn, and Professor Umesh Vazirani.  I
would also like to thank Professor Raymond Chiao for serving as my in
department advisor and Professor Lars Bildsten for serving as my initial
graduate advisor.

Not to forget all my friends who helped me keep my sanity throughout my
graduate career!  Thanks to the QBN crew and especially Ken ``Louie'' Brown and
Simon ``Louie'' Myrgren.  Without the distractions provided by Melissa Hampton,
West Burghardt, Emily Ho, the entire Vegas Tech crew, Ben and Ami Foster, and
the aforementioned Luis de la Fuente and studbud, I would be even crazier than
I actually am!

This thesis would not be possible were it not for the loving support of my
relatives far and wide.  Thanks to my big brother Michael Bacon for not being
too rough on his geeky little brother.  I would especially like to give thanks
for the support of my grandmothers, Norma Morris and Helen Bacon, as well as to
the scientific heritage of my grandfathers, the late Glen Bacon and the late
Lee Morris.

Finally, I would like to thank my sister Catherine Elizabeth Bacon, who is my
greatest inspiration.

\end{acknowledgements}

\end{frontmatter}

\part{Quantum Computation: Decoherence and Control}

\chapter{Philosonomicon}

\begin{quote}
{\em wherein we gently embark on an inquiry into the computational depths of
the physical universe and discover the fragile structure of information with
quantum foundations}
\end{quote}

\section{Prologue}

Our generous universe comes equipped with the ability to
compute\footnote{Blessed be the computational universe which allows this very
thesis\cite{Bacon:01a} to be typed onto a portable computer in the comfort of a
sorted array of pleasant locations.}. By the use of appropriate physical
systems algorithmic tasks can be executed with repeatable results which in turn
allow for the development of our systems of mathematics and physics consistent
with this repeatability. In physics, determination of the {\em allowable}
manipulations of a physical system is of central importance. Computer science,
on the other hand, has arisen in order to {\em quantify} what resources are
needed in order to perform a certain algorithmic function.  For computer
science to be applicable to the real world the quantification of resources
needed to perform a certain algorithmic function should be delimited by what
physics has determined to be allowable manipulations.  Thus we arrive at the
realization that because {\em information is physical}, our understanding of
computer science should be built on primitives which respect our understanding
of the laws of physics.  Terse in expression, ghostly trivial in its conceptual
underpinnings, this mantra that,
\begin{center}
``{Information is physical!}''\cite{Landauer:99a}
\end{center}
nonetheless has deep consequences for both the physicist examining how nature
behaves and the computer scientist attempting to understand the power and
limitations of real world execution of algorithmic tasks.  This very thesis, an
ever-growing body of scientific literature, and an equally expanding community
of scientists (rainbow in its composition of physicists, mathematicians, and
computer scientists), are but a small testament to the usefulness, practical
and abstract, of this one small idea.  ``Information is physical'', we thus
shout, and in this thesis we explore, the consequences of this small idea in
our large and generous universe.

\section{Argument via the inevitability of technology}

It is only through the bright light of hindsight that we can appreciate the
grandeur of scientific achievement during the twentieth century.  Unlike any
previous historical era, this scientific century has erected profound
disciplines from a seeming vacuum of prior consideration, pushing novel
technologies and new understandings in directions inconceivable only a few
years prior.  Among the two most far reaching movements of the twentieth
century's scientific symphony have been the composition of the quantum theory
of nature and the rising crescendo of the computer revolution.  To first
approximation these two fields appear in independent coexistence.  To master
the art of computer programming, knowledge of quantum theory is not
prerequisite.  Likewise, to learn contemporary physical theory, understanding
of modern computational theory is not necessary.

The illusion of separation between computer science and modern physical theory,
however, fades quickly as one's focus on details sharpens. On one hand,
comprehension of modern physical theory does not require computation, but our
understanding of the physical world is sharpened, if not progressed, by the use
of computers in the simulation of physical systems.  Whole realms of physics
would be inaccessible were it not for the use of computers to perform
calculations impossible on a human scale but possible with the calculational
capabilities of modern computers.  Fields like physical chemistry and lattice
quantum field theory now depend on the use of computational power to such a
degree that a growing view among theoretical physicists is to play the cynic
and declare they are ``computer programmers not physicists''\footnote{Posing a
significant retention problem for graduate physics programs!}.

The reverse implication between the two fields also occurs because computers
are physical devices such that quantum theory is essential to understanding
their physical operation.  The modern quantum theory of these devices presents
our best understanding of the physics behind the computer revolution.  Thus,
while there is nothing which is {\em essentially} quantum mechanical about the
algorithmic operation of today's computers, our understanding of the mechanisms
behind the computer architecture is deeply rooted in the quantum theory of the
solid-state.

\begin{figure}[h]
\hspace{2.5cm}  \psfig{figure=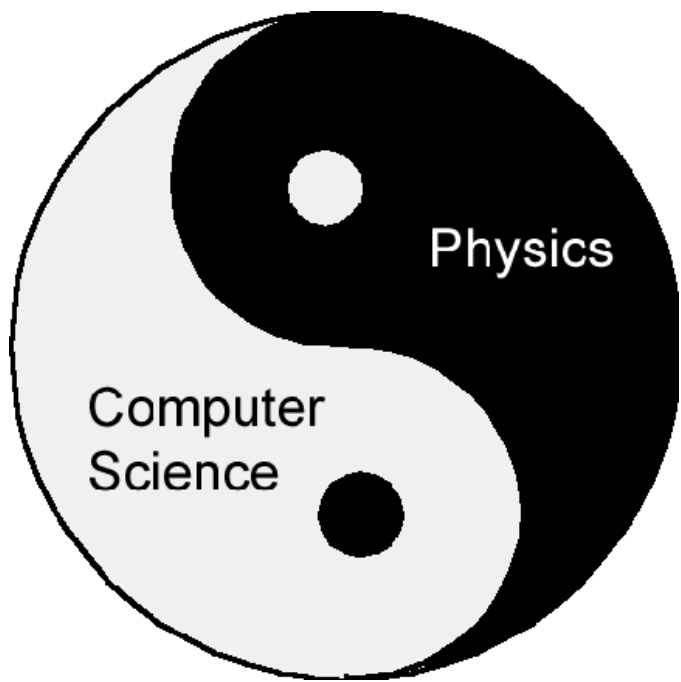,width=3.5in}  \caption{\em Physics
and computer science entangled} \label{fig:tao}
\end{figure}

How far can the two way relationship between computer science and quantum
theory be pressed?  The forward implication asks the question ``what can
computer science tell us about quantum theory?''\cite{Feynman:82a}  This thesis
will not concern itself with this question, and indeed it appears that very
little progress has been made along this line of inquiry (see, however
\cite{PhysComp:82a,Wheeler:89a}).

The reverse implication posses a different query: ``what can quantum theory
tell us about computer science?''  One important difference between this
implication and its inverse lies in the seeming {\em inevitability} of the
relevance of this question.  This inevitability arises from two different
directions.  In 1965 Gordon Moore noticed that the computational power of a
computer doubled approximately every two years\cite{Moore:65a}. A more physical
statement of this principle is that the number of atom's needed to represent
one bit of information will halve approximately every two years.  Since Moore's
1965 observation, Moore's law has continued to hold and been the barometer of
astounding technological progress in computer hardware. As Moore's law moves
into its fortieth year of success, however, a new barrier has arisen on the not
so distant horizon.  If Moore's law continues to hold, around the year 2015
Moore's law predicts that the size of the computational devices constructed
will reach a scale where quantum effects will begin to play a dominant
operational role.  One view of progress maintains that this will be the
essential limit to our current solid-state computer architectures: quantum
effects becoming dominant implies that no more computational power can be
squeezed out of the system.  On the other hand, it is unclear how a computer
operating at this quantum limit will behave. The argument of technological
inevitability leads us to believe that computers operating into the quantum
regime will be built.  Thus it seems technologically relevant to consider how
computers operating with quantum effects dominating will behave.  Quantum
theory can tell us something about how real computers of the future will
function.

A second reason for confidence in the inevitability of the role quantum theory
can play in computer science builds from a long line of experimental progress
in control of quantum systems.  In particular, fields like cavity quantum
electrodynamics\cite{Berman:94a}, ion and neutral atom
trapping\cite{Wineland:98a}, and certain areas of quantum
optics\cite{Gardiner:91a,Scully:97a,Yamamoto:99a}, have made considerable
progress in demonstration of the control of fully quantum degrees of freedom.
These extremely sensitive experimental successes point to a time in which
control over multiple interacting quantum systems will become possible.  From
the computer science prospective, such quantum control will represent
computational devices operating in a quantum regime. Again technological
progress leads us to believe that quantum control will be pressed further and
further until at least small scale computational quantum devices are
constructed.

Inevitably, we are thus led to assume that the relevance of quantum theory to
computational device will grow larger with time.  What, then, are the
consequences of this seemingly inevitable crash between the twentieth centuries
most prolific offspring, quantum theory and modern computation?

\section{The rise of the quantum algorithm}

\begin{quote}
{\em One must solemly affirm one's allegiance to the Quantum God before one may
be admitted to the physics clan.}
\\
\begin{flushright} --Carver A. Mead, {\em Collective Electrodynamics}\cite{Mead:00a}
\end{flushright}

\end{quote}

In the early 1980's Benioff\cite{Benioff:80a, Benioff:82a, Benioff:82b} and
Feynman\cite{Feynman:82a,Feynman:85a} began to consider computers whose {\em
algorithmic} operation was fully quantum mechanical.  Benioff appears to have
been motivated towards such quantum computers via the requirement that
description of quantum theory should be self-consistently described by machines
operating according to quantum theory.  Feynman, on the other hand, had a long
standing interest in the physical limits of computation\cite{Feynman:60a} which
apparently led him towards considering computers with quantum
components\cite{Feynman:96a}. However, while Feynman\cite{Feynman:82a}, and
earlier Manin\cite{Manin:80a}, clearly understood that simulating quantum
systems was in some form a difficult task, it took the ground breaking work of
Deutsch\cite{Deutsch:85a} and Deutsch and Jozsa\cite{Deutsch:92a} to realize
that computers built on quantum principles could perform computational tasks in
an intrinsically more efficient manner than could classical computers.  What
these latter authors showed was that there were circumstances under which
information in a quantum setting manipulated by a {\em quantum computer} had a
different productivity than equivalent classical information manipulated by a
classical computer.   Here, then, was the first interesting answer to the query
``what can quantum physics tell us about computer science?''  The
quantification of resources which is the main thrust of computer science was
shown to be {\em different} when operating in the quantum regime.

The work of Deutsch and Jozsa was followed up by a progression of work
demonstrating increasingly powerful applications of the idea of quantum
computation.  The oracle problem Deutsch and Jozsa investigated (and subsequent
results by Berthiaume and Brassard\cite{Berthiaume:92a,Berthiaume:94a}) was one
in which the amount of resources needed in order to perform the computation on
a quantum computer was exponentially less than a similar {\em exact}
computation performed on a classical computer.  However, a probabilistic
classical computer could solve the problem Deutsch and Jozsa posed with similar
use of resources if the problem output of the algorithm could be wrong with
some vanishingly small probability.  Thus the work of Deutsch and Jozsa alone
did not demonstrate a clear separation between classical and quantum
computation.

Overcoming the exactness requirement of Deutsch and Jozsa, Bernstein and
Vazirani \cite{Bernstein:93a} put forth algorithms which showed a true
superpolynomial resource gap between quantum and classical computation in 1993.
This was followed closely by the work of Simon\cite{Simon:94a} who posed a
problem which required exponentially more resources to solve on a classical
computer than on a quantum computer.  In 1994, following Simon's lead,
Shor\cite{Shor:94a} remarkably demonstrated that quantum computers could factor
numbers and compute a discrete logarithm efficiently.  Much work in complexity
theory has gone into attempting to develop efficient classical algorithms for
these two problems and it is widely believed that such efficient solution on a
classical computer is impossible\cite{Schneier:95a}. In fact, confidence in the
difficulty of these two problems forms the basis for the most widely used
public key cryptography systems\cite{Rivest:78a}.  Further evidence for the
power of quantum computers over classical computers was unveiled when Grover
\cite{Grover:96a,Grover:97a} demonstrated that quantum computers could search
unordered lists quadratically faster than classical computers.

By 1996, a clear separation in productivity between the algorithmic
manipulation of quantum information and classical information had been
established.  Further progress\cite{Boneh:95a,Kitaev:95a}
demonstrated\cite{Jozsa:98a} that Deutsch-Jozsa, Bernstein-Vazirani, Simon, and
Shor's algorithms were all related to a single problem known as the hidden
subgroup problem.  Separate from these Shor-type algorithms, research also
broadened\cite{Boyer:98a,Brassard:98a} and quantified\cite{Bennett:97a} the
algorithm developed by Grover.

A third line of research has shown how to use a quantum computer to efficiently
simulate quantum
systems\cite{Abrams:97a,Boghosian:98a,Lidar:99a,Sornborger:99a,Wiesner:96a,Zalka:98a}.
While there is no general proof that quantum systems are hard to simulate on
classical computers, the vast industry of physicists who have attempted to
provide such efficient simulations have all failed.  Building a quantum
computer would profoundly change the complexity of the quantum models studied
by physicists.

The discovery that quantum algorithms can outperform their classical brethren
is a result which should be fundamentally shocking to all studied computer
scientists.  The computational complexity classes of yesteryear have ethereal
foundations: the true foundations lie in a quantum setting.  Further shock
should also occur to those who use public key cryptosystems based on factoring
and discrete logarithms: the future building of a quantum computer will allow
your encrypted messages to be read!  Like any infant discovery, however, the
true power behind quantum computation is currently unclear.  Past ventures by
humanity in brandishing the skill of foresight--
\begin{quote}
{\em ``I think there is a world market for maybe five computers.'' - Thomas
Watson, chairman of IBM, 1943}
\end{quote}
--give us the confidence and optimism to believe that the field of quantum
algorithms is only beginning to bloom.

\section{Control and the quantum computer}

While the algorithmic speedup promised by quantum computers was being
developed, much work was done defining and understanding the basic question:
what exactly is a quantum computer?

A seminal step in modern computer science was taken when Turing defined the
class of functions now known as recursive or computable
functions\cite{Turing:36a}.  The Church-Turing
thesis\cite{Turing:36a,Church:36a,Church:36b} conjectures that this class of
functions corresponds precisely to what can be computed by an algorithmic
method in the real world.  Thus the Church-Turing thesis provides a fundamental
grounding upon which modern theoretical computer science is built: everything
that is naturally computable by an algorithm is precisely the class of
recursive functions.  Computer scientists are thus assured of job security by
basing their studies on the class of recursive functions.  Furthermore it was
found that a certain class of computers, {\em universal}
computers\cite{Turing:36a}, could be used to efficiently compute a recursive
function. Thus, under the Church-Turing thesis and universality results, a
computer scientist concerned with computation could be myopic to all models of
computation sans a universal computer.  Of particular importance to computer
science is that the Church-Turing thesis and universality results allow for the
development of a quantification\cite{Cook:71a,Karp:72a} of the computational
resources needed to perform a certain algorithmic task which is essentially
robust to the basic model of computation used to perform the task. Modern
computational complexity\cite{Papadimitriou:94a} theory is a house built upon a
frame of universal computers whose structural integrity is encoded in the
robustness claimed by the Church-Turing thesis.

The Church-Turing thesis, however, is not a mathematical proof, however, but an
empirical statement whose validity has withstood over seventy years of testing.
The advent of quantum computation, however, has brought the validity of the
computational complexity models founded upon the Church-Turing thesis into
question and in fact the very basis of computation which is now claimed to be
fundamental in computer science has taken a severe detour into the quantum
realm.  Early research in quantum computation generalized classical models of
computing, the Turing machine and the circuit model, into their quantum
mechanical analogies.  The quantum equivalent of a Turing machine was first
considered by Benioff\cite{Benioff:80a,Benioff:82a,Benioff:82b}.
Deutsch\cite{Deutsch:85a} and Yao\cite{Yao:93a} further developed quantum
Turing machines.  The quantum equivalent of classical circuits was introduced
by Deutsch\cite{Deutsch:89a} and this quantum circuit model (with certain
uniformity constraints)  was shown to be equivalent to the quantum Turing
machine by Yao\cite{Yao:93a}.

In the simplest quantum circuit model a sequence of quantum gates (unitary
evolution) is applied (perhaps in parallel) to an array of quantum mechanical
two-level systems (qubits) with an appropriate initialization and readout of
the quantum information.   One of the first results in quantum computation was
the demonstration that certain sets of quantum gates acting on such an array
are {\em universal} in the sense that any unitary evolution on the array could be
performed by an appropriate sequence of such gates.  Following early results
which required three-body interactions\cite{Deutsch:89a} between quantum
systems it was subsequently realized that two-body
interactions\cite{DiVincenzo:95a} were sufficient to perform universal quantum
computation in the quantum circuit model.

Because quantum interactions are intrinsically analog in nature (interaction
times and coupling strengths are real numbers) the correct description of
universal quantum circuits requires some notion of
approximation\cite{Knill:95a,Knill:95b}.  This is similar to the situation with
probabilistic classical computers.  At first glance it appears that the analog
nature of probabilities may cause unwarranted power due to infinite accuracy in
such classical probabilistic computers. Models which contain bounded accuracy
in their transition probabilities, the real world equivalent to a classical
probabilistic machine, however, can be shown to form a robust computational
class.  Similarly, quantum circuits must be cast within the framework of finite
accuracy.  In particular, discrete sets of quantum gates implemented with a
finite accuracy are the real building blocks of a quantum circuit.  Luckily
such discrete sets were shown to be able to approximate any exact quantum
circuit to within an accuracy $\epsilon$ (defined on some suitable distance
measure) with only $\log^c(\epsilon^{-1})$ computational
overhead\cite{Kitaev:97a,Solovay:95a}. This in turn allows for the
establishment of robust computational complexity classes within the context of
such discrete gate universal quantum computers.

The universality results in the quantum circuit model show that given
sufficient control over quantum systems there is a robust class of computations
based on the quantum circuit model.  Thus sufficient quantum control implies
quantum computation.  But what of the validity of the quantum circuit model as
a real description of quantum systems?  Quantum circuits clearly map to quantum
systems, but how realistic are the assumptions that go into the quantum circuit
model?

\section{The decoherence roadblock}

Unfortunately, the description adopted in the quantum circuit model does not
correspond to the real world in a particularly nasty detail.  The quantum
circuit model describes a quantum computer as a closed quantum system.  The
whole formalism of a quantum circuit assumes that there is a system which
executes the circuit but is completely isolated from the rest of the universe.
In the real world, however, there are no known mechanisms for truly isolating a
quantum system from its environment.  Real quantum systems are open quantum
systems. Open quantum systems couple to their environment and destroy the
quantum information in the system through the process known as
decoherence\cite{Giulini:96a}. When examining the simple evolution of a single
quantum system this system-environment coupling appears to cause errors on the
quantum system's evolution.  The picture of a quantum circuit where only
desired unitary evolution occurs is thus naive.

Decoherence, then, is a direct attack on the physical viability of quantum
computers in the real world\cite{Haroche:96a,Landauer:96a,Unruh:95a}.  Because
quantum information is not easily isolated from its environment, physics
dictates that the quantum information will lose many of the properties that
make the information quantum and not classical.  In fact, much of the infamous
transition from quantum to classical physics has been attributed to the role of
decoherence in physical systems\cite{Zurek:82a,Zurek:91a,Zurek:01a}.  Thus to
the question ``what makes a computer classical and not quantum?'' the answer
``decoherence'' follows. While the algorithmic speedup promised by quantum
computers viewed as a closed system is a profound observation, it is all for
naught if this {\em decoherence problem} cannot be overcome.

There is an analogy here with classical computers operating in noisy
environments.  For example, conventional computers exposed to hard radiation of
space will not function properly due to the errors caused on the computer
hardware by the radiation.  At first glance it would appear that a classical
computer operating in such an environment would be useless.  One mistake in the
calculating the trajectory of a satellite can mean the complete destruction of
the satellite!  Besides the obvious practice of making the computer hardware
resilient to the radiation, perhaps surprisingly, there is another method for
overcoming this problem known a ``fault-tolerant'' computation.  Fault-tolerant
computation is intimately related to the idea of error correcting codes. In
classical error correcting codes, information transmitted through a noisy
channel is made more resistant to the noise by making the information
redundant.  This basic idea, that redundancy can protect information, was
extend by von Neumann\cite{vonNeumann:56a} to provide a method for performing
computations in the presence of noisy environments and imperfect operations.
Thus the question that emerged around 1996 was does there exists a theory of
fault-tolerant quantum computation?

The first step towards solving the decoherence problem was taken in 1995 when
Shor\cite{Shor:95a} (and independently Steane\cite{Steane:96a}) discovered a
quantum analogue of classical error correcting codes.  Shor discovered that by
encoding quantum information, this information could become more resistant to
interaction with its environment.  Following this remarkable discovery a
rigorous theory of quantum error correction was
developed\cite{Bennett:96a,Calderback:96a,Calderback:97a,Ekert:96a,Knill:97a}.
Many different quantum error correcting codes
\cite{Chuang:97b,Gottesman:96a,Knill:96a,Laflamme:96a,Leung:97a,Rains:97a,Steane:96b,Steane:99a}
were discovered and this further led to a theory of fault-tolerant quantum
computation\cite{Aharonov:97a,Gottesman:98a,Kitaev:97b,Knill:98a,Preskill:98a,Shor:96a}.
Fully fault-tolerant quantum computation describes methods for dealing with
system-environment coupling as well as dealing with faulty control of the
quantum computer.  Of particular significance was the discovery of the
threshold theorem for fault-tolerant quantum
computation\cite{Aharonov:97a,Gottesman:97a,Kitaev:97a,Knill:98a,Preskill:98a}.
The threshold theorem states that if the decoherence interactions are of a
certain form and are weaker than the controlling interactions by a certain
ratio, quantum computation to any desired precision can be achieved.  The
threshold theorem for fault-tolerance thus declares a final solution to the
question of whether there are theoretical limits to the construction of robust
quantum computers.

\section{Quantum Gemini: decoherence and control}

The study of information in a quantum setting is beginning to describe an
amazingly rich computational universe.  In this brief introduction we have
learned that quantum algorithms can perform astounding computational feats.
Quantum control can be used to perform these algorithms, while decoherence can
be overcome by this same quantum control.  In spite of these discoveries, the
inevitability of quantum technology remains unclear.  Exactly what physical
systems will be used to build a quantum computer?  There have been a plethora
of proposed physical systems for quantum computation and a few of these have
even moved from the drawing board to small scale
implementation\cite{Turchette:95a,Monroe:95a,Chuang:98a,Nakamura:99a}. Just as
vacuum tubes of the past have been replaced by the silicon wafers of today, the
hardware of future quantum computers, however, is currently far from certain.

Given the state of ignorance as to the suitability of different physical
systems for quantum computation, it is important to provide theoretical
groundwork towards understanding what does and does not make a good quantum
computer. This implies understanding the delicate dance between quantum
computation's twin considerations: decoherence and control.

In Greek mythology, Castor and Pollux were twins born to the same mother but
with different fathers.  Pollux's father was a god while the Castor's father
was a mere mortal.  Thus Pollux was immortal while Castor was mortal.  When
Castor died on the battlefield his brother was so stricken with grief that he
pleaded with Zeus to either send him to the same fate or restore his brother to
life. Zeus was touched by the brotherly love and allowed Castor to spend
alternating days on Olympus with the gods and in the mortal world below the
Earth, Hades. Due to their exemplary example of brotherly love the star
constellation Gemini was placed in the heavens by Zeus in honor of these twins.

In this thesis we venture forth towards understanding a modern day quantum
Gemini.  Quantum control, our Pollux, is the powerful near-immortal master of
quantum computation.  Decoherence, our Castor, pulls quantum computation down
into the mortal real world.  Sufficient quantum control helps pull decoherence
away from real world difficulty and restores the glory of quantum computation.
``Define, clarify, and broaden his brotherly relationship between decoherence
and quantum control'', we thus beseech and in this thesis we explore, ``and
someday quantum computers will move from myth to reality!''

\section{Thesis outline}

This thesis is divided into three main parts.  In part I of the thesis we
introduce the basic notions of decoherence, control and universality in a
quantum computer.  Chapter \ref{ch:qo} discusses the basic formalism of quantum
operators for describing decoherence and presents a non-standard derivation of
a semigroup master equation through the operator-sum representation.  Chapter
\ref{ch:control} then introduces the notion of control of a quantum system.
Necessary and sufficient conditions for interactions which can be used for
control which does not cause decoherence are presented.  The Lie algebraic
structure of control is then discussed along with the important issue of
approximation in quantum control.  Chapter \ref{ch:univ} shifts focus towards
universal quantum computation with a special emphasis on the subsystems nature
of universal quantum computers.  The notion of encoded universality is
introduced with an emphasis on the Lie algebraic structure of such encodings. A
criteria for universal quantum computation is derived which is useful for
deciding when even encoding cannot turn a set of interactions into a universal
set of interactions.

In part II of this thesis we turn to the theory of decoherence-free subspaces
and decoherence-free subsystems.  We begin in Chapter \ref{ch:dfscond} by
deriving necessary and sufficient conditions for the existence of
decoherence-free subspaces and their generalization decoherence-free
subsystems.  The role of the OSR algebra is stressed as a fundamental method
for understanding both decoherence-free subspaces and decoherence-free
subsystems.  The commutant of the OSR algebra is also identified as an
important characterizer of decoherence-free systems.  Decoherence-free
subsystem conditions in the master equation are introduced and the reason why
such conditions are currently only necessary are discussed.  Finally the
connection between symmetrization schemes and decoherence-free subsystems is
discussed.  In Chapter \ref{ch:dfsstab} we discuss the stability of
decoherence-free systems to perturbations.  We show that perturbing
interactions do not destroy the decoherence-free properties.  Chapter
\ref{ch:dfsqc} discusses many of the issues which generically arise when using
a decoherence-free subsystem for quantum computation.  In Chapter~\ref{ch:col}
we introduce an important model of decoherence which supports decoherence-free
subsystems, the collective decoherence model.  Master equations are derived for
both collective dephasing and for collective amplitude damping in order to
better illustrate the conditions under which collective decoherence occurs. The
notion of weak and strong collective decoherence is introduced and the
decoherence-free subsystems for both of these cases is introduced.  In
Chapter~\ref{ch:collectiveuniv} we discuss universal quantum computation on the
weak and strong collective decoherence decoherence-free subsystems.  The use of
only the exchange interaction for quantum computing is discovered and issues of
the explicit use of collective decoherence decoherence-free subsystems for
quantum computation are discussed.  In Chapter~\ref{ch:ion} we discuss
universal quantum computation on an experimentally realized decoherence-free
subspace in ion traps.  Explicit control sequences are identified for such
computation.  Chapter~\ref{ch:exchange} then discusses how solid state
proposals for quantum computation can be simplified and improved by the use of
encoded universality with the exchange interaction.  Finally in
Chapter~\ref{ch:atom} we discuss decoherence-free subspaces in atomic systems.

In part III of this thesis we turn to methods for building a quantum computer
which rely on techniques of robustness due to the energetics of the decoherence
process. In Chapter~\ref{ch:sup} we describe the effect of supercoherence where
quantum information is protected at low environment temperatures.  A
supercoherent system which allows for universal quantum computation is derived
and presented in the context of a solid-state implementation of a quantum
computer. We then present a spin ladder in Chapter~\ref{ch:ladder} which has
many of the properties of supercoherence as well as new error correcting
properties. In Chapter~\ref{ch:lattice} this is taken one step further and a
system with a ground state which is a quantum error correcting is discussed.
This is the first example of such a quantum error correcting ground state which
is fully quantum mechanical and which does not require unreasonable physical
resources. Finally in Chapter~\ref{ch:nft} we discuss the possibility of
naturally fault-tolerant quantum computation. Analogies with the classical
robustness of information are discussed and a general framework for future
natural fault-tolerant quantum computation is provided.

\chapter{The Pain of Isolating Quantum Information: Decoherence} \label{ch:qo}

\begin{quote}
{\em so its quantum computers we want \\ with computational power we can
flaunt\\ well there's a price which we'll have to pay \\ because quantum
coherences rejoice in decay }
\end{quote}

In this chapter we introduce the basic theory of quantum operations for
studying decoherence.  We begin by giving a simple example of how decoherence
can destroy quantum information.  We then introduce decoherence in an abstract
formalism known as the operator-sum representation(OSR).  Shortcomings of this
formalism are illuminated.  We then discuss the physically motivated
approximations of the operator-sum representation known as master equations.  A
mystery in decoherence rates calculated in the operator-sum representations is
presented and solved.

\section{The degradation of quantum information} \label{sec:degrade}

Quantum computation would be a matter of the control of quantum systems (not
itself a completely trivial subject) were it not for the fact that quantum
systems are open systems. The degradation of quantum information due to the
coupling of the system containing the quantum information to the environment is
called {\em decoherence}\footnote{An unfortunate state of nomenclature exists
as to the use of the word decoherence. Early
researchers\cite{Zurek:82a,Zurek:91a} used the word decoherence to refer to
operations which destroyed quantum coherences and transferred information to
the environment in a very specific manner. With the development of quantum
computation many authors loosened the use of this word to refer to any
system-environment couplings, not just those which destroy coherence in a
specific basis or involve specific transfer of information from the system to
the environment. In this thesis we will use the word decoherence to refer to
such generic system-environment couplings.}. Let us begin our understanding of
the degradation of quantum information by examining a simple example.

Suppose we are given a system of consisting a single qubit and an environment
consisting of another qubit.  The Hilbert space of this combined system and
environment is ${\mathcal H}={\mathcal H}_S \otimes {\mathcal H}_E \equiv \CC^2
\otimes \CC^2$. Further suppose that there is a coupling between the system and
the environment given by the Hamiltonian ${\bf H}=\lambda \bmath{\sigma}_z
\otimes \bmath{\sigma}_z$ where $\lambda$ is a fixed coupling constant.

We wish to encode on the system a qubit of quantum information,
$|\psi\rangle=\alpha |0\rangle + \beta |1\rangle$.  If $\lambda=0$ we could
create the state $|\psi\rangle$ and the state of the system would remain
$|\psi\rangle$ for all times after its creation: the quantum information would
be preserved.  If however $\lambda \neq 0$, there is a coupling between the
system and the environment given by the evolution operator
\begin{equation}
{\bf U}(t)=\exp\left[-i \lambda t \bmath{\sigma}_z \otimes
\bmath{\sigma}_z\right]= \cos(\lambda t) {\bf I} -i \sin(\lambda t)
\bmath{\sigma}_z \otimes \bmath{\sigma}_z. \label{eq:exampleu}
\end{equation}
Suppose that the environment is initially in the state $|+\rangle={1 \over
\sqrt{2}} \left(|0\rangle + |1 \rangle \right)$, so that initially the state of
the system plus environment is $|\psi\rangle \otimes |+\rangle$.  At a time $t$
latter, the state of the system plus environment will be
\begin{equation}
|\psi(t)\rangle=\cos(\lambda t) |\psi\rangle \otimes |+\rangle -i\sin(\lambda
t) \left( \bmath{\sigma}_z |\psi\rangle  \right)\otimes |-\rangle,
\end{equation}
where $|-\rangle={1 \over \sqrt{2}} \left(|0\rangle - |1 \rangle \right)$.  The
density matrix of the system at time $t$ is given by
\begin{eqnarray}
\bmath{\rho}_S(t)&=&{\rm Tr}_E\left[|\psi(t) \rangle \langle \psi(t)| \right]=
\cos^2(\lambda t) |\psi \rangle \langle \psi | + \sin^2(\lambda t)
\bmath{\sigma}_z |\psi \rangle \langle \psi | \bmath{\sigma}_z \nonumber \\
&=&\left[\begin{array}{cc}
  |\alpha|^2 & \alpha \beta^* \cos(2 \lambda t) \\
  \alpha^* \beta \cos(2 \lambda t) & |\beta|^2
\end{array} \right]. \label{eq:exampleevo}
\end{eqnarray}
Here ${\rm Tr}_E[\cdot]$ represents tracing over the environment.   The
residual difference between this density matrix and the initial density matrix
is
\begin{equation}
{\delta \bmath{\rho}} =\bmath{\rho}(t) -\bmath{\rho}(0)=\left[\begin{array}{cc}
  0 & \alpha \beta^* (1-\cos(2 \lambda t)) \\
  \alpha^* \beta (1- \cos(2 \lambda t)) & 0
\end{array} \right].
\end{equation}
We here see that as time evolves, the off diagonal elements of the density
matrix oscillate in time.  We thus say that the ``coherence'' between the
$|0\rangle$ and $|1\rangle$ states is being manipulated.  Note that a time
$t={\pi \over \lambda} k$, where $k$ is an integer, the quantum information in
the system is unaffected, $\delta \bmath{\rho}={\bf 0}$. System-environment
coupling alone is not enough to degrade the quantum information.  In addition
to the coupling, {\em an assumption about the inaccessiblity of the
environmental degrees of freedom must be made in order for decoherence to
occur}.  Suppose, for example, that at time $t_0={\pi \over 4 \lambda}$ the
coupling between the system and the environment is turned off and the state of
the environment is made inaccessible to experiment.  At this time the diagonal
elements of the density matrix in the $|0\rangle, |1\rangle$ basis completely
vanish.  Since the environmental degrees of freedom are now, by assumption,
assumed to be inaccessible, the quantum information in the system has been
degraded.  As described in Appendix \ref{apa:tracenorm}, the trace norm between
two density matrices is a good measure of the absolute distinguishability of
the density matrices.  For this example we calculate that
\begin{equation}
D(\bmath{\rho}(0),\bmath{\rho}(t))={1 \over 2} {\rm Tr}|\delta
\bmath{\rho}|=|\alpha| |\beta| |1-\cos(2 \lambda t)|.
\end{equation}
The best measurement to distinguish the initial state from the decohered state
at time $t$ will produce measurements probabilities whose absolute difference
will differ by $|\alpha||\beta||1-\cos(2 \lambda t)|$.  If one is thinking
about using this qubit for some sort of computation, then we see that the
computation will err with a probability of at least this value.

This simple example of decoherence serves to illustrate the basic idea that
coupling between the system and environment can lead to degradation of quantum
information.

\section{Quantum operations}

In this section we describe a basic formalism for understanding open quantum
systems.  In particular we seek to understand the evolution of a system when it
is coupled to an environment as seen from the perspective of the system alone.

\subsection{Derivation} \label{sec:osrderive}

Consider the dynamics of a system $S$ together with the rest of the universe
which we will call the environment $E$.  We will assume that the system $S$
represents full degrees of freedom separate from those of the environment $E$.
The state space of the system plus environment then occupies a Hilbert space
which is the tensor product of the system and environment Hilbert spaces,
${\mathcal H} \equiv {\mathcal H}_S \otimes {\mathcal H}_E$.

Note that this is not the most general definition of a system--it is possible
that the system we are interested in does not have support over a full degree
of freedom.  This is the case, for example, when one is interested in a limited
number of levels of a multi-level atom.  In this situation, probability can
``leak'' in or out of the system from or to the rest of the degree of freedom.
We will develop our formalism for the situation where the system is a full
degree of freedom but note where results can be extended to this more general
definition of a system.

The evolution of the system $S$ plus environment $E$ (which together do from a
closed system by postulate) is unitary with a Hamiltonian given by
\begin{equation}
{\bf H}={\bf H}_S \otimes {\bf I}_E + {\bf I}_S \otimes {\bf H}_E +{\bf
H}_{SE}, \label{eq:hamiltoniandecomp}
\end{equation}
where ${\bf H}_S$ acts on the system degrees of freedom $S$, ${\bf H}_E$ acts
on the environmental degrees of freedom and ${\bf H}_{SE}$ couples these
degrees of freedom.  The evolution of the system and environment is then
governed by the evolution operator ${\bf U}(t)=\exp \left[ -i {\bf H} t
\right]$. When the coupling between the system and the environment is zero,
${\bf H}_{SE}=0$, the evolution of the system plus environment are separately
unitary ${\bf U}(t)=\exp \left[-i {\bf H} t \right] = {\bf U}_S(t) \otimes {\bf
U}_E(t)$ with ${\bf U}_S(t)=\exp \left[-i {\bf H}_S t \right]$ and ${\bf
U}_E(t)=\exp \left[ -i {\bf H}_E t \right]$.  From the perspective of the
system alone, the evolution is therefore strictly unitary {\em independent} of
the possibly entangled initial state of the system and environment:
\begin{equation}
{\bmath{\rho}}_S(t) = {\rm Tr}_E \left[ {\bf U}(t) {\bmath{\rho}}(0) {\bf
U}^\dagger(t) \right] = {\bf U}_S(t) {\bmath{\rho}}_S(0) {\bf U}_S^\dagger(t).
\end{equation}
If this were not true, it would allow for superluminal manipulation of distant
systems.

When, however, ${\bf H}_{SE} \neq 0$, the evolution of the system and the bath
is more complicated.  Let us first examine the situation when the system and
the bath are initially in a tensor products state ${\bmath{\rho}}(0)={\bmath{
\rho}}_S(0) \otimes {\bmath{\rho}}_E(0)$.  From the perspective of the system
the evolution is given by
\begin{equation}
{\bmath{\rho}}_S(t) = {\rm Tr}_E \left[ {\bf U}(t) {\bmath{\rho}}_S(0) \otimes
{\bmath{\rho}}_E(0) {\bf U}^\dagger(t) \right].
\end{equation}
The initial state of the environment can be written in terms of its spectral
decomposition $\bmath{\rho}_E(0) = \sum_\nu p_\nu |\nu \rangle \langle \nu |$
where $|\nu \rangle \in {\mathcal H}_E$ is a complete orthogonal basis for the
environment which diagonalizes $\bmath{\rho}_E(0)$, $0 \leq p_\nu \leq 1$, and
$\sum_\nu p_\nu = 1$. Expanding the trace and using the spectral decomposition
of the environment we find that
\begin{equation}
{\bmath{\rho}}_S(t)= \sum_{\mu,\nu} \langle \mu | {\bf U}(t)
{\bmath{\rho}}_S(0) p_\nu |\nu \rangle \langle \nu | {\bf U}^\dagger(t) |\mu
\rangle,
\end{equation}
or
\begin{equation}
{\bmath{\rho}}_S(t)= \sum_i {\bf A}_i(t) \bmath{\rho}_S(0) {\bf
A}_i^\dagger(t), \label{eq:osr}
\end{equation}
where
\begin{equation}
{\bf A}_{i=(\mu,\nu)}(t)= \sqrt{p_\nu} \langle \mu | {\bf U}(t) |\nu \rangle.
\label{eq:osrexplicit}
\end{equation}
The requirement that ${\bf U}(t)$ is unitary implies that
\begin{equation}
\sum_{i=(\mu,\nu)} {\bf A}_i^\dagger(t) {\bf A}_i(t) ={\bf I}.
\label{eq:osrnorm}
\end{equation}
Eq.~(\ref{eq:osr}) together with the normalization condition
Eq.~(\ref{eq:osrnorm}) form the {\em trace-preserving operator-sum
representation} (OSR).  Notice that the exact form of the OSR operators depends
on the basis $|\mu \rangle$ (not the $|\nu\rangle$ basis which is determined by
the spectral decomposition).  The evolution does not depend on this basis
expansion, but the exact form of the operators ${\bf A}_i(t)$ does depend on
this basis choice.  We will return to this freedom

In fact it can be shown\cite{Kraus:83a} that the most general evolution of a
density matrix, ${\bmath{\rho}}(t)= {\mathcal L}(t) \left[ {\bmath{\rho}}(0)
\right]$ satisfying the requirements
\begin{enumerate}
  \item The map ${\mathcal L}(t)$ takes density matrices to density matrices.
  \item The map ${\mathcal L}(t)$ is a linear map.
  \item The map ${\mathcal L}(t)$ is completely positive.  A completely
  positive map takes positive operators to positive operators when acting as
  identity on an auxiliary space ${\mathcal I} \otimes {\mathcal L}(t) [ {\bf A} ]
  \geq 0$ for ${\bf A} \geq 0$, with ${\mathcal I}$ the identity operator on
  any addition Hilbert space.
\end{enumerate}
must have the form of the OSR.  Every possible OSR has a description in terms
of the action of a unitary operator on a larger Hilbert space.  This allows us
to favor the more concrete derivation of the OSR from the physical perspective
of unitary evolution traced over the environment as opposed to the more
axiomatic approach.

\begin{figure}[h]
\vspace{0.5cm} \hspace{1in} \psfig{figure=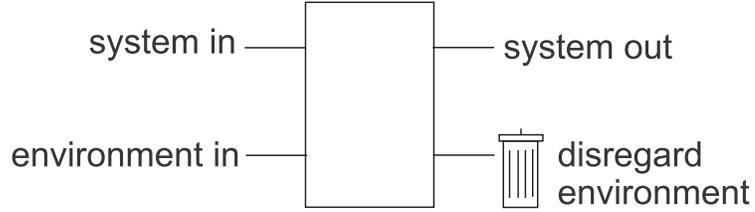,width=4in} \vspace{0.2cm}
\caption{\em Diagram of the operator sum representation} \label{fig:osr}
\end{figure}

\subsection{Fixed basis OSR} \label{sec:fixedbasis}

A tool which we will find useful later in our derivation of master equations is
the fixed basis form of the OSR\cite{Chuang:97a,Bacon:99a}.  Suppose we choose
a fixed basis (see Appendix~\ref{apa:fixedbasis}) for expanding each of the
operators ${\bf A}_i(t)$ in the OSR:
\begin{equation}
{\bf A}_i(t)=\sum_{\alpha} b_{i\alpha}(t) {\bf F}_\alpha. \label{eq:fixedbasis}
\end{equation}
The OSR can then be written as
\begin{equation}
\bmath{\rho}(t)=\sum_i {\bf A}_i(t) \bmath{\rho}(0) {\bf A}_i^\dagger(t) =
\sum_{i \alpha \beta} b_{i\alpha}(t) b_{i\beta}^*(t) {\bf F}_\alpha
\bmath{\rho}(0) {\bf F}_\beta^\dagger= \sum_{\alpha \beta} \chi_{\alpha
\beta}(t) {\bf F}_\alpha \bmath{\rho}(0) {\bf F}_\beta,
\label{eq:fixedbasisosr}
\end{equation}
where
\begin{equation}
\chi_{\alpha \beta}=\sum_i b_{i \alpha }(t) b_{i\beta}^*(t).
\end{equation}
Eq.~(\ref{eq:fixedbasisosr}) is the fixed basis or chi representation of the
OSR. Normalization requires that
\begin{equation}
\sum_{i \alpha \beta} b_{i\alpha}^* b_{i\beta} {\bf F}_\alpha^\dagger {\bf
F}_\beta = \sum_{\alpha \beta} \chi_{\beta \alpha} {\bf F}_\alpha^\dagger {\bf
F}_\beta= {\bf I}. \label{eq:fixedbasisnorm}
\end{equation}
Taking the trace of this equation we find that
\begin{equation}
\sum_{\alpha} \chi_{\alpha \alpha} = d.
\end{equation}
The $\chi_{\alpha \beta}(t)$ matrix is a positive hermitian matrix which
specifies the OSR in a given basis.

Separating out the identity components of Eq.~(\ref{eq:fixedbasisosr}) and
Eq.~(\ref{eq:fixedbasisnorm}) we obtain
\begin{equation}
\bmath{\rho}(t)=\chi_{00}(t) \bmath{\rho}(0) + \sum_{\alpha\neq0} \left[
\chi_{\alpha 0}(t) {\bf F}_\alpha \bmath{\rho}(0) + \chi_{0 \alpha}(t)
\bmath{\rho}(0) {\bf F}_\alpha^\dagger \right] + \sum_{\alpha,\beta \neq 0}
\chi_{\alpha \beta}(t) {\bf F}_\alpha \bmath{\rho}(0) {\bf F}_\beta^\dagger.
\end{equation}
and
\begin{equation}
\chi_{00}(t) {\bf I} + \sum_{\alpha \neq 0} \left[ \chi_{\alpha 0}(t) {\bf
F}_\alpha + \chi_{0 \alpha}(t) {\bf F}_\alpha^\dagger \right] +
\sum_{\alpha,\beta \neq =0} \chi_{\alpha \beta}(t) {\bf F}_\beta^\dagger {\bf
F}_\alpha = {\bf I}.
\end{equation}
Multiplying the second of these equations by ${1 \over 2} \bmath{\rho}(0)$ from
both the left and right, and substituting into the evolution equation, we
obtain
\begin{equation}
\bmath{\rho}(t)-\bmath{\rho}(0) = -i \left[ {\bf S}(t),\bmath{\rho}(0)\right] +
{1 \over 2} \sum_{\alpha,\beta\neq0} \chi_{\alpha \beta}(t) \left( \left[ {\bf
F}_\alpha, \bmath{\rho}(0) {\bf F}_\beta^\dagger \right] + \left[ {\bf
F}_\alpha \bmath{\rho}(0), {\bf F}_\beta^\dagger \right] \right),
\label{eq:newosr}
\end{equation}
where
\begin{equation}
{\bf S}(t)={i \over 2} \sum_{\alpha \neq 0}\left( \chi_{\alpha 0}(t) {\bf
F}_\alpha - \chi_{0 \alpha}(t) {\bf F}_\alpha^\dagger \right).
\end{equation}
This version of the fixed basis OSR will be useful in deriving master
equations.  It is also convenient because any positive $\chi_{\alpha \beta}(t)$
matrix whose trace is $d$ corresponds to an OSR.

\subsection{Example OSR}

As an example of the OSR consider the process described in
Section~\ref{sec:degrade}.  The system-environment evolution operator is given
by Eq.~(\ref{eq:exampleu}) and the initial density matrix of the environment is
$\bmath{\rho}_E(0)=|+\rangle \langle +|$.  In the derivation of the OSR there
are two terms,
\begin{eqnarray}
{\bf A}_1(t)&=&\langle + |_E \left(\cos(\lambda t) {\bf I} -i \sin(\lambda t)
\bmath{\sigma}_z \otimes \bmath{\sigma}_z \right) |+ \rangle_E= \cos(\lambda t)
{\bf I}, \nonumber \\ {\bf A}_2(t)&=&\langle - |_E \left(\cos(\lambda t) {\bf
I} -i \sin(\lambda t) \bmath{\sigma}_z \otimes \bmath{\sigma}_z \right)
|+\rangle_E=-i \sin(\lambda t) \bmath{\sigma}_z.
\end{eqnarray}
Note that ${\bf A}_1^\dagger(t) {\bf A}_1(t)+ {\bf A}_2^\dagger(t) {\bf
A}_2(t)={\bf I}$ as required by unitarity.  The evolution of the initial
density matrix $\bmath{\rho}(0)$ is thus
\begin{equation}
\bmath{\rho}(t)=\cos^2(\lambda t) \bmath{\rho}(0) + \sin^2(\lambda t)
\bmath{\sigma}_z \bmath{\rho}(0) \bmath{\sigma}_z,
\end{equation}
which agrees with Eq.~(\ref{eq:exampleevo}) derived above.

Suppose that instead of the environment being in the initial state $|+\rangle
\langle +|$ it is in the state $|0\rangle \langle 0|$.  In this case, if we us
the basis $|+\rangle$, $|-\rangle$ to calculate the OSR we find that
\begin{eqnarray}
{\bf A}_1(t)&=&\langle + |_E \left(\cos(\lambda t) {\bf I} -i \sin(\lambda t)
\bmath{\sigma}_z \otimes \bmath{\sigma}_z \right) |0 \rangle_E= {1 \over
\sqrt{2}} \left( \cos(\lambda t) {\bf I}-i \sin(\lambda t) \bmath{\sigma}_z
\right), \nonumber
\\ {\bf A}_2(t)&=&\langle - |_E \left(\cos(\lambda t) {\bf I} -i \sin(\lambda
t) \bmath{\sigma}_z \otimes \bmath{\sigma}_z \right) |0\rangle_E= {\bf A}_1(t).
\end{eqnarray}
If we instead use the basis $|0\rangle$, $|1\rangle$ to calculate the OSR, we
find that
\begin{eqnarray}
\tilde{\bf A}_1(t)&=&\langle 0 |_E \left(\cos(\lambda t) {\bf I} -i
\sin(\lambda t) \bmath{\sigma}_z \otimes \bmath{\sigma}_z \right) |0 \rangle_E=
\left( \cos(\lambda t) {\bf I}-i \sin(\lambda t) \bmath{\sigma}_z \right),
\nonumber
\\ \tilde{\bf A}_2(t)&=&\langle 1 |_E \left(\cos(\lambda t) {\bf I} -i \sin(\lambda
t) \bmath{\sigma}_z \otimes \bmath{\sigma}_z \right) |0\rangle_E= {\bf 0}.
\end{eqnarray}
There are two interesting facts about this case.  First, we see how using a
different basis for calculating the OSR gives different operators but the same
evolution:
\begin{equation}
\bmath{\rho}(t)={\bf A}_1(t) \bmath{\rho}(0) {\bf A}_1^\dagger(t) + {\bf
A}_2(t) \bmath{\rho}(0) {\bf A}_2^\dagger(t)= \tilde{\bf A}_1(t)
\bmath{\rho}(0) \tilde{\bf A}_1^\dagger(t).
\end{equation}
Second the evolution of the system is unitary, $\tilde{\bf A}_1^\dagger
\tilde{\bf A}_1={\bf I}$.  Besides demonstrating the non-uniqueness of the OSR,
this example serves to bring up an interesting question: under what conditions
is the evolution of the OSR correspond to unitary evolution?  Since this
question presages future work we will address this question in the next
subsection.

\subsection{OSR and unitary evolution}

The question we pose is under what conditions does
\begin{equation}
\sum_i {\bf A}_i \bmath{\rho} {\bf A}_i^\dagger = {\bf U} \bmath{\rho} {\bf
U}^\dagger, \label{eq:osrunitary}
\end{equation}
where
\begin{equation}
{\bf U}^\dagger {\bf U}= \sum_i {\bf A}_i^\dagger {\bf A}_i = {\bf I},
\label{eq:osrunitarynorms}
\end{equation}
for all $\bmath{\rho}$.  We claim that an iff condition for this to hold is
${\bf A}_i=c_i(t) {\bf U}$ with $\sum_i |c_i|^2 =1$\cite{Duan:99c,Lidar:99b}.

The forward implication is trivial.  Clearly if ${\bf A}_i=c_i {\bf U}$ with
$\sum_i |c_i|^2=1$ then Eq.~(\ref{eq:osrunitary}) and
Eq.~(\ref{eq:osrunitarynorms}) both hold.

To prove the inverse, assume Eq.~(\ref{eq:osrunitary}) and
Eq.~(\ref{eq:osrunitarynorms}) both hold.  It is useful to rewrite
Eq.~(\ref{eq:osrunitary}) as
\begin{equation}
\sum_i {\bf U}^\dagger {\bf A}_i \bmath{\rho} {\bf A}_i^\dagger {\bf U} =
\bmath{\rho},
\end{equation}
and then define $\tilde{\bf A}_i = {\bf U}^\dagger {\bf A}_i$ so that this
becomes
\begin{equation}
\sum_i \tilde{\bf A}_i \bmath{\rho} \tilde{\bf A}_i^\dagger = \bmath{\rho}.
\end{equation}
Since this equation must hold for all $\bmath{\rho}$ it must hold for a
particular choice of $\bmath{\rho}=|\psi \rangle \langle \psi |$.  This
immediately leads to
\begin{equation}
\sum_i |\langle \psi | \tilde{\bf A}_i |\psi \rangle|^2 = 1. \label{eq:osruneq}
\end{equation}
For a given $\tilde{\bf A}_i$, the state $\tilde{\bf A}_i |\psi\rangle$ can be
split into two components $\tilde{\bf A}_i|\psi\rangle = c_i |\psi\rangle +
c_i^\perp |\psi^\perp\rangle$ where $|\psi^\perp\rangle$ is a vector
perpendicular to $|\psi\rangle$.  Eq.~(\ref{eq:osruneq}) then implies
\begin{equation}
\sum_i |c_i|^2 = 1. \label{eq:osrnormeq}
\end{equation}

The normalization condition Eq.~(\ref{eq:osrunitarynorms}) can be recast as
\begin{equation}
\sum_i \tilde{\bf A}_i^\dagger \tilde{\bf A}_i = {\bf I},
\end{equation}
which implies
\begin{equation}
\sum_i |c_i|^2 + |c_i^\perp|^2 = 1.
\end{equation}
Together with Eq.~(\ref{eq:osrnormeq}) this implies that $\sum_i
|c_i^\perp|^2=0$ such that $c_i^\perp=0$ for all $i$.  Thus $|\psi\rangle$ is
an eigenstate of all of the $\tilde{\bf A}_i$, $\tilde{\bf A}_i |\psi\rangle =
c_i |\psi \rangle$ This must hold for all possible $|\psi\rangle$ in the
Hilbert space the OSR operates on and therefore
\begin{equation}
\tilde{\bf A}_i = c_i {\bf I} \Rightarrow {\bf A}_i = c_i {\bf U}.
\end{equation}
Eq.~(\ref{eq:osrunitarynorms}) then implies $\sum_i |c_i|^2=1$.

\subsection{Limits of the OSR}

The OSR is fairly satisfying in terms of describing the evolution of a system
coupled to an environment.  The initial state of the environment together with
a description of the unitary evolution operator on the system and environment
allows for  a description of the evolution of all possible system density
operators in a compact form.  The most troublesome assumption in this
derivation is, perhaps, the assumption that the system and the environment are
initially in a tensor product state ${\bmath{\rho}}(0)={\bmath{\rho}}_S(0)
\otimes {\bmath{\rho}}_E(0)$.

Do there exist situations in which interaction between the system and the
environment cannot be expressed in the OSR?  Consider the situation where the
system and environment are each single qubits and the initial joint state is
$|\psi_1\rangle={1\over\sqrt{2}} \left(|00\rangle + |11\rangle\right)$ or
$|\psi_2\rangle={1 \over\sqrt{2}} \left(|+0\rangle +|-1 \rangle \right)$.
Suppose the system and environment then evolve according to the unitary
evolution ${\bf U}={\bf I} \otimes |0\rangle \langle 0 | + \bmath{\sigma}_x
\otimes |1\rangle \langle 1 |$.  In some sense, the state of the environment is
the same in both of these situations: the density matrices of the environment
for both $|\psi_1\rangle$ and $|\psi_2\rangle$ are both ${1 \over 2} {\bf I}$.
Further, the density matrices of the system for both $|\psi_1\rangle$ and
$|\psi_2\rangle$ are also both ${1 \over 2}{\bf I}$.  After evolution according
to ${\bf U}$, however, the state of the system is different for these two cases
differ
\begin{eqnarray}
\bmath{\rho}_1&=&{\rm Tr}_E \left[ {\bf U} |\psi_1 \rangle \langle \psi_2| {\bf
U}^\dagger \right] = |0\rangle \langle 0| \nonumber \\ \bmath{\rho}_2&=& {\rm
Tr}_E \left[ {\bf U} |\psi_2 \rangle \langle \psi_2 | {\bf U}^\dagger \right] =
{1 \over 2} \left( |0\rangle \langle 0| + |1 \rangle \langle 1| \right).
\end{eqnarray}
Thus we see that the {\em same} density matrix has evolved into two different
density matrices when the environment's density matrices was identical.  Thus
it is clear that there is no hope in deriving an OSR which depends solely on
the initial density matrix of the environment and the system-environment
unitary evolution.  In particular the entangled nature (see
Appendix~\ref{apa:sepdef} for definition) of the system and environment leads
to situations where the OSR fails.

The initial condition of tensor product states for the system and environment
is an assumption that the system and the environment are initially
uncorrelated.  Further we have shown how when the system and the environment
start an entangled state an OSR depending only on the environmental density
matrix and the full evolution is impossible. Let us now show that even when the
system and the environment are classically correlated there are problems in the
derivation of the OSR.  Suppose that the initial state of the system plus
environment can be written in the separable form (see Appendix~\ref{apa:sepdef}
for definition)
\begin{equation}
\rho(0) = \sum_\eta q_\eta {\bmath{\rho}}_{S \eta}(0) \otimes {\bmath{\rho}}_{E
\eta} (0),
\end{equation}
with $0<q_\eta\leq 1$ and $\sum_\eta q_\eta=1$ and ${\bmath{\rho}}_{S
\eta}(0)$, ${\bmath{\rho}}_{E\eta}(0)$ valid density matrices.  The initial
system density matrix is ${\bmath{\rho}}_S(0)={\rm Tr} \left[{\bmath{\rho}}(0)
\right]= \sum_\eta q_\eta {\bmath{\rho}}_{S \eta}(0)$.  Each environmental
density matrix has a spectral decomposition (perhaps over different environment
basis states): ${\bmath{\rho}}_{E\eta}(0)=\sum_{\nu_\eta} p_{\nu \eta}
|\nu_\eta \rangle \langle \nu_\eta |$.  The evolution of the system is then
\begin{equation}
{\bmath{\rho}}_S(t)= \sum_{\mu,\nu,\eta} \langle \mu | {\bf U}(t) {\bf
\rho}_{S\eta}(0)q_\eta p_{\nu \eta} |\nu_\eta \rangle \langle \nu_\eta | {\bf
U}^\dagger(t) |\mu \rangle.
\end{equation}
This can be written in the form
\begin{equation}
{\bmath{\rho}}_S(t)= \sum_\eta \sum_{i=(\mu,\nu)} {\bf A}_{i,\eta}(t) q_\eta
\rho_{S\eta}(0) {\bf A}_{i,\eta}^\dagger(t),
\end{equation}
where
\begin{equation}
{\bf A}_{i=(\mu,\nu),\eta}(t)=\sqrt{p_{\nu \eta}}\langle \mu | {\bf U}(t)
|\nu_\eta \rangle,
\end{equation}
and unitarity requires
\begin{equation}
\sum_{i} {\bf A}_{i,\eta}^\dagger(t) {\bf A}_{i,\eta}(t) = {\bf I}.
\end{equation}
Unless the basis used for each spectral decomposition of the bath is the same
$|\nu_\eta\rangle = |\nu\rangle$ {\em and} the spectral coefficients are the
same $p_{\nu \eta}=p_\nu$, the evolution of the system cannot be expressed as
in the OSR form Eq.~(\ref{eq:osr}).

\section{Master equations}

While the OSR is a convenient formalism for discussing the coupling between the
system and the bath under appropriate initial conditions, it is often too
cumbersome to be used for calculations on real physical system.  One important
reason for this fact is that the environment of real physical systems are often
large complex subsystems whose evolution is difficult to model.  The simplicity
of the system is of little help when dealing with open quantum systems which
require an understanding of environmental degrees of freedom.  Despite this
difficulty, a surprisingly large class of decohering dynamics has been
adequately described by physically derived evolution equations.

A closed quantum system evolves according to the Liouville equation of motion
\begin{equation}
{\partial \bmath{\rho}(t) \over \partial t} = -i [{\bf H},\bmath{\rho}(t)],
\end{equation}
where we have chosen a static Hamiltonian ${\bf H}$ for simplicity.  Oftentimes
it is possible to derive an {\em approximate} evolution equation for an open
quantum system which corresponds to an extra term in this evolution equation:
\begin{equation}
{\partial \bmath{\rho}(t) \over \partial t} = -i [{\bf H},\bmath{\rho}(t)]
 + {\mathcal L}\left[ \bmath{\rho}(t) \right].
\end{equation}
A large class of these approximate evolution equations correspond to {\em
semigroup master equations}.  If we let
$\bmath{\rho}(t)=\Lambda(t)[\bmath{\rho}(0)]$ denote the parameterized map of
the initial density matrix to the density matrix at time $t$, we define a {\em
semigroup master equation} as a map $\Lambda(t)$ which satisfies
\begin{enumerate}
  \item $\Lambda(t)$ is a completely positive linear map continuous in $t$(see
  Section~\ref{sec:osrderive} for the definition of complete positivity).
  \item $\Lambda(t)$ is Markovian: $\Lambda(t) \circ \Lambda(s) =
  \Lambda(s+t)$.
  \item The initial state of the system and environment are in a tensor product
  state.
\end{enumerate}
Gorini, Kossakowski, and Sudarshan\cite{Gorini:76a} and
Lindblad\cite{Lindblad:76a} have shown that any map $\Lambda(t)$ which
satisfies these requirements has an evolution which is generated by the
semigroup master equation (SME)
\begin{equation}
{\partial \bmath{\rho}(t) \over \partial t} = -i [{\bf H},\bmath{\rho}(t)]
 + {1 \over 2} \sum_{\alpha \beta} a_{\alpha \beta} \left(\left[ {\bf F}_\alpha \bmath{\rho}(t), {\bf F}_\beta^\dagger
 \right]+ \left[ [{\bf F}_\alpha,\bmath{\rho}(t) {\bf F}_\beta^\dagger \right], \right)
\end{equation} \label{eq:sme}
where ${\bf F}_\alpha$ are a complete basis for the operators on the Hilbert
space which $\bmath{\rho}$ inhabits and $a_{\alpha \beta}$ is a positive
hermitian matrix.

\subsection{Discrete coarse graining derivation of the SME}

We now show that explicit use of a discrete coarse-graining over time can lead
naturally from the OSR evolution equation, Eq.~(\ref{eq:newosr}) to the SME,
Eq.~(\ref{eq:sme})\cite{Bacon:99a}. A useful form of the fixed basis OSR
Eq.~(\ref{eq:newosr}) is found by taking the derivative of
Eq.~(\ref{eq:newosr}) with respect to time
\begin{equation}
{\partial \bmath{\rho}(t) \over \partial t}= -i \left[ {\partial {\bf S}(t)
\over \partial t} ,\bmath{\rho}(0)\right] + {1 \over 2}
\sum_{\alpha,\beta\neq0} {\partial \chi_{\alpha \beta}(t) \over \partial t}
\left( \left[ {\bf F}_\alpha, \bmath{\rho}(0) {\bf F}_\beta^\dagger \right] +
\left[ {\bf F}_\alpha \bmath{\rho}(0), {\bf F}_\beta^\dagger \right] \right).
\label{eq:newerosr}
\end{equation}

The coarse graining of the evolution will be done with respect to a time
$\tau$.  This time-scale, we will eventually discover, is related to a
environment ``memory'' time scale.  Coarse graining over $\tau$ corresponds to
\begin{equation}
\bmath{\rho}_{j}=\bmath{\rho} (j\tau );\quad \chi _{\alpha \beta ;j}=\chi
_{\alpha \beta }(j\tau );\quad j\in \NN.
\end{equation}
Further, rewriting the OSR Eq.~(\ref{eq:newosr}) as $\bmath{\rho} (t)={\bf
\Lambda } (t)\bmath{\rho} (0)$ and defining $\tilde{{\tt L}}(t)$ through ${\bf
\Lambda } (t)= {\rm T}\exp \left[ \int_{0}^{t}\tilde{{\tt L}}(s)ds\right] $ we
find that
\begin{equation}
{\frac{\partial \bmath{\rho }(t)}{\partial t}}=\tilde{{\tt L}}(t)[\bmath{\rho}
(t)]. \label{eq:osrtrue}
\end{equation}
Defining $\tilde{{\tt L}}_{j}=\int_{j\tau }^{(j+1)\tau }\tilde{{\tt L}}(s)ds$ ,
with $\tau n=t$, we have
\begin{equation}
\int_{0}^{t}\tilde{{\tt L}}(s)ds =\tau \sum_{j=0}^{n-1}\tilde{{\tt L}}_{j}.
\end{equation}
Next we will make the assumption that on the time scale of the environment
$\tau $, the evolution generators $\tilde{{\tt L}}(t)$ commute in the
``average'' sense that $\left[ \tilde{{\tt L}}_{j},\tilde{{\tt L}}_{k}\right]
=0,\forall j,k$. Physically, we imagine this operation as arising from the
``resetting'' of the environment density operator over the time-scale $\tau $.
Under this assumption, the evolution of the system is Markovian when $t\gg \tau
$:
\begin{equation}
{\bf \Lambda}(t)=\prod_{j=0}^{n-1} \exp \left[ \tau \tilde{{\tt L}}_j \right] .
\end{equation}
Under the discretization of the evolution, this product form of the evolution
implies that
\begin{equation}
\bmath{\rho}_{j+1}=\exp \left[ \tau \tilde{{\tt L}}_j \right] [\bmath{\rho}_j]
.
\end{equation}
In the limit of $\tau \ll t$ we expand this exponential, to find that
\begin{equation}
{\frac{\bmath{\rho}_{j+1}-\bmath{\rho}_j }{\tau}}= \tilde{{\tt L}}_j [
\bmath{\rho}_j ]. \label{eq:rhodot}
\end{equation}
This equation is simply a discretization of Eq.~(\ref{eq:osrtrue}) under the
assumption that $\tau \ll \theta$, where $\theta$ is the time-scale of change
for the system density matrix. Notice in particular that the RHS of
Eq.~(\ref{eq:rhodot}) contains the {\it average} value of $\tilde{{\tt L}} (t)
$ over the interval. From the OSR evolution equation Eq.~(\ref{eq:newerosr}),
we know the explicit form of $\tilde{{\tt L } }(t)$ over the first interval
from $0$ to $\tau$. Discretizing over this interval we find that
\begin{eqnarray}
{\frac{\bmath{\rho}_{1}-\bmath{\rho}_0 }{\tau}} &=& - i \left [ \left \langle
{\partial {\bf S}(t) \over \partial t} \right \rangle,\bmath{\rho}_0 \right ] +
{\frac{1 }{2}} \sum_{\alpha,\beta} \left \langle {\partial \chi_{\alpha
\beta}(t) \over \partial t} \right \rangle \left( [{\bf F} _\alpha,
\bmath{\rho}_0 {\bf F}_\beta^\dagger] + [{\bf F} _\alpha \bmath{\rho}(0),{\bf
F} _\beta^\dagger] \right) \nonumber \\ &\equiv&\tilde{{\tt L}}_0 [
\bmath{\rho}_0 ],
\end{eqnarray}
where
\begin{equation}
\left\langle X\right\rangle \equiv {\frac{1}{\tau }}\int_{0}^{\tau }X(s)ds.
\label{eq:t-ave}
\end{equation}
Thus, in the sense of the coarse graining above we have arrived at an explicit
form for $\tilde{{\tt L}}_{0}$.

Consider the evolution beyond this first interval. Deriving an explicit form
for $\tilde{{\tt L}}_{1}$ and for higher terms is now impossible because
Eq.~(\ref{eq:newerosr}) gives the evolution in terms of $\bmath{\rho} (0)$.
However, since we have made the assumption that the environment ``resets'' over
the time-scale $\tau $, we expect the environment to interact with the system
in the same manner over every $\tau $-length coarse-grained interval. This is
equivalent to assuming that $\tilde{{\tt L}} _{i}=\tilde{{\tt L}}_{0},\forall
i$ (which of course is the most trivial way of satisfying the Markovian
evolution condition $[\tilde{{\tt L}}_{i},\tilde{ {\tt L}}_{j}]=0,\forall
i,j$). Then, using Eq.~(\ref{eq:rhodot}), one is led to the form of the
semigroup equation of motion,
\begin{equation}
\frac{\partial \bmath{\rho}(t)}{\partial t}= - i [ \left \langle{\ {\partial
{\bf S}(t) \over
\partial t}} \right \rangle,\bmath{\rho}(t)] + {\frac{1 }{2}} \sum_{\alpha, \beta} \left
\langle {\partial {\chi}_{\alpha \beta}(t) \over \partial t} \right \rangle
\left( [ {\bf F}_\alpha, \bmath{\rho}(t) {\bf F}_\beta^\dagger] + [{\bf
F}_\alpha \bmath{\rho}(t), {\bf F}_\beta^\dagger] \right).
\label{eq:newosrgood}
\end{equation}

We can write this equation of motion in an alternative form which distinguishes
between the system and environment contributions to the evolution.  Since
Eq.~(\ref{eq:newerosr}) is linear in the $\chi_{\alpha \beta}(t)$ matrix, one
can calculate $\chi _{\alpha \beta }^{(0)}(t)$ for the isolated system and
hence define the new terms which come about from the coupling of the system to
the environment:
\begin{equation}
\chi _{\alpha \beta }(t)=\chi _{\alpha \beta }^{(0)}(t)+\chi _{\alpha \beta
}^{(1)}(t).
\end{equation}
The terms which correspond to the isolated system will therefore produce a
normal $-i[{\bf H},\bmath{\rho} (t)]$ Liouville term in
Eq.~(\ref{eq:newosrgood}). Thus Eq.~(\ref{eq:newosrgood}) can be rewritten as
\begin{equation}
\frac{\partial \bmath{\rho} (t)}{\partial t}=-i \left[ {\bf H} +\left\langle
{{\partial{\bf S}^{(1)}(t) \over \partial t}}\right\rangle ,\bmath{\rho}
(t)\right] +{\frac{1}{ 2}}\sum_{\alpha,\beta} \left\langle {\partial
{\chi}_{\alpha \beta }^{(1)}(t) \over \partial t} \right\rangle \left( [{\bf
F}_{\alpha },\bmath{\rho} (t){\bf F}_{\beta }^{\dagger }]+[{\bf F}_{\alpha
}\bmath{\rho} (t),{\bf F}_{\beta }^{\dagger }]\right) , \label{eq:newosrham}
\end{equation}
which with the identification of $\left\langle {\partial {\chi}_{\alpha \beta
}(t) \over \partial t} \right\rangle $ with $a_{\alpha \beta }$ is equivalent
to Eq.~({\ref{eq:sme}), except for the presence of the second term derived from
$\left\langle {\partial {\bf S}^{(1)}(t) \over \partial t}\right\rangle $ in
the Liouvillian.  This second term induces unitary dynamics on the system,
$\left\langle {\partial {\bf S}^{(1)}(t) \over \partial t}\right\rangle $, is
referred to as the {\em Lamb shift}.  This term explicitly describes an unitary
effect which the environment has on the system.  It is often implicitly assumed
to be present in Eq.~(\ref{eq:sme}).

We have shown how coarse-graining the evolution over the environment time-scale
$\tau $ allows one to understand the connection between the OSR and the
semigroup evolution.  The assumptions which went into this derivation are
explicitly
\begin{enumerate}
    \item The time-scale for the evolution of the system density matrix is much
larger than the time-scale for the resetting of the environment ($\tau \gg
\theta$).
    \item The evolution of the system should be Markovian ($[\tilde{{\tt L}}_{i},\tilde{
{\tt L}}_{j}]=0,\forall i,j$)
    \item The environment resets to the same state so that the system evolution is the same over every coarse graining ($\tilde{{\tt
L}}_i=\tilde{{\tt L}}_0,\forall i$).
    \item The system and the environment start in a tensor product state.
\end{enumerate}
The importance of Eq.~(\ref {eq:newerosr}) lies in the fact that it allows one
to pinpoint the exact point at which the assumption of Markovian dynamics are
made and further, due to the general likeness of its form to the SME, provides
an easily translatable connection when going from the non-Markovian OSR to the
Markovian SME. Notice also that the assumption of Markovian dynamics introduces
an arrow of time in the evolution of the system through the ordering of the
environmental states: the system evolves through time in the direction of each
successive resetting of the environment.

A detailed study of this coarse graining procedure on a specific model has been
carried out in \cite{Lidar:01a} where the authors examine the application of
this procedure to a spin-boson model.  Amazingly at low order in perturbation
theory the coarse grain procedure described above provides an accurate
description of the open system dynamics.

The use of Markovian master equations in physics has a long and storied
history.  From the early study of phenomenological models
\cite{Bloch:46a,Wangsness:53a}, to more rigorous derivations
\cite{Kossakowski:72a,Davies:74a,Davies:76a,Lindblad:76a}, and the saturation
of master equations in the quantum optics community\cite{Carmichael:93a},
master equations are a useful tool for modeling the behavior of many different
physical systems.  It has even been suggested that instead of an approximation
of the full unitary dynamics, the SME is a fundamental evolution equation for
nature (for a good discussion of this matter, and why it fails to solve the
``measurement problem'', see \cite{Giulini:96a}).  What we have provided in
this section is a different manner of understanding how the SME can arise as an
approximate evolution of a system.  Eq.~(\ref{eq:newosr}) and
Eq.~(\ref{eq:newerosr}) provide an path between the exact OSR and the
approximate SME via our specific coarse graining procedure.

\subsection{Resolving a mystery in decoherence rates}

Decoherence rates
\begin{equation}
{1 \over \tau_n}= \left\{{\rm Tr}\left[\bmath{\rho}(0) \bmath{\rho}^{(n)}(0)
\right] \right\}^{1 \over n},
\end{equation}
(see Appendix~\ref{apa:rates} for the motivation behind this definition) can be
used to understand the time scales of a decoherence process. Interestingly,
under the SME, first order decoherence rates ($1/\tau_1$) are finite while in
the OSR these decoherence rates vanish.

One can see the vanishing of the first order decoherence rate in the OSR by
directly substituting in the pre-OSR Hamiltonian dynamics and using the
cyclical nature of the trace operation,
\begin{eqnarray}
{1 \over \tau_1}&=&{\rm Tr}_S \left[ \bmath{\rho}_S(0) \left( {\partial \over
\partial t} {\rm Tr}_E \left[ {\bf U}_{SE}(t) \bmath{\rho}_S(0) \otimes
\bmath{\rho}_E(0) {\bf U}_{SE}^\dagger(t)  \right] \right)_{t=0} \right]
\nonumber \\ &=& {\rm Tr}_S \left[ \bmath{\rho}_S(0) {\rm Tr}_E \left[ -i{\bf
H}_{SB} \bmath{\rho}_S(0) +i \bmath{\rho}_S(0) {\bf H}_{SB} \right] \right]=0.
\end{eqnarray}
The only possible manner in which this vanishing of this trace could not occur
would be to play some tricks with limits of infinite matrices.

However, in the SME, the first order decoherence rate does not vanish.
Explicitly, in the SME, we find that (in the absence of a system evolution
${\bf H}_S=0$),
\begin{equation}
{1 \over \tau_1}= {\rm Tr}_S \left[ \bmath{\rho}_S(0) \left( {1 \over 2}
\sum_{\alpha,\beta\neq0} \left( [{\bf F}_\alpha \bmath{\rho}(0),{\bf
F}_\beta^\dagger] + [{\bf F}_\alpha,\bmath{\rho}(0){\bf F}_\beta^\dagger]
\right) \right) \right],
\end{equation}
which in general does not vanish (see for example
\cite{Zanardi:98a,Zanardi:98b}).

Now lets present a bit (or more precisely a qubit!) of a paradox.  Consider the
often quoted example of phase damping of a qubit. In this case, it would appear
that there is a finite first order decoherence rate.  Yet, phase damping of a
qubit is often presented within the OSR\cite{Chuang:97a,Knill:97a,Nielsen:00a},
which, as we have just shown above, would predict {\em zero} first order
decoherence rates for any non-singular Hamiltonian. In this example, the OSR
operators are given by{\cite{Chuang:97a} }
\begin{eqnarray}
{\bf A}_0(t)=\left(
\begin{array}{cc}
1 & 0 \\ 0 & e^{-\lambda t}
\end{array}
\right), \quad {\rm and} \quad {\bf A}_1(t)= \left(
\begin{array}{cc}
0 & 0 \\ 0 & \sqrt{1 - e^{-2\lambda t}}
\end{array}
\right) ,  \label{eq:osrphase}
\end{eqnarray}
and a simple calculation using these operators yields a minimum first order
decoherence rate of $1/\tau_1=-\lambda/2$. How can this be?  In particular we
know that every OSR corresponds to some Hamiltonian dynamics on a larger
Hilbert space and we have previously showed that first order decoherence rates
vanish in the OSR.  Yet, here is an example of an OSR where the first order
decoherence rate does not vanish!

We can resolve this apparent paradox by examining the coarse graining procedure
used to derive the SME from the OSR.

Using Eq.~(\ref{eq:newerosr}) the first order decoherence rate in the OSR
becomes
\begin{eqnarray}
{1 \over\tau _{1}}&=&-i{\rm Tr}\left[ \bmath{\rho} (0) \left[ \left({\partial
{\bf S}(t) \over \partial t} \right)_{t=0},\bmath{\rho} (0)\right] \right] \\
\nonumber &+&{\rm Tr} \left[{1 \over 2}\sum_{\alpha ,\beta \neq1}
\left({\partial {\chi} _{\alpha \beta }(t) \over
\partial t} \right)_{t=0} \left( [{\bf F}_{\alpha },\bmath{\rho} (0){\bf
F}_{\beta }^{\dagger }]+[{\bf F}_{\alpha }\bmath{\rho} (0),{\bf F}_{\beta
}^{\dagger }]\right)  \right] . \label{eq:osrrate}
\end{eqnarray}
Using the decomposition of the OSR operators, Eq.~(\ref{eq:osrexplicit}), and
knowing that ${\bf U}(0)={\bf I}_{S}\otimes {\bf I}_{B}$, we find that ${\bf
A}_{i}(0)=\sqrt{\nu }{\bf I}_{S}\delta _{i,(\nu ,\nu )}$. Thus, since the $
{\bf F}_{\alpha }$'s form a linearly independent basis, it follows, using
Eq.~(\ref{eq:fixedbasis}), that the expansion coefficients must be
\begin{equation}
b_{i\alpha}(0)= \delta_{\alpha 0} \sqrt{\nu d} \delta_{i,(\nu,\nu)}.
\end{equation}
where $d$ is dimension of the system Hilbert space.  By direct evaluation,
\begin{equation}
\left( {\partial {\chi}_{\alpha \beta }(t) \over \partial t}
\right)_{t=0}=\sum_{\nu d}\sqrt{\nu d }\left[ \left( \delta _{\alpha 0}\left({
\partial {b}_{(\nu ,\nu ),\beta }^{\ast }(t) \over \partial t} \right)_{t=0}+
\delta _{\beta 0}\left({\partial {b} _{(\nu ,\nu ),\alpha }(t) \over \partial t
} \right)_{t=0}\right) \right] , \label{eq:chicond}
\end{equation}
which implies the vanishing (as long as $\left( {\partial {b}_{(\nu ,\nu
),\alpha }(t) \over \partial t} \right)_{t=0}$ remains finite) in
Eq.~(\ref{eq:osrrate}) of every term except ${\rm Tr}\left[\bmath{\rho}
(0)\left[\left({
\partial{\bf S}(t) \over \partial t} \right)_{t=0},\bmath{\rho} (0) \right] \right]$. However, this in turn vanishes by cyclic
permutation of the trace. Thus we see as claimed, that the OSR first order
decoherence rate vanishes.

We can now use our coarse graining derivation of the SME to understand how
first order decoherence rates appear in the SME.  Examination of our derivation
of the SME, Eqs.~(\ref{eq:newosrgood}) and (\ref{eq:newosrham}), now shows how
non-zero first order decoherence rates can arise when the evolution is
considered to be Markovian.  In  the derivation of the semigroup equation in
the Markovian limit we made the assumption that the matrices $\left( {\partial
{\chi}_{\alpha \beta }(t) \over \partial t} \right)_{t=0}$ can be identified
with the constant matrices $a_{\alpha \beta }$ of the semigroup equation,
Eq.~(\ref {eq:sme}). However, when this is done, the matrix elements $\left(
{\partial {\chi} _{\alpha \beta }(t) \over \partial t } \right)_{t=0}$ in
Eq.~(\ref{eq:osrrate}) are replaced by their time-averaged values, for which
the relation Eq.(\ref{eq:chicond}) no longer applies. Hence, in general, the
first order decoherence rates are necessarily not zero when the Markovian
coarse-graining is applied. For a finite total Hamiltonian ${\bf H}_{SB}$,
non-zero first order rates are therefore seen to be an artifact of the
Markovian assumption, and their appearance emphasizes the delicate nature of
the transition to the Markovian regime.

We have seen how the first order decoherence rate can not vanish in the
transition from the OSR to the SME, but we are still left with the paradox of a
first order decoherence rate in the OSR for the phase damping example.  To
resolve this dichotomy, we consider how the above phase damping OSR operators
are generated from the unitary dynamics of a qubit system $S$ and a qubit bath
$B$. The evolution operator
\begin{equation}
{\bf U}(t)=\left(
\begin{array}{cccc}
1 & 0 & 0 & 0 \\ 0 & e^{-\lambda t} & 0 & \sqrt{1 - e^{-2 \lambda t}} \\ \ 0 &
0 & 1 & 0 \\ 0 & -\sqrt{1-e^{-2\lambda t}} & 0 & e^{-\lambda t}
\end{array}
\right)
\begin{array}{c}
|\! \downarrow 0 \rangle \\ |\! \downarrow 1 \rangle \\ |\! \uparrow 0 \rangle
\\ |\! \uparrow 1 \rangle
\end{array},
\end{equation}
(where the first qubit represents the bath ($|\! \uparrow \rangle, |\!
\downarrow \rangle$) and the second represents the system ($|0
\rangle,|1\rangle$) as denoted in the columns above) with the bath initially in
the state $|\! \downarrow \rangle$, immediately gives the OSR operators of
Eq.~(\ref{eq:osrphase}). It is easy to calculate the Hamiltonian which
generates this evolution, (using ${\bf H}_{SB}(t)=i \hbar {\frac{d{\bf U}(t)
}{dt}} {\bf U}^\dagger(t)$):

\begin{equation}
{\bf H}_{SB}(t)= \left(
\begin{array}{cccc}
0 & 0 & 0 & 0 \\ 0 & 0 & 0 & -g(t) \\ 0 & 0 & 0 & 0 \\ 0 & g(t) & 0 & 0
\end{array}
\right),
\end{equation}
where
\begin{equation}
g(t)=i\hbar {\frac{\gamma e^{-\gamma t} }{\sqrt{1 - e^{-2 \gamma t}}}}.
\end{equation}
However, we see that as $t \rightarrow 0$, $g(t) \rightarrow \infty$. Thus, in
this simple example, we find that at $t=0$, the Hamiltonian becomes singular.
This illustrates our claim that first order decoherence rates in the OSR are
the result of an infinite Hamiltonian, and do not contradict the general OSR
result of zero rates for finite Hamiltonians.

\subsection{Diagonal form of the SME}

In the SME, Eq.~(\ref{eq:sme}), we have selected used a specific full basis
${\bf F}_\alpha$.  This choice of basis is, of course, arbitrary.  A different
basis, ${\bf G}_\alpha$ could have been selected and this new basis will be
related to the old basis via
\begin{equation}
{\bf F}_\alpha= \sum_{\beta\neq0} g_{\alpha \beta} {\bf G}_\beta,
\label{eq:basisnew}
\end{equation}
for $\alpha \neq0$.  If we require the new basis to maintain the trace inner
product, then
\begin{equation}
{\rm Tr}\left[ {\bf F}_\alpha^\dagger {\bf F}_\beta \right] = {\rm
Tr}\left[\sum_{\mu,\nu} g_{\alpha \mu}^* g_{\beta \nu} {\bf G}_{\mu}^\dagger
{\bf G}_{\nu} \right]= \sum_\nu g_{\alpha \nu}^* g_{\beta \nu} = \delta_{\alpha
\beta}.
\end{equation}
Thinking about $g_{\alpha \nu}^*$ as a matrix, this implies that $g_{\alpha
\nu}^*$ is a unitary matrix.

The non-Hamiltonian generator of the SME, Eq.~(\ref{eq:sme}) is defined as
\begin{equation}
{\tt L}\left[ \bmath{\rho} \right] = {1 \over 2} \sum_{\alpha,\beta \neq 0}
a_{\alpha \beta} \left( [{\bf F}_\alpha \bmath{\rho} ,{\bf F}_\beta^\dagger] +
[{\bf F}_\alpha ,\bmath{\rho} {\bf F}_\beta^\dagger] \right).
\end{equation}
The change of basis, Eq.~(\ref{eq:basisnew}), transforms this generator to
\begin{equation}
{\tt L}\left[ \bmath{\rho} \right] = {1 \over 2} \sum_{\alpha,\beta,\nu,\mu
\neq 0} a_{\alpha \beta} g_{\alpha \nu} g_{\beta \mu}^*\left( [{\bf G}_\nu
\bmath{\rho}, {\bf G}_\mu^\dagger] + [{\bf G}_\nu, \bmath{\rho} {\bf
G}_\mu^\dagger] \right).
\end{equation}
This is a new generator for a SME
\begin{equation}
{\tt L}\left[ \bmath{\rho} \right] = {1 \over 2} \sum_{\nu,\mu \neq 0} a_{\mu
\nu}^\prime\left( [{\bf G}_\nu \bmath{\rho}, {\bf G}_\mu^\dagger] + [{\bf
G}_\nu, \bmath{\rho} {\bf G}_\mu^\dagger] \right),
\end{equation}
where
\begin{equation}
a_{\mu \nu}^\prime = \sum_{\alpha,\beta\neq=0} a_{\alpha \beta} g_{\alpha \nu}
g_{\beta \mu}^*.
\end{equation}
Since $g_{\alpha \nu}^*$ can be any unitary matrix, and $a_{\alpha \beta}$ is a
hermitian matrix, we can choose $g_{\alpha\nu}^*$ such that this matrix
diagonalizes $a_{\mu \nu}^\prime$.  In this case, the generator of the SME is
given by
\begin{equation}
{\tt L}\left[ \bmath{\rho} \right] = {1 \over 2} \sum_{\nu \neq 0}
a_{\nu}\left( [{\bf G}_\nu \bmath{\rho}, {\bf G}_\nu^\dagger] + [{\bf G}_\nu,
\bmath{\rho} {\bf G}_\nu^\dagger] \right),
\end{equation}
which we can rescale such that the SME becomes
\begin{equation}
{\partial \bmath{\rho}(t) \over \partial t} = -i[{\bf H},\bmath{\rho}(t)] +{1
\over 2} \sum_{\alpha \neq 0} \left( [{\bf L}_\alpha \bmath{\rho}(t), {\bf
L}_\alpha^\dagger] + [{\bf L}_\alpha, \bmath{\rho}(t) {\bf L}_\alpha^\dagger ]
\right).
\end{equation}
The operators ${\bf L}_\alpha$ are called the Lindblad operators after
\cite{Lindblad:76a} and this form of the SME is called the Lindblad diagonal
form.

\section{Decoherence}

In the previous two sections we have developed formalisms for understanding the
coupling of a system to its environment.  Along the way we have encountered
assumptions which allowed us to make formal progress in modeling the
decoherence.  Much of the justification for the formalisms of the OSR and SME
must come from the empirical evidence in favor of these descriptions.  Barring
this justification, one must resort back to the {\em fully} Hamiltonian
description of the system plus environment in order to make progress in
understanding a particular decoherence process.  Thus, while the decoherence
formalisms of the OSR and SME allow a nice description of decoherence, there is
much to be said for thinking about decoherence from a purely Hamiltonian system
plus environment viewpoint.  In this thesis we will have the chance to work
with all three of these approaches, the OSR, the SME, and the full Hamiltonian
formulation of decoherence.

\chapter{Quantum Control} \label{ch:control}

\begin{quote}
{\em Two questions: \\
 1. What does it mean to control the evolution of a quantum system? \\
 2. Given some control, what can be done? }
\end{quote}

In this chapter, we introduce the notion of control of a quantum system.  The
role of control which does not cause decoherence is emphasized.  Various
formalisms are developed to understand when such non-decohering control is
possible.  This formalism is then applied to the case of control of a qubit
when coupled via a Jaynes-Cummings Hamiltonian to a coherent state of the
electromagnetic field.  Finally, work regarding what can be done with a given
amount of control is reviewed with the role of the Lie algebraic structure
being emphasized.

\section{Control and measurement}

Suppose one is given a quantum system $S$ and some means of controlling this
system.  By a quantum system $S$, we mean a system which experiment has showed
can produce effects whose description obeys quantum mechanics or at least some
semi-classical quantum principles.  In general, it seems that there are two
forms of interactions which an external system can influence on a quantum
system: control and measurement.

In control one manipulates a {\em controlling apparatus} whose state controls
the unitary evolution of the system.  In order for this manipulation to be a
valid {\em quantum control}, the evolution of the system should not become
entangled with the controlling apparatus.  Another way of stating this is that
the {\em act of control should not induce decoherence on the system}.

In contrast to control, in measurement a {\em measuring apparatus} interacts
with the system in such a way that the state of the measuring apparatus becomes
entangled or correlated in such a way that the state of the apparatus provides
information about the system.

\section{Conditions for control}

Let us try to quantify exactly what is meant for a control mechanism to be a
good control mechanism which does not cause decoherence on the system.  We will
model the problem in a manner which we think reasonably captures a large number
of experimental methods for classical control of quantum systems.

Suppose we are given a quantum system $S$ and an apparatus $A$.   We will
assume that there is some constant coupling Hamiltonian between the system $S$
and the apparatus $A$, ${\bf H}_{SA}$. There are two objections to this
assumption.  The first objection claims that this is not a good assumption
because it is possible to take an apparatus and remove it across the room such
that the apparatus no longer interacts with the system. The resolution of this
objection is two-fold.  First of all it seems to always be possible to model
the removal of apparatus from interaction {\em within} the Hamiltonian ${\bf
H}_{SA}$ and the apparatus evolution ${\bf I}_S \otimes {\bf H}_A$.  The reason
for this is our fundamental belief that quantum mechanics is obeyed by all
physical laws. Thus once we have defined our system, there can only be
Hamiltonian coupling to an outside quantum system. The second reason this
objection is not well founded is the experimental reality that almost all
control of quantum systems some component of the apparatus in contact with the
system.  Thus, for example, if one is manipulating the electronic state of an
atom with a laser, the atom is in constant contact with the electromagnetic
mode which will be used for control. The second objection to the model of a
constant ${\bf H}_{SA}$ is that it disallows a possibly time dependent ${\bf
H}_{SA}$.  Much of what we will derive can easily be extended to the case of a
time dependent Hamiltonian and our assumption of time-independence in this
respect is merely a convenience in order to simplify our discussion.  We thus
start from a full system-apparatus evolution Hamiltonian of
\begin{equation}
{\bf H}= {\bf H}_S \otimes {\bf I}_A + {\bf I}_S \otimes {\bf H}_A + {\bf
H}_{SA},
\end{equation}
with the corresponding unitary evolution of ${\bf U}_{SA}(t)=\exp\left[ -i {\bf
H} t\right]$.

\subsection{Orthogonal pure state stationary control}

\begin{quote}
{\em When sufficient is easy, necessary is almost always difficult.}
\end{quote}

Given the assumption of a constant system-apparatus coupling, let us examine a
simple general model for classical control.
\begin{definition}
{\em (Orthogonal pure state stationary control) } Suppose we are given an
orthogonal set ${\mathcal A}$ of pure states $|a\rangle$ of the apparatus A.
Orthogonal pure state stationary control is then defined as the situation where
for every input $|a\rangle$ into the apparatus (defined as the situation where
the density matrix of the system plus apparatus is $\bmath{\rho}_S(0) \otimes
|a\rangle \langle a|$) the evolution of the system is unitary with some
generating Hamiltonian ${\bf H}_a$ and the state of the apparatus is
$|a\rangle$ at all times during the evolution.
\end{definition}
The condition of orthogonal pure state stationary control is therefore
\begin{equation}
{\bf U}_{SA}(t) \left(\bmath{\rho}_S(0) \otimes |a\rangle \langle a |
\right){\bf U}_{SA}^\dagger(t)  = {\bf U}_a (t) \bmath{\rho}_S(0) {\bf
U}_a^\dagger(t) \otimes |a\rangle \langle a| \quad \forall t, \forall a \in
{\mathcal A}, \label{eq:psscexact}
\end{equation}
where ${\bf U}_a(t) =\exp \left[ -i {\bf H}_a t \right]$.  The question we now
seek to answer is whether there is a succinct method for determining whether a
given Hamiltonian ${\bf H}$ can be used to perform orthogonal pure state
stationary control?

Let us begin by expressing the system-apparatus evolution as an expansion over
the system tensor apparatus operators.  In particular we will choose a complete
hermitian basis ${\bf F}_\alpha$ (see Appendix~\ref{apa:fixedbasis}) for the
expansion over the system component of Hamiltonian
\begin{equation}
{\bf H}=\sum_\alpha {\bf F}_\alpha \otimes {\bf A}_\alpha = {\bf I}_S \otimes
{\bf H}_A + \sum_{\alpha \neq 0} {\bf F}_\alpha \otimes {\bf A}_\alpha,
\label{eq:saexpansion}
\end{equation}
where we have conveniently expanded out the identity component of this
expansion.  Hermiticity of the Hamiltonian implies that the ${\bf A}_\alpha$
operators can be chosen to be Hermitian as well. Then, a sufficient condition
for pure state stationary control to hold is
\begin{eqnarray}
{\bf A}_\alpha |a \rangle &=& c_{\alpha,a} |a\rangle \nonumber \\ {\bf H}_A
|a\rangle &=& \lambda_a |a\rangle,  \label{eq:pssccond}
\end{eqnarray}
for all $|a\rangle$.  We can check that this is sufficient by direct evaluation
\begin{eqnarray}
&&{\bf U}_{SA}(t)\left(\bmath{\rho}_S(0) \otimes |a\rangle \langle a |
\right){\bf U}_{SA}^\dagger(t)  \nonumber \\&&= \exp \left[-i \sum_\alpha {\bf
F}_\alpha \otimes {\bf A}_\alpha t \right] \left(\bmath{\rho}_S(0) \otimes
|a\rangle \langle a | \right) \exp \left[ i \sum_\beta {\bf F}_\beta \otimes
{\bf A}_\beta t \right] \nonumber \\ &&= \exp \left[-i \sum_\alpha c_{\alpha,a}
{\bf F}_\alpha t \right ] \bmath{\rho}_S(0) \exp \left[ i \sum_\alpha
c_{\alpha,a}^* {\bf F}_\alpha t \right] \otimes |a\rangle \langle a|,
\end{eqnarray}
which we can easily see by using the Taylor expansion of the exponential,
evaluating the apparatus operators and regrouping.  Thus if
Eq.~(\ref{eq:pssccond}) holds then the evolution is that of orthogonal pure
state stationary control with the controlled Hamiltonians ${\bf H}_a =\lambda_a
{\bf I}_S+ \sum_{\alpha \neq 0} c_{\alpha,a} {\bf F}_\alpha$.

Let us now show that Eq.~(\ref{eq:pssccond}) is also a necessary condition for
orthogonal pure state stationary control.  Differentiating the orthogonal pure
state stationary control condition, Eq.~(\ref{eq:psscexact}), with respect to
time $t$ and evaluating this equation at $t=0$ we find that
\begin{equation}
[{\bf H}_{SA} , \bmath{\rho}_S(0) \otimes |a\rangle \langle a| ] = [{\bf
H}_a,\bmath{\rho}_S(0)] \otimes |a\rangle \langle a|. \label{eq:psscham}
\end{equation}
Expanding
\begin{equation}
\bmath{\rho}_S(0)= \sum_\alpha \rho_\alpha {\bf F}_\alpha = {1 \over d } {\bf
I} + \sum_{\alpha \neq 0} \rho_\alpha {\bf F}_\alpha,
\end{equation}
and tracing over the system we find that this implies
\begin{equation}
\sum_{\alpha \neq 0} \rho_\alpha \left[ {\bf A}_\alpha , |a\rangle \langle a|
\right] + \left[{\bf H}_A,|a\rangle \langle a | \right] ={\bf 0}.
\label{eq:psscalmost}
\end{equation}
This must hold for all $\bmath{\rho}_S(0)$.  The $\rho_\alpha$, $\alpha \neq 0$
form a convex set in the vector space $\RR^{d^2-1}$ where $d$ is the dimension
of the system Hilbert space.  This convex set contains the origin (which
corresponds to $\rho_S(0)={1 \over d}{\bf I}$) and an open ball of dimension
$d^2-1$ around the origin\cite{Zyczkowski:98a}.  This in turn implies that each
of the terms in the expansion of Eq.~(\ref{eq:psscalmost}) must vanish
separately
\begin{eqnarray}
 \left[ {\bf A}_\alpha , |a\rangle \langle a|
\right] &=&{\bf 0} \nonumber \\ \left[ {\bf H}_A , |a\rangle \langle a| \right]
&=&{\bf 0}.
\end{eqnarray}
This in turn directly implies our claimed condition Eq.~(\ref{eq:pssccond}) and
must hold for all $|a\rangle$

\subsection{Commuting mixed state stationary control}

In orthogonal pure state stationary control, we assumed that the system was in
one of an orthogonal set of states $|a\rangle$.  Our choice of the {\em
orthogonal} input state $|a\rangle$ was made in order to satisfy in a nebulous
manner some requirement that our apparatus is a {\em classical} control device.
A more satisfying requirement would be to loosen our apparatus to start in a
mixed state.  In this case, the more appropriate choice of classicality is that
the different possible controlling mixed states commute(see \cite{Barnum:96a}
for a good motivation for this choice).  Thus we define:
\begin{definition}
{\em (Commuting mixed state stationary control)}  Suppose we are given a
commuting set ${\mathcal M}$ of mixed-states $\bmath{\rho}_a$ of the apparatus
A. Commuting mixed state stationary control is then defined as the situation
where for every input $\bmath{\rho}_a$ into the apparatus (defined as the
situation where the density matrix of the system plus apparatus is
$\bmath{\rho}_S(0) \otimes \bmath{\rho}_a$) the evolution of the system is
unitary with some generating Hamiltonian ${\bf H}_a$ and the state of the
apparatus is $\bmath{\rho}_a$ at all times during the evolution.
\end{definition}
The condition of commuting mixed state stationary control is therefore
\begin{equation}
{\bf U}_{SA}(t) \left(\bmath{\rho}_S(0) \otimes \bmath{\rho}_a \right){\bf
U}_{SA}^\dagger(t) = {\bf U}_a (t) \bmath{\rho}_S(0) {\bf U}_a^\dagger(t)
\otimes \bmath{\rho}_a \quad \forall t, \forall \bmath{\rho}_a \in {\mathcal
M}. \label{eq:msscexact}
\end{equation}

We claim that a necessary and sufficient condition for commuting mixed state
stationary control is
\begin{eqnarray}
 {\bf A}_\alpha \bmath{\rho}_a &=& c_{\alpha,a} \bmath{\rho}_a \label{eq:msscacond}
\\
{\bf H}_A \bmath{\rho}_a  &=&\lambda_{a} \bmath{\rho}_a, \label{eq:msschcond}
\end{eqnarray}
for all of the commuting $\bmath{\rho}_a$.  Note that the orthogonal pure state
stationary control condition is contained within this condition.

We prove this claim via the time honored tradition of using purification to map
this onto the problem we already know how to solve: the orthogonal pure state
stationary control condition.

Let us introduce an auxiliary system $R$ such that the purifications of the
commuting $\bmath{\rho}_a$ are orthogonal:
\begin{equation}
\bmath{\rho}_a= {\rm Tr}_R \left[ |a\rangle \langle a | \right],
\end{equation}
where the $|a\rangle$ are orthogonal.  It is always possible to perform such an
orthogonal purification when the $\bmath{\rho}_a$ commute (but not possible
always possible when they do not commute).  The commuting mixed state
stationary control condition Eq.~(\ref{eq:msscexact}) then becomes
\begin{equation}
 {\bf U}_{SA}(t) \otimes {\bf I}_R \left(\bmath{\rho}_S(0)
\otimes |a\rangle \langle a | \right){\bf U}_{SA}^\dagger(t) \otimes {\bf I}_R
 = {\bf U}_a (t) \bmath{\rho}_S(0) {\bf U}_a^\dagger(t) \otimes |a\rangle
 \langle a|
\quad \forall t,a,
\end{equation}
which we can express as
\begin{equation}
 {\bf U}_{SAR}(t)  \left(\bmath{\rho}_S(0)
\otimes |a\rangle \langle a | \right){\bf U}_{SAR}^\dagger(t)
 = {\bf U}_a (t) \bmath{\rho}_S(0) {\bf U}_a^\dagger(t) \otimes |a\rangle
 \langle a|
\quad \forall t,a,
\end{equation}
where ${\bf U}_{SAR}(t) = \exp\left[-i{\bf H} \otimes {\bf I}_R t \right]$.  A
necessary and sufficient condition for this is just the orthogonal pure state
stationary control conditions from above with an identity tensored onto the
operators
\begin{eqnarray}
{\bf A}_\alpha \otimes {\bf I}_R |a \rangle &=& c_{\alpha,a} |a\rangle
\nonumber \\ {\bf H}_A \otimes {\bf I}_R |a\rangle &=& \lambda_a |a\rangle.
\end{eqnarray}
Forming the operators ${\bf A}_\alpha \otimes {\bf I}_R |a\rangle \langle a |$
and ${\bf H}_A \otimes {\bf I}_R |a\rangle \langle a|$ and tracing over $R$ we
then arrive at the claimed necessary and sufficient conditions
Eq.~(\ref{eq:msscacond}) and Eq.~(\ref{eq:msschcond}) holding for all of the
commuting $\bmath{\rho}_a$.

\subsection{Non-stationary control and throwing the switch}

Throughout our derivation of the control equations we have required that the
input and the output of apparatus remain the same.  Thus the adjective
``stationary'' was appended to all of our derivations of control.  In general,
it seems likely that a more general condition allows no entanglement between
the system and the apparatus but allows the state of the apparatus to change.
What we are {\em not} talking about here is the situation where the apparatus
and the system are entangled at some midway point and then at some final time
the state is no longer entangled.  This latter case is an example of quantum
control via a quantum apparatus because maintenance of the quantum nature of
the apparatus is necessary in order to perform the operation without
decoherence on the system.

One of the potential problems with non-stationary control is the fact that
observation of the apparatus as the state changes can lead to entanglement of
the apparatus with an external observer which then induces decoherence on the
system.  Knowing the state of the apparatus will provides information about how
far along a certain evolution on the system has progressed and when different
observations are made at different times, decoherence can result.  It is an
interesting open question, then, to understand non-stationary control of
quantum system.

Along similar lines of thought, the model we have presented for control assumes
that there is a manner in which the state of the apparatus can be rapidly
changed between the different controlling states $\bmath{\rho}_a$.  The reason
rapid control is needed in this model is that if the state of the apparatus
gets caught in either a superposition or mixture of two controlling
Hamiltonians which produce different evolution this will cause decoherence from
the perspective of the system.  Thus the model we present is one in which the
state of the apparatus can be efficiently manipulated on time scales shorter
than the time scale of the controlled dynamics on the quantum system.

\section{Control examples}

Here we examine two simple control examples.  One of these allows control while
the other does not allow for control. In both examples the systems $S$ and the
apparatus $A$ are single qubits.

The first example is a trivial example where pure state stationary control is
possible.  Consider the system-apparatus Hamiltonian
\begin{equation}
{\bf H}= \lambda \left(\bmath{\sigma}_z \otimes {\bf I} +\bmath{\sigma}_x
\otimes \bmath{\sigma}_z + {\bf I} \otimes \bmath{\sigma}_z \right).
\end{equation}
The ${1 \over \sqrt{2}}\bmath{\sigma}_\alpha$ are a good fixed operator basis
for the system Hilbert space, we therefore find in the fixed-basis expansion
Eq.~(\ref{eq:saexpansion}) that
\begin{eqnarray}
{\bf H}_A&=& \lambda \bmath{\sigma}_z \nonumber \\ {\bf A}_z &=& \lambda
\sqrt{2} {\bf I} \nonumber \\ {\bf A}_x &=& \lambda \sqrt{2} \bmath{\sigma}_z.
\end{eqnarray}
Clearly the eigenstates of $\bmath{\sigma}_z$, $|0\rangle$ and $|1\rangle$,
satisfy the orthogonal pure state stationary control conditions
Eq.~(\ref{eq:pssccond}).  In particular we see that if the apparatus is in the
state $|0\rangle$, then the evolution of the system is according to the
Hamiltonian ${\bf H}_0 = \lambda \left( \bmath{\sigma}_z + \bmath{\sigma}_x +
\bf{I} \right)$.  If, on the other hand, the apparatus is in the state
$|1\rangle$, then the evolution is governed by the Hamiltonian ${\bf H}_1 =
\lambda \left( \bmath{\sigma}_z - \bmath{\sigma}_x + {\bf I} \right)$.

Next we present an example of a system-apparatus Hamiltonian which does not
allow for control.  Consider the system-apparatus Hamiltonian
\begin{equation}
{\bf H} = \lambda \left( \bmath{\sigma}_z \otimes {\bf I} + \bmath{\sigma}_x
\otimes \bmath{\sigma}_ x + {\bf I} \otimes \bmath{\sigma}_z \right).
\end{equation}
Again, using the ${1 \over \sqrt{2} } \bmath{\sigma}_z$ as the fixed-basis for
the system one finds the terms in expansion Eq.~(\ref{eq:saexpansion}),
\begin{eqnarray}
{\bf H}_A&=& \lambda \bmath{\sigma}_z \nonumber \\ {\bf A}_z &=& \lambda
\sqrt{2} {\bf I} \nonumber \\ {\bf A}_x &=& \lambda \sqrt{2} \bmath{\sigma}_x.
\end{eqnarray}
There are no states which are the eigenstates of all three of these operators
(this would contradict $\left[ \bmath{\sigma}_z, \bmath{\sigma}_x \right] = 2i
\bmath{\sigma}_ y$).  Thus there are no states which satisfy the orthogonal
pure state stationary control conditions Eq.~(\ref{eq:pssccond}).

\section{Control with coherent states}

As a more physically relevant application of the orthogonal pure state
stationary control condition, let us consider the control of a two level system
via coupling to a boson field mode.  We assume that the system is a qubit and
the apparatus is a boson field mode with creation and annihilation operators
${\bf a}^\dagger$ and ${\bf a}$ respectively.  We will consider the evolution
of the system and apparatus as dominated by the post-rotating wave
approximation Jaynes-Cumming Hamiltonian exactly at resonance\cite{Jaynes:63a},
\begin{equation}
{\bf H}=g \bmath{\sigma}_- \otimes {\bf a}^\dagger + g^* \bmath{\sigma}_+
\otimes {\bf a}, \label{eq:jcham}
\end{equation}
where $\bmath{\sigma}_\pm = \bmath{\sigma}_x \pm i \bmath{\sigma}_y$.  Using
the ${1 \over \sqrt{2}} \bmath{\sigma}_\alpha$ as a basis for the system
operators, we obtain the expansion
\begin{equation}
{\bf H}=\bmath{\sigma}_x \otimes (g {\bf a}^\dagger + g^* {\bf a} ) +
\bmath{\sigma}_y \otimes i(-g {\bf a}^\dagger + g^* {\bf a} ).
\end{equation}
The issue of whether this Hamiltonian can be used for stationary control is
therefore reduced to whether the operators
\begin{eqnarray}
{\bf A}_x &=& \sqrt{2} \left(g {\bf a}^\dagger + g^* {\bf a} \right) \nonumber
\\ {\bf A}_y &=&i \sqrt{2}( -g {\bf a}^\dagger + g^* {\bf a})
\end{eqnarray}
have simultaneous eigenstates.  First we will show why these operators do not
have simultaneous eigenstates but then we will show how in a certain limit
these operators can have a approximate simultaneous eigenstates.

Suppose that ${\bf A}_x$ and ${\bf A}_y$ had a simultaneous eigenstate
$|\psi\rangle$ with eigenvalue $a_x$ and $a_y$ respectively.  Since ${\bf A}_x$
and ${\bf A}_y$ are both Hermitian, $a_x$ and $a_y$ are both real.  The
commutator between ${\bf A}_x$ and ${\bf A}_y$ is
\begin{equation}
\left[ {\bf A}_x, {\bf A}_y \right] = -4i |g|^2  \left[{\bf a},{\bf a}^\dagger
\right]= -4i|g|^2 {\bf I}. \label{eq:axay}
\end{equation}
The fact that $|\psi\rangle$ is a simultaneous eigenstate of the ${\bf A}_x$
and ${\bf A}_y$ operator implies
\begin{equation}
\langle \psi | \left[ {\bf A}_x, {\bf A}_y \right] | \psi \rangle = a_x a_y -
a_y a_x =0.
\end{equation}
However on the right hand side of Eq.~(\ref{eq:axay}) we find the $\langle \psi
| (-4i|g|^2 {\bf I}) | \psi \rangle=-4i|g|^2$.  This is a contradiction and
therefore ${\bf A}_x$ and ${\bf A}_y$ cannot have simultaneous eigenstates.
Thus the Jaynes-Cummings Hamiltonian Eq.~(\ref{eq:jcham}) cannot be used for
orthogonal pure state stationary control.

Let us show, however, despite the fact that the Jaynes-Cummings Hamiltonian
cannot be used for exact control, that with a suitable approximation the
Jaynes-Cummings Hamiltonian can indeed be used for control.

The coherent state $|\alpha\rangle$ where $\alpha \in \CC$ is defined in terms
of the number states $|n\rangle$ as\cite{Scully:97a}
\begin{equation}
|\alpha\rangle= e^{-{|\alpha|^2 \over 2}} \sum_{n=0}^\infty {\alpha^n \over
\sqrt{n!} } |n\rangle.
\end{equation}
and is an eigenstate of the annihilation operator ${\bf a}|\alpha \rangle =
\alpha |\alpha \rangle$.  If the bosonic field we are considering is an
electromagnetic field, then lasers produce coherent states with very high
fidelity.

We next find that
\begin{equation}
{\bf A}_x |\alpha \rangle = \sqrt{2} \left( g^* \alpha |\alpha \rangle + g
e^{-|\alpha|^2/2} \sum_{n=0}^\infty {\sqrt{n+1} \alpha^n \over \sqrt{n!}}
|n+1\rangle \right).
\end{equation}
Defining the normalized state
\begin{eqnarray}
|\psi_\alpha \rangle &=& {1 \over \sqrt{\sum_{n=0}^\infty {(n+1) (|\alpha|^2)^n
\over n!} }}\sum_{n=0}^\infty {\sqrt{n+1} \alpha^n \over \sqrt{n!}} |n+1\rangle
\nonumber \\ &=& {e^{-|\alpha|^2/2} \over \sqrt{1+|\alpha|^2}}
\sum_{n=0}^\infty {\sqrt{n+1} \alpha^n \over \sqrt{n!}} |n+1\rangle.
\end{eqnarray}
We find that
\begin{eqnarray}
{\bf A}_x |\alpha \rangle &=& \sqrt{2} \left( g^* \alpha |\alpha \rangle + g
e^{-|\alpha|^2/2} \sqrt{\sum_{n=0}^\infty {(n+1) (|\alpha|^2)^n \over n!} }
|\psi_\alpha \rangle \right) \nonumber \\ &=& \sqrt{2} \left( g^* \alpha
|\alpha \rangle + g \sqrt{1+|\alpha|^2} |\psi_\alpha \rangle \right).
\end{eqnarray}
Now $|\psi_\alpha \rangle$ is nearly $|\alpha\rangle$ for large $|\alpha|$
\begin{equation}
|\langle \alpha | \psi_\alpha \rangle| =  \left|{e^{-|\alpha|^2} \over \sqrt{1+
|\alpha|^2}} \sum_{n=0}^\infty {(|\alpha|^2)^n \alpha^* \sqrt{n+1} \over \sqrt{
n! (n+1)!}} \right|={|\alpha| \over \sqrt{1+|\alpha|^2}}.
\end{equation}
In particular we find that
\begin{equation}
{\bf A}_x |\alpha \rangle \approx \sqrt{2} \left( g^* \alpha + g \alpha^*
\right) |\alpha\rangle,
\end{equation}
for $|\alpha| \gg 1$.  Similarly it can be shown that
\begin{equation}
{\bf A}_y |\alpha \rangle \approx  \sqrt{2} i\left( g^* \alpha - g \alpha^*
\right) |\alpha\rangle,
\end{equation}
for $|\alpha| \gg 1$.  Thus we have shown that $|\alpha\rangle$ is nearly an
eigenstate of ${\bf A}_x$ and ${\bf A}_y$ with eigenvalues $\sqrt{2} \left(g^*
\alpha +g \alpha^* \right)$ and $\sqrt{2} i \left(g^*\alpha - g \alpha^*
\right)$ respectively.

We have therefore shown that a system interacting with an apparatus which is in
the coherent state $|\alpha\rangle$ will, to a high degree of approximation,
produce an evolution on the system when $|\alpha| \gg 1$.

\section{The unitary control question}

We have examined the conditions under which control of a quantum system is
possible.  Now suppose that one is given some control over a quantum system. In
this section we address the issue of what can be done given the ability to
exercise some specified control.  For discussions in this section, we assume
ideal control conditions (no decoherence, perfect control of the controlling
apparatus and related couplings).  In the section following this one we deal
with the issue of approximation within the issue of control although we will
touch on the subject briefly in this section.  Another shortcoming of our
discussion is the fact that we ignore the effect which measurements can have
for controlling a system evolution. Thus what we are really asking is a
question of {\em unitary} control.

The most generic manner of posing the question of control is to assume that a
set of unitary evolutions ${\bf U}_i \in {\mathcal U}$ can be enacted on the
system via some controlling apparatus.  Given the ability to perform each of
these evolutions ${\bf U}_i$ a sequence of control can then be enacted like
\begin{equation}
{\bf U}_{i_1} {\bf U}_{i_2} {\bf U}_{i_3} \cdots {\bf U}_{i_p} \quad {\rm
where} \quad {\bf U}_{i_k} \in {\mathcal U}. \label{eq:controlseq}
\end{equation}
We will call such an evolution a {\em control sequence}.

\subsection{Densely filled group}

It might seem obvious that the control sequences form a group, but it turns out
this is not true in an exact sense.  The reason for this is that the control
sequences are {\em finitely} generated.  Let us demonstrate a trivial example
of a control sequence which does not form a group.  Suppose there is only one
${\bf U}_1 \in {\mathcal U}$ which acts on a single qubit as ${\bf
U}_1=|0\rangle \langle 0| + e^{i \gamma} |1\rangle \langle 1 |$ for some
$\gamma \in \RR$.  Then the control sequences we can generate are ${\bf
U}_1^p=|0\rangle \langle 0 | + e^{ip\gamma} |1\rangle \langle 1|$ for $p \in
\NN^+$.  In order for ${\bf U}_1$ to have an inverse (and hence form a group)
there must exist a $p$ such that ${\bf U}_1^p={\bf I}$ or $\exp[ip\gamma]=1$.
The only way in which this can be true for a finite $p$ is for $\gamma$ to be a
rational number.  Thus the control sequences do not always exactly form a
group.

However it is easy to see that to some degree of approximation, the control
sequences do form a group.  Writing a given ${\bf U}_i$ in the diagonal form
$\sum_\alpha e^{i \theta_\alpha } |\alpha \rangle \langle \alpha|$ we see that
for a given $\alpha$ $e^{i\theta_\alpha p}$, $p \in \NN^+$ are reachable by
repeated application of ${\bf U}_i$  For a given $\theta_\alpha$, either $e^{i
\theta_\alpha p}=1$ for some finite $p$ or $e^{i \theta_\alpha p}$ densely
fills $e^{i x}, \forall x \in \RR$ and as such densely fills the neighborhood
around $e^{ix}=1$.  Thus there always exists a finite $p$ such that ${\bf
U}_i^p \approx {\bf I}$ where the approximation is in the sense of deviation
from each $e^{i \theta_\alpha p}$ from $1$.  Thus we see that all control
sequences densely fill a group: every group element can be arbitrarily
accurately approximated by some control sequence.

The question of what can be generated by a control sequence is therefore
generally answered with {\em a group}, with the understanding that this answer
hinges on the densely filling structure of the control sequences.

\subsection{Hamiltonian control}

In most experimental control of a quantum system, instead of being given a set
of ${\bf U}_i \in {\mathcal U}$ which can be implemented, one usually
encounters the situation where evolution according to some set of Hamiltonians
${\bf H}_i \in {\mathcal O}$ can be achieved.  We will make the assumption that
the achievable control for such a set of Hamiltonians is given by all
evolutions of the form $\exp[-i {\bf H}_i t] \forall {\bf H}_i \in {\mathcal
O}, \forall t\in \RR^+$. This is an ideal assumption whose validity in the real
world is lacking due to (1) infinite precision in $t$ and (2) inability to
perform extremely fast turning on and off of a Hamiltonian. Problem (1) is true
for any control sequence and is addressed in the next section.  Problem (2),
however, also doesn't pose a huge problem because repeated application of
$\exp[-i {\bf H}_i t_0]$ for a fixed $t_0$ can be used to densely fill the
torus of all $\exp[-i {\bf H}_i t]$. Thus it is a generally good assumption
that control of a quantum system will allow for the implementation of given
$\exp[i{\bf H}_i t]$ for all real values of $t$.  We call this case of control
{\em Hamiltonian control}. In Hamiltonian control, one asks the question what
can be achieved via a Hamiltonian control sequence
\begin{equation}
e^{-i {\bf H}_{i_1} t_1 } e^{-i {\bf H}_{i_2} t_2} \cdots e^{-i {\bf H}_{i_p}
t_p} \quad {\rm where} \quad {\bf H}_{i_k} \in {\mathcal O}, t_i \in \RR^+.
\label{eq:hamcont}
\end{equation}

\subsection{Lie structure of Hamiltonian control} \label{sec:liestructure}

Lloyd\cite{Lloyd:95a} and Deutsch, Barenco, and Ekert\cite{Deutsch:95a},
building on work hinted at by DiVincenzo\cite{DiVincenzo:95a}, were the first
to raise and answer the question of what can be done with Hamiltonian control
within the context of quantum computation.  We have seen in the previous
section how control sequences form a group.  In the case of Hamiltonian
control, the group which is generated is a Lie group.  In particular, the
Hamiltonian control sequences Eq.~(\ref{eq:hamcont}) generate a continuously
parameterized group with nice smoothness and continuity properties over the
parameterization.

When we refer to the Lie group structure of the Hamiltonian control sequences,
we are just referring to the abstract group multiplication law between elements
of the Hamiltonian control sequences $g(\alpha) g(\beta) = g(\delta)$ where
$\alpha, \beta, \delta$ are the parameters of the group elements $g(\alpha),
g(\beta), g(\delta)$. Our Hamiltonian control sequences, however, have an
explicit representation as unitary linear operators on a Hilbert space
${\mathcal H}$, ${\bf g}(\alpha)$.  This explicit representation is called a
{\em unitary representation} of the Lie group.  A representation of a Lie group
is said to the reducible if it has an invariant proper subspace, by which we
mean that the action of any group element ${\bf g}(\alpha)$ on a vector in the
subspace remains in the subspace.  A representation which is not reducible is
irreducible.  All of the Lie groups generated by Eq.~(\ref{eq:hamcont}) are
completely reducible Lie groups.  This means that the representations we deal
with can always be written as the direct product of irreducible representations
(irreps),
\begin{equation}
{\bf g}(\alpha) = {\bf g}_1 (\alpha) \oplus {\bf g}_2(\alpha) \oplus \cdots
\oplus {\bf g}_k(\alpha),
\end{equation}
where each ${\bf g}_i(\alpha)$ is an irrep parameterized by $\alpha$.

Decomposing the action of the Hamiltonian control sequences into completely
reducible form tells us a lot about what can be done with such sequences.  It
doesn't give use direct access to what sort of computation (something we
haven't even introduced, but the meaning should be clear) can be performed on
the quantum system because we haven't defined an input, output relationship on
the system.  On the other hand, specifying the completely reducible form of a
Lie group describes exactly the limits of what can be done with a given
Hamiltonian control sequence.  The completely reducible form of a given
Hamiltonian control sequence succinctly describes all possible unitary actions
which can be performed on a controlled system.

Lets also point out how just knowing which Lie group one is dealing with is not
enough to pin down what can be done with a given Hamiltonian control sequence.
One needs to also know which dimensional representation one is dealing with.
The easiest example for illustrating this is to examine the one-dimensional
representation of $SU(2)$
\begin{equation}
{\bf g}_1(\alpha_1,\alpha_2,\alpha_3) = [ 1 ],
\end{equation}
and compare this to the two-dimensional representation of $SU(2)$,
\begin{equation}
{\bf g}_2(\alpha_1,\alpha_2,\alpha_3) = \exp \left[-i \vec{\alpha} \cdot
\vec{\bmath{\sigma}} \right],
\end{equation}
where $\vec{\bmath{\sigma}}= \left( \bmath{\sigma}_1, \bmath{\sigma}_2 ,
\bmath{\sigma}_3 \right)$ is the vector of the two-dimensional Pauli matrices.
Clearly the action of these two operators are very different.  One does
absolutely nothing, while the other manipulates a two-dimensional quantum
system in a non-trivial manner.  Thus just knowing what Lie group one has
control over is not enough--information about which irrep is also needed.

Every Lie group has a corresponding Lie algebra which we can use to good effect
to understand what can be done with a given Hamiltonian control sequence. Given
the ability to enact the Hamiltonians ${\bf H}_i \in {\mathcal O}$, every
Hamiltonian which can be generated from these Hamiltonians via the following
two actions can be physically enacted:
\begin{enumerate}
    \item Real linear combination of elements: $a {\bf H}_\alpha + b {\bf
    H}_\beta$ where $a,b \in \RR$.
    \item Lie commutation of elements: $i[{\bf H}_\alpha , {\bf H}_\beta]$.
\end{enumerate}
The reason Hamiltonian control sequences with Hamiltonians generated by this
set of operators are reachable follow from the two identities
\begin{eqnarray}
&& \lim_{n \rightarrow \infty} \left( \exp\left[-i {{\bf H}_\alpha t \over
\sqrt{n}} \right] \exp\left[i {{\bf H}_\beta t \over \sqrt{n}} \right]
\exp\left[i {{\bf H}_\alpha t \over \sqrt{n}} \right]\exp\left[-i {{\bf
H}_\beta t \over \sqrt{n}} \right] \right)^n=\exp \left[ [{\bf H}_\alpha , {\bf
H}_\beta ]  t \right] \nonumber
\\ && \lim_{n\rightarrow \infty} \left( \exp\left[-i {a {\bf H}_\alpha t \over
n} \right] \exp \left [-i {b {\bf H}_\beta t \over n} \right] \right) ^n = \exp
\left[ -i \left( a {\bf H}_\alpha + b {\bf H}_\beta \right) t \right].
\label{eq:liestructure}
\end{eqnarray}
In fact, we know from the famous theorem of Lie that the reachable operators
are exactly those which can generated via these two processes.  Thus the Lie
algebra generated by the $i {\bf H}_\alpha$ describes the Hamiltonians which
can be enacted by a Hamiltonian control sequence.

Again, just knowing the Lie algebraic structure of the Hamiltonians, however,
does not tell everything about the Lie group generated by the Hamiltonians.
Here there is an even further complication in that isomorphic Lie algebras may
correspond to different Lie groups.  Thus the abstract specification of the Lie
algebra is not enough to understand what can be done with a Hamiltonian control
sequence.  In spite of this fact, which just means that we can't look at the
abstract nature of the Lie algebra and jump to conclusions, if we completely
reduce a Lie algebra this will tell us everything about what can be done with a
given Hamiltonian control sequence.

\section{Control and approximation}

An important notion in control of quantum systems is how badly executed
operations influence the outcome of a control sequence.  Bernstein and
Vazirani\cite{Bernstein:97a} were the first to discuss how a sequence of poorly
approximated quantum operations influence the outcome of a particular control
sequence.  We follow the discussion of Nielsen and Chuang\cite{Nielsen:00a} on
the issue of approximating control sequences.

\subsection{Approximate unitary evolution} \label{sec:approxu}

Suppose we start a quantum system in the state $|\psi\rangle$ and then execute
a single unitary evolution ${\bf U}$ on the system and then perform a
measurement with POVM elements ${\bf M}_\alpha$ (see Appendix~\ref{apa:povm}).
How do the probabilities of these measurements differ if instead of enacting
${\bf U}$, the evolution operator ${\bf V}$ was executed?  Outcome $\alpha$
occurs with probability $\langle \psi | {\bf U}^\dagger {\bf M}_\alpha {\bf U}|
\psi \rangle$ if ${\bf U}$ is executed but occurs with probability $\langle
\psi | {\bf V}^\dagger {\bf M}_\alpha {\bf V}| \psi \rangle$ if ${\bf V}$ is
executed. The absolute value of the difference in these probabilities is
\begin{eqnarray}
\delta P_\alpha &=& \left|\langle \psi | {\bf U}^\dagger {\bf M}_\alpha {\bf
U}| \psi \rangle - \langle \psi | {\bf V}^\dagger {\bf M}_\alpha {\bf V}| \psi
\rangle \right|\nonumber \\ &=& \left| \langle \psi | {\bf U}^\dagger {\bf
M}_\alpha |\Delta\rangle + \langle \Delta |{\bf M}_\alpha {\bf V} |\psi \rangle
\right|,
\end{eqnarray}
where $|\Delta\rangle=\left( {\bf U} - {\bf V} \right) |\psi\rangle$.  Using
Cauchy-Schwarz we find that
\begin{eqnarray}
\delta P_\alpha  &\leq& \left| \langle \psi | {\bf U}^\dagger {\bf M}_\alpha
|\Delta\rangle\right| +  \left|\langle \Delta |{\bf M}_\alpha {\bf V} |\psi
\rangle \right| \nonumber \\ &\leq& 2 \left\| |\Delta \rangle \right\|
\nonumber \\ &\leq& 2E({\bf U},{\bf V}),
\end{eqnarray}
where
\begin{equation}
E({\bf U},{\bf V}) \equiv \max_{|\psi\rangle} \| \left( {\bf U} - {\bf V}
\right) |\psi\rangle \|.
\end{equation}
Therefore $E({\bf U},{\bf V})$ gives a quantification of how different a
measurement outcome can be if the two different evolutions ${\bf U}$ or ${\bf
V}$ are executed.  We will thus call $E({\bf U},{\bf V})$ the error between the
evolutions ${\bf U}$ and ${\bf V}$.

An important class of error which can occur in an evolution occur when the
variation of the controlled Hamiltonian is negligible while there are problem
executing the evolution for a precise time $t$.  In this case the error is
\begin{eqnarray}
E \left(\exp \left[-i {\bf H} t \right] , \exp \left[-i {\bf H}(t + \delta t)
\right] \right)&=& \max_{|\psi \rangle}  \left \| \left(\exp \left[-i {\bf H} t
\right] - \exp \left[-i {\bf H}(t + \delta t) \right] \right) |\psi\rangle
\right\| \nonumber \\ &=& \max_{|\psi \rangle}  \left \| \left( {\bf I}- \exp
\left[-i {\bf H}\delta t \right] \right) |\psi\rangle \right\| \nonumber
\\ &=& 1- \exp\left[-i \left(\max_{|\psi\rangle} \left \|{\bf H} |\psi \rangle \right\|
\right) \delta t\right].
\end{eqnarray}
For small $\delta t$, the error is thus
\begin{equation}
E \left(\exp \left[-i {\bf H} t \right] , \exp \left[-i {\bf H}(t + \delta t)
\right] \right) \approx \delta t \max_{|\psi\rangle} \left \| {\bf H} |\psi
\rangle \right \|.
\end{equation}

Suppose we are attempting to execute a control sequence ${\bf U}_{i_1} {\bf
U}_{i_2} \cdots {\bf U}_{i_p}$.  Due to inaccuracies, however, the control
sequence ${\bf V}_{i_1} {\bf V}_{i_2} \cdots {\bf V}_{i_p}$  was enacted.  The
error between these two control sequences is then
\begin{equation}
E({\bf U}_{i_1} {\bf U}_{i_2} \cdots {\bf U}_{i_p},{\bf V}_{i_1} {\bf V}_{i_2}
\cdots {\bf V}_{i_p}).
\end{equation}
It turns out that the error caused by such a sequence is at most the sum of the
errors of the individual operations
\begin{equation}
E({\bf U}_{i_1} {\bf U}_{i_2} \cdots {\bf U}_{i_p},{\bf V}_{i_1} {\bf V}_{i_2}
\cdots {\bf V}_{i_p})= \sum_{j=1}^p E\left( {\bf U}_{i_j},{\bf V}_{i_j}
\right).
\end{equation}
This can be proved via induction.  For $p=2$, we can use the triangle
inequality to show that
\begin{eqnarray}
E\left({\bf U}_{i_1} {\bf U}_{i_2} , {\bf V}_{i_1} {\bf V}_{i_2} \right) &=&
\left \| \left( {\bf U}_{i_1} {\bf U}_{i_2} - {\bf V}_{i_1} {\bf V}_{i_2}
\right) |\psi \rangle \right\| \nonumber \\ &=& \left\| \left( {\bf U}_{i_1}
{\bf U}_{i_2} - {\bf V}_{i_2} {\bf U}_{i_2} \right) |\psi\rangle + \left( {\bf
V}_{i_2} {\bf U}_{i_1} - {\bf V}_{i_2} {\bf V}_{i_1} \right) |\psi\rangle
\right\| \nonumber
\\ &\leq& \left\| \left( {\bf U}_{i_2}- {\bf V}_{i_2} \right) {\bf U}_{i_1}
|\psi \rangle \right\| + \left \|  {\bf V}_{i_2}\left({\bf U}_{i_1} - {\bf
V}_{i_1} \right) |\psi \rangle \right \| \nonumber \\ &\leq& E\left({\bf
U}_{i_1},{\bf U}_{i_2} \right) + E\left( {\bf U}_{i_2}, {\bf V}_{i_2} \right).
\end{eqnarray}
The general case for $p>2$ then quickly follows from induction.

Thus we have seen how $E({\bf U},{\bf V})$ quantifies the notion of how close
two unitary operators are in terms of difference in measurement outcomes
following the different unitary operators.  Further for a sequence of unitary
evolutions, the total error is bounded by the sum of the individual errors.
This latter property will be important when we discuss the relationship between
probabilities and computation.

\subsection{Approximate OSR evolution}

Is there an equivalent definition of an error distance between two OSR
evolutions $\{ {\bf A}_i \}$ and $\{ {\bf B}_i \}$?  As above, we can examine
the absolute difference in a POVM outcome measurement probability given the
input state $|\psi\rangle$, but now after the OSR evolutions via OSR operators
$\{ {\bf A}_i \}$ and $\{ {\bf B}_i \}$,
\begin{eqnarray}
\delta P_\alpha &=&\left| \sum_i {\bf A}_i |\psi \rangle \langle \psi | {\bf
A}_i^\dagger {\bf M}_\alpha - \sum_i {\bf B}_i |\psi \rangle \langle \psi |
{\bf B}_i^\dagger {\bf M}_\alpha \right| \nonumber \\ &=& \sum_i \left| \langle
\psi | \left( {\bf A}_i^\dagger {\bf M}_\alpha {\bf A}_i - {\bf B}_i^\dagger
{\bf M}_\alpha {\bf B}_i \right) |\psi \rangle \right| \nonumber \\ &=& \sum_i
\left | \langle \psi | {\bf A}_i^\dagger {\bf M}_\alpha |\Delta_i \rangle +
\langle \Delta_i | {\bf M}_\alpha  {\bf B}_i | \psi \rangle \right|,
\end{eqnarray}
where $|\Delta_i\rangle=\left( {\bf A}_i - {\bf B}_i \right) |\psi\rangle$.
Cauchy-Schwarz then implies
\begin{eqnarray}
\delta P_\alpha &\leq& \sum_i \left( |\langle \psi | {\bf A}_i^\dagger {\bf
M}_\alpha |\Delta_i\rangle| + |\langle \Delta_i | {\bf M}_\alpha {\bf B}_i
|\psi \rangle | \right).
\end{eqnarray}
We can now use the trick of recalling that the OSR comes from unitary evolution
on a larger space.  If the environment starts in the state $|0\rangle$ and the
OSR operators $\{ {\bf A}_i \}$ and $\{ {\bf B}_i \}$ come from the unitary
evolution ${\bf U}_A$ and ${\bf U}_B$ respectively, we find that
\begin{eqnarray}
\delta P_\alpha &\leq& \sum_i \left( |\langle 0 | \langle \psi | {\bf
U}_A^\dagger {\bf M}_\alpha |\Delta_i\rangle |i \rangle| + |\langle i |\langle
\Delta_i | {\bf M}_\alpha {\bf U}_B |\psi \rangle |0 \rangle | \right)
\nonumber
\\ &\leq& 2 \sum_i \left \| |\Delta_i \rangle |i \rangle \right \| \leq 2
\sum_i \left \| |\Delta_i\rangle \right \| \leq 2 E\left( \{ {\bf A}_i \}, \{
{\bf B}_i \} \right),
\end{eqnarray}
where we define the error between the two OSR evolutions $\{ {\bf A}_i \}$ and
$\{ {\bf B}_i \}$ as
\begin{equation}
E\left( \{ {\bf A}_i \}, \{ {\bf B}_i \} \right) =  \max_{|\psi\rangle} \|
\sum_i \left( {\bf A}_i - {\bf B}_i \right) |\psi\rangle \|.
\end{equation}
Thus we see that OSR evolutions have a similar notion of error to those of
unitary evolution.

\section{Control}

In this chapter we have seen how to define what is and what is not good
control.  We have begun to explored what can be done with this control and
understood how approximate control can be given a quantitative basis.  Later in
this thesis we will discuss the use of control sequences for quantum
computation.  Oftentimes it will be useful to work in the perfect control arena
even though the validity of this assumption is certainly not realized in
experiment.

Consider this state of affairs from the perspective of the status of classical
computers in the 1940's.  At that time it was unclear that machines could {\em
reliably} execute computations, and indeed early computers were prone to
breaking down.  Even today, hardware errors in computers can occur but the
probability of such errors occurring is extremely small (due in part to the
largess of Avogadro's number, see Chapter~\ref{ch:nft}).  The myth of perfect
control for classical computers is a good but only approximate truth.  The
question for quantum computation, of course, is whether it will ever be
possible to achieve the low probability of failure for a given control. Of
particular note in this quest is the demonstration of fault-tolerant quantum
computation\cite{Aharonov:97a,Gottesman:98a,Kitaev:97b,Knill:98a,Preskill:98a,Shor:96a}
where, even with imperfect control, nearly perfect control is achievable
without a drastic increase in resources.  On the other hand, there is no good
reason to believe that there do not exist systems which are naturally
fault-tolerant (see Chapter~\ref{ch:nft}).  The issues of control we have
raised in this chapter then, are the central language which will motivate our
quest for reliable quantum computation.

\chapter{Universal Quantum Computation} \label{ch:univ}

\begin{quote}
{\em```Mechanical process' is supposed to be a metaphor, Alan\dots''} \\
\begin{flushright} --Niel Stephenson, {\em Cryptonomicon}\cite{Stephenson:99a}~
\end{flushright}
\end{quote}

In the previous two chapters we have seen how to understand the evolution, both
desired (in the form of controllable evolution) and undesired (in the form of
decoherence) of a quantum system.  In this chapter we address the issue of how
to put the controllable evolution to use to perform quantum computation.  We
begin with a discussion of the notion of quantum subsystems.  The fundamental
localizable subsystems of modern physical theories then allow us to define and
make a case for the quantum circuit model as a valid model of quantum
computation.  The notion of a universal gate set is then introduced and two
important lemmas are presented which simplify the identification of universal
gate sets.  The most commonly cited universal gate set is then shown to be
universal.  In order to put the field of quantum computational complexity on
solid footing, the Kitaev-Solovay theorem is presented and the connection
between discrete and Hamiltonian control is discussed.  We then present an
example of a gate set which is not fully universal.  This leads to a discussion
of the concept of encoded universality wherein one uses the fungible nature of
quantum information to make a gate set universal.  An example of an encoded
universal gate set is presented.  An open question about the relationship of
representation theory of Lie algebras to quantum computation is presented and a
discussion of different dimensional irreducible representations of $SU(2)$ is
shown to give a broad leeway into the question of what is a qubit.  Finally,
the growth function of a Lie algebra is defined and shown to be a powerful tool
in showing when a gate set is {\em not} universal.

\section{Quantum subsystems}

In our discussion of decoherence we divided the universe up into a system and
an environment.  We made the assumption that this division was such that the
Hilbert space factorized as ${\mathcal H}= {\mathcal H}_S \otimes {\mathcal
H}_E$.  This was an assumption that the universe could be divided up into {\em
subsystems}: the system (perhaps a poorly planned nomenclature in hindsight!)
and the environment.  Similarly when we discussed decoherence-free control we
had the two subsystems, the system and the apparatus.  What dictates the
subsystem structure of quantum systems?

Let us first examine the notion of subsystems from an abstract mathematical
point of view.  The simplest concept of a subsystems structure is the one which
we most frequently encounter in nature: {\em full tensor product} subsystems.
These are subsystems in which the full Hilbert space ${\mathcal H}$ can be
divided up into a tensor product of $n$ subsystems, ${\mathcal H}=
\bigotimes_{i=1}^n {\mathcal H}_i$ where each ${\mathcal H}_i$ is a Hilbert
space corresponding to a subsystem. Note, however, that this is not the most
generic notion of a subsystem.  In particular it is possible that there are
{\em subspace tensor product} subsystems.  This means that instead of the full
tensor product structure there is a tensor product structure over restricted
subspaces of the Hilbert space ${\mathcal H}= \bigoplus_{j=1}^p \left(
\bigotimes_{i=1}^{n_p} {\mathcal H}_{ij} \right)$.  Here $j$ labels a subspace
of the global Hilbert space ${\mathcal H}$ and $i$ labels the $i$th subsystem
over this subspace.  We thus see that the most general notion of a subsystems
is one which act within different subspaces of the global Hilbert space
${\mathcal H}$.  Note that one could take one of the ${\mathcal H}_{ij}$
Hilbert space and further decompose this Hilbert space into a subsystem
structure.  If this is done, however, one can always express this subsystem
structure as in the subspace tensor product structure.  Thus the subspace
tensor product structure is the most general subspace tensor product structure
possible.

Of course from a mathematical point of view, we can always view any global
Hilbert space ${\mathcal H}$ as having any subsystem structure (full tensor
product or subspace tensor product) we desire. What is needed in order to make
progress in understanding subsystems is to ask {\em how physics} dictates a
subsystem structure.  In particular, {\em the notion of subsystems is a
empirically derived concept}.  The basic postulates of quantum systems {\em do
not} dictate the subsystem structure of quantum systems.

How, then, does the notion of subsystems arise in quantum systems?  Subsystems
arise due to the empirically motivated physical theories which we paste onto
the basic postulates of quantum systems.  The physical theories provide
Hamiltonians ${\bf H}_i$ which dictate the evolution of quantum systems and the
manner in which these Hamiltonians act on the system provide the notion of
subsystems. Of particular significance is the realization that currently all
empirically verified fundamental physical theories carry with them the
requirement of locality.  The notion of locality establishes a causal structure
on the evolution of quantum systems in spacetime: the Hamiltonians of these
theories establish a subsystem structure corresponding to the idea of local
subsystems. The basic postulate of locality thus leads to physical theories
which contain localizable subsystems.

The fundamental physical theories thus provided a {\em fundamental subsystem
structure} on a quantum system.  This can be separated from an {\em induced
subsystem structure} which takes the fundamental subsystem structure and builds
up subsystems from the fundamental physical subsystems.  For example, the
notion of individual atomic systems as a being separate subsystems is an
induced subsystem structure arising from the more fundamental physical
subsystem structure of quantum electrodynamics (and to a lesser degree the
quantum theories of the weak and strong forces).  It is a basic conjecture of
modern physics that
\begin{conjecture}
All empirical induced subsystems arise from localizable fundamental subsystems.
\end{conjecture}
Since induced subsystems are derived from fundamental subsystems, we are
therefore motivated to consider fundamental localizable subsystems as the basic
notion of quantum subsystems.

Let us be more concrete in our description of what we mean by localizable
subsystems.  In particular we will not address the issue of what does or does
not constitute a localizable subsystem but instead we will present a model of a
localizable subsystem which we claim captures the notion of locality in most
modern theories.  Suppose we are given a $d$ dimensional hypercubic lattice
with vertices ${\mathcal V}$ and edges ${\mathcal E}$.  We associate with each
of the vertices $v \in {\mathcal V}$ in this lattice a subsystem ${\mathcal
H}_v$ such that the global Hilbert space factors with a full tensor product
${\mathcal H}=\bigotimes_{v \in {\mathcal V}} {\mathcal H}_v$.  Local physical
theories produce nonvanishing Hamiltonians only when the Hamiltonians act as
single-body interactions on individual subsystem (${\bf H}_{v_i}$ on a given
vertices $v_i$ Hilbert space ${\mathcal H}_{v_i}$ tensored with identity on all
other subsystems) or between individual subsystems which are neighbors(${\bf
H}_{v_i,v_j}$ acting nontrivially on the combined Hilbert space ${\mathcal
H}_{v_i} \otimes {\mathcal H}_{v_j}$ where $v_i$ and $v_j$ are neighbors on the
hypercubic lattice tensored with identity on all other subsystem).  We claim
that this model of local subsystems can be used as the basis for all modern
quantum physical theories.  Of course in modern field theory, the subsystem
structure is really over a continuum of subsystems, so what we are really
claiming is that the continuum model of quantum field theory can be well
approximated by our basic model.

Further, as we have emphasized, the local subsystem structure of quantum
systems is really an empirical question for physical theories within the
framework of quantum principles.  Of special note on this subject is the
collected work of Kitaev, Freedman, and
coworkers\cite{Bravyi:00a,Freedman:00a,Freedman:00b,Freedman:00c,Freedman:00d,Kitaev:00a}
who have examined different physical theories of nature in terms of their local
subsystem structure.  For instance, these workers have described how some
modern topological field theories can be cast within a local subsystem
structure. The vigilant theorist, therefore, should take an interest in new
theories of nature which do not appear to provide a local subsystem
structure--if these theories turn out to have an empirical basis and a
non-local subsystem structure, the basis of quantum computation which local
subsystems provide may need updating!

We have introduced the notion of subsystems here because our future work in
this thesis will deal with induced subsystems of a nontrivial nature.  A
crucial role of a quantum computer will be the ability to simulate the
fundamental local subsystem structure with this induced subsystem structure.
This is the motivation which makes this section fundamentally important to the
understanding of what makes a quantum computer.

\section{The quantum circuit model}

We will now introduce the quantum circuit model of quantum computation.  This
model was first introduced by Deutsch\cite{Deutsch:89a} with more rigorous
theory being presented by Yao\cite{Yao:93a}.

We know from the previous section that modern physical theories are well
described by localizable quantum subsystems.  We would like to build a model of
quantum computation which, in the spirit of a modern day Church-Turing thesis,
provides a good model for what can be computed using quantum systems in the
real world.  For concreteness we will introduce the {\em qubit} quantum circuit
model and then describe how this model fits in with the more general notion of
a quantum computer.

The qubit quantum circuit model on $n$ qubits is built of a collection of $n$
two state systems.  Given $n$ qubits, the subsystem structure of this system is
${\mathcal H}=\bigotimes_{i=1}^n \CC^2$.  We endow the qubits with a
computational basis $|0\rangle$ and $|1\rangle$ which are the $\pm 1$
eigenstates of $\bmath{\sigma}_z$.  The {\em input} to the quantum circuit is a
basis vector $|i\rangle=|i_1\rangle |i_2\rangle \cdots |i_n\rangle$ where each
qubit is in a particular basis state $|0\rangle$ or $|1\rangle$.  The input
represents a prepared state upon which the quantum computation will act.  The
evolution of the system once the input has been prepared is then described by a
series of local control operators known as local quantum gates.  A quantum gate
acting on $k$ qubits is a unitary $2^k \times 2^k$ evolution matrix which
describes the effect of some evolution on the prescribed $k$ qubits.  We will
assume that the quantum gates whose evolution we can implement are all two or
one qubit gates, but we will allow parallel operation of such gates (see
\cite{Aharonov:96a} for our motivation for allowing parallel operators).  We
also restrict our quantum computer to have some realistic localized subsystem
structure and only allow operators which operate nontrivially between local
subsystems.  A quantum circuit is then a specification of the gates which will
operate upon the quantum system. Upon execution of the evolution the qubits are
measured in the computational basis and the output will then be a computational
basis output state $|j\rangle$. The outcome of the circuit will, in general, be
probabilistic.

The qubit quantum circuit model is clearly a restricted class of a much larger
class which we will label the {\em subsystems quantum circuit model}.  In the
subsystems quantum circuit model, one is given a system with some subsystem
structure ${\mathcal H}=\bigotimes_{i=1}^n {\mathcal H}_i$ (if it is not a full
tensor product structure, then we will examine a full tensor product structure
over some subspace of a subspace tensor product).  Preparation now corresponds
to preparing an input state which for each of the subsystems. The quantum gates
now correspond to operators on the subsystems and between local subsystems.
Finally, measurement is now a complete projective measurement on each of the
subsystem.  In further discussion, we will refer to the subsystems quantum
circuit model as the quantum circuit model unless needed.

A quantum circuit is a specification of the gates which will perform a quantum
computation on the input which results in an output with some probability.  In
this definition, it will turn out (see below) that every possible manipulation
of an input to an output is a quantum circuit.  For a fixed $n$, then, it is
possible to construct every possible quantum circuit.  However, the notion of
simply being a quantum circuit is not enough to capture the notion of an
algorithm: algorithms tell us how to work with inputs of varying length in a
{\em uniform} manner.  In particular there should be some method for
constructing a quantum circuit corresponding to some algorithm for all possible
input sizes.

To resolve this inadequacy of the quantum circuit model we must introduce the
notion of {\em uniform quantum circuit families}.  A quantum circuit family is
a set of circuits ${\mathcal C}$ whose elements are circuits $C_m$ indexed by a
label $m$ which describes the number of input bits into the given circuit. Each
of these circuits can be augmented by any number of extra work bits and the
output may also have any number of extra output bits (i.e. possible output
greater than $m$ bits). On an input string $|i\rangle$ with $m$ qubits, the
circuit labeled by $m$ produces an output $C_m(|i\rangle)$.  We require that
the circuits in ${\mathcal C}$ be consistent in that $C_m(|0\rangle \otimes
|i\rangle)=C_n(|i\rangle)$ where $m>n$ and $i$ is an $n$ bit input. Furthermore
we must require that there is some procedure for constructing the circuit for a
given input $|i\rangle$.  We say that a circuit family ${\mathcal C}$ is {\em
uniform} is there is a (classical) Turing machine which, given the input $i$
generates a description of the circuit $C_m$ which will act on the input
$|i\rangle$.  We will not delve into the definition of a classical Turing
machine--for our purposes we can just substitute our intuitive notion of a
modern classical computer (which is a Turing machine (almost: today's computers
do not have unlimited memory!)).

We have thus seen that the notion of a quantum algorithm can be recast into the
notion of uniform quantum circuit families.  The quantum circuit model itself
consisted of three major procedures: preparation, evolution due to quantum
gates, and measurement.  The quantum circuit model was also endowed with a
specific subsystem structure and certain localized limits on the actions which
could be performed on this subsystem structure.  In order to make the quantum
circuit model correspond to some notion of an algorithmic task we have had to
introduce the notion of uniform quantum circuit families.  The qubit quantum
circuit model with uniform quantum circuit families is a specific realization
of what is meant to carry out an algorithm on a quantum system.

\section{Universality}

In the previous section we have defined a quantum algorithm as a quantum
circuit family acting on a qubit quantum circuit model.  The circuit family
${\mathcal C}$ will contain an algorithm for constructing a given circuit $C_m$
for input on $m$ bits.  The output of the classical algorithm describing as
specific $C_m$ for a given input $|i\rangle$ will contain a set of instructions
${\mathcal I}(i)$ for building the circuit $C_m$.  In particular these
instructions ${\mathcal I}(i)$ will describe what gates should be implemented,
how they should be implemented (between which qubits), and the order in which
the gates should be executed.  The execution of a specific circuit family
${\mathcal C}$ requires that certain specific quantum gates are executed in the
fashion described by the instructions ${\mathcal I}(i)$.  Thus it would seem
that different circuit families might require different possible quantum gates.
This would be a torrid state of affairs for quantum computation if every
quantum circuit family required a complete reengineering of the quantum
hardware.

We would thus like to ask the question of whether there exist some set of
elementary quantum gates which can be used to build our algorithms such that
once we have access to this set of gates, we can in principle build up any
quantum circuit desired.  Actually what we want is a little less restrictive
because, as we discussed in Chapter~\ref{ch:control}, we must deal with some
sort of approximate evolution.  We want to ask if there is a set of gates which
we can use to approximate any quantum circuit.  An important note in this
definition is that because the quantum circuit description coming from a
classical computer is finite, the set of quantum gates we will use must also be
finite.

Let us therefore begin by making the following definition:
\begin{definition} {\em ($n$-qubit universal gate set)}
A set of quantum gates ${\mathcal G}$ acting on a qubit quantum circuit model
with $n$ qubits is defined to be a {\em $n$-qubit universal gate set} if, for
any $\epsilon>0$ a sequence of gates from this set ${\mathcal G}$ can be used
to approximate any unitary evolution on all $n$ qubits to accuracy $\epsilon$
(if ${\bf G}$ represents the evolution due to a sequence of such gates then we
require $E({\bf U},{\bf G})<\epsilon$ where $E$ is defined as in
Section~\ref{sec:approxu}).
\end{definition}
We can loosen this definition a bit if we allow some finite number of ancilla
qubits to be acted upon:
\begin{definition} {\em ($n$-qubit universal gate set augmented by $m$ ancilla
qubits)} A set of quantum gates ${\mathcal G}$ acting on a qubit quantum
circuit model with $n+m$ qubits is defined to be a {\em $n$-qubit universal
gate set augmented by $m$ ancilla qubits} if, for any $\epsilon>0$ a sequence
of gates from this set ${\mathcal G}$ can be used to approximate any unitary
evolution on $n$ qubits to accuracy $\epsilon$.
\end{definition}
Further we can extend the notion of a universal set of quantum gates to the
subsystem quantum circuit model via simply substituting subsystem quantum
circuit model for qubit quantum circuit model.  We then say that a set of gates
is a {\em $n$-subsystem universal gate set}.

The first to demonstrate a universal set of gates was Deutsch\cite{Deutsch:89a}
in 1989.  The universal gate set obtained by Deutsch, however, consisted of
operators on three qubits.  The three-body interactions necessary to produce
such a gate, however, are extremely difficult if not impossible to
experimentally realize. DiVincenzo was the first to demonstrate a universal set
of gates which required only two-body interactions\cite{DiVincenzo:95a}.
Perhaps the most widely cited universal gate set consists of the controlled-not
combined with a finitely generated group dense in single qubit
rotations\cite{Barenco:95b}.

Another important result states that a generic gate together with the ability
to permute qubits is universal\cite{Deutsch:95a,Lloyd:95a}. Here generic is the
rather limited notion of a gate with no inherit symmetry drawn from the space
of all possible gates.  While this result is of great existential value, as it
allows maintains that generically universality is not hard to achieve, in
practice this result has few applications.  The reason for this is that
physical interactions tend to have symmetries in their interactions.  Such
symmetries confine the given Hamiltonian to a lower dimensional space than the
full space of all operators on a given space.  Nature, in general, does not
uphold a mathematician's generic.

Below we assemble a list of important universal gate sets.

\newpage

\begin{table}[h]
\begin{tabular}{|l|c|}

 \hline Gate set & Reference \\ \hline \hline

  ${\bf Q}(\alpha) \equiv({\bf P}_{|00\rangle} + {\bf P}_{|01\rangle} + {\bf P}_{|11\rangle} ) \otimes {\bf I} +
  {\bf P}_{|11\rangle}\otimes ie^{-i \alpha \bmath{\sigma}_x}$,
  & Deutsch, 1989\cite{Deutsch:89a} \\
  and Permutations & \\ \hline
  $\bmath{\sigma}_x$, $e^{\pm i\sqrt{\alpha}}{\bf P}_{|00\rangle} + {\bf P}_{|01\rangle} +
  {\bf P}_{|10\rangle}+{\bf P}_{|11\rangle}$,  &
  DiVincenzo, 1995\cite{DiVincenzo:95a} \\
  ${\bf P}_{|0\rangle} \otimes {\bf I} + {\bf P}_{|1\rangle}
  ie^{\pm i \sqrt{\alpha} \bmath{\sigma}_x}$, and Permutations  &  \\ \hline
  Any single ${\bf P}_{|0\rangle} \otimes {\bf I} + {\bf P}_{|1\rangle} \otimes e^{i\sum_{i=1}^3 \alpha_i
  \bmath{\sigma}_i}$,
  & Barenco, 1995\cite{Barenco:95a} \\
  and Permutations & \\ \hline
  Almost any single two-qubit gate, and & Deutsch {\em et.al}, 1995\cite{Deutsch:95a} \\
  Permutations & Lloyd, 1995\cite{Lloyd:95a} \\ \hline
  ${\bf  ^C X}\equiv{\bf P}_{|0\rangle} \otimes {\bf I} + {\bf P}_{|1\rangle} \otimes
  \bmath{\sigma}_x$, and any gate set which
  & Barenco {\em et.al}, 1995\cite{Barenco:95b} \\
  densely generates single qubit gates &   \\ \hline
  ${\bf H}$, ${\bf P}$, ${\bf ^C X}$, ${\bf Q}(\pi/2)$
  & Shor, 1996\cite{Shor:96a} \\ \hline
  ${\bf P}$, ${\bf ^C X}$, ${\bf ^C P} \equiv {\bf P}_{|00\rangle} + {\bf P}_{|01\rangle}
  +{\bf P}_{|10\rangle}+i {\bf P}_{|11\rangle}$, & Knill {\em
  et.al}, 1998\cite{Knill:98b} \\
  and the ability to prepare $|\pm\rangle={1 \over \sqrt{2}} \left( |0\rangle \pm |1\rangle \right)$ & \\ \hline
  ${\bf H}$, ${\bf P}$, ${\bf ^C X}$, ${\bf ^C P}$ & Knill {\em
  et.al,} 1998\cite{Knill:98b} \\ \hline
  ${\bf H}$, ${\bf P}$, ${\bf ^C X}$, and the ability to prepare & Knill {\em
  et.al}, 1998\cite{Knill:98b,Knill:98a} \\
   $|\pi/8\rangle
  = \cos(\pi/8)|0\rangle + \sin(\pi/8) |1\rangle$ &  \\ \hline

  ${\bf ^C X}$, ${\bf Q}(\pi/2)$, plus the ability to  & Gottesman, 1998\cite{Gottesman:98a} \\
  measure
  $\bmath{\sigma}_i$, $i\in\{x,y,z\}$ & \\ \hline
     Single qubit operations, Bell measurements, & Gottesman and
    Chuang\cite{Gottesman:99a} \\
    and GHZ states & \\ \hline
  $\exp\left[{-i \alpha \sum_i \bmath{\sigma}_i \otimes \bmath{\sigma}_i} \right]$ &
  Bacon {\em et.al}, 2000\cite{Bacon:00a} \\
   & Kempe {\em et.al}, 2001\cite{Kempe:01a} \\ \hline
   $\bmath{\sigma}_x \otimes \bmath{\sigma}_x + \bmath{\sigma}_y \otimes
   \bmath{\sigma}_y$ between qubits & Kempe, 2001\cite{Kempe:01a} \\
    & Wu and Lidar, 2001\cite{Wu:01a} \\ \hline
    Linear optics elements, single photon sources, & Knill, Laflamme, \\
    single photon detectors & and Milburn, 2001\cite{Knill:01a} \\ \hline
    Entangled cluster state and single measurements & Raussendorf \\ & and Briegel,
    2001\cite{Raussendorf:01a} \\ \hline

\end{tabular}
\caption{\em Universal gate constructions}

Universal sets of gates.  Many of these gates are elements in the normalizer of
the Pauli group (see Appendix~\ref{apa:pauli}).  $\alpha$, and $\alpha_i$ are
irrational multiples of $\pi$. ${\bf P}_{|\psi\rangle}=|\psi\rangle \langle
\psi|$ not to be confused with ${\bf P}=|0\rangle \langle 0| + i |1\rangle
\langle 1|$. Permutations means the ability to permute the wires of the qubits.
\end{table}

\subsection{The inductive subsystem lemma}

Before proceeding, it is useful to introduce the following tools for
universality proofs.

\begin{lemma} \label{lem:subspace} \cite{Aharanov:99a}
Suppose one is given two sets ${\mathcal O}_1$ and ${\mathcal O}_2$ of
operators which densely act as $SU(d_A)$ and $SU(d_B)$ on two subspaces
${\mathcal H}_A$ (dimension $d_A$) and ${\mathcal H}_B$ (dimension $d_B$) of a
larger Hilbert space ${\mathcal H}$. If ${\mathcal H}_A$ and ${\mathcal H}_B$
are not disjoint then the set of operators which can be achieved by combining
these operators is $SU(d)$ acting on the union of these Hilbert spaces
${\mathcal H}_A \cup {\mathcal H}_B$ (dimension $d$)
\end{lemma}
Proof: See \cite{Aharanov:99a}.

Following this lemma, the inductive subsystem lemma follows
\begin{lemma} \label{lem:subsystem}
Suppose one is given a Hilbert space ${\mathcal H}$ with a subsystem structure
${\mathcal H}=\bigoplus_{i=1}^n {\mathcal H}_i$, each subsystem ${\mathcal
H}_i$ of dimension $d_i$.  We say that two subsystems ${\mathcal H}_a$ and
${\mathcal H}_b$ of dimensions $d_a$ and $d_b$ are {\em computationally
connected} if operators on the combined Hilbert space ${\mathcal H}_a \otimes
{\mathcal H}_b$ densely generate $SU(d_ad_b)$.  Let $G$ be the graph of
computationally connected subsystems for a given gate set.  If this graph is
connected, then the gate set can densely generate $SU(\prod_{i=1}^n d_i)$.
\end{lemma}
Proof: Follows simply from induction using Lemma~\ref{lem:subspace}.

\subsection{Universal gate set example}

As a quick example of a universal gate set, we give here the example of the
controlled-not ${\bf ^C X}$ plus finitely generated dense single qubit
gates\cite{Barenco:95b}.

By postulate, this gate set generates any single qubit operation to any desired
accuracy.  In particular it generates an approximation to a single qubit
$\bmath{\sigma}_x$ rotation near identity, $\exp\left[ i \delta
\bmath{\sigma}_x \right]$.  Sandwiching this operation in between two
controlled not operators, we find that on two qubits
\begin{equation}
{\bf ^C X} \exp\left[ i \delta \bmath{\sigma}_x \otimes {\bf I} \right] {\bf ^C
X} = \exp \left[ i \delta \bmath{\sigma}_x \otimes \bmath{\sigma}_x \right].
\end{equation}
We recall (see Appendix~\ref{apa:pauli}) that elements of the single qubit
Pauli normalizer act as an automorphism on the single qubit Pauli operators.
This implies that there are single qubit operators which when conjugated about
${\bmath \sigma}_x \otimes {\bmath \sigma}_x$ produce the Pauli operators
${\bmath \sigma}_\alpha \otimes {\bmath \sigma}_\beta$.  Thus our gate set can
produce
\begin{equation}
{\bf N}_1^\dagger \otimes {\bf N}_2^\dagger \exp \left[ i \delta
\bmath{\sigma}_x \otimes \bmath{\sigma}_x \right] {\bf N}_1 \otimes {\bf N}_2 =
\exp \left[ i \delta \bmath{\sigma}_\alpha \otimes \bmath{\sigma}_\beta
\right],
\end{equation}
where ${\bf N}_i$ are elements of the single qubit Pauli normalizer.  We have
therefore shown that the gate set can produce, to a given accuracy any
infinitesimal generator of the $SU(4)$ over two qubits:
\begin{eqnarray}
&&\exp \left[i  \vec{\delta} \cdot \vec{\bmath{\sigma}} \otimes {\bf I } \delta
\right] \quad {\rm postulate} \nonumber \\ &&\exp \left[i {\bf I} \otimes
\vec{\delta} \cdot \vec{\bmath{\sigma}} \right] \quad {\rm postulate} \nonumber
\\
 && \exp \left[ i \delta \bmath{\sigma}_\alpha \otimes \bmath{\sigma}_\beta
 \right] \quad {\rm above}.
\end{eqnarray}
Because we have the infinitesimal generators of $SU(4)$, we can therefore
produce any gate in the $SU(4)$ of two qubits: we have shown how to produce all
two qubit unitary gates.

Next we can use Lemma~\ref{lem:subsystem}.  Because we can generate $SU(4)$
between local qubits, we can therefore produce any $SU(2^n)$ on $n$ qubits.
Thus we have shown how the controlled not plus local single qubit gates can be
used to enact any possible quantum circuit.

\subsection{The Kitaev-Solovay theorem}

We have defined a universal gate set such that any quantum circuit can be
constructed to any desired accuracy from this gate set.  Now, suppose one is
given a quantum circuit family.  The circuits in this circuit family will come
with descriptions of the quantum gates to be executed in the quantum algorithm.
There are many ways to place the cost on implementing such a circuit: the
breadth of the circuit, the depth of the circuit, the total number of gates
used, etc.  The field of quantum computational
complexity\cite{Bernstein:93a,Bernstein:97a,Berthiaume:92a} seeks to understand
how these resources grow for different quantum algorithms.  The cost function
which is perhaps most important is the depth of the circuit.  This depth
corresponds in some fashion to the total running time of the circuit.  In view
of the universality results for quantum circuits, it would be nice to know that
different universal gate sets do not lead to radically different assessments of
the complexity of different circuits.

That this is ostensibly true is guaranteed by a theorem due to
Kitaev\cite{Kitaev:97a} and independently
Solovay\cite{Solovay:95a,Solovay:00a}.
\begin{theorem}{(\em Solovay-Kitaev)}
Let ${\mathcal G}$ be a finite set of quantum gates which contains each gate's
inverse and which densely generates $SU(d)$.  For $\epsilon>0$, there is
sequence of gates of length $l$ which is within $\epsilon$ of every element of
$SU(d)$ (using trace distance, Section~\ref{sec:approxu}) where
$l=O\left(\log^c \left(1 \over \epsilon \right) \right)$ where $c$ is some
fixed constant which depends on $d$.
\end{theorem}
Proof: See \cite{Nielsen:00a}.  It is interesting to note the connections
between this theorem and the study of ``geometric group
theory''\cite{delaHarpe:00a}.

The Solovay-Kitaev theorem indicates that a universal set of quantum gates can
be used to approximate another universal set of gates with only a
polylogarithmic overhead in the depth of the circuit.  Consider two sets of
gates ${\mathcal G}_1$ and ${\mathcal G}_2$.  Because each of these gate sets
are universal, every gate in ${\mathcal G}_1$ can be approximated by a sequence
of gates in ${\mathcal G}_2$ and vice versa.  The content of the Solovay-Kitaev
theorem tells us the sequence of gates from one set used to approximate a gate
from the other set requires $O\left(\log^c \left( {1 \over \epsilon} \right)
\right)$ gates.  Thus the depth difference between circuits constructed with
different universal sets of gates to an accuracy $\epsilon$ is only
$O\left(\log^c \left( {1 \over \epsilon} \right) \right)$.

\subsection{Discrete versus Hamiltonian control}

In Chapter \ref{ch:control} we introduce the notion of a Hamiltonian control
sequence given a set ${\mathcal O}$ of implementable Hamiltonians
\begin{equation}
e^{-i {\bf H}_{i_1} t_1 } e^{-i {\bf H}_{i_2} t_2} \cdots e^{-i {\bf H}_{i_p}
t_p} \quad {\rm where} \quad {\bf H}_{i_k} \in {\mathcal O}, t_i \in \RR^+.
\end{equation}
How do Hamiltonian control sequences, which we are most likely to encounter in
real quantum control situations, mesh with the idea of universal quantum gate
sets?

Universal sets of quantum gates are a finite set of gates which can be
implemented with a certain accuracy while Hamiltonian control sequences are a
continuum of gates which can be implemented with a certain accuracy.  In
practice, one would take the set of Hamiltonian control sequences and make
these operators a discrete set in order to use the control sequence as a
universal set of gates. There is a simplification, however, in describing the
universality properties of Hamiltonian control sequences which often makes
determining their universality properties simple.  Given a subsystem structure
and a Hamiltonian control sequence, the universality properties follow directly
from analysis of the Lie algebra generated by the control Hamiltonians.

Representation theory of the Lie algebra for a set of control Hamiltonian
$\mathcal O$ describes exactly what can be done with a Hamiltonian control
sequence. Combined with a description of the accuracy with which a given
Hamiltonian control sequence can be implemented, this information describes how
every element of a Lie group corresponding to the Lie algebra can be obtain to
within some accuracy given a Hamiltonian control sequence.  It is important to
realize that just understanding what can be done with some given control is
insufficient for resolving questions about universality. A mapping from the
quantum system to a subsystems quantum circuit model must also be made.
Analysis of Lie algebra alone does not give a complete understanding of
universality properties.

\subsection{Example use of Lie algebraic structure} \label{sec:lieexamp}

Suppose we are given a linear array of $n$ qubits where $n$ is odd.  Each
individual qubit has an energy ${\bf H}_0=\epsilon \sum_{i=1}^n
\bmath{\sigma}_z^{(i)}$ which is always present and we have no control of the
energy spacing $\epsilon$. Between neighboring qubits there is an interaction
$\tilde{\bf H}_i= \bmath{\sigma}_x^{(i)}  \bmath{\sigma}_x^{(i+1)}$ over which
we have complete control.  Thus at the set of implementable Hamiltonians we can
achieve is $\bf{H}_i={\bf H}_0+ \tilde{\bf H}_i$ and ${\bf H}_0$.

Let us describe the Lie algebra achievable with these interactions.  Clearly we
can start with ${\bf H}_0/\epsilon$ and $\tilde{\bf H}_i$ as our interactions
because the first is given up to scaling and the second can be obtain via
subtracting this first from ${\bf H}_i$.  Taking the commutator yields
\begin{equation}
{\bf H}_i^{(1)}={i \over 2}\left[   {\bf H}_0, \tilde{\bf H}_i  \right] =
\bmath{\sigma}_y^{(i)}  \bmath{\sigma}_x^{(i+1)} + \bmath{\sigma}_x^{(i)}
\bmath{\sigma}_y^{(i+1)}.
\end{equation}
Taking the commutator of this operation with ${\bf H}_0$ or $\tilde{\bf H}_i$
yields
\begin{eqnarray}
{\bf H}_i^{(2)}&=&{i \over 2} \left[ {\bf H}_i^{(1)} , {\bf H}_0\right]= -2
\bmath{\sigma}_x^{(i)}  \bmath{\sigma}_x^{(i+1)} + 2 \bmath{\sigma}_y^{(i)}
\bmath{\sigma}_y^{(i+1)} \nonumber \\ {\bf H}_i^{(3)}&=&{i \over 2} \left[ {\bf
H}_i^{(1)}, \tilde{\bf H}_i \right] = \bmath{\sigma}_z^{(i)} +
\bmath{\sigma}_z^{(i+1)}.
\end{eqnarray}
Because $n$ is odd, the last of these commutators implies that
$\bmath{\sigma}_z^{(i)}$ is in the Lie algebra generated by ${\mathcal O}$.  At
this point it is clear that the Lie algebra generated by ${\mathcal O}$ is the
same as the Lie algebra generated by ${\mathcal O}^\prime= \{
\bmath{\sigma}_z^{(i)} , \bmath{\sigma}_x^{(i)} \bmath{\sigma}_x^{(i+1)} \}$.

Suppose we wanted to use these interactions for a quantum circuit model on all
$n$ qubits, i.e. with subsystem structure $\bigotimes_{i=1}^n \CC^2$. We can
show that this is not possible, i.e. that it is not possible to generate
$SU(2^n)$ with the operators in ${\mathcal O}^\prime$.

Consider the elements of the Lie algebra generated by ${\mathcal O}^\prime$.
All of these elements will be linear combinations of commutators of the
generators in ${\mathcal O}^\prime$.  We will now show that the parity, defined
as the eigenvalue of $\bigotimes_{i=1}^n \bmath{\sigma}_z^{(i)}$, cannot be
changed by the any element in the Lie algebra generated by ${\mathcal
O}^\prime$.  First note that all of the elements of ${\mathcal O}^\prime$
commute with $\bigotimes_{i=1}^n \bmath{\sigma}_z^{(i)}$.  This in turn implies
that all commutators formed from the generators of ${\mathcal O}^\prime$
commute with $\bigotimes_{i=1}^n \bmath{\sigma}_z^{(i)}$.  Thus
$\bigotimes_{i=1}^n \bmath{\sigma}_z^{(i)}$ commutes with all elements of the
Lie algebra generated by ${\mathcal O}^\prime$.  Since $\bigotimes_{i=1}^n
\bmath{\sigma}_z^{(i)}$ commutes with all of the elements of the Lie algebra
generated by ${\mathcal O}^\prime$ there are elements of $SU(2^n)$ which are
not in this Lie algebra.

In this example we have shown how a specific control mechanism fails to be
fully universal.  It is impossible to use the control of local
$\bmath{\sigma}_x^{(i)} \bmath{\sigma}_x^{(i+1)}$ with each qubit having a
constant energy to produce every unitary evolution on $n$ qubits.

\section{Encoded universality}

\subsection{The fungible nature of quantum information}

An important property of classical information which carries over to the
quantum regime is the fungible nature of information\cite{Bacon:01c}.  A
resource is fungible if interchanging it with another resource does not destroy
the value of the resource.  Whether we represent a classical bit by the
presence or absence of a chad on a punch-card \cite{Supremecourt:00-949} or in
the orientation of a billion electron spins, the intrinsic value of the
information (the value of the bit) is untouched.  Information does not depend
upon the medium in which it is represented.  The fungible nature of information
has been key to the exponential growth of the computer revolution.  The fact
that it does not matter that the information is being confined to smaller and
smaller components on silicon chips has been central to the continuing success
of Moore's law\cite{Moore:65a}.  So, too, goes quantum information: the
plethora of experimentally proposed systems from which a quantum computer could
be built is made possible by the fungibility of quantum information.  Whether
we store quantum information in the electronic levels of an atomic system or in
the spin of a single electron impurity in a solid-state system, the information
is still quantum and can be used for the basis of building a quantum computer.

One central aspect of the fungible nature of quantum information is that the
information can be encoded in some highly non-trivial manner.  Of course when
we represent a qubit in the spin of an electron or in the hyperfine levels of a
ion, we are essentially encoding the qubit into a given Hilbert space. However,
it is important that this notion can be considerably extended.  In particular,
given multiple quantum subsystem, quantum information can be stored in highly
entangled states between these subsystems.  The fact that quantum information
can be encoded is essential to the development of the theory of quantum error
correcting codes.  By choosing a particular encoding of the quantum
information, quantum error correcting codes provide a method for identifying
and correcting the effect of quantum errors on the code.  In part II of this
thesis, we will explore how certain encodings of quantum information can be
used to perfectly isolate the quantum information from particular forms of
decoherence.

The fungible nature of quantum information is a warning sign on the path
towards building a quantum computer.  While a gaggle of labs quest to develop a
particular system for quantum computation, the fungible nature of quantum
information tells us that a successful architecture for quantum computing may
look nothing like the currently envisioned system.  Because information can be
encoded, it is unclear exactly where we will store the information that makes
up a future quantum computer.  An optimistic viewpoint of the fungible nature
of quantum information, then, tells us that the quest for physical systems
which can quantum compute is far from a closed deal.  We will return to this
issue in Chapter \ref{ch:nft}.

\subsection{Encoded universality constructions}

Encoding of quantum information can also be of use in the construction of
universal gate sets\cite{Bacon:01c}.  There are two complementary ways of
looking at this problem.  On the one hand, because quantum information can be
encoded, certain interactions which were not universal over the entire Hilbert
space can be made universal on a particular encoded space.  At the other end of
the spectrum, it is common in quantum computing to develop a particular
encoding (for error correction, due to physical constraints, etc.) and then to
ask: what manipulations are needed to compute on such an encoded space.  Of
course, these viewpoints are complementary to each other.  In this section, we
will discuss the first of these viewpoints: how encoding can make gate sets
universal over an encoding.

Suppose one is given a gate set ${\mathcal G}$.  As we have previously argued,
the universal properties of this gate set is really a question of the relation
of this gate set to representations of Lie groups.  In fact the notion of an
irreducible representation directly contains our notion of encoded
universality.  Thus the idea of encoded universality is nothing more than the
observation that the power of a set of gates is described by the irreducible
representations and these irreducible representations may act on some encoded
space.  We refer to a set of gates which acts on some encoded subsystem
structure in a universal manner as a universal set of gates on a quantum
circuit model with encoded subsystems.

Notice that the notion of encoded universality changes the rules not only for
the manipulation of the quantum information, but also for the preparation and
measurement procedures of a quantum circuit model with encoded subsystems.  Of
particular importance here it to note that the preparation procedure should not
be overly inefficient.  We will return to these questions when we address a
specific example in Chapter~\ref{ch:collectiveuniv}.

Let us define what is needed in order to present an encoded universality
quantum circuit model which can be used to construct uniform quantum circuit
families:
\begin{itemize}
\item A particular subsystem structure on the Hilbert space must be described
which maps onto a subsystem structure of an unencoded quantum circuit model.
This is perhaps the most stringent of the requirements for an encoded quantum
computer.  Without a subsystems structure, uniformity cannot be enforced and
the very nature of a scalable architecture is violated.
\item The ability to prepare the subsystems into a particular initial state.
Here there is more leeway.  It is, even for the standard quantum circuit model,
never necessary to obtain perfect preparation.  An important issue for encoded
universality constructions is the fact that it is possible for quantum
information to ``leak'' out of the encoded subsystem.
\item Operations which act in a universal manner on the encoded subsystem
structure.
\item The ability to extract information from the encoded subsystems.  Again,
perfect measurement is not necessary.  The ability to extract even a little bit
of information is often sufficient for quantum computation.
\end{itemize}

A particularly interesting class of encoded universality constructions are what
we will term {\em few subsystems} encoded universality.  While in practice,
given a gate set, ${\mathcal G}$ one can analyze the action of this gate set on
ever larger numbers of qubits, the uniformity condition of the quantum circuit
model puts a condition on the encoding such that there should be a some map
onto a subsystem structure which grows proportional to the number of qubits
added.  Thus the typical manner in which encoded universality will be used is
to take a constant number of subsystems and the encode a basic subsystem into
this constant number of subsystems.  For example one may find that taking
triads of qubits allows for an encoded qubit which can be robustly manipulated
(prepared, measured, unitarily controlled).  Then one proceeds to take the
encoded subsystems, map it onto a quantum subsystem circuit model and
(hopefully) show universality on this encoded subsystem structure.

\subsection{Few subsystems encoded universality example}

Suppose one is given a spin chain of $n$ qubits with interactions as shown in
Figure~\ref{fig:encodechain} below. In particular assume that the Hamiltonians
which can be enacted on this spin chain come from the set
\begin{equation}
{\mathcal S}=\left\{ \bmath{\sigma}_x^{(i)} \bmath{\sigma}_x^{(i+1)},
\bmath{\sigma}_z^{(i)} + \bmath{\sigma}_z^{(i+1)}, \bmath{\sigma}_z^{(i)}
\bmath{\sigma}_z^{(i+1)} \right\}.
\end{equation}
\begin{figure}[h]
\hspace{2cm}  \psfig{figure=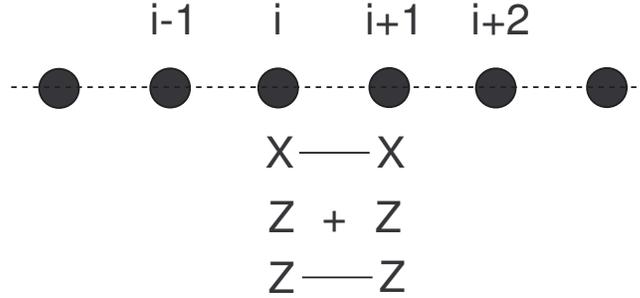,width=3.5in} \vspace{0.2cm}
\caption{\em Example encoded universality spin chain} \label{fig:encodechain}
\end{figure}
To see that full universality on all $n$ qubits (the ability to implement
$SU(2^n)$ on the system) is not possible, note that $\prod_{i=1}^n
\bmath{\sigma}_z^{(i)}$ commutes with all of the elements of ${\mathcal S}$ and
thus, via the same argument of Section~\ref{sec:lieexamp}, the Lie algebra
generated by these Hamiltonians is not the full $SU(2^n)$.

Let us examine the action of the Hamiltonians listed above on pairs of qubits.
Notice that these have the following Lie algebraic structure
\begin{eqnarray}
&&\left[  \bmath{\sigma}_z^{(i)}
+\bmath{\sigma}_z^{(i+1)},\bmath{\sigma}_x^{(i)} \bmath{\sigma}_x^{(i+1)}
\right]=2i \left( \bmath{\sigma}_x^{(i)} \bmath{\sigma}_y^{(i+1)} +
\bmath{\sigma}_y^{(i)} \bmath{\sigma}_x^{(i+1)}\right) ~~+~~ {\rm
cyclic~permutations} \nonumber \\ &&\left[ \bmath{\sigma}_z^{(i)}
\bmath{\sigma}_z^{(i+1)}, \bmath{\sigma}_x^{(i)}
\bmath{\sigma}_x^{(i+1)}\right]= \left[\bmath{\sigma}_z^{(i)}
\bmath{\sigma}_z^{(i+1)}, \bmath{\sigma}_z^{(i)}+
\bmath{\sigma}_z^{(i+1)}\right]=0 \nonumber \\ && \left[ \bmath{\sigma}_z^{(i)}
\bmath{\sigma}_z^{(i+1)}, \left( \bmath{\sigma}_x^{(i)}
\bmath{\sigma}_y^{(i+1)} + \bmath{\sigma}_y^{(i)}
\bmath{\sigma}_x^{(i+1)}\right) \right] =0.
\end{eqnarray}
The first of these commutation relations indicates an $SU(2)$ structure for
these operators and the final three commutation relations indicate that
$\bmath{\sigma}_z^{(i)} \bmath{\sigma}_z^{(i+1)}$ is an abelian subalgebra over
these two qubits.  In particular we note that over the subspace with basis
states $|00\rangle$ and $|11\rangle$ (for the $i$ and $i+1$th qubit), the
operators $\bmath{\sigma}_x^{(i)} \bmath{\sigma}_x^{(i+1)}, \left(
\bmath{\sigma}_x^{(i)} \bmath{\sigma}_y^{(i+1)} + \bmath{\sigma}_y^{(i)}
\bmath{\sigma}_x^{(i+1)}\right), \bmath{\sigma}_z^{(i)}
+\bmath{\sigma}_z^{(i+1)}$ act as $\bmath{\sigma}_x, \bmath{\sigma}_y,
\bmath{\sigma}_z$ on this encoded space, while $\bmath{\sigma}_z^{(i)}
\bmath{\sigma}_z^{(i+1)}$ acts as identity on this subspace.  Thus we can use
these operators to enact $SU(2)$ on the encoded subspace spanned by the logical
qubits $|0_L\rangle = |00\rangle $ and $|1_L\rangle=|11\rangle$.  Also note
that the subspace $|01\rangle$ and $|10\rangle$ is not acted upon by these
operators in a non-commuting manner and hence over these operators cannot be
used as a qubit.

Having shown that there is an encoding over two qubits for which the operators
in ${\mathcal S}$ act as $SU(2)$, we then hope to extend this encoding to a
full quantum circuit model.  In particular we take our encoded qubits (the
subsystems) to be two physical qubits with the encoding of
$|0_L\rangle=|00\rangle$ and $|1_L\rangle=|11\rangle$.  We have already shown
that any single qubit operation is possible on this encoded space and thus it
is sufficient to show that we can implement a non-trivial two body encoded
operation between the qubits in order to produce a encoded universal quantum
circuit.  If we take the encoding between the $1$st and $2$nd, $3$rd and $4$th,
etc. qubits then the operation $\bmath{\sigma}_z^{(2k)}
\bmath{\sigma}_z^{(2k+1)}$ with $1 \leq k \leq n/2$ provides this coupling.  In
particular note that
\begin{eqnarray}
\bmath{\sigma}_z^{(2k)} \bmath{\sigma}_z^{(2k+1)} |0_L\rangle |0_L\rangle &=&
\bmath{\sigma}_z^{(2k)} \bmath{\sigma}_z^{(2k+1)} |00\rangle |00\rangle =
|00\rangle |00\rangle \nonumber \\ \bmath{\sigma}_z^{(2k)}
\bmath{\sigma}_z^{(2k+1)} |0_L\rangle |1_L\rangle &=& \bmath{\sigma}_z^{(2k)}
\bmath{\sigma}_z^{(2k+1)} |00\rangle |11\rangle = -|00\rangle |11\rangle
\nonumber
\\ \bmath{\sigma}_z^{(2k)} \bmath{\sigma}_z^{(2k+1)} |1_L\rangle |0_L\rangle
&=& \bmath{\sigma}_z^{(2k)} \bmath{\sigma}_z^{(2k+1)} |11\rangle |00\rangle =-
|11\rangle |00\rangle \nonumber \\ \bmath{\sigma}_z^{(2k)}
\bmath{\sigma}_z^{(2k+1)} |1_L\rangle |1_L\rangle &=& \bmath{\sigma}_z^{(2k)}
\bmath{\sigma}_z^{(2k+1)} |11\rangle |11\rangle = |11\rangle |11\rangle.
\end{eqnarray}
Thus we see that this operation acts like an encoded $\bmath{\sigma}_z \otimes
\bmath{\sigma}_z$ between encoded qubits.  When this operation is enacted as a
{\em Hamiltonian}, combined with single encoded qubit operators this allows for
universal control of the encoded qubits.

In this example we have seen how pairing the qubits together we can obtain an
encoding such that there is a mapping from encoded two-qubit states to encoded
qubits and universal quantum computation can be obtained on this encoding.

\section{What to do with the strange irreducible representations}

In constructing encoded universal gate sets from Hamiltonian control sequences,
one can perform an analysis of the Lie algebra structure Hamiltonians to get a
hold of how these Hamiltonians can be used for encoded universality.  Luckily
the analysis of the Lie algebras we will deal with on a quantum computer have
long ago been identified and classified!  We will not deal with this issue here
but instead refer the reader to the standard texts of
Cornwell\cite{Cornwell:97a} and Georgi\cite{Georgi:99a}.

However, we would like to bring up two points related to representation theory
of Lie algebras which are important but have not received extensive discussion
in the quantum computing literature.

\subsection{What to do with Lie Algebra X?}

All of the universality constructions to date have shown how a suitable $SU(k)$
can be executed on a given circuit model.  But there is more under the sun than
the Lie group $SU(k)$!  In particular there are other Lie groups with differing
Lie algebras which, in theory, can arise or be simulated on a quantum system.
An interesting open question is if these Lie algebras have anything to do with
quantum computing.
\begin{open}
Do Lie algebras other than $su(k)$ play any role in the realm of quantum
computing?
\end{open}
An important point in this discussion is that there are only four infinite
families of Lie algebras in the classification scheme ($A_n$, $B_n$, $C_n$, and
$D_n$ in the standard notation.)  These would appear to be the Lie algebras
which are most likely to support some sort of computation because they satisfy
the requirement of allowing for the notion of the power of the computer growing
with the number of subsystems added.

\subsection{What is a qubit?}

The notion of encoded universality also raises some particularly interesting
questions.  One particular issue which has been raised in the literature is the
notion of ``what is a qubit?''  Viola, Knill, and Laflamme \cite{Viola:01a}
present models of qubits encoded into different spaces.  These authors rightly
take an operational view of a qubit.  A qubit is defined by how one can access
the information in the qubit: both in control and in measurement of the qubit.
Unfortunately the authors only present qubits where they have operators which
act on the qubits which satisfy both the commutation and anti-commutation
relations of the standard Pauli matrices:
\begin{eqnarray}
\left[ \bmath{\sigma}_\alpha, \bmath{\sigma}_\beta \right] &=& 2i
\epsilon_{\alpha \beta \gamma} \bmath{\sigma}_\gamma \nonumber \\ \left \{
\bmath{\sigma}_\alpha, \bmath{\sigma}_\beta \right \} &=& 2 \delta_{\alpha
\beta} {\bf I}.
\end{eqnarray}
This, however, is a limited notion of a qubit from the point of view of the
representation theory of quantum information.  To see this, consider three
different irreps of $SU(2)$: i.e. the Lie algebra which satisfies the
commutation relations above, but not necessarily the anti-commutation
relations. The first representation is the one-dimensional irreducible
representation. In this representation, the $SU(2)$ operators all act as $0$
\begin{equation}
\bmath{\sigma}_x^{[1]} \cong \left[0 \right], \quad \bmath{\sigma}_y^{[1]}
\cong \left[ 0 \right], \quad \bmath{\sigma}_z^{[1]} \cong \left[0 \right].
\end{equation}
The second representation is the two-dimensional irreducible representation:
\begin{eqnarray}
 \bmath{\sigma}_x^{[2]} \cong \left[\begin{array}{cc} 0 & 1 \\ 1 & 0
\end{array}\right], \quad \bmath{\sigma}_y^{[2]} \cong\left[\begin{array}{cc} 0 & -i \\ i & 0
\end{array}\right], \quad \bmath{\sigma}_z^{[2]} \cong \left[\begin{array}{cc} 1 & 0 \\ 0 & -1
\end{array}\right].
\end{eqnarray}
Finally we examine the three-dimensional irreducible representation
\begin{eqnarray}
\bmath{\sigma}_x^{[3]} \cong \sqrt{2} \left[ \begin{array}{ccc} 0 & 1 & 0 \\ 1
& 0 & 1
\\ 0 & 1 & 0 \end{array} \right], \quad \bmath{\sigma}_y^{[3]} \cong \sqrt{2} \left[
\begin{array}{ccc} 0 & -i & 0 \\ i & 0 & -i \\ 0 & i & 0 \end{array} \right],
\quad \bmath{\sigma}_z^{[3]} \cong 2 \left[ \begin{array}{ccc} 1 & 0 & 0 \\ 0 &
0 & 0
\\ 0 & 0 & -1 \end{array} \right].
\end{eqnarray}

Clearly the one-dimensional irrep of $SU(2)$ is useless.  This irrep can, in no
manner, be considered a qubit.  And, of course, the two-dimensional irrep of
$SU(2)$ is what we normally think of as a qubit.  But what about the
three-dimensional irrep?  Let us show that any manipulation of the
two-dimensional irrep can be mimicked by the three-dimensional irrep.  Let
$|a\rangle$, $|b\rangle$, and $|c\rangle$ denote the vectors upon which the
three-dimension irrep acts and $|0\rangle$, $|1\rangle$ denote the vectors upon
with the two-dimensional irrep acts.

In the two-dimensional irrep, a state of the system can be written as
\begin{equation}
\bmath{\rho}^{[2]}(\vec{n})={1 \over 2} \left({\bf I} + \vec{n} \cdot
\vec{\bmath{\sigma}}^{[2]} \right).
\end{equation}
Let us map the state of the two-dimensional irrep described by $\vec{n}$ to the
three-dimensional state
\begin{equation}
\bmath{\rho}^{[3]}(\vec{n})={1 \over 3} {\bf I} + {1 \over 6} \vec{n} \cdot
\vec{\bmath{\sigma}}^{[3]}.
\end{equation}
Notice that a pure state in the two-dimensional irrep is {\em not} mapped onto
a pure state in the three-dimensional irrep.

An observable on the two-dimensional irrep is given by
\begin{equation}
{\bf H}^{[2]}(m_0,\vec{m})=m_0 {\bf I} + \vec{m} \cdot \bmath{\sigma}^{[2]}.
\end{equation}
The expectation of this observable is
\begin{equation}
{\rm Tr} \left[ \bmath{\rho}^{[2]}(\vec{n}) {\bf H}(m_0,\vec{m}) \right] =m_0 +
\vec{m} \cdot \vec{n}.
\end{equation}
For the three dimensional irrep we can define the equivalent observable
\begin{equation}
{\bf H}^{[3]}(m_0,\vec{m})=m_0 {\bf I} + {3 \over 4} \vec{m} \cdot
\vec{\bmath{\sigma}},
\end{equation}
such that the expectation value of this observable is identical
\begin{equation}
{\rm Tr} \left[ \bmath{\rho}^{[3]}(\vec{n}) {\bf H}^{[3]}(m_0,\vec{m}) \right]
= m_0+ \vec{m} \cdot \vec{n}.
\end{equation}

Finally note that evolution on the two-dimensional irrep
\begin{equation}
{\bf U}^{[2]}(\vec{v},t) =\exp \left[-it \vec{v} \cdot
\vec{\bmath{\sigma}}^{[2]} \right],
\end{equation}
can be directly mapped onto evolution of the three-dimensional irrep
\begin{equation}
{\bf U}^{[3]}(\vec{v},t)=\exp \left[-it \vec{v} \cdot
\vec{\bmath{\sigma}}^{[3]} \right],
\end{equation}
such that the evolution of the density matrix has the same effect
\begin{eqnarray}
{\bf U}^{[2]}(\vec{v},t) \bmath{\rho}^{[2]}(\vec{n}) {\bf
U}^{[2]\dagger}(\vec{v},t)&=&\bmath{\rho}^{[2]}(\vec{n}^\prime) \nonumber \\
{\bf U}^{[3]}(\vec{v},t) \bmath{\rho}^{[3]}(\vec{n}) {\bf
U}^{[3]\dagger}(\vec{v},t)&=&\bmath{\rho}^{[3]}(\vec{n}^\prime).
\end{eqnarray}

Thus we have seen that there is a mapping between density matrices,
observables, and evolutions of the two and three-dimensional irreps which
perfectly preserves the structure of a qubit.  In general $d>1$ dimensional
irreps of $SU(2)$ can be used in a similar manner to construct a valid qubit. A
qubit is more than just the two-dimensional irreducible representations of
$SU(2)$!

\section{Subsystem growth of a Lie algebra and quantum computation}

As important as the question of when a given Hamiltonian control sequence has
universal control is the negative of this question.  Here we present a useful
criteria for detecting Lie algebras which are not universal.

Suppose one is given a set of Hamiltonians ${\mathcal S}_n$ which can be
implemented in a Hamiltonian control sequence on $n$ subsystems.  Let
${\mathcal L}_n$ denote the Lie algebra which can be generated by ${\mathcal
S}_n$ and let $g(n)$ denote the number of linearly independent operators in
${\mathcal L}_n$. We call $g(n)$ the subsystems growth function.
\begin{theorem} \label{th:growth}
A growth function $g(n)$ which is polynomial in $n$ is not universal on a
quantum circuit model.
\end{theorem}
Proof: The basic idea behind this theorem is to note that a quantum circuit
model on $n$ subsystems has a state space which grows exponentially in $n$ and
therefore performing unitary operators on this space is equivalent to
generating elements of an exponentially growing Lie algebra.

Return now to the example presented in Section \ref{sec:lieexamp} where we
examined the power of Hamiltonian control sequences generated by Hamiltonians
in the set
\begin{equation}
{\mathcal O}^\prime= \{ \bmath{\sigma}_z^{(i)} , \bmath{\sigma}_x^{(i)}
\bmath{\sigma}_x^{(i+1)} \}.
\end{equation}
We will now show that, even with the help of encoding, this set of Hamiltonians
is not universal.

Define the operation ${\bf M}_{jk,\alpha,\beta}= \bmath{\sigma}_\alpha^{(j)}
\prod_{i=j+1}^{k-1} \bmath{\sigma}_z^{(i)} \bmath{\sigma}_\beta^{(k)}$ where
$j<k$ and $\alpha,\beta \in \{x,y\}$.  We claim that the operators in the Lie
algebra generated by ${\mathcal O}^\prime$ are all linear combinations of the
form ${\bf M}_{jk,\alpha,\beta}$ plus the single qubit
$\bmath{\sigma}_z^{(i)}$.  Notice that this is true for $n=2$.  We will prove
the result by induction.  First we note that because our generators are made up
of Pauli operators, we need not worry about linear combinations of operators,
but only need to worry about the operators which can be generated by
commutation. Let ${\mathcal L}^n$ denote the Lie algebra on $n$ qubits
generated by taking commutators in ${\mathcal O}^\prime$.  For example
${\mathcal L}_2=\{{\bf M}_{12,x,x},{\bf M}_{12,x,y},{\bf M}_{12,y,x},{\bf
M}_{12,y,y}, \bmath{\sigma}_z^{(1)}, \bmath{\sigma}_z^{(2)} \}$ as claimed
above.  Assume that ${\mathcal L}_n= \{ {\bf M}_{jk,\alpha,\beta}, 1\leq
j<k\leq n,\alpha,\beta \in \{x,y\} \} \cup \{ \bmath{\sigma}_z^{(i)}, 1\leq i
\leq n \}$.  First notice that taking commutators of elements of ${\mathcal
L}_n$ and $\bmath{\sigma}_z^{(i)}$ only produces elements in ${\mathcal L}_n$:
the only elements which $\bmath{\sigma}_z^{(i)}$ do not commute with are ${\bf
M}_{ik,\alpha,\beta}$ and ${\bf M}_{ki,\alpha,\beta}$ and this commutation only
serves to flip the value of $\alpha$ or $\beta$.  Finally, note that taking the
commutator between elements of ${\mathcal L}_n$ and $\bmath{\sigma}_x^{(i)}
\bmath{\sigma}_x^{(i+1)}$ can only generate elements in ${\mathcal L}_{n+1}$.
To see this, first note that the only nontrivial commutators are those which
occur with the $\bmath{\sigma}_z^{(i)}$ operators, which just produce elements
in ${\mathcal L}_2$.  Further commutators between $\bmath{\sigma}_x^{(i)}
\bmath{\sigma}_x^{(i+1)}$ and ${\bf M}_{jk,\alpha,\beta}$ only create ${\bf
M}_{j^\prime k^\prime,\alpha^\prime,\beta^\prime}$ which are one qubit larger
or smaller.  Thus we have proved that the Lie algebra generated by elements of
${\mathcal O}^\prime$ are spanned by the set of linearly independent operators
in ${\mathcal L}_n= \{ {\bf M}_{jk,\alpha,\beta}, 1\leq j<k\leq n,\alpha,\beta
\in \{x,y\} \} \cup \{ \bmath{\sigma}_z^{(i)}, 1\leq i \leq n \}$.

Let us count the operators in ${\mathcal L}_n$.  There are $n$
$\bmath{\sigma}_z^{(i)}$ operators and $4 {n \choose 2}$ ${\bf
M}_{jk,\alpha,\beta}$ operators.  Thus the growth function for this Lie algebra
is $g(n)=n+4 \left( {n(n-1) \over 2} \right) = 2n^2-n$.  This growth function
is polynomial in $n$ and thus via Theorem \ref{th:growth} this set  of operator
is not universal.

\section{Universal quantum computation}

Universality is one of the most powerful concepts to arise from the theoretical
study of computer science.  In this chapter we have dealt with the ideal
conditions needed for universal quantum computation.  This ideal model presents
an abstract connection to the question of exactly what is a quantum computer.
Real world realization of a universal quantum computer, however, must deal with
decoherence, faulty operations, faulty preparation, and faulty measurements.
Luckily the theory of fault-tolerant quantum computation has been developed
which deals directly with these
issues\cite{Aharonov:97a,Gottesman:98a,Kitaev:97b,Knill:98a,Preskill:98a,Shor:96a}
and a theorem which basically states that if these problems are not too severe,
the ideal model can nearly ideally be
obtained\cite{Aharonov:97a,Gottesman:97a,Kitaev:97a,Knill:98a,Preskill:98a}.

If Alan Turing were to return from the dead and see the modern classical
computer, he would surely be shocked by the technological progress achieved in
the past fifty years.  However, if one explained to Turing how the modern
computer works, he would surely recognize the manner in which the modern
computer attains universal computation (no slouch, that Turing: he could work
in a base-32 notation that others had to convert to decimal to understand!) One
of the main motivations for studying the theory of universal quantum
computation is simply the realization that we do not know exactly what a future
quantum computer will look like, but we have some notion of what is required in
order to obtain universality.  The unknowable future, then, has already given
way to novel proposals for quantum computation, and, in the end, may present
the ultimate road towards building a quantum computer.

\part{Decoherence-Free Quantum Computation}

\chapter{Decoherence-Free Conditions} \label{ch:dfscond}

\begin{quote}
{\em wherein the demons of decoherence are first shown the door\\and the door
is described by necessary and sufficient conditions}
\end{quote}

In this chapter we introduce the basic conditions for decoherence-free
subspaces and decoherence-free subsystems.  We begin with a simple classical
example of a subsystem which withstands a classical error process.  The
fundamental algebraic theorem of decoherence is then derived and the concept of
the OSR algebra is defined.  Decoherence-free subspaces are then introduced and
an iff condition for such subspaces is derived.  A simple example of a
decoherence-free subspace is presented and how to handle system specific
evolution is discussed.  Decoherence-free subsystems are then defined and with
the help of a basic theorem of the representation theory of complex associative
$\dagger$-closed algebras and an iff condition for such subsystems is derived.
An example of a decoherence-free subsystem is examined and the role of the a
nontrivial commutant is introduced.  Finally decoherence-free conditions for
master equations are presented.

\section{Protecting information by encoding}

Two parties, Alice and Bob, wish to communicate an important message.  At their
disposal is a classical communication channel.  Alice and Bob can send two
classical bits at a time down this channel.  Unfortunately this classical
channel has a devilish manner of distorting the information sent down the
channel.  When the parties send their two bits down the channel, there is a
noise process in the channel which will flip the value of {\em both} bits. Thus
if Alice sends $00$ down the line, Bob will either receive the undisturbed $00$
or the flipped $11$.  If Alice sends $01$ through the channel, Bob will either
receive the undisturbed $01$ or the flipped $10$.  Clearly, if Alice and Bob
wish to communicate using the full capacity of the classical channel (meaning
each using both bits of the classical channel) they will fail.

Let $(x_1,x_2)$ denote the classical bits sent down two bit channel.  All of
the information which is in the pair $(x_1,x_2)$ is also in the pair $(x_1
\oplus x_2 , x_2)$ where $\oplus$ is the exclusive-or of the two bits ($x_1
\oplus x_2=x_1+x_2~{\rm mod}~2$).  To see this, simply note that this is a map
which is one-to-one: $00 \rightarrow 00$, $01 \rightarrow 11$, $10 \rightarrow
10$, $11 \rightarrow 01$. The pair $(x_1 \oplus x_2,x_2)$ is a particular {\em
encoding} of the classical information.  What is interesting to Alice and Bob
about this encoding is that the first bit $x_1 \oplus x_2$ is unchanged by the
error process of the channel.  If the channel does not act on the bits, then of
course nothing happens to $x_1 \oplus x_2$.  If the channel flips both of the
bits, then $x_1\oplus x_2 \rightarrow \bar{x}_1 \oplus \bar{x}_2=x_1 \oplus
x_2$ where $\bar{x}$ indicates the negation operation and we have used the fact
that the exclusive-or of two negated bits is the same as the exclusive-or of
the unnegated bits.  On the other hand the second bit in the encoding $x_2$ is
unprotected from the action of the channel.

Alice and Bob can thus use this noisy channel to communicate with {\em perfect}
fidelity by using the encoding $(x_1 \oplus x_2,x_2)$.  If Alice wants to send
an encoded bit down the channel, she encodes her bit into the parity of the two
bits (choosing either of the two possible $(x_1,x_2)$ for a given choice of
parity) and sends these two bits to Bob.  The channel cannot change the parity
of the two bits and thus Bob can decode the bit which Alice encoded by
examining the parity of two bits he receives.

There are two morals from this simple example.  The first moral is that
information can be protected from disturbance via an appropriate encoding.  The
second moral comes from the observation that the reason an appropriate encoding
exists which perfectly protects the information is due to a {\em symmetry} of
the noise process.  In particular the errors which the channel induce on the
two bits are identical on each individual bit.  This is the symmetry of the
error process which allows for encoding into parity which perfectly preserves
the information.

The above example demonstrates how classical information can be perfectly
protected from noise via an appropriate encoding of the information.  In this
part of the thesis we will be examining similar constructions, but now in the
context of quantum information.

\section{The OSR Algebra}

Consider the evolution of a system $S$ and an environment $E$ with joint
Hilbert space ${\mathcal H}={\mathcal H}_S \otimes {\mathcal H}_E$ which
evolves according to some Hamiltonian ${\bf H}$.  Consider the expansion of the
Hamiltonian into system and environment operators
\begin{equation}
{\bf H}=\sum_{\alpha=0}^A {\bf S}_\alpha \otimes {\bf B}_\alpha,
\label{eq:osrexpansion}
\end{equation}
where ${\bf S}_\alpha$ (${\bf B}_\alpha$) acts on ${\mathcal H}_S$(${\mathcal
H}_B$).  The expansion we give above is, of course, always possible.  The
expansion, on the other hand is not unique.  We will place the requirement on
this expansion that the ${\bf B}_\alpha$ are a complete fixed basis which are
linearly independent, hermitian and have an inner product ${\rm Tr}\left[ {\bf
B}_\alpha^\dagger {\bf B}_\beta \right]= \delta_{\alpha \beta}$. Such a basis
can always be chosen over the environment Hilbert space (see
Appendix~\ref{apa:fixedbasis}).  We will often refer to the ${\bf S}_\alpha$ as
the {\em system operators} and the ${\bf B}_\alpha$ as the {\em environment
operators}.

Let us recall that evolution of the system plus environment which initially
starts in a tensor product state $\bmath{\rho}(0)=\bmath{\rho}_S(0) \otimes
\bmath{\rho}_E(0)$ is given by the OSR evolution
\begin{equation}
\bmath{\rho}(t) = \sum_i {\bf A}_i(t) \bmath{\rho}_S(0) {\bf A}_i^\dagger (t).
\end{equation}
We now claim that a basis for the OSR operators ${\bf A}_i(t)$ corresponding to
the Hamiltonian ${\bf H}$ in Eq.~(\ref{eq:osrexpansion}) is given by the {\em
complex associative algebra} $\tt A$ generated by the ${\bf S}_\alpha$ plus the
identity operator ${\bf I}$.
\begin{definition} {\em (Complex associative algebra)} \cite{Knill:00a,Landsman:98a}
The {\em complex associative algebra} ${\tt A}$ generated by the set of
operators ${\bf S}_\alpha$ is the set of operators which can be constructed
from the operators ${\bf S}_\alpha$ via the processes of linear combination
over $\CC$ and simple operator multiplication.
\end{definition}
We claim that
\begin{lemma} \label{lem:osralgebra}  \cite{Knill:00a}
Consider the evolution of a system plus environment due to a Hamiltonian ${\bf
H}$ with expansion Eq.~(\ref{eq:osrexpansion}).  The OSR operators ${\bf
A}_i(t)$ corresponding to evolution due to ${\bf H}$ are elements of the
complex associative algebra $\tt A$ generated by the ${\bf S}_\alpha$ in
Eq.~(\ref{eq:osrexpansion}) plus identity ${\bf I}$.
\end{lemma}
Proof: The Taylor expansion of the full system-environment evolution operator
is given by
\begin{eqnarray}
{\bf U}(t)&=& \sum_{n=0}^\infty{  \left( -it \sum_{\alpha=0}^A {\bf S}_\alpha
\otimes {\bf B}_\alpha \right)^n \over n!} \nonumber \\ &=& \sum_{n=0}^\infty {
(-it)^n \over n!} \left[\sum_{\alpha_1=0}^A \cdots \sum_{\alpha_n=0}^A
\left({\bf S}_{\alpha_1} \cdots {\bf S}_{\alpha_n}\right) \otimes  \left( {\bf
B}_{\alpha_1} \cdots {\bf B}_{\alpha_n} \right) \right].
\end{eqnarray}
A fixed basis form over the environment of the evolution operator can be
obtained by expanding
\begin{eqnarray}
{\bf B}_{\alpha_1} \cdots {\bf B}_{\alpha_n} &=& \sum_{\alpha=0}^A
b_\alpha(\alpha_1,\dots,\alpha_n) {\bf B}_\alpha \nonumber \\
b_\alpha(\alpha_1,\dots,\alpha_n) &=& {\rm Tr} \left[ {\bf B}_\alpha^\dagger
\left( {\bf B}_{\alpha_1} \cdots {\bf B}_{\alpha_n} \right) \right],
\end{eqnarray}
such that
\begin{equation}
{\bf U}(t)= \sum_{n=0}^\infty { (-it)^n \over n!} \left[\sum_{\alpha_1=0}^A
\cdots \sum_{\alpha_n=0}^A \left({\bf S}_{\alpha_1} \cdots {\bf
S}_{\alpha_n}\right) \otimes  \left( \sum_{\alpha=0}^A
b_\alpha(\alpha_1,\dots,\alpha_n) {\bf B}_{\alpha} \right) \right].
\end{equation}
Using the definition of the OSR operators, Eq.~(\ref{eq:osrexplicit}) we find
that for an initial evolution of $\rho_E(0)=\sum_\nu p_\nu |\nu \rangle \langle
\nu|$
\begin{eqnarray}
{\bf A}_{i=(\mu \nu)}&=&\sqrt{p_\nu} \sum_{n=0}^\infty { (-it)^n \over n!}
\left[\sum_{\alpha_1=0}^A \cdot \cdot \sum_{\alpha_n=0}^A \left({\bf
S}_{\alpha_1} \cdot \cdot {\bf S}_{\alpha_n}\right)  \left( \sum_{\alpha=0}^A
b_\alpha(\alpha_1,\dots,\alpha_n) \langle \mu |{\bf B}_{\alpha}| \nu \rangle
\right)\right] \nonumber \\ &=& \sqrt{p_\nu} \sum_{n=0}^\infty
\sum_{\alpha_1=0}^A \cdots \sum_{\alpha_n=0}^A \sum_{\alpha=0}^A { (-it)^n
\over n!} b_\alpha(\alpha_1,\dots,\alpha_n) B_{\alpha,\mu\nu} ({\bf
S}_{\alpha_1} \cdots {\bf S}_{\alpha_n}), \label{eq:aexpand}
\end{eqnarray}
where $B_{\alpha,\mu\nu}=\langle \mu | {\bf B}_\alpha |\nu \rangle$.  Thus we
see that each OSR operator ${\bf A}_i$ is a complex linear combination of
products of the ${\bf S}_\alpha$'s plus identity ${\bf I}$ for $n=0$.  Thus the
${\bf A}_i$ are elements of the complex associative algebra $\tt A$ generated
by the ${\bf S}_\alpha$ plus identity ${\bf I}$ as claimed.

Suppose we are given a complex associative algebra $\tt A$ generated by the
elements ${\bf S}_\alpha$ plus identity ${\bf I}$.  Let ${\bf F}_\beta^{\tt A},
\beta=1 \dots B$ denote a complete basis for the operators in this algebra
which has an identical span as the elements of ${\tt A}$. Then we can expand
the elements of $\tt A$ as
\begin{eqnarray}
{\bf S}_{\alpha_1} \cdots {\bf S}_{\alpha_n} &=& \sum_{\beta=1}^B
s_{\beta}(\alpha_1,\dots,\alpha_n) {\bf F}_\beta^{\tt A} \nonumber \\
s_\beta(\alpha_1,\dots,\alpha_n) &=& {\rm Tr} \left[ {\bf F}_\beta^{\tt A
\dagger} \left( {\bf S}_{\alpha_1}\cdots{\bf S}_{\alpha_n} \right) \right].
\end{eqnarray}
Expanding Eq.~(\ref{eq:aexpand}) in terms of ${\bf F}_\beta^{\tt A}$,
\begin{eqnarray}
{\bf A}_{i=(\mu,\nu)}&=&\sqrt{p_\nu} \sum_{n=0}^\infty \sum_{\alpha_1=0}^A
\cdots \sum_{\alpha_n=0}^A \sum_{\alpha=0}^A  \sum_{\beta=0}^B { (-it)^n \over
n!} b_\alpha(\alpha_1,\dots,\alpha_n) B_{\alpha,\mu\nu}
s_\beta(\alpha_1,\dots,\alpha_n) {\bf F}_\beta^{\tt A} \nonumber \\ &=&
 \sum_{\beta=0}^B a_\beta {\bf F}_\beta^{\tt A},
\end{eqnarray}
where
\begin{equation}
a_\beta = \sqrt{p_\nu} \sum_{n=0}^\infty \sum_{\alpha_1=0}^A \cdots
\sum_{\alpha_n=0}^A \sum_{\alpha=0}^A  { (-it)^n \over n!}
b_\alpha(\alpha_1,\dots,\alpha_n) B_{\alpha,\mu\nu}
s_\beta(\alpha_1,\dots,\alpha_n).
\end{equation}
For general initial environment initial conditions, we can use this expression
to show that there are evolutions such that $a_\beta$ is non-vanishing for some
time $t>0$.  Consider the $k$th derivative of $a_\beta$ with respect to time
evaluated at $t=0$,
\begin{equation}
\left.{\partial^k a_\beta(t) \over \partial t^k} \right|_{t=0} = \sqrt{p_\nu}
\sum_{\alpha_1=0}^A \cdots \sum_{\alpha_n=0}^A \sum_{\alpha=0}^A  {(-i)^k \over
k!} b_\alpha(\alpha_1,\dots,\alpha_k) B_{\alpha,\mu\nu}
s_\beta(\alpha_1,\dots,\alpha_k).
\end{equation}
For general environmental initial conditions, we can choose an initial
environmental condition and basis $|\mu\rangle$ such that $\sqrt{p_\nu}
B_{\alpha,\mu\nu} \neq 0$ for any $\alpha$.  Further, $b_\alpha(\alpha_1,
\dots,\alpha_k) \neq 0$ for at least one $\alpha$ for any
$\alpha_1,\dots,\alpha_n$.  Finally, because the ${\bf F}_\beta^{\tt A}$ have
an identical span to the complex associative algebra generated by the ${\bf
S}_\alpha$ plus identity ${\bf I}$, there must exist a $k$ and
$\alpha_1,\dots,\alpha_k$ such that $s_\beta(\alpha_1,\dots,\alpha_n) \neq0$.
Thus we have shown that $a_\beta \neq 0$ for some $t>0$ for every $\beta$.

Together with Lemma \ref{lem:osralgebra} this implies an extremely important
theorem in the study of decoherence.  Let us first define the OSR algebra
\begin{definition} {\em (OSR algebra)}
The {\em OSR algebra} is the complex associative algebra generated by (i) the
${\bf S}_\alpha$ operators in the expansion of a system-environment Hamiltonian
${\bf H}=\sum_{\alpha=0}^A {\bf S}_\alpha \otimes {\bf B}_\alpha$ where the
${\bf B}_\alpha$ are linearly independent operators and (ii) the identity ${\bf
I}$.
\end{definition}
\begin{theorem} \label{th:fundamental}
{\em (Fundamental algebraic theorem of decoherence) }\cite{Knill:00a} Suppose a
system and environment evolve according to the Hamiltonian ${\bf
H}=\sum_{\alpha=0}^A {\bf S}_\alpha \otimes {\bf B}_\alpha$ where the ${\bf
B}_\alpha$ are linearly independent.  The OSR evolution operators ${\bf
A}_i(t)$ are in the OSR algebra and the span of the ${\bf A}_i(t)$ for generic
environment initial conditions is identical to the OSR algebra.
\end{theorem}
The significance of this theorem is that it reduces the study of system
evolution with generic environment initial conditions to the study of the
algebraic structure of the corresponding OSR algebra.  Thus if one wishes to
understand the effects and OSR operator can have, it is enough to examine the
span of the system operators of a system-environment expansion of the
Hamiltonian.  It provides an iff connection between the OSR operators and the
OSR algebra under generic environmental initial conditions.

\section{Decoherence-free subspaces}

In the example from the first section of this chapter we saw that information
could be protected from an environment via a suitable encoding of the
information.  In this section we present the first discovered and simplest
condition under which similar protection can be endowed to a quantum system.

Consider a system $S$ with Hilbert space ${\mathcal H}_S$ which evolves
according to some joint Hamiltonian ${\bf H}=\sum_{\alpha=0}^A {\bf S}_\alpha
\otimes {\bf B}_\alpha$ with linearly independent environment operators ${\bf
B}_\alpha$.  Corresponding to this evolution and given an environmental initial
condition there are a set of OSR operators ${\bf A}_i(t)$ for the evolution of
the system.  We say that a system density matrix $\bmath{\rho}_S$ is {\em
invariant} under the OSR operators ${\bf A}_i(t)$ if
\begin{equation}
\sum_i {\bf A}_i(t) \bmath{\rho}_S {\bf A}_i^\dagger(t) = \bmath{\rho}_S.
\label{eq:invariant}
\end{equation}
Clearly an invariant density matrix does not evolve even though the system and
environment may have some non-trivial coupling.
\begin{definition} {\em (Decoherence-free subspace)}
A subspace ${\mathcal S}$ of a system's Hilbert space ${\mathcal H}_S$ is
called a {\em decoherence-free subspace} (DFS) with respect to a
system-environment coupling if every pure state from this subspace is invariant
under the corresponding OSR evolution for any possible environment initial
condition:
\begin{equation}
\sum_i {\bf A}_i(t) |j\rangle \langle j| {\bf A}_i^\dagger(t) = |j\rangle
\langle j|, \quad \forall |j\rangle \in {\mathcal S} \quad {\rm and} \quad
\forall \bmath{\rho}_E(0).
\end{equation}
\end{definition}
A decoherence-free subspace is a perfect quantum memory.  The fungible nature
of quantum information tells us that quantum information encoded into a
subspace has the same fundamental value as any other representation of quantum
information.  Thus while the fundamental value of the quantum information is
unchanged by encoding into a DFS, an important property of the way in which
this information interacts with its environment is changed by encoding into a
DFS.

While we have defined a DFS in terms of invariant pure states, mixed states
fall nicely within the protection of a DFS as well.  In particular a mixed
state which has support only over a the pure states of a DFS will be also be
invariant and hence protected from decoherence.

\subsection{Decoherence-free subspace condition}

Let us describe a necessary and sufficient condition for the existence of a DFS
given a specific system-environment coupling as in Eq.~(\ref{eq:osrexpansion}).
\begin{theorem} \label{th:dfsubspace} {\em (Decoherence-free subspace Hamiltonian
criteria)}\cite{Zanardi:97a} A subspace ${\mathcal S}$ is a DFS iff the system
operators ${\bf S}_\alpha$ act proportional to identity on the subspace:
\begin{equation}
{\bf S}_\alpha|j\rangle = c_\alpha |j\rangle \quad \forall |j\rangle \in
{\mathcal S}. \label{eq:dfsubspacecond}
\end{equation}
\end{theorem}
Proof: First we prove sufficiency.  Suppose ${\bf S}_\alpha|j\rangle =
c_\alpha|j\rangle$. Due to the fundamental algebraic theorem of decoherence,
all OSR operators are elements of the complex associative algebra generated by
${\bf S}_\alpha$. Thus
\begin{equation}
{\bf A}_i(t)|j\rangle = c_i(t) |j\rangle,
\end{equation}
where $c_i$ is some complex constant which is a complex combination of the
$c_\alpha$'s.  Therefore
\begin{equation}
\sum_i {\bf A}_i(t) |j\rangle \langle j | {\bf A}_i^\dagger(t) = \sum_i
|c_i(t)|^2 |j\rangle \langle j|.
\end{equation}
The normalization condition $\sum_i {\bf A}_i^\dagger(t) {\bf A}_i(t) = {\bf
I}$ implies $\sum_i |c_i(t)|^2=1$.  Thus if ${\bf
S}_\alpha|j\rangle=c_\alpha|j\rangle$ for all $|j\rangle$ in a subspace
${\mathcal S}$, the ${\mathcal S}$ is a DFS.  Next we prove necessity.  For
every ${\bf A}_i(t)$ acting on a state $|j\rangle$, we can separate the
resulting state into a component which is along $|j\rangle$ and a component
which is perpendicular to $|j\rangle$, $|j_\perp(i)\rangle$ which depends on
${\bf A}_i(t)$:
\begin{equation}
{\bf A}_i(t)|j\rangle=a_i(t) |j\rangle + b_i(t) |j_\perp(i) \rangle.
\end{equation}
The invariant condition on $|j\rangle$, Eq.~(\ref{eq:invariant})
\begin{equation}
\sum_i \left( a_i(t) |j\rangle +b_i(t) |j_\perp(i)\rangle \right) \left(
\langle j | a_i^*(t) + \langle j_\perp(i)| b_i^*(t) \right) = |j\rangle \langle
j|,
\end{equation}
or
\begin{equation}
\sum_i |a_i(t)|^2 =1.
\end{equation}
The OSR normalization condition $\sum_i {\bf A}_i^\dagger(t) {\bf A}_i(t) =
{\bf I}$ implies
\begin{equation}
\sum_i |a_i(t)|^2 + |b_i(t)|^2=1.
\end{equation}
Thus $\sum_i |b_i(t)|^2 =0$ which implies $b_i(t)=0$ for all $i$ and $t$.  Thus
we see that ${\bf A}_i(t)|j\rangle= a_i(t) |j\rangle$.  We can now invoke the
fundamental algebraic theorem of decoherence.  The span of the ${\bf A}_i(t)$
for generic environmental initial conditions is identical to the complex
associative algebra generated by the ${\bf S}_\alpha$.  Thus ${\bf
A}_i(t)|j\rangle = a_i(t)|j\rangle$ implies ${\bf S}_\alpha |j\rangle =
c_\alpha |j\rangle$.

We have seen how the condition Eq.~(\ref{eq:dfsubspacecond}) is an iff
condition for the existence of a DFS.  Stated succinctly, a DFS is the
degenerate common eigenspace of the ${\bf S}_\alpha$ system operators.  Perhaps
the most important aspect of the DFS criteria is to notice how degeneracy is
essential to the definition.  A system-environment coupling which is degenerate
cannot distinguish between the degenerate states.

\subsection{Example decoherence-free subspace}

Let us examine a particularly simple example of decoherence-free subspace.
Suppose two qubits are coupled to a harmonic oscillator environment via the
Hamiltonian
\begin{equation}
{\bf H}= g \left( \bmath{\sigma}_z \otimes {\bf I} + {\bf I} \otimes
\bmath{\sigma}_z \right) \otimes \left( {\bf a} + {\bf a}^\dagger \right),
\end{equation}
where ${\bf a}$ (${\bf a}^\dagger$) is the destruction (creation) operator for
the harmonic oscillator.  The OSR algebra is then the complex associative
algebra generated by the operators ${\bf I}$ and $\left( \bmath{\sigma}_z
\otimes {\bf I} + {\bf I} \otimes \bmath{\sigma}_z \right)$.  The second of
these operators has eigenstates $|00\rangle$, $|01\rangle$, $|10\rangle$,
$|11\rangle$ with eigenvalues $2$, $0$, $0$, and $-2$ respectively.  Thus there
is a subspaces spanned by $|01\rangle$ and $|10\rangle$ which satisfies the DFS
condition Eq.~(\ref{eq:dfsubspacecond}).  We can now directly see how
superpositions of these two basis states do not decohere
\begin{equation}
\exp[-i {\bf H} t] \left( \alpha |01\rangle + \beta |10\rangle \right) \otimes
|\psi_{env}\rangle = \left( \alpha |01\rangle + \beta |10\rangle \right)
\otimes |\psi_{env} \rangle,
\end{equation}
because $\bmath{\sigma}_z \otimes {\bf I} + {\bf I} \otimes \bmath{\sigma}_z$
annihilates each basis state $|01\rangle$ and $|10\rangle$.  In general the
Hamiltonian operator will not annihilate the states, but will act as a constant
on the states.  This implies that a global phase to the subspace will be
applied.  However a global phase does not cause decoherence on the system.

\subsection{System Hamiltonian and the DFS} \label{sec:syshamdfs}

We have defined a decoherence-free subspace as a subspace for which the OSR
evolution produces no evolution of the quantum information stored in the
subspace.  Clearly such a subspace would be useless for quantum {\em
computation} because the information stored in the subspace does not evolve!

A system-environment Hamiltonian can be expanded as ${\bf H}={\bf H}_S \otimes
{\bf I} + {\bf I} \otimes {\bf H}_E + {\bf H}_{SE}$ where the all of the
nontrivial coupling between the system and the environment is included in the
${\bf H}_{SE}$ term.  For time independent Hamiltonians, all of the decoherence
for general environmental initial conditions comes from the ${\bf H}_{SE}$
coupling.  Thus the decoherence-free subspace condition should be applied to an
expansion of the ${\bf H}_{SE}$.  The unitary evolution will then be due to
${\bf H}_S$.  This evolution should {\em preserve} the DFS.  By this we mean
that the evolution due to ${\bf H}_S$ should not take states with support over
the subspace to states with support outside of the subspace.
\begin{definition}
{\em (Subspace preserving Hamiltonian)} A Hamiltonian ${\bf H}$ preserves a
subspace ${\mathcal S}$ if the spectral decomposition of the Hamiltonian ${\bf
H}=\sum_i h_i |i\rangle \langle i|$ can be expanded such that the states
$|i\rangle$ which are entirely within the subspace ${\mathcal S}$ or entirely
outside of the subspace ${\mathcal S}$.
\end{definition}
Notice that because the spectral decomposition is not unique when there is a
degeneracy of the system Hamiltonian, the subspace preserving condition must be
defined in terms of {\em if} the spectral decomposition can be made {\em such
that} the subspace is preserved.

If a system Hamiltonian ${\bf H}_S$ preserves a subspace and that subspace is a
DFS with respect to the system-environment coupling ${\bf H}_{SE}$, then the
evolution of the DFS will be entirely unitary.  We will refer to a DFS which
evolves unitarily by the term DFS unless a distinction is needed and then we
will refer to a unitarily evolving DFS.

Another way in which the presence of a system Hamiltonian can be dealt with is
to work in the interaction picture.  In the interaction picture, the evolution
of a system due to the Hamiltonian ${\bf H}={\bf H}_0+{\bf V}$ is recast into
examining the evolution of $\tilde{\bmath{\rho}}(t) = {\bf U}_0^\dagger(t)
\bmath{\rho}(t) {\bf U}_0(t)$ where ${\bf U}_0(t)=\exp[-i {\bf H}_0 t]$.  The
evolution of $\tilde{\bmath{\rho}}(t)$ is given Schr\"{o}dinger equation
evolution under the interaction Hamiltonian $\tilde{\bf V}(t)={\bf
U}_0^\dagger(t) {\bf V} {\bf U}_0(t)$.  If ${\bf H}_0$ consists only of
separate system and environment evolution (i.e. no system-environment coupling)
then a state $|\psi\rangle$ which is invariant with respect to $\tilde{\bf
V}(t)$ will evolve unitarily.  To see this simply note that if a state is
invariant in the interaction picture, then $\tilde{\bmath{\rho}(t)} =
|\psi\rangle \langle \psi | \Rightarrow \bmath{\rho}(t)= {\bf U}_0^S(t)
|\psi\rangle \langle \psi| {\bf U}_0^S(t)^\dagger$ where ${\bf U}_0^S(t)$
represents the system evolution operator alone.

\subsection{Decoherence-free subspaces and quantum error correction}

The theory of quantum error correction (see, for example,
\cite{Knill:97a,Gottesman:98a,Nielsen:00a}) provides a method of preserving
quantum coherence by actively manipulating the quantum information.  In this
theory, one can show that certain encodings of quantum information can be
arranged such certain error processes can be detected and corrected on this
encoding without destroying the coherence between the encoded quantum
information.  Suppose ${\bf E}_\alpha$ are a set of {\em error operators} which
act on a given system.  These errors are usually taken from an expansion of the
OSR algebra, but are not necessarily a complete basis for the OSR algebra.
These errors usually represent ``the largest'' component of the OSR operators
often on some short time expansion of the OSR operators.

A necessary and sufficient condition for there to exist a subspace which can
detect and correct these errors is given by \cite{Bennett:96a,Knill:97a}
\begin{equation}
\langle i| {\bf E}_\alpha^\dagger {\bf E}_\beta |j\rangle = C_{\alpha \beta}
\delta_{ij}, \label{eq:qecc}
\end{equation}
for the basis states $|i\rangle$ and $|j\rangle$ in the subspace and for all
errors ${\bf E}_\alpha$ and ${\bf E}_\beta$.  The intuition behind this
criteria is that the errors should take the basis states to distinguishable
subspaces so that these errors can be diagnosed and then corrected.

How do DF subspaces fit in with the theory of quantum error correction?  If we
identify the error operators ${\bf E}_\alpha$ with the OSR algebra, then the DF
subspace characterizing Theorem~\ref{th:dfsubspace}, implies that a DF subspace
necessarily satisfies the condition
\begin{equation}
\langle i| {\bf E}_\alpha^\dagger {\bf E}_\beta  |j\rangle = c_\alpha^* c_\beta
\delta_{ij},
\end{equation}
where $|i\rangle$ and $|j\rangle$ are both in the DF subspace.  Since
$c_\alpha^* c_\beta$ is a rank one matrix, it is possible to choose a basis for
the error operators ${\bf E}_\alpha$ such that $\langle i| \tilde{\bf
E}_\alpha^\dagger {\bf E}_\beta |j\rangle= c \delta_{ij}$.  In the theory of
quantum error correcting codes, the rank of the $C_{\alpha \beta}$ matrix in
Eq.~(\ref{eq:qecc}) is known as the degeneracy of the code\cite{Gottesman:98a}.
Thus we are lead to the characterization\cite{Lidar:99b,Duan:99b}
\begin{lemma}
A decoherence free subspace ${\mathcal S}$ from some OSR algebra is a fully
degenerate quantum error correcting code for all elements of the OSR algebra.
\end{lemma}

For more discussion of the relationship between quantum error correction and
decoherence-free subspaces the reader is referred to \cite{Lidar:99b,Duan:99b}.

\section{Decoherence-free subsystems}

In the previous section we have seen how information can be encoded into a
subspace of the system's Hilbert space such that the information does not
decohere.  For constructing a quantum computer, however, this condition is not
the most general condition under which quantum information can be stored in a
decoherence-free manner.  The basic reason for this is that quantum information
is stored most generally in subsystems and not necessarily subspaces.

To define a decoherence-free subsystem, we must first define an operation which
we will call the subsystem trace operator.
\begin{definition}
Suppose a Hilbert space ${\mathcal H}$ has a general subsystem structure
${\mathcal H}= \bigoplus_{j=1}^p \left( \bigotimes_{i=1}^{n_p} {\mathcal
H}_{ij} \right)$.  Let $|k^{(j)}\rangle$ denote a basis for the subspace
defined by $j$ in this expansion.  $|k^{(j)}\rangle$ is in the Hilbert space
$\bigotimes_{i=1}^{n_p} {\mathcal H}_{ij}$ and a basis over this tensor product
structure is given by $|k_1^{(j)}\rangle \otimes |k_2^{(j)}\rangle \otimes
\cdots |k_{n_j}^{(j)}\rangle$.  We define the subsystem trace operator over the
subsystem ${\mathcal H}_{ij}$ as
\begin{equation}
{\rm Tr}_{ij}\left[ {\bf O} \right] = \sum_{k_i^{(j)}} \langle k_i^{(j)} | {\bf
O} | k_i^{(j)} \rangle.
\end{equation}
\end{definition}
We say that information $\bmath{\rho}_I$ has been encoded into a subsystem
${\mathcal H}_{ij}$ when the density matrix of the full Hilbert space
$\bmath{\rho}$ satisfies
\begin{equation}
{\rm Tr}_{1j} [ \cdots {\rm Tr}_{(i-1)j} [ {\rm Tr}_{(i+1)j} [ \cdots {\rm
Tr}_{n_j j} [ \bmath{\rho}] \cdots ] ] \cdots ] =\bmath{\rho}_I.
\end{equation}
Let us define the above operator as the ${ij}$ subsystem extractor,
\begin{equation}
\Upsilon_{ij}\left[ {\bf O} \right] ={\rm Tr}_{1j} [ \cdots {\rm Tr}_{(i-1)j} [
{\rm Tr}_{(i+1)j} [ \cdots {\rm Tr}_{n_j j} [ {\bf O} ] \cdots ] ] \cdots ].
\end{equation}

This allows us to define what it means to be decoherence-free when information
is encoded into a subsystem.
\begin{definition}{\em (Decoherence-free subsystem (DFS))}
Given a system Hilbert space with a general subsystem decomposition ${\mathcal
H}= \bigoplus_{j=1}^p \left( \bigotimes_{i=1}^{n_p} {\mathcal H}_{ij} \right)$.
A subsystem ${\mathcal H}_{ij}$ is said to be a {\em decoherence-free
subsystem} (DFS) with respect to a system-environment coupling if every pure
state encoded into this subsystem is invariant with respect to this subsystem
under the corresponding OSR evolution for any possible environment initial
condition.  If $\bmath{\rho}_\phi$ denotes the situation where the pure state
$|\phi\rangle$ has been encoded in the ${\mathcal H}_{ij}$ subsystem, then this
condition is equivalent to
\begin{equation}
\Upsilon_{ij}\left[  \sum_k {\bf A}_{k}(t) \bmath{\rho}_\phi {\bf
A}_k^\dagger(t)  \right]= |\phi\rangle \langle \phi|.
\end{equation}
\end{definition}
We use the abbreviation DFS for both decoherence-free subspaces and
decoherence-free subsystems.  We can see from the above definition of a
decoherence-free subsystem, decoherence-free subspaces are examples of
decoherence-free subsystems.  In particular decoherence-free subspaces occur
when the matrix algebra ${\tt M}_{d_J}$ is one dimensional $d_J=1$ and hence
all of the operators act as a constant on a subspace.  Unless we need to
distinguish between the subsystem and subspace definitions, we will refer to
both as DFSs.

What is the difference between storing information in a subspace and storing
information in a subsystem?  This question often leads to confusion, so let us
address this by examining an illuminating example.  Consider encoding a single
qubit of information into a two qubit system.  One particular way to encode a
qubit into the four dimensional Hilbert space of two qubits is to encode the
information into a subspace spanned by two orthogonal states.  Thus for
instance we can encode the information of a qubit $\alpha |0\rangle + \beta |1
\rangle$ as $\alpha|01\rangle + \beta |10\rangle$:
\begin{equation}
\alpha|0\rangle + \beta |1\rangle \rightarrow \alpha |01\rangle + \beta
|10\rangle.
\end{equation}
We then say that the information has been encoded into a subspace of the two
qubit Hilbert space. Suppose we simply encode the information into the first
qubit of the two qubits.  It doesn't matter, then, what the state of the second
qubit is
\begin{equation}
\alpha|0\rangle +\beta |1\rangle \rightarrow \left(\alpha |0\rangle + \beta
|1\rangle \right) \otimes |\phi\rangle.
\end{equation}
Notice that this map is a one-to-many mapping from the quantum information in
one qubit to a two qubit Hilbert space.  For a particular mapping to a
$|\phi\rangle$, the mapping is the same as the mapping from a qubit to a
subspace of the two qubit Hilbert space.

\subsection{Representation theory for the OSR algebra} \label{sec:reposr}

\begin{quote}
{\em ``The universe is an enormous direct product of representations of
symmetry groups.''}
\\
\begin{flushright} --Steve Weinberg (as quoted in \cite{Gallian:92a})
\end{flushright}
\end{quote}

We now present a theorem which exactly delineates where quantum information can
be stored decoherence-free in a quantum system.  First we note that the OSR
algebra is a $\dagger$-closed algebra.  A $\dagger$-closed algebra is an
algebra that satisfies the requirement that if ${\bf S}$ is the algebra, then
${\bf S}^\dagger$ is also in the algebra.  For the OSR algebra, this follows
from the hermiticity of the system-environment Hamiltonian.

The theorem we want is a basic theorem from representation theory of complex
associative algebras which are $\dagger$-closed (see, for example,
\cite{Landsman:98a})
\begin{theorem}{\em (Basic representation theorem of $\dagger$-closed complex associative algebras)}
Let $\tt A$ be a complex associate algebra which is $\dagger$-closed acting on
a Hilbert space ${\mathcal H}$ and which contains the identity operator. In
general $\tt A$ will be a reducible subalgebra of the full algebra over
${\mathcal H}$.  In particular the algebra $\tt A$ is isomorphic to a direct
sum of full matrix algebras
\begin{equation}
{\tt A} \cong \bigoplus_{J\in {\mathcal J}} {\tt I}_{n_J} \otimes {\tt
M}_{d_J}. \label{eq:algebrareduc}
\end{equation}
Here ${\tt I}_d$ is the $d$ dimensional identity algebra (which just consists
of the $d$ dimensional identity operator) and ${\tt M}_d$ is the $d$
dimensional complex associative algebra corresponding to all general linear
operators on the $d$ dimensional space.
\end{theorem}
${\mathcal J}$ is a set describing the different irreducible representations
and $n_J$ is referred to as the degeneracy of the $J$th irreducible
representation (irrep). This theorem implies that there is a basis such that
the operation of every operator ${\bf S}$ in a $\dagger$-closed complex
associative algebra acts on the Hilbert space as
\begin{equation}
{\bf S}= \bigoplus_{J \in {\mathcal J}} {\bf I}_{n_J} \otimes {\bf S}_{d_J},
\label{eq:subsystemdecomp}
\end{equation}
where ${\bf S}_{d_J}$ is a $d_J$ dimensional operator and ${\bf I}_{n_J}$ is
the $n_J$ dimensional identity operator.  Because ${\tt M}_d$ is the $d$
dimensional complex associative algebra corresponding to all general linear
operators on a $d$ dimensional space, the ${\bf S}_{d_J}$ span the entire space
of $d_J$ dimensional operators.

Corresponding to the decomposition in Eq.~(\ref{eq:subsystemdecomp}) we can
construct a basis which will simplify our notation considerably.  Let
$|J,\lambda,m\rangle$ denote the basis where $J$ labels the subspace of the
irrep, $\lambda$ labels the degenerate component of the decomposition and $m$
labels the component of the composition which is acted upon non-trivially by
the irrep.  This basis implies that the decomposition
Eq.~(\ref{eq:subsystemdecomp}) can be written as
\begin{equation}
{\bf S}= \sum_{J \in {\mathcal J}} \sum_{\lambda,\lambda^\prime=1}^{n_J}
\sum_{m,m^\prime=1}^{d_J} S_{m,m^\prime} |J,\lambda,m \rangle \langle
J,\lambda^\prime,m^\prime|. \label{eq:reposr}
\end{equation}
Another useful operator to define is the operator which performs the subsystem
trace over the matrix algebra component of a given irrep.  Define
\begin{equation}
\Upsilon_{J} \left[ {\bf O} \right] = \sum_m \langle J, m | {\bf O} | J,
m\rangle,
\end{equation}
where we implicitly use the subsystem structure for a given $J$,
$|J,\lambda,m\rangle= |J,m\rangle \otimes |m\rangle$ in this sum.

The basic representation theorem of complex associative $\dagger$-closed
algebras describes a subsystem structure which we previously identified as a
subspace tensor product structure.  Given a $J \in {\mathcal J}$ for an algebra
$\tt A$ there is a subspace over which the operators act.  One manner in which
quantum information can be encoded with respect to the algebra $\tt A$ is to
encode the information into the subspace for a given $J$.  Information encoded
in this manner is acted upon non-trivially by the operators in the algebra $\tt
A$.  Over the subspace for a given $J$, there is a two-fold tensor product
structure.  Information which is encoded into the subspace corresponding to a
particular $J$ can then be encoded such that it respects this tensor product
structure.  Thus information in the subspace can be encoded into the degenerate
degrees of freedom corresponding to the ${\tt I}_{n_J}$ algebra or the
information can be encoded into the degrees of freedom corresponding to the
${\tt M}_{d_J}$ algebra.  We say that information has been encoded into the
degeneracy of the $J$th irrep if that information has support only over the
degrees of freedom of the ${\tt I}_{n_J}$ algebra. Notice that because we are
encoding into a subsystem, for a given $J$ information which is encoded into
the degeneracy will be accompanied by information encoded into the degrees of
freedom of the ${\tt M}_{d_J}$ algebra.

\subsection{Decoherence-free subsystem condition}

The basic representation theorem of complex associative algebras combined with
the fundamental algebraic theorem of decoherence together form an excellent iff
description of decoherence on a quantum system.  Given an OSR algebra generated
by the system operators ${\bf S}_\alpha$, we can decompose this algebra as in
Eq.~(\ref{eq:subsystemdecomp}).  This in turn implies that we can represent the
OSR operators as
\begin{equation}
{\bf A}_i(t) = \bigoplus_{J \in {\mathcal J}} {\bf I}_{n_J} \otimes ({\bf
A}_{d_J})_i(t), \label{eq:osrdecomp}
\end{equation}
where $({\bf A}_{d_J})_i(t)$ are $d_J$ dimensional OSR operators.  The span of
the $({\bf A}_{d_J})_i(t)$ are the entire space of all $d_J$ linear operators
and the $({\bf A}_{d_J})_i(t)$ are themselves valid OSR operators which satisfy
the completeness relation $\sum_i ({\bf A}_{d_J})_i^\dagger(t) ({\bf
A}_{d_J})_i(t)= {\bf I}_{d_J}$.

\begin{theorem} {\em (Decoherence-free subsystem Hamiltonian criteria)}
\cite{Knill:00a} A subsystem is a decoherence-free subsystem iff this subsystem
is encoded in the degeneracy of single irrep $\bar{J}$ from the OSR algebra
${\tt A} \cong \bigoplus_{J \in {\mathcal J}} {\tt I}_{n_J} \otimes {\tt
M}_{d_J}$.
\end{theorem}
Proof: First, sufficiency.  Suppose that the pure state $|j\rangle$ is encoded
into the degeneracy of a single irrep $\bar{J}$ of the OSR algebra ${\tt
A}\cong \bigoplus_{J \in {\mathcal J}} {\tt I}_{n_J} \otimes {\tt M}_{d_J}$.
This means that
\begin{equation}\bmath{\rho}=\bigoplus_{J \in {\mathcal J}} \left\{
\begin{array}{cc}
{\bf 0}_{n_J} \otimes {\bf 0}_{d_J} & {\rm if} J \neq \bar{J} \\ |j\rangle
\langle j| \otimes \bmath{\rho}_{d_{\bar{J}}} & {\rm if } J=\bar{J}
\end{array} \right.
\end{equation}
where $\rho_{d_{\bar{J}}}$ is any arbitrary $d_{\bar{J}}$ dimensional density
matrix and ${\bf 0}_d$ is the $d$ dimensional zero matrix.  We will represent
this state as
\begin{equation}
\bmath{\rho} = |j\rangle \langle j| \otimes \bmath{\rho}_{d_{\bar{J}}},
\end{equation}
where the support of the density matrix is taken to be only over the $\bar{J}$
irrep.  The OSR operators then act on the this states as
\begin{eqnarray}
\sum_i {\bf A}_i(t) \bmath{\rho} {\bf A}_i^\dagger(t) &=& \sum_i {\bf
I}_{n_{\bar{J}}} \otimes \left({\bf A}_{d_{\bar{J}}}\right)_i(t)
\left(|j\rangle \langle j | \otimes \bmath{\rho}_{d_{\bar{J}}} \right) {\bf
I}_{n_{\bar{J}}} \otimes \left({\bf A}_{d_{\bar{J}}}\right)_i^\dagger(t)
\nonumber \\ &=& |j\rangle \langle j| \otimes \sum_i \left( {\bf
A}_{d_{\bar{J}}}  \right)_i(t) \bmath{\rho}_{d_{\bar{J}}} \left( {\bf
A}_{d_{\bar{J}}}  \right)^\dagger_i(t).
\end{eqnarray}
Thus we see that the pure state $|j\rangle$ encoded into the degeneracy of the
algebra is invariant with respect to the subsystem.  Next we prove necessity.
Notice that the decoherence-free subsystem can be thought of as a
decoherence-free subspace after a certain subsystem reduction has been
performed.  We can thus use the necessary component of the decoherence-free
subspace condition after we trace over the appropriate subsystem.  Suppose the
information was encoded into a subsystem which does not correspond to the
degeneracy of a given irrep $J$ in Eq.~(\ref{eq:osrdecomp}). This information
will not be acted upon proportional to identity because (i) a component of the
information is encoded into the ${\tt M}_{d_J}$ algebras, (ii) a component of
the information is encoded into different irreps labeled by $J$, or (iii) both
(i) and (ii).  In case (i), the information will be acted upon nontrivially
over the subsystem because the ${\tt M}_{d_J}$ are the full matrix algebra over
such a space.  In case (ii), the information will be acted upon by differing
algebras. This allows for differing global factors between the OSR operators.
Finally, in case (iii) the information is infected by both of these problems.
The decoherence-free subspace iff condition then implies that the information
must be encoded into the degeneracy of the a single irrep of the OSR algebra.

\subsection{The commutant of the OSR algebra and DFSs} \label{sec:osrcommutant}

Given an OSR algebra $\tt A$, how does one know whether there is a degeneracy
over which information can be encoded in a degenerate manner?  The easiest way
to examine this question is to examine the {\em commutant} of the OSR algebra.
The commutant of an algebra $\tt A$ is denoted by ${\tt A}^\prime$ and is the
set of all operators which commute with {\em all} of the elements of $\tt A$.
The commutant of the commutant of an algebra is itself the algebra $({\tt
A}^\prime)^\prime={\tt A}$.  If the algebra $\tt A$ is reducible to the form
\begin{equation}
{\tt A} \cong \bigoplus_{J\in {\mathcal J}} {\tt I}_{n_J} \otimes {\tt
M}_{d_J},
\end{equation}
as in Eq.~(\ref{eq:algebrareduc}), then the commutant of the algebra is
reducible to the form
\begin{equation}
{\tt A}^\prime \cong \bigoplus_{J \in {\mathcal J}} {\tt M}_{n_J} \otimes {\tt
I}_{d_J}.
\end{equation}
Thus the existence of a non-trivial commutant of the OSR algebra implies the
existence of a DFS for the OSR algebra.

\subsection{Example decoherence-free subsystem} \label{sec:dfsubsystemex}

Here we consider a simple example of a decoherence-free subsystem. This example
is not motivated physically (we will return to physically motivated examples
later) but serves as a good illustration of a decoherence-free subsystem.
Consider a three qubit system coupled to a bath via the Hamiltonian
\begin{equation}
{\bf H}=\bmath{\sigma}_x \otimes \bmath{\sigma}_x \otimes \bmath{\sigma}_x
\otimes {\bf B}_x + \bmath{\sigma}_z \otimes \bmath{\sigma}_z \otimes
\bmath{\sigma}_z \otimes {\bf B}_z, \label{eq:dfsexham}
\end{equation}
where ${\bf B}_\alpha$ are linearly independent bath operators.  The OSR
algebra ${\tt A}$ for this Hamiltonian is generated by the set of operators $\{
{\bf I}, \bmath{\sigma}_x \otimes \bmath{\sigma}_x \otimes \bmath{\sigma}_x,
\bmath{\sigma}_z \otimes \bmath{\sigma}_z \otimes \bmath{\sigma}_z \}$. Closing
this algebra $\tt A$ we see that the OSR algebra $\tt A$ is spanned by the
operators $\{{\bf I}, \bmath{\sigma}_x \otimes \bmath{\sigma}_x \otimes
\bmath{\sigma}_x,\bmath{\sigma}_y \otimes \bmath{\sigma}_y \otimes
\bmath{\sigma}_y,\bmath{\sigma}_z \otimes \bmath{\sigma}_z \otimes
\bmath{\sigma}_z \}$.

A complete set of commuting observables for the three qubit system is given by
the operators ${\bf Z}_{12}=\bmath{\sigma}_z \otimes \bmath{\sigma}_z \otimes
{\bf I}$, ${\bf Z}_{23}={\bf I} \otimes \bmath{\sigma}_z \otimes
\bmath{\sigma}_z$, and ${\bf Z}_{123}=\bmath{\sigma}_z \otimes \bmath{\sigma}_z
\otimes \bmath{\sigma}_z$.  Define the basis labeled by the eigenvalues of
these operators via $|z_{12},z_{23},z_{123}\rangle$.  Expressing these basis
states in terms of the standard computational basis we find that
\begin{eqnarray}
&&|+1,+1,+1\rangle = |000 \rangle, \quad |+1,+1,-1\rangle=|111\rangle, \quad
|+1,-1,+1\rangle = |110 \rangle, \nonumber \\ &&|+1,-1,-1\rangle = |001\rangle,
\quad |-1,+1,+1\rangle=|011\rangle, \quad |-1,+1,-1 \rangle = |100\rangle
\nonumber \\
 &&|-1,-1,+1\rangle=|101\rangle, \quad |-1,-1,-1\rangle = |010\rangle.
\end{eqnarray}
Next notice how the operators in the OSR algebra $\tt A$ only affect the
$z_{123}$ index:
\begin{eqnarray}
\bmath{\sigma}_z \otimes \bmath{\sigma}_z \otimes \bmath{\sigma}_z
|z_{12},z_{23},z_{123} \rangle &=& z_{123}|z_{12},z_{23},z_{123} \rangle
\nonumber \\ \bmath{\sigma}_x \otimes \bmath{\sigma}_x \otimes \bmath{\sigma}_x
|z_{12},z_{23},z_{123} \rangle &=& |z_{12},z_{23},-z_{123} \rangle \nonumber
\\
\bmath{\sigma}_y \otimes \bmath{\sigma}_y \otimes \bmath{\sigma}_y
|z_{12},z_{23},z_{123} \rangle &=& -iz_{123}|z_{12},z_{23},-z_{123} \rangle,
\end{eqnarray}
i.e. we could have written this such that only the $z_{123}$ subsystem is
affected
\begin{equation}
\bmath{\sigma}_\alpha \otimes \bmath{\sigma}_\alpha \otimes
\bmath{\sigma}_\alpha |z_{12},z_{23},z_{123} \rangle = |z_{12},z_{23}\rangle
\otimes {\bf O}_\alpha |z_{123}\rangle.
\end{equation}
Thus two qubits of information can be stored in $z_{12}$ and $z_{23}$ which
will not decohere under the coupling Hamiltonian Eq.~(\ref{eq:dfsexham}).
Notice how the OSR algebra can and does affect the $z_{123}$ quantum number,
but this coupling does not destroy the information in the quantum numbers
$z_{12}$ and $z_{23}$.  This is the essential component of a decoherence-free
subsystem which differs from a decoherence-free subspace.  In the subspace
case, the information in an subspace does not evolve while in the subsystem
case, degrees of freedom other than those of the subsystem evolve.  In terms of
the OSR algebra $\tt A$ we see that the algebra is reducible to the form
\begin{equation}
{\tt A} \cong {\tt I}_4 \otimes {\tt M}_2.
\end{equation}

We also could have seen that information can be encoded into a DFS by examining
the commutant of the OSR algebra.  The commutant of the OSR algebra is
generated by the operators $\{ \bmath{\sigma}_z \otimes \bmath{\sigma}_z
\otimes {\bf I}, {\bf I} \otimes \bmath{\sigma}_z \otimes \bmath{\sigma}_z,
\bmath{\sigma}_x \otimes \bmath{\sigma}_x \otimes {\bf I}, {\bf I} \otimes
\bmath{\sigma}_x \otimes \bmath{\sigma}_x \}$.  The algebra generated this set
of operators is a two-fold degenerate $4$ dimensional matrix algebra:
\begin{equation}
{\tt A}^\prime \cong {\tt M}_4 \otimes {\tt I}_2.
\end{equation}

\section{Master equation decoherence-free conditions}

We recall that the diagonal form of the semigroup master equation (SME) is
given by
\begin{equation}
{\partial \bmath{\rho}(t) \over \partial t} = -i[{\bf H},\bmath{\rho}(t)] +{1
\over 2} \sum_{\alpha \neq 0} \left( [{\bf L}_\alpha \bmath{\rho}(t), {\bf
L}_\alpha^\dagger] + [{\bf L}_\alpha, \bmath{\rho}(t) {\bf L}_\alpha^\dagger ]
\right),
\end{equation}
where the ${\bf L}_\alpha$ are the Lindblad operators.  Here we present
conditions for decoherence-free evolution for the SME.  We will ignore the
evolution due to the system Hamiltonian (see Section~\ref{sec:syshamdfs}),
${\bf H}=0$.

\begin{theorem} {\em (Decoherence-free subspace master
equation criteria)}\cite{Zanardi:98a,Lidar:98a} A subspace ${\mathcal S}$ of
the system Hilbert space ${\mathcal H}_S$ is a decoherence-free subspace when
evolving due to a semigroup master equation  iff the Lindblad operators all
satisfy ${\bf L}_\alpha |j\rangle =c_\alpha |j\rangle$ for every $|j\rangle \in
{\mathcal S}$.
\end{theorem}
Proof: Sufficiency follows via the fact that if ${\bf L}_\alpha|j\rangle=
|j\rangle$, then $ \left[ {\bf L}_\alpha, |j\rangle \langle j| \right]=0$. Thus
if the initial state is $|j\rangle \langle j|$, then the decoherence term
vanishes:
\begin{equation}
{1 \over 2} \sum_{\alpha \neq 0} \left( [{\bf L}_\alpha |j\rangle \langle j|,
{\bf L}_\alpha^\dagger] + [{\bf L}_\alpha, |j\rangle \langle j| {\bf
L}_\alpha^\dagger ] \right) =0.
\end{equation}
Thus ${\partial \bmath{\rho}(t) \over \partial t}=0$ and the state is a DFS. To
show the necessity of this condition, note that ${\partial \bmath{\rho}(t)
\over \partial t}=0$ implies that
\begin{equation}
\sum_{\alpha \neq 0} \left( \left| \langle j | {\bf L}_\alpha |j\rangle \right
|^2 - \langle j | {\bf L}_\alpha^\dagger {\bf L}_\alpha |j\rangle \right)=0.
\end{equation}
If each ${\bf L}_\alpha$ acts on the states $|j\rangle$ as ${\bf L}_\alpha =
a_\alpha |j\rangle + b_\alpha |j_\perp(\alpha) \rangle$, then this implies
\begin{equation}
\sum_{\alpha \neq 0} |b_\alpha|^2 =0,
\end{equation}
and hence each $b_\alpha=0$.  Thus the condition is also necessary.

For subsystems, the situation, unfortunately is not quite as easy.  First let
us define the SME algebra
\begin{definition}{\em (SME algebra)}
Suppose one is given a semigroup master equation with diagonal Lindblad
operators ${\bf L}_\alpha$.  The SME algebra is the complex associative algebra
generated by the Lindblad operators ${\bf L}_\alpha$, their adjoints ${\bf
L}_\alpha^\dagger$ and identity ${\bf I}$.
\end{definition}
Having defined the SME algebra, we can present a sufficient condition for
decoherence-free evolution under the SME.
\begin{theorem}{\em (Decoherence-free subsystem semigroup master equation
criteria)} A sufficient criteria for the existence of a decoherence-free
subsystem under the evolution of a semigroup master equation is that the
subsystem is acted upon as identity by the corresponding SME algebra.
\end{theorem}
Proof: The Lindblad operators and their adjoints along with the identity act as
the reducible complex associative algebra such that we can express these
operators as
\begin{equation}
{\bf L}_\alpha =\sum_{J \in {\mathcal J}} \sum_{\lambda,\lambda^\prime=1}^{n_J}
\sum_{m,m^\prime=1}^{d_J} L_{m,m^\prime} |J,\lambda,m \rangle \langle
J,\lambda^\prime,m^\prime|.
\end{equation}
If we encode into the degeneracy of irrep $J$, then the initial density matrix
of the system will be
\begin{equation}
\bmath{\rho}(0) = |j\rangle \langle j| \otimes \bmath{\rho}_{d_{J}},
\end{equation}
such that $\Upsilon_J \left[\bmath{\rho} \right] = |j\rangle \langle j|$.  It
is then easy to check that
\begin{equation}
{\partial \Upsilon_J \left[\bmath{\rho}(t) \right] \over \partial t} =
\Upsilon_J \left[ {\partial\bmath{\rho}(t)  \over \partial t}
\right]=\Upsilon_J  \left[ {1 \over 2} \sum_{\alpha \neq 0} \left( [{\bf
L}_\alpha \bmath{\rho}(t), {\bf L}_\alpha^\dagger] + [{\bf L}_\alpha,
\bmath{\rho}(t) {\bf L}_\alpha^\dagger ]\right) \right]=0.
\end{equation}

Let us also give an example of why the above condition cannot be also a
necessary condition\cite{Kempe:01a}.  Suppose we are given a single qubit which
is subjected to a semigroup master equation with only one non-zero ${\bf
L}_1=\gamma \bmath{\sigma}_- = \gamma \left(\bmath{\sigma}_x - i
\bmath{\sigma}_y \right)$. This situation corresponds to a single two-level
system subject to spontaneous decay.  Clearly there is a stationary state of
the system $|0\rangle$ which does not evolve.  However, if one examine the
algebra generated by ${\bf L}_1$, ${\bf L}_1^\dagger$ and ${\bf I}$, one can
easily see that any linear operator over the two-qubits can be found in this
algebra.  Then according to the above criteria, there would be no states which
do not evolve.  But this is a contradiction to our earlier observation. Thus
the condition is sufficient but not necessarily necessary.  Only in the
subspace regime does the above condition become necessary and sufficient.

Why did the condition we used in the OSR fail for the SME?  The main reason for
this is that the SME represents evolution which is a semigroup (a group without
the requirement that every element have an inverse).  What is needed in order
to obtain a necessary and sufficient condition is representation theory for
semigroups.  We will not delve into this subject but in the finite dimensional
case there should be no difficulty applying representation theory of semigroups
to the decoherence-free problem.

\section{Inducing decoherence-free conditions}

A final topic which we would like to address is the issue of inducing or
symmetrizing the evolution of a system such that the system-environment
coupling exhibits a certain symmetry which supports a decoherence-free
condition.  Viola and Lloyd\cite{Viola:98a,Viola:98b} were the first to suggest
that it might be possible to use ultra-fast system evolution to reduce
decoherence in the context of quantum computation (see also
\cite{Duan:99a,Duan:99d,Vitali:99a}). Viola, Lloyd, and
Knill\cite{Viola:99a,Viola:99b} and independently Zanardi
\cite{Zanardi:99b,Zanardi:99c,Zanardi:00a} then developed the mathematical
theory behind these symmetrization schemes and demonstrated how universal
quantum computation could also be performed on these systems.  With the seminal
paper of Knill, Laflamme, and Viola\cite{Knill:00a} which introduced the notion
of decoherence-free subsystems, it was quickly realized by Viola, Knill, and
Lloyd\cite{Viola:00a} and by Zanardi\cite{Zanardi:01a} that there is an
intimate relationship between the ideas of symmetrized evolutions and
decoherence-free subsystems.  It is this relationship which we will now briefly
described.

Suppose that one has the ability to perform ultra-fast gates and their inverses
on a system from some the unitary representation of some finite group
${\mathcal G}$.  Let ${\bf G}_g$ be represent of the element $g$ of this group.
If one applies the operation ${\bf G}_g$, then allows the system to evolve
according to some Hamiltonian ${\bf H}$, and finally applies the operation
${\bf G}_g^\dagger$, the resulting operation is
\begin{equation}
{\bf G}_g^\dagger e^{-i{\bf H} t} {\bf G}_g = e^{-it {\bf G}_g^\dagger{\bf H}
{\bf G}_g }, \label{eq:Heff}
\end{equation}
so that the system effectively evolves according to the Hamiltonian ${\bf
H}_{eff}={\bf G}_g^\dagger {\bf H} {\bf G}_g$.  Suppose that all elements of
the groups are applied in this fashion to an evolution due to some Hamiltonian
${\bf H}$.  The evolution is then approximately
\begin{equation}
{\bf G}_{g_1}^\dagger e^{-i{\bf H}\Delta t} {\bf G}_{g_1} {\bf G}_{g_2}^\dagger
e^{-i{\bf H}\Delta t} {\bf G}_{g_2} \cdots {\bf G}_{g_p}^\dagger e^{-i {\bf H}
\Delta t} {\bf G}_{g_p} \approx e^{-i{\bf H}_{eff} |{\mathcal G}| \Delta t},
\end{equation}
where $|{\mathcal G}|$ is the order of the group ${\mathcal G}$ and
\begin{equation}
{\bf H}_{eff}= {1 \over |{\mathcal G}|} \sum_{g \in {\mathcal G}} {\bf
G}_{g}^\dagger {\bf H} {\bf G}_g.
\end{equation}
Define the ${\mathcal G}$-symmetrizing operator
\begin{equation}
\Pi_{\mathcal G}[ {\bf X} ] = {1 \over |{\mathcal G}|} \sum_{g \in {\mathcal
G}} {\bf G}_{g}^\dagger {\bf X} {\bf G}_g.
\end{equation}
Notice that $Pi_{\mathcal G}[{\bf X}]$ commutes with all of the element of
${\mathcal G}$
\begin{equation}
\Pi_{\mathcal G}[ {\bf X} ] {\bf G}_h ={1 \over |{\mathcal G}|} \sum_{g \in
{\mathcal G}} {\bf G}_{g}^\dagger {\bf X} {\bf G}_{gh}={1 \over |{\mathcal G}|}
\sum_{g \in {\mathcal G}} {\bf G}_{gh^{-1}}^\dagger {\bf X} {\bf G}_{g}={\bf
G}_h \Pi_{\mathcal G}[ {\bf X}].
\end{equation}
Define $\CC {\mathcal G}$ as the complex associative algebra generated by
elements of the group ${\mathcal G}$.  We call this algebra the group algebra.
This algebra is reducible, $\CC {\mathcal G} \cong \bigoplus_{J \in {\mathcal
J}} {\tt I}_{n_J} \otimes {\tt M}_{d_J}$.  Operators acted on by $\Pi_G$,
$\Pi_{\mathcal G} [{\bf X}]$ are all in the commutant $\CC {\mathcal G}^\prime$
of the group algebra.  This algebra is reducible to the form $\CC {\mathcal G}
\cong \bigoplus_{J \in {\mathcal J}} {\tt M}_{n_J} \otimes {\tt I}_{d_J}$.

Suppose the ${\mathcal G}$-symmetrizing procedure is applied to the system
component of a system-environment coupling ${\bf H}_{SB}=\sum_{\alpha} {\bf
S}_\alpha \otimes {\bf B}_\alpha$.  Then if this procedure is applied fast
enough \cite{Viola:98a}, the evolution of the system and environment will be
governed by the effective Hamiltonian
\begin{equation}
\Pi_G \left[ {\bf H}_{SB} \right] = \sum_\alpha \Pi_G \left[ {\bf S}_\alpha
\right] \otimes {\bf B}_\alpha.
\end{equation}
One can now apply the decoherence-free conditions to the symmetrized system
operators
\begin{equation}
{\bf S}_{\alpha,eff} = {1 \over |{\mathcal G}|} \sum_{g \in {\mathcal G}} {\bf
G}_{g}^\dagger {\bf S}_\alpha {\bf G}_g.
\end{equation}
As we described above, the ${\bf S}_{\alpha,eff}$ can be reduced because they
are in the commutant of the group algebra.  More generally, if the ${\bf
S}_\alpha$ for a complete basis for the full matrix algebra of the system's
Hilbert space, then the ${\bf S}_{\alpha,eff}$ will exactly realize the entire
commutant $\CC {\mathcal G}^\prime$ of the group algebra $\CC {\mathcal G}$.
\begin{equation}
{\bf S}_{\alpha,eff} = \bigoplus_{J \in {\mathcal J}} {\bf S}_{\alpha,n_J}
\otimes {\bf I}_{d_J},
\end{equation}
where ${\bf S}_{\alpha,n_J}$ are $n_J$ dimensional operators which act on the
degeneracy of the $J$th irrep.

Thus we see that there is an intimate connection between the symmetrization
procedure described above and decoherence-free condition.  By symmetrizing the
evolution, a symmetry in the system-environment coupling can be induced and a
decoherence-free subspace or subsystem can be used to store protected quantum
information.

\section{A brief history of decoherence-free conditions}

Decoherence-free subspaces are somewhat related to pointer
bases\cite{Zurek:82a,Zurek:91a}.  In particular decoherence-free subspaces can
be thought of as degenerate pointer basis: the kind of pointer basis which
would cause fits for the measurement problem interpretation usually attached to
environment-induced pointer basis selection.  Also related are the Dicke states
of optics \cite{Dicke:54a}.  Both of these example, however, degeneracy was
nothing more than a theoretical hindrance.

The first indication of states which are resistant to decoherence as applied to
quantum computation was the work of Palma, Suominen, and Ekert\cite{Palma:96a}
as well as the work of Chuang and Yamamoto\cite{Chuang:95a,Chuang:97c}.  These
authors made observations of specific dephasing based DF subspaces and noted
the consequences of these states for quantum computation.

Work on concrete realizations of DF subspaces was presented in a series of
papers by Duan and
Guo\cite{Duan:97c,Duan:97d,Zanardi:97d,Duan:98a,Duan:98b,Duan:98c,Duan:98d,Duan:98e,Duan:99c}.
These authors derived different physical conditions under which DF subspaces
could exist.  Most of this work was presented in the context of semigroup
master equations and focused solely on the theory the existence of such
subspaces for specific examples.

The first work to mathematically put down the DF subspace condition were the
seminal papers of Zanardi and
Rasetti\cite{Zanardi:97a,Zanardi:97b,Zanardi:97c}.  In these papers Zanardi and
Rasetti put forth the Hamiltonian iff criteria for the existence of a DF
subspace.  Zanardi and Rossi also developed proposals for physical realization
of DF subspaces in quantum dots\cite{Zanardi:98b,Zanardi:99a}.  The semigroup
master equation iff criteria for DF subspaces was then derived by
Zanardi\cite{Zanardi:98a} and by Lidar, Chuang, and Whaley\cite{Lidar:98a}.

Developing, at first independently from DF conditions, Viola and Lloyd
presented the notion that symmetrization could be used to avoid
decoherence\cite{Viola:98a,Viola:98b,Viola:99a,Viola:99b}.  Duan and Guo also
examined pulsed control of decoherence\cite{Duan:99a,Duan:99d}.  Zanardi
developed the general mathematical theory of such
symmetrization\cite{Zanardi:99b}.  Both Viola, Lloyd and Knill\cite{Viola:00a}
along with Zanardi\cite{Zanardi:99c} also presented methods for performing
computation on such symmetrized evolutions.

In an important generalization of the DF subspace notion, the work on dynamical
induced symmetrization led Knill, Laflamme, and Viola to introduce the notion
of DF subsystems\cite{Knill:00a}.  DF subsystems were also derived,
independently, by de Filipo. This was the first derivation of the DF subsystem
criteria. Zanardi\cite{Zanardi:01a,Zanardi:00a} and Viola, Knill, and
Laflamme\cite{Viola:00a} then discussed the dynamical generation of coherence
preserving evolutions and the general theory of DF subsystems.

\section{Decoherence-free conditions}

DF subspaces and their generalization DF subsystems offer a method for avoiding
specific symmetric decoherence mechanism.  In the next few chapters we explore
the stability of DFSs and how DFSs fit in with the notion of a quantum computer
before turning to a concrete physical realization of a DFS.  It should be
mentioned, however, that a recent
experiment\cite{Kielpinski:01a,Kielpinski:01b} using ion traps has demonstrated
the existence of a DF subspace.  Thus this work is not just a matter of wishful
theorizing: there is experimental evidence that the notion of DFSs will play an
important role in a future quantum computer.

\chapter{Stability of Decoherence-Free Systems} \label{ch:dfsstab}

\begin{quote}
{\em Does poking and prodding decoherence-free systems remove the free from
decoherence-free?}
\end{quote}

In this chapter we address the issue of the stability of decoherence-free
systems to additional perturbing decoherence processes.   We first give a
simple example of the stability of a decoherence-free subspace.  The stability
of a DFS with respect to a memory fidelity is treated within both the OSR and
SME. If the strength of the perturbation is $\epsilon$, the decoherence-rates
to all orders are shown to vary as $O(\epsilon^2)$.  Finally, the issue of the
dynamical stability of a DFS is addressed.

\section{Stability example for a decoherence-free subspace}

Suppose one has a decoherence-free subsystem corresponding to some
system-environment coupling.  This coupling may be extremely strong and thus it
is not unreasonable to think that a perturbing non-decoherence-free supporting
interaction could couple with this strong evolution yielding a decoherence-free
subsystem which is highly unstable.  In this chapter we will concern ourselves
with understanding the stability of such a situation.  Before we proceed to the
mathematically messy calculation, however, it is useful to present the simplest
example of such stability.  This analysis was first presented by Lidar, Chuang
and Whaley in \cite{Lidar:98a}.

Consider the addition of a perturbing interaction to that of a DFS supporting
evolution in the SME:
\begin{eqnarray}
{\partial \bmath{\rho}(t) \over \partial t} &=& \sum_{\alpha,\beta}
a_{\alpha\beta} \left( [{\bf F}_\alpha, \bmath{\rho}(t) {\bf F}_\beta^\dagger]
+ [{\bf F}_\alpha \bmath{\rho}(t), {\bf F}_\alpha^\dagger] \right) \nonumber
\\ &&+ \epsilon a_{\alpha \beta}^\prime  \left( [{\bf G}_\alpha, \bmath{\rho}(t)
{\bf F}_\beta^\dagger] + [{\bf G}_\alpha \bmath{\rho}(t), {\bf
F}_\beta^\dagger] \right) \nonumber \\   &&+   \epsilon a_{\alpha
\beta}^{\prime \prime}\left( [{\bf F}_\alpha, \bmath{\rho}(t) {\bf
G}_\beta^\dagger] + [{\bf F}_\alpha \bmath{\rho}(t), {\bf G}_\beta^\dagger]
\right).
\end{eqnarray}
We are interested here in the first-order decoherence rate (see Appendix
~\ref{apa:rates})
\begin{equation}
{1 \over \tau_1} = {\rm Tr} \left[ \bmath{\rho}(0) {\partial \bmath{\rho} \over
\partial t} (0) \right].
\end{equation}

If we encoded into a DF subspaces, $\bmath{\rho}(0)=|\psi \rangle \langle
\psi|$, then ${\bf F}_\alpha \bmath{\rho}(0) = \bmath{\rho}(0) {\bf F}_\alpha =
c_\alpha \bmath{\rho}(0)$ where we have made the simplifying assumption that
${\bf F}_\alpha$ is hermitian and is a basis with the same algebraic structure
as that of the Lindblad operators and their adjoints (the SME algebra). This in
turn implies that
\begin{equation}
{\rm Tr} \left[ \bmath{\rho}(0) \left( [{\bf G}_\alpha, \bmath{\rho}(0) {\bf
F}_\beta^\dagger] + [{\bf G}_\alpha \bmath{\rho}(0), {\bf F}_\beta^\dagger]
\right) \right] =0,
\end{equation}
by using the cyclical property of the trace.  This holds for the other
perturbing term as well.  Thus we see that the first-order decoherence rate
vanishes to order $\epsilon$ under a $\epsilon$ strong perturbation.  In the
following sections we expand this result to higher orders and work in both the
OSR and the SME.  Furthermore we also generalize this result to the subsystems
situation.  This extends the subspace analysis originally presented in
\cite{Bacon:99a}.

\section{Stability under the operator-sum representation}

Consider the addition to a DFS supporting Hamiltonian of new perturbing terms
in the interaction Hamiltonian: ${\bf H}_{SB}^{\prime }={\bf H}_{SB}+\epsilon
{\bf H }_{I}^{\prime }$.  The new full evolution operator is given by
\begin{eqnarray}
{\bf U}^{\prime }(t) &=& \exp\left[ -i {\bf H}_{SB}^\prime t
\right]=\sum_{n=0}^\infty \frac{(-it)^{n}}{n!} \left( {\bf H}_{SB} + \epsilon
{\bf H}_{I}^{\prime } \right)^n  \nonumber \\ &=& {\bf U}(t) +
\sum_{k=1}^\infty \epsilon^k \sum_{n=k}^{\infty }\frac{ (-it)^{n}}{n!}
f^{(k)}_{n}({\bf H}_{SB},{\bf H}_{I}^{\prime }),
\end{eqnarray}
where
\begin{eqnarray}
f_{1}^{(1)}({\bf H}_{SB},{\bf H}_{I}^{\prime }) &=&{\bf H}_{I}^{\prime }
\nonumber \\ f_{2}^{(1)}({\bf H}_{SB},{\bf H}_{I}^{\prime }) &=&{\bf
H}_{SB}{\bf H} _{I}^{\prime }+{\bf H}_{I}^{\prime }{\bf H}_{SB}  \nonumber \\
f_{3}^{(1)}({\bf H}_{SB},{\bf H}_{I}^{\prime }) &=&{\bf H}_{SB}^{2}{\bf H}
_{I}^{\prime }+{\bf H}_{SB}{\bf H}_{I}^{\prime }{\bf H}_{SB}+{\bf H}
_{I}^{\prime }{\bf H}_{SB}^{2}  \nonumber \\ f_{2}^{(2)}({\bf H}_{SB},{\bf
H}_{I}^{\prime }) &=&{{\bf H}_{I}^{\prime }} ^{2}  \nonumber \\
f_{3}^{(2)}({\bf H}_{SB},{\bf H}_{I}^{\prime }) &=&{\bf H}_{SB}{{\bf H}
_{I}^{\prime }}^{2}+{\bf H}_{I}^{\prime }{\bf H}_{SB}{\bf H}_{I}^{\prime }+{ \
\ {\bf H}_{I}^{\prime }}^{2}{\bf H}_{SB},
\end{eqnarray}
etc.  Here ${\bf U}(t)=\exp[-i{\bf H}_{SB} t]$ is the unperturbed evolution
operator.  In this chapter we will concern ourselves with the correction due of
the evolution to first order in the perturbing parameter $\epsilon$:
\begin{equation}
{\bf U}^\prime(t) ={\bf U}(t) + \epsilon \sum_{n=1}^{\infty }\frac{
(-it)^{n}}{n!} f^{(1)}_{n}({\bf H}_{SB},{\bf H}_{I}^{\prime }) + O(\epsilon^2).
\end{equation}
Corresponding to this evolution operator are the OSR operators
\begin{equation}
{\bf A}_{i=(\mu,\nu)}^\prime(t) ={\bf A}_i(t) +   \epsilon \sqrt{p_\nu}
\sum_{n=1}^{\infty }\frac{ (-it)^{n}}{n!} \langle \mu |f^{(1)}_{n}({\bf
H}_{SB},{\bf H}_{I}^{\prime }) |\nu\rangle + O(\epsilon^2).
\end{equation}
Expand the unperturbed OSR operators and the perturbing terms about different
fixed basis (see Section \ref{sec:fixedbasis}) ${\bf F}_\alpha$ and ${\bf
G}_\alpha$:
\begin{eqnarray}
{\bf A}_i(t) &=& \sum_\alpha b_{i\alpha}(t) {\bf F}_\alpha \nonumber \\
 \sqrt{p_\nu}
\sum_{n=1}^{\infty }\frac{ (-it)^{n}}{n!} \langle \mu |f^{(1)}_{n}({\bf
H}_{SB},{\bf H}_{I}^{\prime }) |\nu\rangle &=& \sum_\alpha c_{i\alpha}(t) {\bf
G}_\alpha,
\end{eqnarray}
such that the evolution operator to first order in $\epsilon$ is
\begin{equation}
{\bf A}_i^\prime(t)= \sum_\alpha b_{i\alpha}(t) {\bf F}_\alpha + \epsilon
c_{i\alpha}(t) {\bf G}_\alpha+O(\epsilon^2).
\end{equation}
The evolution due to this OSR is thus
\begin{equation}
\bmath{\rho}(t)= \sum_{\alpha \beta} \left( \chi_{\alpha \beta}^{ff}(t) {\bf
F}_\alpha \bmath{\rho}(0) {\bf F}_\beta^\dagger + \epsilon \chi_{\alpha
\beta}^{fg} {\bf F}_\alpha \bmath{\rho}(0) {\bf G}_\beta^\dagger + \epsilon
\chi_{\alpha \beta}^{gf} {\bf G}_\alpha \bmath{\rho}(0) {\bf F}_\beta^\dagger
\right) + O(\epsilon^2),
\end{equation}
where $\chi_{\alpha \beta}^{ff} = \sum_i b_{i\alpha} b_{i\beta}^*$,
$\chi_{\alpha \beta}^{fg} = \sum_i b_{i\alpha} c_{i\beta}^*$,and $\chi_{\alpha
\beta}^{gf} = \sum_i c_{i\alpha} b_{i\beta}^*$.  The normalization condition is
\begin{equation}
\sum_{\alpha \beta} \left( \chi_{\alpha \beta}^{ff} {\bf F}_\beta^\dagger {\bf
F}_\alpha + \epsilon \chi_{\alpha \beta}^{fg} {\bf G}_\beta^\dagger {\bf
F}_\alpha + \epsilon \chi_{\alpha \beta}^{gf} {\bf F}_\beta^\dagger {\bf
G}_\alpha \right) + O(\epsilon^2)={\bf I}.
\end{equation}
As in Section \ref{sec:fixedbasis} we can separate out the identity components
of the evolution and normalization conditions and obtain
\begin{eqnarray}
\bmath{\rho}(t) - \bmath{\rho}(0)&=& -i \left[ {\bf S}^{ff}(t)+ \epsilon {\bf
S}^{fg}(t) + \epsilon{\bf S}^{gf}(t) ,\bmath{\rho}(0) \right]  \nonumber \\ &&+
{\mathcal L}^{ff}(t) \left[ \bmath{\rho}(0) \right] + \epsilon {\mathcal
L}^{fg}(t) \left[ \bmath{\rho}(0) \right] + \epsilon {\mathcal L}^{gf}(t)
\left[ \bmath{\rho}(0) \right]+O(\epsilon^2),
\end{eqnarray}
where
\begin{eqnarray}
{\bf S}^{ff}(t) &=& { i \over 2} \sum_{\alpha \neq 0} \chi_{\alpha 0}^{ff}(t)
{\bf F}_\alpha - \chi_{0 \alpha}^{ff}(t) {\bf F}_\alpha^\dagger \nonumber \\
 {\bf S}^{fg}(t) &=& {i \over 2} \sum_{\alpha \neq 0} \chi_{\alpha 0}^{fg}(t)
 {\bf F}_\alpha - \chi_{0 \alpha}^{fg}(t) {\bf G}_\alpha^\dagger \nonumber \\
 {\bf S}^{gf}(t) &=& {i \over 2} \sum_{\alpha \neq 0} \chi_{\alpha 0}^{gf}(t)
 {\bf G}_\alpha - \chi_{0 \alpha}^{gf}(t) {\bf F}_\alpha^\dagger,
\end{eqnarray}
and
\begin{eqnarray}
{\mathcal L}^{ff}(t) \left[ \bmath{\rho}(0) \right] &=& \sum_{\alpha,\beta \neq
0} \chi_{\alpha \beta}^{ff}(t) \left( \left[ {\bf F}_\alpha \bmath{\rho}(0),
{\bf F}_\beta^\dagger \right] + \left[ {\bf F}_\alpha, \bmath{\rho}(0) {\bf
F}_\beta^\dagger \right] \right) \nonumber \\ {\mathcal L}^{fg} (t)\left[
\bmath{\rho}(0) \right] &=& \sum_{\alpha,\beta \neq 0} \chi_{\alpha
\beta}^{fg}(t) \left( \left[ {\bf F}_\alpha \bmath{\rho}(0), {\bf
G}_\beta^\dagger \right] + \left[ {\bf F}_\alpha, \bmath{\rho}(0) {\bf
G}_\beta^\dagger \right] \right) \nonumber \\ {\mathcal L}^{gf} (t)\left[
\bmath{\rho}(0) \right] &=& \sum_{\alpha,\beta \neq 0} \chi_{\alpha
\beta}^{gf}(t) \left( \left[ {\bf G}_\alpha \bmath{\rho}(0), {\bf
F}_\beta^\dagger \right] + \left[ {\bf G}_\alpha, \bmath{\rho}(0) {\bf
F}_\beta^\dagger \right] \right).
\end{eqnarray}

Suppose that quantum information $|j\rangle$ is encoded into the degeneracy of
the $J$th irrep of the OSR algebra of the unperturbed OSR evolution:
$\Upsilon_J[\bmath{\rho}(0)]=|j\rangle \langle j|$, or
$\bmath{\rho}(0)=|j\rangle \langle j| \otimes \bmath{\rho}_{d_J}$  The $p$th
order decoherence rate $\tau_p$ for this evolution is given by (see Appendix
\ref{apa:rates})
\begin{equation}
\left( {1 \over \tau_p} \right)^p = {\rm Tr} \left[ \Upsilon_J \left[
\bmath{\rho}(0)\right] \Upsilon_J \left[ \bmath{\rho}^{(p)}(0) \right] \right]
= \langle j | \Upsilon_J \left[ \bmath{\rho}^{(p)}(0) \right] |j \rangle,
\end{equation}
where $\rho^{(p)}(0)$ is the $p$th time derivative of the evolved density
matrix $\bmath{\rho}(t)$ evaluated at $t=0$. Recall that
\begin{equation}
\Upsilon_J \left[ {\bf X} \right] = \sum_m \langle J,m | {\bf X} |J,m\rangle,
\end{equation}
transforms an operator ${\bf X}$ on the full Hilbert space to an operator which
acts only on the degeneracy of the $J$th irrep.  Now, explicitly, we can
calculate that
\begin{eqnarray}
\bmath{\rho}^{(p)}(0)&=&  -i \left[ {\bf S}^{ff (p)}(0)+ \epsilon {\bf
S}^{fg(p)}(0) + \epsilon{\bf S}^{gf(p)}(0) ,\bmath{\rho}(0) \right]  \nonumber
\\ &+& {\mathcal L}^{ff(p)}(0) \left[ \bmath{\rho}(0) \right] + \epsilon {\mathcal
L}^{fg(p)}(0) \left[ \bmath{\rho}(0) \right] + \epsilon {\mathcal L}^{gf(p)}(0)
\left[ \bmath{\rho}(0) \right]+O(\epsilon^2).
\end{eqnarray}
Using this expression we can evaluate the contribution of each term to $p$th
order decoherence rate.  First we find that
\begin{eqnarray}
&&\langle j | \Upsilon_J \left[ -i \left[ {\bf S}^{ff (p)}(0)+ \epsilon {\bf
S}^{fg(p)}(0) + \epsilon{\bf S}^{gf(p)}(0) ,|j\rangle \langle j| \otimes
\bmath{\rho}_{d_J}(0) \right] \right] |j\rangle \nonumber \\ &&=-i \langle j|
\sum_m \langle J,m|  \left[ {\bf S}^{ff (p)}(0)+ \epsilon {\bf S}^{fg(p)}(0) +
\epsilon{\bf S}^{gf(p)}(0) ,|j\rangle \langle j| \otimes \bmath{\rho}_{d_J}(0)
\right] | J,m \rangle |j\rangle \nonumber \\ &&=-i \langle j| \sum_m \langle
J,m| \left( {\bf S}^{ff (p)}(0)+ \epsilon {\bf S}^{fg(p)}(0) + \epsilon{\bf
S}^{gf(p)}(0) \right) \bmath{I}_{n_J} \otimes \bmath{\rho}_{d} |J,m\rangle
|j\rangle \nonumber \\ &&+i \langle j| \sum_m \langle J,m| \bmath{I}_{n_J}
\otimes \bmath{\rho}_{d} \left( {\bf S}^{ff (p)}(0)+ \epsilon {\bf
S}^{fg(p)}(0) + \epsilon{\bf S}^{gf(p)}(0) \right)
 |J,m\rangle |j\rangle \nonumber \\ &&=0,
\end{eqnarray}
where we have used the fact
\begin{equation}
\Upsilon_J \left[ \left({\bf I}_{n_J} \otimes {\bf X}_{d_J}\right) {\bf
Y}\right] = \Upsilon_J \left[ {\bf Y} \left({\bf I}_{n_J} \otimes {\bf
X}_{d_J}\right) \right]. \label{eq:upsiloncyclic}
\end{equation}

Next, because the unperturbed evolution is a DFS,
\begin{equation}
\langle j | \Upsilon_J \left[ {\mathcal L}^{ff(p)}(t) \left[\bmath{\rho}(0)
\right] \right] |j\rangle =0.
\end{equation}
To show that the final two traces vanish, we recall that basic representation
theory of complex associative algebras tells us that the expansion operators
${\bf F}_\alpha$ can be taken to have the same reducible structure as the OSR
algebra
\begin{equation}
{\bf F}_\alpha = \bigoplus_{K \in {\mathcal J}} {\bf I}_{n_K} \otimes \left(
{\bf F}_{d_K} \right)_\alpha.
\end{equation}
Thus we find that
\begin{eqnarray}
&&\Upsilon_J \left[{\mathcal L}^{fg}\left[ \bmath{\rho}^{(p)}(0) \right]
\right] = \Upsilon_J \left[ \sum_{\alpha,\beta \neq 0} \chi_{\alpha
\beta}^{fg(p)}(t)\left( 2 {\bf F}_\alpha \bmath{\rho}(0) {\bf G}_\beta^\dagger
- {\bf G}_\beta^\dagger {\bf F}_\alpha \bmath{\rho}(0) - \bmath{\rho}(0) {\bf
G}_\beta^\dagger {\bf F}_\alpha \right) \right] \nonumber \\ &&= \Upsilon_J
\left[ \sum_{\alpha,\beta \neq 0} \chi_{\alpha \beta}^{fg(p)}(t)\left( 2
\bigoplus_{K \in {\mathcal J}} {\bf I}_{n_K} \otimes \left( {\bf F}_{d_K}
\right)_\alpha \bmath{\rho}(0) {\bf G}_\beta^\dagger - {\bf G}_\beta^\dagger
\bigoplus_{K \in {\mathcal J}} {\bf I}_{n_K} \otimes \left( {\bf F}_{d_K}
\right)_\alpha \bmath{\rho}(0) \right. \right. \nonumber \\ &&- \left .\left.
\bmath{\rho}(0) {\bf G}_\beta^\dagger \bigoplus_{K \in {\mathcal J}} {\bf
I}_{n_K} \otimes \left( {\bf F}_{d_K} \right)_\alpha\right) \right],
\end{eqnarray}
and using the fact that $\Upsilon_J$ pulls out only the $J$th irrep,
\begin{eqnarray}
&\Upsilon_J&  \left[{\mathcal L}^{fg}\left[ \bmath{\rho}^{(p)}(0) \right]
\right]= \sum_{\alpha,\beta \neq 0} \Upsilon_J \left[   \chi_{\alpha
\beta}^{fg(p)}(t) \left(2 \left[ |j\rangle \langle j| \otimes \left( ({\bf
F}_{d_J})_\alpha \bmath{\rho}_{d_J}(0) \right) \right] {\bf G}_\beta^\dagger
\right. \right. \nonumber \\ &&- \left. \left. {\bf G}_\beta^\dagger \left[
|j\rangle \langle j| \otimes \left( ({\bf F}_{d_J})_\alpha
\bmath{\rho}_{d_J}(0) \right) \right] -  \left[ |j\rangle \langle j| \otimes
\left( ({\bf F}_{d_J})_\alpha \bmath{\rho}_{d_J}(0) \right) \right] {\bf
G}_\beta^\dagger \right) \right].
\end{eqnarray}
Finally, using the cyclic property of $\Upsilon_J$,
Eq.~(\ref{eq:upsiloncyclic}), this implies that
\begin{equation}
\langle j| \Upsilon_J \left[{\mathcal L}^{fg}\left[ \bmath{\rho}^{(p)}(0)
\right] \right]|j\rangle = 0.
\end{equation}
A similar calculation finds that
\begin{equation}
\langle j| \Upsilon_J \left[{\mathcal L}^{gf}\left[ \bmath{\rho}^{(p)}(0)
\right] \right]|j\rangle = 0.
\end{equation}

Thus, we have shown that, to first order in $\epsilon$, the decoherence rates
of a DFS on the OSR for a perturbing interaction of strength $\epsilon$
vanishes
\begin{equation}
\left( {1 \over \tau_p} \right)^p= 0 + O(\epsilon^{2}).
\end{equation}
This result implies that perturbing interactions can indeed be treated as
perturbing.  A priori one can worry that a strong DFS supporting interaction
could produce effects that scale like $\epsilon g$ where $g$ is the coupling
strength of the unperturbed interaction.  The above calculation shows that one
must go to order $\epsilon^2$ before such interactions can destroy the
decoherence-free nature of the DFS.

Finally we note that we have worked with a pure input state $|j\rangle \langle
j|$.  Notice, however, that a mixed state which evolves according to the OSR
can be thought of as a convex combination of the pure state evolutions
\begin{eqnarray}
\bmath{\rho}(t)&=&\sum_i {\bf A}_i(t) \bmath{\rho}(0) {\bf A}_i^\dagger(t) =
\sum_i \sum_j {\bf A}_i(t) p_j |j \rangle \langle j| {\bf
A}_i^\dagger(t)\nonumber \\ &=& \sum_j p_j \left( \sum_i {\bf A}_i(t) |j\rangle
\langle j| {\bf A}_i^\dagger(t) \right).
\end{eqnarray}
This implies that the above perturbative analysis carries over to the initial
mixed state case.

\section{Stability under the semigroup master equation}

Consider in addition to a DFS supporting SME
\begin{eqnarray}
 {\partial \bmath{\rho}(t) \over \partial t}  &=& {\mathcal L}_{{\bf
F}_\alpha,{\bf F}_\beta} \left[ \bmath{\rho}(t) \right] \nonumber \\ {\mathcal
L}_{{\bf F}_\alpha,{\bf F}_\beta}\left[ \bmath{\rho}(t) \right] &=& {1 \over 2}
\sum_{\alpha,\beta} a_{\alpha \beta} \left( \left[ {\bf F}_\alpha
\bmath{\rho}(t), {\bf F}_\beta^\dagger \right]+ \left[ {\bf F}_\alpha,
\bmath{\rho}(t) {\bf F}_\beta^\dagger \right] \right),
\end{eqnarray}
and additional $\epsilon$ perturbing term
\begin{equation}
{\partial \bmath{\rho}(t) \over \partial t}  =  {\mathcal L}\left[
\bmath{\rho}(t) \right]={\mathcal L}_{{\bf F}_\alpha,{\bf F}_\beta} \left[
\bmath{\rho}(t) \right] + {\mathcal L}_{{\bf F}_\alpha,\epsilon{\bf G}_\beta}
\left[ \bmath{\rho}(t) \right]+{\mathcal L}_{\epsilon{\bf G}_\alpha,{\bf
F}_\beta} \left[ \bmath{\rho}(t) \right],
\end{equation}
where
\begin{eqnarray}
{\mathcal L}_{{\bf F}_\alpha,\epsilon{\bf G}_\beta}\left[ \bmath{\rho}(t)
\right]  &=& {\epsilon \over 2} \sum_{\alpha,\beta} b_{\alpha \beta} \left(
\left[ {\bf F}_\alpha \bmath{\rho}(t), {\bf G}_\beta^\dagger \right]+ \left[
{\bf F}_\alpha, \bmath{\rho}(t) {\bf G}_\beta^\dagger \right] \right) \nonumber
\\ {\mathcal L}_{\epsilon{\bf G}_\alpha,{\bf F}_\beta}\left[ \bmath{\rho}(t)
\right] &=& {\epsilon \over 2} \sum_{\alpha,\beta} c_{\alpha \beta} \left(
\left[ {\bf G}_\alpha \bmath{\rho}(t), {\bf F}_\beta^\dagger \right]+ \left[
{\bf G}_\alpha, \bmath{\rho}(t) {\bf F}_\beta^\dagger \right] \right).
\end{eqnarray}
The $p$ time derivative of $\bmath{\rho}(t)$ evaluated at $t=0$ is then given
by
\begin{equation}
\bmath{\rho}^{(p)}(0) = {\mathcal L}^{(p-1)} \left[ \bmath{\rho}(0) \right],
\end{equation}
where
\begin{equation}
{\mathcal L}^{(p-1)}[{\bf X}]= \underbrace{{\mathcal L}[ {\mathcal L}[ \cdots
{\mathcal L} [{\bf X}] \cdots ] ] }_{p-1 \quad {\mathcal L}{\rm 's}}.
\end{equation}

Now suppose that quantum information $|j\rangle$ is encoded into the degeneracy
of the $J$th irrep of the SME algebra: $\bmath{\rho}(0)= |j\rangle \langle j|
\otimes \bmath{\rho}_{d_J}(0)$.  The decoherence rates are then, as in the
previous section, given by
\begin{equation}
\left( {1 \over \tau_p} \right)^p =\langle j | \Upsilon_J \left[
\bmath{\rho}^{(p)} (0) \right] |j\rangle = \langle j | \Upsilon_J \left[
{\mathcal L}^{(p-1)} \left[ \bmath{\rho}(0) \right] \right] |j\rangle.
\end{equation}
To first order in $\epsilon$, the only nonvanishing terms in ${\mathcal
L}^{(p-1)} \left[ |j\rangle \langle j| \otimes \bmath{\rho}_{d_J} (0) \right]$
are
\begin{equation}
{\mathcal L}^{(p-1)} \left[ |j\rangle \langle j| \otimes \bmath{\rho}_{d_J} (0)
\right]= \underbrace{{\mathcal L}_{{\bf F}_\alpha,{\bf F}_\beta} [ \cdots
{\mathcal L}_{{\bf F}_\alpha,{\bf F}_\beta} [ }_{p-2 \quad {\mathcal L}_{{\bf
F}_\alpha,{\bf F}_\beta} {\rm 's}} \left( {\mathcal L}_{{\bf F}_\alpha,\epsilon
{\bf G}_\beta } +{\mathcal L}_{\epsilon{\bf G}_\alpha,{\bf F}_\beta } \right)
[|j\rangle \langle j| \otimes \bmath{\rho}_{d_J} (0)]]\cdots],
\end{equation}
because ${\mathcal L}_{{\bf F}_\alpha,{\bf F}_\beta} \left[ |j\rangle \langle
j| \otimes \bmath{\rho}_{d_J} (0) \right]=0$.  We can now expand the ${\bf
F}_\alpha$'s in terms of the SME algebra:
\begin{equation}
{\bf F}_\alpha = \bigoplus_{K \in {\mathcal J}} {\bf I}_{n_K} \otimes ({\bf
F}_{d_J} )_{\alpha}.
\end{equation}
It is useful here to notice that
\begin{equation}
\langle j| \Upsilon_J \left[ {\bf X} \right]  |j\rangle= \sum_m {\rm Tr} \left[
{\bf P}_{J,j,m} {\bf X} {\bf P}_{J,j,m} \right],
\end{equation}
where ${\bf P}_{J,j,m}=|j\rangle \langle j| \otimes |J,m\rangle \langle J,m|$.
Because we can choose ${\bf F}_\alpha$ to be hermitian, ${\bf P}_{J,j,m} {\bf
F}_\alpha = {\bf F}_\alpha {\bf P}_{J,j,m}$ so that
\begin{eqnarray}
&&\langle j| \Upsilon_J\left[ {\mathcal L}^{(p-1)} \left[ |j\rangle \langle j|
\otimes \bmath{\rho}_{d_J} (0) \right] \right]|j\rangle = \sum_m {\rm Tr}
\left[ {\bf P}_{J,j,m} {\mathcal L}^{(p-1)} \left[ |j\rangle \langle j| \otimes
\bmath{\rho}_{d_J} (0) \right] {\bf P}_{J,j,m} \right] \nonumber \\ &&= \sum_m
{\rm Tr} \left[ \underbrace{{\mathcal L}_{{\bf F}_\alpha,{\bf F}_\beta} [ \cdot
\cdot {\mathcal L}_{{\bf F}_\alpha,{\bf F}_\beta} [ }_{p-2 \quad {\mathcal
L}_{{\bf F}_\alpha,{\bf F}_\beta} {\rm 's}} {\bf P}_{J,j,m} \left({\mathcal
L}_{{\bf F}_\alpha,\epsilon {\bf G}_\beta } + {\mathcal L}_{\epsilon{\bf
G}_\alpha, {\bf F}_\beta } \right) [|j\rangle \langle j| \otimes
\bmath{\rho}_{d_J} (0)] {\bf P}_{J,j,m}] \cdot \cdot] \right]. \nonumber \\
\end{eqnarray}
Now
\begin{eqnarray}
&&\sum_m {\bf P}_{J,j,m} {\mathcal L}_{{\bf F}_\alpha,\epsilon {\bf G}_\beta }
\left[ |j\rangle \langle j| \otimes \bmath{\rho}_{d_J}(0) \right] {\bf
P}_{J,j,m} \nonumber \\ &&= \sum_m {\epsilon \over 2} \sum_{\alpha,\beta \neq
0} b_{\alpha \beta} {\bf P}_{J,j,m} \left[ 2 {\bf F}_\alpha \left( |j\rangle
\langle j| \otimes \bmath{\rho}_{d_J}(0) \right){\bf G}_\beta^\dagger -  {\bf
G}_\beta^\dagger {\bf F}_\alpha \left( |j\rangle \langle j| \otimes
\bmath{\rho}_{d_J}(0) \right) \right. \nonumber \\
 &&- \left.
 \left( |j\rangle
\langle j| \otimes \bmath{\rho}_{d_J}(0) \right){\bf G}_\beta^\dagger {\bf
F}_\alpha \right] {\bf P}_{J,j,m}.
\end{eqnarray}
The $\sum_m {\bf P}_{J,j,m} \cdots {\bf P}_{J,j,m}$ is essentially a trace
operator over $m$ and since only ${\bf G}_\beta^\dagger$ acts non-trivially
over the decomposition, we see that we can cycle the operators such that this
term vanishes
\begin{equation}
\sum_m {\bf P}_{J,j,m} {\mathcal L}_{{\bf F}_\alpha,\epsilon {\bf G}_\beta }
\left[ |j\rangle \langle j| \otimes \bmath{\rho}_{d_J}(0) \right] {\bf
P}_{J,j,m}=0.
\end{equation}
A similar conclusion holds for the ${\mathcal L}_{{\bf F}_\alpha,\epsilon {\bf
G}_\beta }$ term.

Thus we see that, as is the case for the OSR, in the SME
\begin{equation}
\left( {1 \over \tau_p} \right)^p =0 + O(\epsilon^2).
\end{equation}

\section{Dynamical stability}

The results derived in the previous Section imply that DFSs are robust to small
perturbations when the DFS is operating as a quantum {\em memory}.  In order to
address what happens when perturbations are made on the system as it evolves
according to some desired quantum {\em computation}, we have to first define an
analog of the mixed-state memory fidelity for an evolving system. This is
\begin{equation}
F_{{\rm d}}(t)={\rm Tr}[\bmath{\rho}_U(t) \bmath{\rho}(t)] ,,
\end{equation}
where $\bmath{\rho}_U(t)$ is the desired unitary evolution,
\begin{equation}
\bmath{\rho} _{U}(t)={\bf U}_{S}(t)\bmath{\rho} (0){\bf U}_{S}^{\dagger
}(t),\quad {\rm with} \quad {\bf U}_{S}(t)=\exp \left[ -i{\bf H}_{S}t\right] .
\end{equation}
Here ${\bf H}_{S}$ is the system Hamiltonian.  This {\em dynamical} fidelity is
a good measure of the difference between the desired evolution of the system
and the actual, noisy evolution.  Thus, $0\leq F_{ {\rm d}}(t)\leq 1$, with
$F_{{\rm d}}(t)=1$ if and only if the evolution is perfect, i.e., $\bmath{\rho}
(t)=\bmath{\rho}_{U}(t)$. The decoherence rates for the dynamical fidelity are
defined in the same manner as for the memory fidelity:
\begin{eqnarray}
F_{{\rm d}}(t)=\sum_{n}\frac{1}{n!}\left( \frac{t}{\bar{\tau}_{n}}\right)
^{n}:\quad \frac{1}{\bar{\tau}_{n}}=\left\{ {\rm Tr}[\left\{ \rho _{U}(t)\rho
(t)\right\} ^{(n)}]\right\} ^{1/n}.
\end{eqnarray}
In \cite{Bacon:99a}, it was shown that $\left({1 \over \bar{\tau}_1} \right)=0$
for both the OSR and the SME.  The interested reader is referred to this
article for more information on this result.

\section{Stability}

We assemble in Table \ref{tab:pert} all of the perturbation results

\begin{table}[h]
\begin{tabular}{|l|c|c|} \hline
& SME & OSR \\ \hline General & $1/\tau_{1}\neq 0$ & $1/\tau_{1}=0 $
\\
& $1/\tau_{n}\neq 0$, $n\geq 2$ & $1/\tau_{n}\neq 0$, $n\geq 2$  \\ \hline DFSs
& $1/\tau_{n}=0$ & $1/\tau_{n}=0$ \\ \hline memory fidelity for $\epsilon
$-perturbed DFSs & $1/\tau_n=0+O(\epsilon^2)$ & $ 1/\tau_n=0+O(\epsilon^2)$
\\ \hline dynamical fidelity for $\epsilon$-perturbed DF & $1/\bar{\tau}
_{1}=0$ & $1/\bar{\tau}_{1}=0$ \\ subspaces  & $1/\bar{\tau}_{n}\neq 0$, $n\geq
2$  & $1/\bar{\tau}_{n}\neq 0$, $n\geq 2$  \\\hline
\end{tabular}
\caption{\em Perturbed decoherence-rates}\label{tab:pert}
\end{table}
The first indications of the stability of a DF subspace can be found in the
numerical simulations done by Zanardi in \cite{Zanardi:98a}.  Lidar, Chuang,
and Whaley\cite{Lidar:98a} then presented the general memory stability
condition of ${1 \over \tau_1}=0$ in the context of DF subspaces.  The general
memory stability results to all orders in time for DF subspaces were derived by
Bacon, Lidar, and Whaley in \cite{Bacon:99a}.  In this chapter, we have
broadened these stability results from the arena of DF subspaces to DF
subsystems.

The stability of DFSs to perturbations is a particularly nice result for using
DFSs as a stable quantum memory.  It is unlikely that absolutely perfect DF
conditions will exist in nature and therefore it is important to understand how
perturbing interactions change the DF nature of the system.  The above
perturbation results indicate that one can treat perturbing interactions on a
DFS as independent of the DF condition.  This will later turn out to be an
important issue when one thinks about how to use DFSs within the context of
fault-tolerant quantum error correction.

\chapter{Decoherence-Free Subsystems and the Quantum Computer} \label{ch:dfsqc}

\begin{quote}
{\em To compute or not to compute, that is the question.}
\end{quote}

In this chapter we address the issue the relationship between DFSs and quantum
computation.  We begin by describing how DFSs can be used in a concatenated
manner and how this fits in with the idea of fault-tolerant quantum
computation.  A particularly important aspect of DFSs necessary for their use
in quantum computation is the ability to perform universal quantum computation
on the encoded information.  The special, but not unique, role of the commutant
of the OSR or SME algebra for universal quantum computation is outlined.  The
issue of measurement on a DFS and leakage errors is also introduced with
application to more concrete models put off until the following chapters.

\section{Quantum computation and decoherence-free subspaces}

In the previous two chapters we have introduced the notion of decoherence-free
subsystem and examined the stability of such DFSs to perturbing interactions.
We have seen that it is possible to perform an encoding of quantum information
such that the information is protected from a certain source of decoherence.
Let us now discuss how such decoherence-free subsystems can be put to use
towards building a quantum computer.

One of the common misconceptions about decoherence-free subsystems is that they
were intended as an {\em ultimate} solution towards building a quantum
computer.  There are two main reasons why such a future is highly unlikely to
unfold.

The first reason why DFSs are not the ultimate solution arises from the fact
that the symmetries necessary for maintaining a decoherence-free condition will
almost certainly not be perfectly realized in the physical world.  Encoding
into a DFSs, should be thought of in the context of eliminating a {\em
particular} decoherence mechanism.  This decoherence mechanism may be the
dominant mechanism or it may be on equal footing with other non-DFS supporting
decoherence mechanisms.  This is not to deemphasize the importance of DFSs
towards building a quantum computer: elimination of a particular decoherence
mechanism should not be brushed under the rug and dismissed.

The second reason why DFSs are not the ultimate solution to building a quantum
computer is because the concept of a decoherence-free subsystem does nothing to
address the issue of fault-tolerant quantum computation.  Suppose one has,
miraculously, found a system whose only decoherence mechanism supports a DFS
and it is possible to encode and make suitable measurements on the DFS.  In
order to use such DFSs for quantum computation, one must be able to perform
operations on the DFS which manipulate the quantum information.  These
operations will most likely be faulty: it will not be possible to perfectly
execute an operation on the DFS perhaps.  The condition of being
decoherence-free says nothing about the faulty operation of gates on the
encoded DFS.

So, in general, the theory of DFSs must be cast within the broader quest
towards building a quantum computer.  The most likely usefulness of the DF idea
in quantum computation is to work alongside the theory of fault-tolerant
quantum error correction.  The theory of fault-tolerant quantum computation
\cite{Aharonov:97a,Gottesman:98a,Kitaev:97b,Knill:98a,Preskill:98a,Shor:96a}
describes how faulty operations and decoherence, if the effects of both are
sufficiently weak and sufficiently non-pathological, can be used to perform
quantum computation to any desired accuracy with only a polynomial slowdown in
the quantum computation.  The idea of putting DFSs to work in quantum
computation is the elimination of a particular decoherence mechanism such that
the threshold for fault-tolerant quantum computation can be achieved.  This
philosophy is perhaps best summarized by the saying ``use symmetry first!''
implying that symmetries in the system-environment coupling should be used to
first eliminate bothersome decoherence mechanisms before quantum error
correction is then applied to build a reliable quantum computer.

\section{DFSs for quantum computation}

To build a quantum computer, we must make a mapping to the quantum subsystem
circuit model.  The most straightforward manner of achieving this goal in the
context of DFSs is to take individual DFSs as the subsystems of the quantum
subsystem circuit model.  The idea is something like that depicted in Figure
\ref{fig:subcat}.
\begin{figure}[h]
\hspace{1cm}
 \psfig{figure=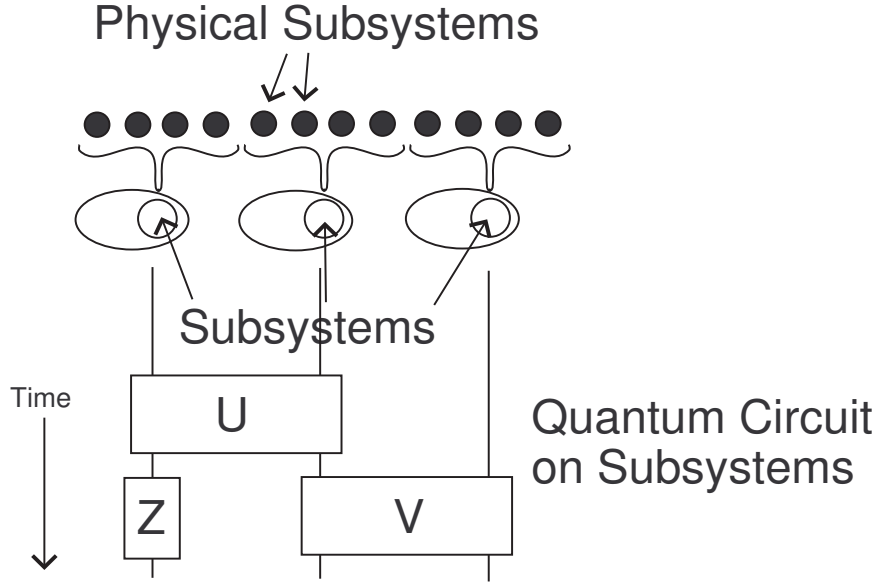,width=5in} \vspace{0.2cm} \caption{\em
Decoherence-free subspace and the quantum circuit model} \label{fig:subcat}
\end{figure}
Information in physical subsystems is encoded into a DFS which may span several
physical subsystems.  These encoded subsystems then will become the building
blocks of the quantum subsystem circuit model.  Viewed from the lens of coding
theory what we are doing is using the DFSs as a code from which the quantum
subsystems circuit model is constructed.

Suppose one has two OSR or SME algebras ${\tt A}$ and ${\tt B}$ which have
representations ${\tt A}\cong \bigoplus_{J\in {\mathcal J}} {\tt I}_{n_J}
\otimes {\tt M}_{d_J}$ and ${\tt B}\cong \bigoplus_{K\in {\mathcal K}} {\tt
I}_{n_K} \otimes {\tt M}_{d_K}$.  Now suppose that these two algebra act on two
separate subsystems of a Hilbert space ${\mathcal H}={\mathcal H}_A \otimes
{\mathcal H}_B$.  The algebra on this conjoined space then acts as ${\tt A}
\otimes {\tt B}$.  Using the reducible representation of each algebra we find
that
\begin{eqnarray}
{\tt A} \otimes {\tt B} &\cong& \left( \bigoplus_{J\in {\mathcal J}} {\tt
I}_{n_J} \otimes {\tt M}_{d_J} \right) \otimes \left(  \bigoplus_{K\in
{\mathcal K}} {\tt I}_{n_K} \otimes {\tt M}_{d_K} \right) \nonumber \\ &\cong&
\bigoplus_{J \in {\mathcal J}, K \in {\mathcal K}} {\tt I}_{n_J} \otimes {\tt
I}_{n_K} \otimes {\tt M}_{d_J} \otimes {\tt M}_{d_K} \nonumber \\ &\cong&
\bigoplus_{J \in {\mathcal J}, K \in {\mathcal K}} ({\tt I}_{n_J} \otimes {\tt
I}_{ n_K}) \otimes {\tt M}_{d_J d_K},
\end{eqnarray}
where we have used the fact that the tensor product of two full matrix algebras
is the full matrix algebra on the tensor product state ${\tt M}_{d_1} \otimes
{\tt M}_{d_2} \cong {\tt M}_{d_1 d_2}$.  This result implies that if we store
build a quantum subsystem circuit out of subsystems which are each individually
DFSs, then the conjoined subsystems will still be DF.  This, however, will not
always be the situation.  Due to the particular symmetry involved in a DFS, it
may be possible that conjoining two DFSs produces an OSR or SME algebra which
is larger than the simple tensor product structure of the above decomposition.

Here it should be pointed out that, much like the case of the quantum
subsystems circuit model itself, there is some arbitrariness in the how we map
from a DFS to the quantum subsystem circuit model.  In the above description,
and as depicted in Figure~\ref{fig:subcat} each encoded subsystem is in
one-to-one correspondence with a set of physical qubits.  The tensor product
between these sets of physical qubits then becomes the tensor product between
the encoded subsystems in the quantum circuit model.  This model, of course, is
not the most general.  We will, however concentrate on this model as it appears
to be the most physically relevant model.

DFS encoded subsystems will be used in the quantum subsystem circuit model to
perform fault-tolerant quantum computation. Placing quantum information into a
DFS presents extra challenges for the theory of fault-tolerant quantum
computation.  Let us enumerate the ways in which a DFS fits in with the
standard model of fault-tolerant quantum computation.

\begin{enumerate}
\item Preparation.  There should be some manner to create states with support over a DFS
with a certain fidelity of preparation.  In the theory of fault-tolerant
quantum error correction there are often cases where preparation of a
particular state is desired.  Specific DFS models then will require these
special preparation steps.
\item Measurement.  Closely tied with the issue of preparation, it should be
possible to extract information via a measurement which makes some distinction
between different encoded information.  Of course it would be highly desirable
to perform any possible measurement, but much of the theory of fault-tolerant
quantum computation can be adapted to models where only minimal information
extraction is possible.  We will address this issue in
Section~\ref{sec:dfsmeas}.
\item Universality.  A set of interactions must be possible which act on the
information encoded in the DFS.  If the DF condition is to maintained there is
an important restriction here that does not appear in normal quantum
computation: the interactions should always act within the protected subsystem.
This issue is addressed in Section~\ref{sec:dfsuniv}.
\item Noise models.  The threshold theorem for fault-tolerant quantum
computation deals with noise models of a specific form.  This means that the
perturbing noise (i.e. non-decoherence-free) on DFSs should fit within these
noise models.  Of non-trivial significance in this context is the problem that
information which has been encoded can leak out of the encoding.  A
particularly useful technique for attacking leakage in the context of error
correction is given in Section~\ref{sec:dfsleak}
\end{enumerate}

\section{The commutant and universal quantum computation} \label{sec:dfsuniv}

Using DFSs to construct a quantum subsystems circuit model one must be able to
perform computation on the encoded subsystems.  In a given physical setup where
one is attempting to couple or act on single subsystems, there will be a given
OSR or SME algebra ${\tt A}$ which is relevant when the evolution on the
encoded subsystem is being acted upon.  The operation which is enacted on the
encoded subsystem may be on a single encoded subsystem or between multiple
subsystems, but in both cases, there will be a relevant OSR or SME algebra
${\tt A}$ describing the DFS.  In particular because leaving a DFS may be
disastrous to the encoded quantum information, we require that the evolution of
the system never cause information in the DFS to leak out of the DFS.  What are
the necessary and sufficient conditions for a Hamiltonian dynamics ${\bf H}$ to
maintain the DF condition?

We recall that every algebra ${\tt A}$ has a commutant ${\tt A}^\prime$ which
is the set of all operators which commute with the elements of the algebra
${\tt A}$.  These operators have a dual reducible structure (recall Section
\ref{sec:osrcommutant})
\begin{eqnarray}
{\tt A} &\cong& \bigoplus_{J \in {\mathcal J}} {\tt I}_{n_J} \otimes {\tt
M}_{d_J} \nonumber \\ {\tt A}^\prime &\cong& \bigoplus_{J \in {\mathcal J}}
{\tt M}_{n_J} \otimes {\tt I}_{d_J}. \label{eq:dual}
\end{eqnarray}
In particular we see that elements of the commutant act to preserve the
reducible structure of the algebra ${\tt A}$.  This leads us the following
sufficient condition for a Hamiltonian ${\bf H}$ to act only on information
encoded in a particular irrep.
\begin{lemma} \label{lem:commuteosr}
Suppose one is given an OSR or SME algebra ${\tt A} \cong \bigoplus_{J \in
{\mathcal J}} {\tt I}_{n_J} \otimes {\tt M}_{d_J}$ and information has been
encoded into the degeneracy of the $K$th irrep.  A Hamiltonian ${\bf H}$ which
commutes with all of the elements of ${\tt A}$ will act on this encoded
information and will not take this encoded information out of the $K$th irrep.
\end{lemma}
Proof:  Trivial application of the idea of the commutant of the algebra ${\tt
A}$.  If ${\bf H}$ commutes with ${\tt A}$, then it is in ${\tt A}^\prime$ and
therefore has the decomposition described in Eq.~(\ref{eq:dual}) which
preserves the information encoded into the degeneracy.

\subsection{Example of the commutant condition}

Recall in Section~\ref{sec:dfsubsystemex} we found that the OSR algebra ${\tt
A}$ spanned by  $\{ {\bf I}, \bmath{\sigma}_x \otimes \bmath{\sigma}_x \otimes
\bmath{\sigma}_x,\bmath{\sigma}_y \otimes \bmath{\sigma}_y \otimes
\bmath{\sigma}_y , \bmath{\sigma}_z \otimes \bmath{\sigma}_z \otimes
\bmath{\sigma}_z \}$ supported a four dimensional DFS.  The commutant of the
OSR algebra ${\tt A}$ is generated by the operators $\{ \bmath{\sigma}_z
\otimes \bmath{\sigma}_z \otimes {\bf I} ,
 {\bf I} \otimes \bmath{\sigma}_z \otimes \bmath{\sigma}_z , \bmath{\sigma}_x
 \otimes \bmath{\sigma}_x \otimes {\bf I},\bmath{\sigma}_x ,  {\bf I} \otimes \bmath{\sigma}_x \otimes
 \bmath{\sigma}_x, {\bf I} \}$.  Let us discuss how this commutant can be used
 to enact interactions on the DFS.

The DFS corresponding to ${\tt A}$ is four dimensional.  We can therefore think
of this subsystem as composed of two qubits.  The splitting of the space into
two qubits is, as always, arbitrary in where we place the tensor product
structure.  We recall that a complete set of commuting observables for the
three qubit Hilbert space of this DFS is given by ${\bf
Z}_{12}=\bmath{\sigma}_z \otimes \bmath{\sigma}_z \otimes {\bf I}$, ${\bf
Z}_{23}={\bf I} \otimes \bmath{\sigma}_z \otimes \bmath{\sigma}_z$, and ${\bf
Z}_{123}=\bmath{\sigma}_z \otimes \bmath{\sigma}_z \otimes \bmath{\sigma}_z$
with a corresponding basis $|z_{12},z_{23},z_{123}\rangle$.    The DFS
information is encoded into the subsystem spanned by the
$|z_{12},z_{23}\rangle$.  We will take our two qubits to be the $|z_{12}\rangle
\otimes |z_{23}\rangle$ subsystem structure.  We therefore see that the
operation ${\bf Z}_{12}$ acts non-trivially on this encoded information
\begin{eqnarray}
{\bf Z}_{12} |+1,z_{23},z_{123}\rangle &=& +|+1,z_{23},z_{123}\rangle \nonumber
\\
{\bf Z}_{12} |-1,z_{23},z_{123}\rangle &=& -|-1,z_{23},z_{123}\rangle.
\end{eqnarray}
Similarly for ${\bf Z}_{23}$ acts only on the $z_{23}$ component of
$|z_{12},z_{23},z_{123}\rangle$.  The ${\bf Z}_{12}$ and ${\bf Z}_{23}$ act as
an encoded $\bmath{\sigma}_z$ on each of the encoded qubits.  Similarly we find
that
\begin{eqnarray}
{\bf I} \otimes \bmath{\sigma}_x \otimes \bmath{\sigma}_x
|+1,z_{23},z_{123}\rangle &=& |-1,z_{23},z_{123}\rangle \nonumber
\\
{\bf I} \otimes \bmath{\sigma}_x \otimes \bmath{\sigma}_x
|-1,z_{23},z_{123}\rangle &=& |+1,z_{23},z_{123}\rangle,
\end{eqnarray}
such that this operator acts as an encoded $\bmath{\sigma}_x$ on the
$|z_{12}\rangle$ qubit.

What should now be clear this that these operators which are in the commutant
of ${\tt A}$ act as single qubit Pauli operators on the encoded subsystems.
Let us denote the two qubits via $a$ and $b$.  Then the encoded operators on
the these subsystems are enacted by the encoded Pauli operators
\begin{eqnarray}
\bmath{\sigma}_x^{(a)}={\bf I} \otimes \bmath{\sigma}_x \otimes
\bmath{\sigma}_x, \quad \bmath{\sigma}_y^{(a)}=  \bmath{\sigma}_z \otimes
\bmath{\sigma}_y \otimes \bmath{\sigma}_x , \quad \bmath{\sigma}_z^{(a)} =
\bmath{\sigma}_z \otimes \bmath{\sigma}_z \otimes {\bf I}, \nonumber \\
\bmath{\sigma}_x^{(b)}= \bmath{\sigma}_x \otimes \bmath{\sigma}_x \otimes {\bf
I} , \quad \bmath{\sigma}_y^{(b)}=  \bmath{\sigma}_x \otimes \bmath{\sigma}_y
\otimes \bmath{\sigma}_z , \quad \bmath{\sigma}_z^{(b)} = {\bf I} \otimes
\bmath{\sigma}_z \otimes \bmath{\sigma}_z.
\end{eqnarray}
Thus if one wishes to perform a single qubit rotation on the qubit $a$, one can
use a Hamiltonian of the form
\begin{equation}
{\bf H}^{(a)}=\vec{n} \cdot \vec{\bmath{\sigma}}^{(a)} = n_x {\bf I} \otimes
\bmath{\sigma}_x \otimes \bmath{\sigma}_x+ n_y \bmath{\sigma}_z \otimes
\bmath{\sigma}_y \otimes \bmath{\sigma}_x+ n_z \bmath{\sigma}_z \otimes
\bmath{\sigma}_z \otimes {\bf I}.
\end{equation}

Similarly we can construct the operators which act between qubits $a$ and $b$.
For example the operator which acts as $\bmath{\sigma}_x \otimes
\bmath{\sigma}_x$ on the encoded qubits is given by
\begin{equation}
\bmath{\sigma}_x^{(a)} \bmath{\sigma}_x^{(b)} = \left( {\bf I} \otimes
\bmath{\sigma}_x \otimes \bmath{\sigma}_x \right) \left( \bmath{\sigma}_x
\otimes \bmath{\sigma}_x \otimes {\bf I} \right)= \bmath{\sigma}_x \otimes {\bf
I} \otimes \bmath{\sigma}_x.
\end{equation}
This operator and all the other similar two qubit operators is, like the
$\bmath{\sigma}_\alpha^{(a,b)}$, in the commutant of ${\tt A}$.

Thus we see how examining the commutant of an algebra $\tt A$ can allow us to
find operators which perform interactions on the encoded subsystem.

\subsection{Why the commutant condition is sufficient but not necessary}

While Lemma \ref{lem:commuteosr} describes a sufficient condition for a
Hamiltonian to preserve the information encoded into a DFS subsystem, the
condition is not necessary.  To see this, one recalls that information will be
stored in a particular $J^\prime\in {\mathcal J}$ irrep of the algebra ${\tt A}
\cong \bigoplus_{J \in \mathcal{J}} {\tt I}_{n_J} \otimes {\tt M}_{d_J}$.
Information which has been encoded into the degeneracy of a particular single
$J^\prime$ irrep is unaffected by what happens in the other irreps. Elements
which commute with operators in ${\tt A}$ preserve {\em every} $J$ irrep.
Information which is encoded in a particular single degeneracy can be acted
upon by operators which are not in the commutant of ${\tt A}$ and which still
preserve the DF encoded information.

Using the criteria that a Hamiltonian commute with the OSR or SME algebra,
then, is not a necessary condition for preserving information encoded into the
corresponding DFS.  We will find, however, that while the criteria is not
necessary it is often sufficient for our needs.

\subsection{Representation theory and the commutant}

A further area which often causes confusion in describing computation on a DFS
is the difference between a complex associate algebra and a Lie algebra.
Suppose one is given the ability to enact a set of Hamiltonians which generate
the commutant of the OSR or SME algebra ${\tt A}$ for a certain DFS.  The
ability to enact the Hermitian generators of the commutant is not enough to
guarantee that every operation on the encoded DFS can be enacted.  The reason
for this is that the generators we have specified are generators in the sense
of a complex associative algebra (multiplication, linear combination) and not
in the sense of a Lie algebra (Lie bracket, linear combination).

Let us given an illustrative example of this situation to clarify the problem.
Suppose we are given the ability to enact a three-dimensional irrep of the Lie
algebra $su(2)$,
\begin{eqnarray}
\bmath{\sigma}_x^{[3]} \cong \sqrt{2} \left[ \begin{array}{ccc} 0 & 1 & 0 \\ 1
& 0 & 1
\\ 0 & 1 & 0 \end{array} \right], \quad \bmath{\sigma}_y^{[3]} \cong \sqrt{2} \left[
\begin{array}{ccc} 0 & -i & 0 \\ i & 0 & -i \\ 0 & i & 0 \end{array} \right],
\quad \bmath{\sigma}_z^{[3]} \cong 2 \left[ \begin{array}{ccc} 1 & 0 & 0 \\ 0 &
0 & 0
\\ 0 & 0 & -1 \end{array} \right].
\end{eqnarray}
It is easy to check that the complex associative algebra generated by
multiplication and linear combination is the entire space of linear operators
on the three-dimensional space.  However, the Lie algebra generated by these
operators is just the three operators $\bmath{\sigma}_\alpha^{[3]}$ which do
not span the space of linear operators on the three-dimensional space. Elements
like $\left(\bmath{\sigma}_x^{[3]}\right)^2$ are in the complex associative
algebra generated by the $\bmath{\sigma}_\alpha^{[3]}$, but are not in the Lie
algebra generated by the $\bmath{\sigma}_\alpha^{[3]}$.

The correct way to state Lemma \ref{lem:commuteosr} in terms of the generators
of a Lie algebra is then
\begin{lemma}
Suppose one is given an OSR or SME algebra ${\tt A} \cong \bigoplus_{J \in
{\mathcal J}} {\tt I}_{n_J} \otimes {\tt M}_{d_J}$ and information has been
encoded into the degeneracy of the $K$th irrep.  A set of Hamiltonians
${\mathcal S}$ each of which commutes with all of the elements of ${\tt A}$
will act on this encoded information without taking this encoded information
out of the $K$th irrep.  Furthermore this set of Hamiltonians ${\mathcal S}$
will generate a Lie algebra ${\mathcal L}$ which has a reducible structure of
the form
\begin{equation}
{\mathcal L} \cong \bigoplus_{J \in {\mathcal J}} {\mathcal L}_{n_J} \otimes
{\mathcal I}_{d_J},
\end{equation}
where ${\mathcal L}_{n_J}$ is a (perhaps further reducible) $n_J$ dimensional
Lie algebra and ${\mathcal I}_{d_J}$ represents the identity action on a $d_J$
dimensional space.
\end{lemma}

\subsection{Existential universality on a DFS} \label{sec:existuniv}

It is important to realize that universal sets of gates always {\em exist} for
any given subsystem structure mapped onto a quantum circuit
model\cite{Lloyd:95a,Zanardi:99c}. This is to say that it is always possible to
construct a given set of interactions between subsystems.  However, for a
specific DFS, there are important limitations which prevent this existential
result from holding any weight.  In particular, the set of operators which can
be enacted on the DFS often is from a limited set of physically viable
operators.  In most systems, more that two-body interactions will be very
difficult to enact on the system.  Thus existentially there are always
universal gate sets, but under most conditions, these existential results are
not of use.

Suppose, for example, that one has encoded on qubit of information into $5$
physical qubits in terms of the basis states $|0_L\rangle = |00000\rangle$ and
$|1_L\rangle=|11111\rangle$.  Clearly there is a single qubit encoded
$\sigma_z$ between these qubits which is given by $|0_L\rangle \langle 0_L| -
|1_L\rangle \langle 1_L|= |00000\rangle \langle 00000| - |11111\rangle\langle
11111|$.  Notice however, that this is a five-qubit interaction which we would
not expect to be easily implementable on a system.  On the other hand, one can
also see that a single Pauli $\bmath{\sigma}_z$ acting on a single qubit of
this encoding produces an encoded $\bmath{\sigma}_z$: $\bmath{\sigma}_z^{(1)}
|00000\rangle =|00000\rangle$ and $\bmath{\sigma}_z^{(1)} |11111\rangle =
-|11111\rangle$.

\section{Measurement on DFSs for quantum computation} \label{sec:dfsmeas}

Suppose we are trying to extract information via a measurement which has been
encoded into the degeneracy of the $J^\prime$th irrep of some OSR or SME
algebra ${\tt A}$.  Clearly the measurement of an operator which is this OSR
will not yield any information about the information encoded into the
degeneracy.  This is because these operators all act as identity on the encoded
information and measuring identity gives no information about the encoded
information.

Suppose we wish the measurement operators to preserve the DFS structure of the
encoded information.  In this case, the nontrivial elements of the commutant of
${\tt A}$ provide operators which preserve the DFS structure and return
information about the encoded information.
\begin{lemma}
Let ${\bf M}$ be a hermitian observable which is a member of the commutant of
the OSR or SME algebra ${\tt A}$.  Information which has been encoded into a
irrep of the algebra ${\tt A}$ will remain in the irrep after a measurement of
${\bf M}$.
\end{lemma}

The issue of measurements, however, is again far from contained within elements
of the commutant only.  Just like in the unitary manipulation of DF encoded
information, there are measurements which are not in the commutant which still
preserve the encoded information.

\section{Making leakage into noise}\label{sec:dfsleak}

Finally we would like to address the issue of noise models on a DFS.  In the
standard theory of error correction, one works with operators ${\bf E}$ which
are called the {\em errors}, and represent the action of major component of the
OSR algebra on system evolution.

An important form of noise on a DFS is a leakage error\cite{Lidar:99b}. If we
encode information into the $J^\prime$th irrep of some algebra ${\tt A}$, then
we can classify three types of errors.
\begin{enumerate}
\item Errors which act on the DFS information but preserve the subsystem structure.
These errors act on the $J^\prime$th irrep in a non-trivial manner. If we are
using a given DFS for fault-tolerant error correction, these errors will be the
standard errors which the fault-tolerant error correction serves to fix.
\item Error which preserve the DFS information but act nontrivially otherwise.
These are errors like those generated by the OSR algebra ${\tt A}$.
\item Errors which do not preserve the DFS information.  These errors take
information in a subsystem and leak the information to outside of the
subsystem.  For example information in the $J^\prime$th irrep may be moved to
the $J^{\prime \prime}$th irrep.
\end{enumerate}
If the subsystem structure of the algebra ${\tt A}$ corresponds to
\begin{center}
\begin{tabular}{c|cc|}
\cline{2-3} & & \\
 &
 \begin{tabular}{cc}
 \begin{tabular}{|c|}
 \hline
 ${\tt I}_{n_{J^\prime}}$ \\
 \hline
 \end{tabular} $\otimes$
  &
  \begin{tabular}{|c|}
 \hline
 ${\tt M}_{d_{J^\prime}}$ \\
 \hline
 \end{tabular}
 \end{tabular} &  \begin{tabular}{|c|} \hline 0 \\ \hline \end{tabular}\\
${\tt A} \cong$  & &  \\
 & \begin{tabular}{|c|} \hline 0 \\ \hline \end{tabular} &
 \begin{tabular}{|c|} \hline $\bigoplus_{J\neq J^\prime} {\tt I}_{n_J} \otimes {\tt M}_{d_J}$ \\ \hline \end{tabular} \\
 & & \\
 \cline{2-3}
\end{tabular}
\end{center}
Then the errors detailed above correspond to operators with nonvanishing
element in the following locations
\begin{center}
\begin{tabular}{c|cc|}
\cline{2-3} & & \\
 &
 \begin{tabular}{cc}
 \begin{tabular}{|c|}
 \hline
 1 \\
 \hline
 \end{tabular} $\otimes$
  &
  \begin{tabular}{|c|}
 \hline
 2 \\
 \hline
 \end{tabular}
 \end{tabular} &  \begin{tabular}{|c|} \hline 3 \\ \hline \end{tabular}\\
${\bf E}\cong$  & &  \\
 & \begin{tabular}{|c|} \hline 3 \\ \hline \end{tabular} &
 \begin{tabular}{|c|} \hline 1,2,3 \\ \hline \end{tabular} \\
 & & \\
 \cline{2-3}
\end{tabular}
\end{center}
Of these errors, those in 3 are the most troublesome in the use of a DFS
concatenated within a fault-tolerant quantum error correction procedure.  These
``leakage'' errors, however, do not pose a {\em fundamental} problem for the
theory of fault-tolerant quantum computation\cite{Preskill:98a,Gottesman:98a}.
A particularly nice technique for deal with leakage errors is to simply make
these errors type 1/2 errors.  To do this one makes a measurement which
distinguishes between states in the DFS and states outside of the DFS and then
depending on the outcome takes states outside of the DFS back into states in
the DFS.  Thus it is possible to convert errors which leak out of the subsystem
and make these errors which occur on the subsystem.  For a specific example of
this technique applied to a DFS/quantum error correction scheme see
\cite{Lidar:99b}.

\section{Decoherence-free subsystems as components of a quantum computer}

The purpose of this chapter was to address some of the issues which occur when
attempting to use DFSs in conjunction with the theory of fault-tolerant quantum
computation.  There are no fundamental difficulties in such a melding of DFSs
and fault-tolerant quantum computation.  Much like in the theory of
fault-tolerant quantum computation, however, specific application to a specific
physical system which supports a DFS poses different challenges in melding DFSs
with fault-tolerance.  In the next few chapters we will have the opportunity to
examine a specific physically relevant model of a DFS and thus the results in
this chapter will be directly addressed for this physical model.

\chapter{Collective Decoherence} \label{ch:col}

\begin{quote}
{\em Where not being able to distinguish subsystems is a symmetry}
\end{quote}

In this chapter we introduce an important physical model of decoherence which
supports decoherence-free evolution: collective decoherence.  This model is, in
some sense, a generic model and demonstrates an important symmetry which can be
realized in suitable natural quantum systems.  Due to the physical relevance of
this model, it will be the subject of this thesis in the following four
chapters.  We begin with a non-rigorous discussion of the conditions which lead
to collective decoherence.  We then turn to the example of collective dephasing
and present models of this decoherence process in the Hamiltonian and master
equation formulations.  Specific conditions for collective dephasing are
derived.  We then discuss collective amplitude damping and the conditions under
which such a process occurs.  Finally, we categorize the three different types
of collective decoherence as weak collective decoherence, strong collective
decoherence, and collective amplitude damping.  The DFS structure of each of
these models is then given.

\section{Collective coupling to an environment}

Consider two physical qubits which are situated in close proximity to each
other.  When we think about the environment of these qubits, we are generally
thinking about the environment as the rest of the universe.  Thus even when the
qubits are not in close proximity, the entity of the environment is really the
same for each qubit.  However, as the qubits are brought from close proximity
to large separation, the environments with which each qubit most strongly acts
separate out into two local environments for each qubit.  Physical assumptions
then usually allow us to consider each qubit as coupling strongly to a local
environment and weakly or vanishingly to the other qubits' environment.
Conversely, when the two qubits are situated close together, the environment
which each qubit interacts with is essentially the same environment.

In most physical situations it is impossible to put two physical qubits on top
of each other--especially without these qubits interacting with each other--but
let us imagine for the moment that this is possible. In the limit of qubits on
top of each other and not interacting, we expect each qubit to couple to the
environment in an identical manner.  Now suppose we increase the physical
separation between these qubits.  Clearly the identical manner in which the
qubits couple to the system will now no longer be identical.  The coupling to
the same environment, however, for small enough separation, should still be the
main mechanism of decoherence for these closely spaced qubits. This is {\em
exactly} analogous to the reasoning behind distant qubits having separate local
environments.  We will refer to the situation where each qubit couples in an
identical manner to individual quantum subsystems as the case of {\em
collective decoherence}.

Another way to metaphorically codify the idea of collective coupling is to
think about decoherence as a spying process on the system.  Decoherence is the
process through which the environment becomes entangled with the system and
some of the quantum information of the system is transferred to a joint
system-environment state.  Viewed in this manner, the decoherence process is
the manner in which the environment observer the system.  Now consider the case
of two closely spaced qubits which are being observed by an environment. Since
the qubits are closely spaced, the environment may not be able to distinguish
between each of the qubits when the environment observers (interacts) with the
two qubits.  The inability of the environment to distinguish two or more
closely spaced physical qubits is exactly the case of collective decoherence.

\section{Collective dephasing} \label{sec:coldephase}

Consider the evolution of a system of $n$ qubit coupled to an environment.
These qubits have a natural energy levels and the process of dephasing is the
mechanism through which the populations of these levels do not change but the
coherence between the levels do change.  This setup is most generally
characterized by a system-environment Hamiltonian of the form
\begin{equation}
{\bf H} = \underbrace{\sum_i \omega_i \bmath{\sigma}_z^{(i)}}_{{\bf H}_S} +
{\bf H}_E + \underbrace{ \sum_i \bmath{\sigma}_z^{(i)} \otimes {\bf B}_i}_{{\bf
H}_{SE}},
\end{equation}
where ${\bf H}_E$ is some environment Hamiltonian.  If these energy levels are
identical then $\omega_i=\omega$.  The case of collective dephasing corresponds
to the situation when ${\bf B}_i={\bf B}$ and the energy levels are identical.
This then corresponds to the Hamiltonian
\begin{equation}
{\bf H}= 2\omega {\bf S}_z + {\bf H}_E + {\bf S}_z \otimes 2{\bf B},
\end{equation}
where we have defined
\begin{equation}
{\bf S}_z= \sum_i \bmath{\sigma}_z^{(i)}.
\end{equation}
In the collective dephasing setup, the OSR algebra will be generated by ${\bf
I}$ and ${\bf S}_z$ (and thus consists of all higher powers of ${\bf S}_z$.) We
will discuss the DFSs generated by the collective dephasing model in
Section~\ref{sec:weakdfs}.

Let us introduce a less generic model of the dephasing of qubits which we can
use to make arguments about the situations under which collective decoherence
in the form of dephasing should occur.  Consider a system of $n$ identical
qubits ${\mathcal H}_S=\bigotimes_{i=1}^n \CC^2$ coupled to a quantized field
expressed as a set of harmonic oscillator modes which are the environment
${\mathcal H}_E=\bigotimes_{k} {\mathcal H}_K$ via the Hamiltonian
\begin{equation}
{\bf H}= \sum_i \omega_0 \bmath{\sigma}_z^{(i)} + \sum_k \omega_k {\bf
a}_k^\dagger {\bf a}_k +  \sum_{ik}  \bmath{\sigma}_z^{(i)} \left( g_{ik} {\bf
a}_k + g_{ik}^* {\bf a}_k^\dagger \right),
\end{equation}
where ${\bf a}_k$ (${\bf a}_k^\dagger$) is the annihilation (creation) operator
for the $k$th mode.  The coupling constant $g_{ik}$ will, in general, depend on
the location of the $i$th system.  In many situations it may be possible to
make approximations directly on the coupling constants $g_{ik}$.  The situation
corresponding to collective dephasing is then when the coupling between the
system and the environment is identical for each qubit $g_{ik}=g_k$.  In this
case the Hamiltonian is given by
\begin{equation}
{\bf H}= 2 \omega_0 {\bf S}_z  + \sum_k \omega_k {\bf a}_k^\dagger {\bf a}_k +
 2 {\bf S}_z  \sum_{k} \left( g_{k} {\bf a}_k + g_{k}^* {\bf a}_k^\dagger
 \right).
\end{equation}
To give an idea of when $g_{ki}=g_k$ we recall that the spatial dependence of
$g_{ki}$ is given by a normal mode expansion of the field.  Thus $g_{ki} =
g_k(r_i)$ where $r_i$ is the location of the $i$th qubit and $g_k(r)$ describes
the spatial variation of the $k$th mode.  The condition of $g_{ki}=g_k$ then
corresponds to $g_k(r_i)=g_k(r_j)$ for all $i$ and $j$.  In other words, when
the spacing between the qubits is small enough that the normal mode $k$ does
not vary significantly over the positions of these qubits, collective dephasing
will dominate.  If the normal mode, for example, is a plane wave $g_{ki}=g_k
e^{i \vec{k} \cdot \vec{r}_i}$ and the spacing between the qubits is much less
than the wavelength of this plane wave, $\vec{k} \cdot (\vec{r}_i-\vec{r}_j)
\ll 1$, then $e^{i\vec{k} \cdot \vec{r}_i} \approx e^{i \vec{k} \cdot
\vec{r}_j}$ or $g_{ki} \approx g_k$.

\subsection{Master equation collective dephasing}

In order to obtain the collective dephasing regime, it is necessary that there
be a reason why modes which distinguish between different qubits contribute
little to the dynamics of the system-environment evolution.  In order to
clarify the role of this assumption, we present a derivation of a semigroup
master equation for this Hamiltonian which can help clarify under what
conditions this assumption is a good assumption. This is the semigroup master
equation formulation of collective dephasing\cite{Palma:96a,Duan:98a}.

The first step in the derivation of the master equation is to move into the
interaction picture.  Define ${\bf H}_0=\omega_0\sum_i \bmath{\sigma}_z^{(i)} +
{\omega_k} \sum_k {\bf a}_k^\dagger {\bf a}_k$ and ${\bf H}_I= \sum_{ik}
\bmath{\sigma}_z^{(i)} \left( g_{ik} {\bf a}_k + g_{ik}^* {\bf a}_k^\dagger
\right)$.  Then the interaction picture Hamiltonian is given by
\begin{equation}
{\bf H}_I^\prime(t) = \sum_{ik} \bmath{\sigma}_z^{(i)} \left( g_{ik} {\bf a}_k
e^{-i \omega_k t } + g_{kl}^* {\bf a}_k^\dagger e^{i \omega_k t} \right).
\end{equation}
The $\bmath{\sigma}_z^{(i)}$ each individually commute with this Hamiltonian
and thus the populations of each qubit will be unaffected by the evolution due
to this Hamiltonian.  Under the approximation, the general form of a master
equation is given by
\begin{equation}
{\partial \bmath{\rho} \over \partial t} = -i {\rm Tr}_E \left( \left[ {\bf
H}_I^\prime(t), \bmath{\rho} \otimes \bmath{\rho}_E(0) \right] \right) -
\int_0^t {\rm Tr}_E \left( \left[ {\bf H}_I^\prime(t), \left[ {\bf
H}_I^\prime(\tau), \bmath{\rho} \otimes \bmath{\rho}_E(0) \right] \right] d\tau
\right). \label{eq:master}
\end{equation}
We will make the assumption that the environment is in thermal equilibrium at
temperature $T$ (we set $k_B=1$, $\beta={1 \over T}$).  When we make this
assumption we will refer to the environment as the bath. The bath density
matrix is thus given by \cite{Gardiner:91a}
\begin{equation}
\bmath{\rho}_E(0)={1 \over {\rm Tr} \left[\exp \left( - \beta {\bf H}_B \right)
\right] } \exp \left( - \beta  {\bf H}_B  \right)=\bigotimes_k \int d^2
\alpha_k {1 \over \pi \left<N_{\omega_k} \right>} \exp \left( -
{|\alpha_{\omega_k}|^2 \over \left< N_{\omega_k} \right> } \right) |\alpha_k
\rangle \langle \alpha_k |,
\end{equation}
where $\left< N_{\omega_k} \right>$ is the mean occupation number for mode $k$,
\begin{equation}
\left< N_{\omega_k} \right>= {1 \over \exp \left( \beta \omega_k \right) - 1},
\end{equation}
and $|\alpha_k\rangle$ is a coherent state for the $k$th mode.

The first term in the master equation, Eq.~(\ref{eq:master}), is given by
\begin{eqnarray}
-i {\rm Tr}_E \left( \left[ {\bf H}_I^\prime(t), \bmath{\rho} \otimes
\bmath{\rho}_E(0) \right] \right) &=& -i \sum_{ik} \left[
\bmath{\sigma}_z^{(i)} , \bmath{\rho} \right] \left(  \left< g_{ik} {\bf a}_k
e^{-i \omega_k t} + g_{ik}^*  {\bf a}_k^\dagger  e^{i \omega_k t} \right>_E
\right) \nonumber \\ &=&-i \sum_{ik} \nu_{ik} \left[ \bmath{\sigma}_z^{(i)}
,\bmath{\rho} \right],
\end{eqnarray}
where $\left< {\bf O} \right>_E= {\rm Tr} \left({\bf O} \bmath{\rho}_E(0)
\right)$ and
\begin{equation}
\nu_{ik}= \left< g_{ik} {\bf a}_k e^{-i \omega_k t} + g_{ik}^* {\bf
a}_k^\dagger  e^{i \omega_k t} \right>_E.
\end{equation}
This term vanishes identically for a bath in equilibrium $\nu_{ik}=0$.

The second term in the master equation, Eq.~(\ref{eq:master}), is given by
\begin{eqnarray}
-\int_0^t {\rm Tr}_E \left( \left[ {\bf H}_I^\prime(t), \left[ {\bf
H}_I^\prime(\tau), \bmath{\rho} \otimes \bmath{\rho}_E(0) \right] \right] d\tau
\right)&&=\sum_{ijkk^\prime} \left( \Gamma_{ijkk^\prime}^{(1)} \left[
\bmath{\sigma}_z^{(j)} \bmath{\rho}, \bmath{\sigma}_z^{(i)} \right]  \right.
\nonumber
\\ && \left. + \Gamma_{ijkk^\prime}^{(2)} \left[ \bmath{\sigma}_z^{(i)},\bmath{\rho} \bmath{\sigma}_z^{(j)} \right]
\right),
\end{eqnarray}
where
\begin{eqnarray}
\Gamma_{ijkk^\prime}^{(1)}&=&\int_0^\tau \left <  \left( g_{ik} {\bf a}_k e^{i
\omega_k t} + g_{ik}^* {\bf a}_k^\dagger e^{-i\omega_kt} \right) \left(
g_{jk^\prime} {\bf a}_{k^\prime} e^{i \omega_{k^\prime} \tau} + g_{jk^\prime}^*
{\bf a}_{k^\prime}^\dagger e^{-i\omega_{k^\prime} \tau} \right) \right>_E d
\tau \nonumber \\ \Gamma_{ijkk^\prime}^{(2)}&=&\int_0^t \left <   \left(
g_{jk^\prime} {\bf a}_{k^\prime} e^{i \omega_{k^\prime} \tau} + g_{jk^\prime}^*
{\bf a}_{k^\prime}^\dagger e^{-i\omega_{k^\prime} \tau} \right) \left( g_{ik}
{\bf a}_k e^{i \omega_k t} + g_{ik}^* {\bf a}_k^\dagger e^{-i\omega_kt} \right)
\right>_E d \tau . \nonumber \\
\end{eqnarray}
Using the thermal equilibrium density matrix it is easy to calculate that
\begin{eqnarray}
\Gamma_{ijkk^\prime}^{(1)}&=&\int_0^t \delta_{kk^\prime} \left( g_{ik} g_{jk}^*
\left(\left< N_{\omega_k}\right> +1\right)  e^{i \omega_k(t- \tau)} + g_{ik}^*
g_{jk} \left< N_{\omega_k} \right> e^{-i \omega_k (t-\tau)}\right) d \tau
\nonumber \\
 \Gamma_{ijkk^\prime}^{(2)}&=&\int_0^t \delta_{kk^\prime}
\left( g_{jk} g_{ik}^* \left(\left< N_{\omega_k}\right> +1\right)  e^{-i
\omega_k (t-\tau)} + g_{jk}^* g_{ik} \left< N_{\omega_k} \right> e^{i \omega_k
(t-\tau)}\right) d \tau.
\end{eqnarray}
The evolution is therefore given by
\begin{eqnarray}
{\partial \bmath{\rho} \over \partial t} &=&  \sum_{ij} \Gamma_{ij} \left(
\left[ \bmath{\sigma}_z^{(i)} \bmath{\rho}, \bmath{\sigma}_z^{(j)} \right]+
\left[ \bmath{\sigma}_z^{(i)},\bmath{\rho} \bmath{\sigma}_z^{(j)} \right]
\right)  \label{eq:dephasemaster} \\ \Gamma_{ij}&=& \sum_k \int_0^t \left(
g_{jk} g_{ik}^* e^{-i \omega_k (t-\tau)} + g_{jk}^* g_{ik} e^{i \omega_k
(t-\tau)} \right) \left(2\left< N_{\omega_k}\right> +1\right)  d \tau.
\nonumber
\end{eqnarray}
This dephasing master equation shows how the coefficient matrix $\Gamma_{ij}$
contains information about the correlation of decoherence between different
qubits.

There are two important limits to Eq.~(\ref{eq:dephasemaster}).  In the first
limit, $\Gamma_{ij}=\delta_{ij} \Gamma_i$.  In this case the master equation
can be written as a sum of two Lindblad operators on each qubit
\begin{eqnarray}
{\partial \bmath{\rho} \over \partial t} &=&  \sum_{i} \Gamma_i \left( \left[
\bmath{\sigma}_z^{(i)} \bmath{\rho}, \bmath{\sigma}_z^{(i)} \right]+ \left[
\bmath{\sigma}_z^{(i)},\bmath{\rho} \bmath{\sigma}_z^{(i)} \right] \right).
\end{eqnarray}
This is the case of {\em independent} dephasing.  Each qubit evolves
independent of the evolution of the other qubit.  The other important limit is
when $\Gamma_{ij}$ is constant, $\Gamma_{ij}=\Gamma/4$.  In this case the
master equation contains just one Lindblad operator which acts on all qubits
\begin{eqnarray}
{\partial \bmath{\rho} \over \partial t} &=&  \sum_{i} \Gamma \left( \left[
{\bf S}_z \bmath{\rho}, {\bf S}_z \right]+ \left[ {\bf S}_z,\bmath{\rho} {\bf
S}_z \right] \right).
\end{eqnarray}
This is the case of {{\em collective} dephasing.

In the continuum model where the bath corresponds to some quantized field, we
can make the substitution $\sum_k \rightarrow {V_r \over (2\pi)^r} \int d^r k$
where $r$ is the dimension of the field.  We will examine the case of $r=1$.
The other dimensional cases follow similar lines of investigation.

We assume the coefficients $g_{ki}$ have a spatial relationship $g_{ki}=g(k)
e^{ikr_i}$ where $r_i$ is the position of the $i$th qubit.  Then if the
quantized region is length $L$,
\begin{eqnarray}
\Gamma_{ij}&=&{L \over 2 \pi} \int dk |g(k)|^2 \int_0^t \left(   e^{ i k
(r_j-r_i)-i \omega_k (t-\tau)} +  e^{ik (r_i-r_j)+ i \omega_k (t-\tau)} \right)
\left(2\left< N_{\omega_k}\right> +1\right)  d \tau \nonumber \\ &=&
 {L \over \pi} \int dk { |g(k)|^2 \over \omega_k}     \sin\left[  k
(r_j-r_i)- \omega_k t \right] \left(2\left< N_{\omega_k}\right> +1\right).
\end{eqnarray}
We would like to see what conditions the length scale at which the
approximation $\Gamma_{ij}=\Gamma$ occurs.  Moving into the frequency domain,
we find that
\begin{equation}
\Gamma_{ij}={L \over \pi} \int d\omega {d k \over d \omega} { |g(\omega)|^2
\over \omega} \sin\left[ k(\omega) (r_j-r_i)- \omega t \right] \left(2\left<
N_{\omega}\right> +1\right),
\end{equation}
where we have dropped the superfluous $k$ index on $\omega_k$.  Define the
envelope function  $f(\omega)={d k \over d \omega} { |g(\omega)|^2 \over
\omega}  \left(2\left< N_{\omega}\right> +1\right)$ such that $\Gamma_{ij}= {L
\over \pi} \int d \omega f(\omega) \sin\left[ k(\omega) (r_j-r_i)- \omega t
\right]$.  The function $f(\omega)$ determines which $\omega$ modes contribute
maximally to this integral.  We can split $f(\omega)$ into two contributions
\begin{eqnarray}
f(\omega)&=&f_{T}(\omega)+ f_{V}(\omega) \nonumber \\
 f_T(\omega)&=&2 {d k \over d \omega} { |g(\omega)|^2 \over
\omega}  \left< N_{\omega}\right> \nonumber \\
 f_V(\omega)&=&{d k \over d \omega} {
|g(\omega)|^2 \over \omega}.
\end{eqnarray}
$f_T(\omega)$ represents the thermal contribution to $f(\omega)$ while
$f_V(\omega)$ comes from the vacuum fluctuation contribution to $f(\omega)$.
The thermal contribution to the $f(\omega)$ has a natural cutoff frequency
given by the thermal frequency
\begin{equation}
f_T(\omega)=2 {d k \over d \omega} { |g(\omega)|^2 \over \omega}  {1 \over
e^{\beta \omega} -1}.
\end{equation}
Thus for $\omega \gg T$, $f_T(\omega)$ is exponentially suppressed.  Assuming a
linear dispersion relation ${d k \over d \omega}=c$, if the qubits are spaced
such that $|r_i-r_j| \ll {c \over T}$, then the integral is not exponentially
suppressed in the region where $(r_i-r_j) k \approx 0$.  Thus, if the qubits
are spaced closer than the thermal spacing $l_T={c \over T}$, the thermal
contribution to $f(\omega)$ will contribute $\Gamma_{ij}=\Gamma_T$ independent
of $i$ and $j$.  For a given temperature, there is a spectrum of bath modes
which are occupied.  The temperature then determines the longest wavelength
which has non-negligible occupation and this wavelength then determines the
spacing needed in order to achieve collective dephasing.

The vacuum contribution to $f(\omega)$ however, does not have such an
exponential suppression except as given by the field theory which provides a
coupling constant with a cutoff frequency $g(\omega) \propto \omega^n e^{\omega
\over \omega_c}$.  If the bath field is a phonon field, the natural cut-off can
be identified with the Debye frequency.  In this case an identical argument to
the thermal case gives a characteristic vacuum spacing $l_V={c \over
\omega_c}$.  Qubits spaced closed that this vacuum spacing will dephase
collectively due to the vacuum contribution $f_V(\omega)$.

\section{Collective amplitude damping} \label{sec:colamp}

In the previous section we investigated the situation where no population
transfer occurred on the system's qubits but the phase of the qubits state was
affected.  Let us now examine the situation where population transfer does
occur.

Consider the situation of $n$ qubits coupled to a radiation field.  In the
interaction picture and under the rotating wave approximation, the Hamiltonian
for this system plus environment is given by
\begin{equation}
{\bf H}_I(t)= \sum_{i=1}^n \sum_{\vec{k}} \left[ g_{\vec{k}} e^{-i \vec{k}
\cdot \vec{r}_i -i (\omega_{\vec{k}} - \omega_0) t } \bmath{\sigma}_+^{(i)}
{\bf a}_k + g_{\vec{k}}^* e^{i \vec{k} \cdot \vec{r}_i +i (\omega_{\vec{k}} -
\omega_0) t } \bmath{\sigma}_-^{(i)} {\bf a}_k^\dagger \right],
\label{eq:radcouple}
\end{equation}
where $\bmath{\sigma}_\pm = \bmath{\sigma}_x \pm i\bmath{\sigma}_y$ and
$\omega_0$ is the energy spacing of each qubit.  Under the assumption of
$\vec{k} \cdot (\vec{r}_i-\vec{r}_j) \ll 1$, this Hamiltonian becomes
\begin{equation}
{\bf H}_I(t)= \sum_{i=1}^n \sum_{\vec{k}} \left[ \tilde{g}_{\vec{k}} e^ {-i
(\omega_{\vec{k}} - \omega_0) t } \bmath{\sigma}_+ {\bf a}_k +
\tilde{g}_{\vec{k}}^* e^{i (\omega_{\vec{k}} - \omega_0) t } \bmath{\sigma}_-
{\bf a}_k^\dagger \right],
\end{equation}
where $\tilde{g}_{\vec{k}} = e^{i \vec{k} \cdot r_1} g_{\vec{k}}$.  Notice that
the system operators couple collectively to the bath
\begin{equation}
{\bf H}_I(t)= 2 \sum_{\vec{k}} \left[ \tilde{g}_{\vec{k}} e^ {-i
(\omega_{\vec{k}} - \omega_0) t } {\bf S}_+ {\bf a}_k + \tilde{g}_{\vec{k}}^*
e^{i (\omega_{\vec{k}} - \omega_0) t } {\bf S}_- {\bf a}_k^\dagger \right],
\end{equation}
where
\begin{equation}
{\bf S}_\pm = {1 \over 2} \sum_i (\bmath{\sigma}_x \pm i\bmath{\sigma}_y).
\label{eq:spmdef}
\end{equation}
The OSR algebra for this Hamiltonian under the assumption $\vec{k} \cdot
(\vec{r}_i-\vec{r}_j) \ll 1$ is therefore generated by ${\bf I}$ and ${\bf
S}_\pm$.  We will later return to this situation, which we will label strong
collective decoherence.

\subsection{Master equation collective amplitude damping}

Let us examine the evolution due to the pre-approximated (except the
rotating-wave approximation) Hamiltonian
Eq.~(\ref{eq:radcouple})\cite{Duan:98a}. Using the master equation
Eq.~(\ref{eq:master}) and the assumption that the environment modes are all in
the vacuum state, we can easily obtain the master equation in the interaction
picture as
\begin{equation}
{\partial \bmath{\rho} \over \partial t}= -i \sum_{ij} \epsilon_{ij} \left[
\bmath{\sigma}_+^{(j)} \bmath{\sigma}_-^{(i)} , \bmath{\rho}
\right]+\alpha_{ij} \left(\left[\bmath{\sigma}_-^{(i)}
\bmath{\rho},\bmath{\sigma}_+^{(j)}\right]+\left[\bmath{\sigma}_-^{(i)},
\bmath{\rho}\bmath{\sigma}_+^{(j)}\right] \right),
\end{equation}
where
\begin{eqnarray}
\epsilon_{ij} &=& {1 \over 4} \sum_{\vec{k}} |g_{\vec{k}}|^2 {1 \over \omega_0
- \omega_{\vec{k}}} e^{i \vec{k} \cdot \left(\vec{r}_i - \vec{r}_j \right) }
\nonumber \\
 \alpha_{ij}&=& \sum_{\vec{k}} \pi |g_{\vec{k}}|^2 \delta(\omega_0 -
 \omega_{\vec{k}}) e^{i \vec{k} \cdot (\vec{r}_i -\vec{r}_j) }.
\end{eqnarray}
In the continuum limit, the main contribution to these terms occur at
$\omega_{\vec{k}} = \omega_0$.  Thus in order to attain a collective regime,
the requirement is that
\begin{equation}
\vec{k}_0 \cdot (\vec{r}_i-\vec{r}_j) \ll 1,
\end{equation}
where $\vec{k}_0$ is the wavenumber where $\omega_{\vec{k}_0}=\omega_0$.  Due
to the resonance condition, the conditions for collective amplitude damping are
much easier to describe than those of collective dephasing.  The main pathway
for amplitude damping is exchange of $\omega_0$ energy with the bath and
therefore this dominant pathway provides the condition for collective amplitude
damping.

In the collective regime, the master equation reduces to
\begin{equation}
{\partial \bmath{\rho} \over \partial t}= -i  \epsilon \left[ {\bf S}_+ {\bf
S}_-, \bmath{\rho} \right]+\alpha \left(\left[{\bf S}_- \bmath{\rho},{\bf S}_+
\right]+\left[{\bf S}_-, \bmath{\rho}{\bf S}_+\right] \right),
\end{equation}
where ${\bf S}_\pm$ are defined as in Eq.~(\ref{eq:spmdef}).  Notice that the
Lindblad operator ${\bf S}_-$ here does not include an equivalent ${\bf
S}_-^\dagger={\bf S}_+$ Lindblad operator.  The case of collective amplitude
damping, then, is a case where the SME algebra may give differing DFS
structures than the actual DFS for the master equation.

In both collective dephasing and collective amplitude damping, the fundamental
requirement to enter into these regimes is that the spacing of the qubits be
sufficiently small that the important wavelengths of the interacting baths
cannot distinguish the qubits.  There are other natural situations where
collective decoherence will dominate.  For example if both qubits are coupled
to another quantum system external to the two qubits, the wavelength criteria
need not be met, but only the fact that the two qubits couple identically to
the states of the other system is needed.  The models we have presented in
Sections~\ref{sec:coldephase} and~\ref{sec:colamp} are meant to serve as guides
to finding systems where collective decoherence is exhibited.

\section{Collective decoherence}

In the last two sections we have examined models which exhibit collective
coupling of a system to the environment.  There are three relevant arenas for
this collective coupling which we will label weak collective decoherence,
strong collective decoherence, and collective amplitude damping.  For
completeness, we recall our definition of the collective operators on $n$
qubit,
\begin{equation}
{\bf S}_\alpha=\sum_{i=1}^n {1 \over 2} \bmath{\sigma}_\alpha^{(i)}.
\end{equation}
where $\alpha \in \{x,y,z,+,-\}$.  When we need to refer to the collective
operators on a specific number of qubits, we will do this with a superscript
${\bf S}_\alpha^{[n]}$ is the collective operator on $n$ qubits
\begin{equation}
{\bf S}_\alpha^{[n]}=\sum_{i=1}^n {1 \over 2} \bmath{\sigma}_\alpha^{(i)}.
\end{equation}
These operators form a representation of the Lie algebra $su(2)$, meaning they
satisfy the commutation relations
\begin{equation}
\left[ {\bf S}_\alpha, {\bf S}_\beta \right] = i\epsilon_{\alpha,\beta,\gamma}
{\bf S}_\gamma \quad \alpha,\beta,\gamma \in \{x,y,z\}.
\end{equation}

The three cases of collective decoherence are then specified by
\begin{definition}{\em (Weak collective decoherence)} When the OSR or SME
algebra consists of only a single $\sum_\alpha n_\alpha {\bf S}_\alpha$ ($
n_\alpha \in \RR$, $n_x^2+n_y^2+n_z^2=1$) and identity we call this decoherence
mechanism {\em weak collective decoherence}.  Single qubit rotations are always
possible which take this operator to the operator ${\bf S}_z$.  We will assume
that this has been done and thus weak collective decoherence for our purposes
will be when the OSR or SME algebra consists only of ${\bf S}_z$ and ${\bf I}$.
\end{definition}
\begin{definition}{\em (Strong collective decoherence)} When the OSR or SME
algebra contains all ${\bf S}_\alpha$, $\alpha \in \{x,y,z\}$ and the identity
we call this decoherence mechanism {\em strong collective decoherence}.
\end{definition}
\begin{definition}{\em (Collective amplitude damping)} When the SME contains
only the Lindblad operator ${\bf S}_-$ and a Hamiltonian term ${\bf S}_+ {\bf
S}_-$, we call this decoherence mechanism {\em collective amplitude damping}.
Notice that collective amplitude damping when extended to the full SME algebra
is strong collective decoherence.
\end{definition}

\section{Weak collective decoherence DFSs} \label{sec:weakdfs}

In weak collective decoherence on $n$ qubits, the only nontrivial error
operator is ${\bf S}_z$.  This error operator thus forms an abelian algebra
${\tt A}$ with elements spanned by the set $\{{\bf S}_z^0, {\bf S}_z^1, {\bf
S}_z^2, \dots {\bf S}_z^n\}$.  Due to the fact that ${\bf S}_z$ is hermitian,
there is not difference between the DFSs in the Hamiltonian/OSR treatment and
the SME treatment.  In the first case (OSR) ${\bf S}_z$ will be the system
operator and in the second case (SME) ${\bf S}_z$ will be the sole Lindblad
operator.  Furthermore, because the algebra for the weak collective decoherence
is abelian, the DF structure will be that of DF subspaces.  This is because
abelian algebras all have irreps which are one-dimensional and one-dimensional
irreps simply correspond to DF subspaces (note that the converse is does not
hold. There can be DF subspaces when the algebra is non-abelian.  The algebra
will be abelian over the subspaces, but over the entire space it can be
non-abelian.)

The easiest way to understand the weak collective decoherence DFS is to work in
the basis where ${\bf S}_\alpha$ is diagonalized.  This basis is just the
standard computational basis $|i\rangle=|i_1\rangle \otimes |i_2\rangle \otimes
\cdots \otimes  |i_n\rangle$,
\begin{equation}
{\bf S}_z |i_1\rangle \otimes |i_2\rangle \otimes \cdots \otimes |i_n\rangle =
{1 \over 2} \left( \sum_{l=1}^n (-1)^{i_l}\right) |i_1\rangle \otimes
|i_2\rangle \otimes \cdots \otimes |i_n\rangle.
\end{equation}
Let $H(i)$ denote the Hamming length of number $i$ in binary: $H(i)$ is the
number of $1$'s in the binary expression of $i$.  Then this is just
\begin{equation}
{\bf S}_z |i\rangle = {1 \over 2} \left(n-2H(i)\right)|i\rangle.
\end{equation}
Notice that for a given Hamming distance $H(i)$, the action of ${\bf S}_z$ on
all states with this Hamming distance $H(i)$ is identical.  The DF subspace
criteria is the ${\bf S}_z|\psi\rangle = c|\psi\rangle$ for each of the states
$|\psi\rangle$ in the subspace.  Thus in our case we see that the DF subspaces
correspond to states with equal Hamming weight.
\begin{definition}{\em (Weak collective decoherence DF subspace DFS$_n(H)$)}
Weak collective decoherence on $n$ qubits supports DF subspaces labeled by the
integer $0 \leq h \leq n$, DFS$_n(h)$.  DFS$_n(h)$ is spanned by basis states
in the computational basis $|i\rangle$ which have Hamming weight $H(i)$ equal
to $h$.
\end{definition}
This result follows directly from the DF subspace Hamiltonian and semigroup
master equation criteria.

The dimension of a given DFS$_n(h)$ is given by the number of ways a $n$ bit
number can be written which has a Hamming distance $h$.  This is given by
\begin{equation}
dim( {\rm DFS}_n(h)) =n_h= {n \choose h}. \label{eq:weakdim}
\end{equation}
The largest DFS for a fixed number of qubits then corresponds to the case when
$h={n \over 2}$ when $n$ is even, or $h={n\pm1 \over 2}$ when $n$ is odd.

\subsection{The weak DFS basis}

A complete set of commuting observables for the weak collective decoherence DFS
on $n$ qubits is given by the set of operators $\{{\bf S}_z^{[1]}, {\bf
S}_z^{[2]}, {\bf S}_z^{[3]}, \dots, {\bf S}_z^{[n]}\}$\cite{Kempe:01a}.  The
corresponding basis is then denoted by
$|S_z^{[1]},S_z^{[2]},\dots,S_z^{[n]}\rangle$.  This basis is especially nice
because it allows a for a graphical representation of the DFSs and their basis
states.  We will call this basis the weak DFS basis.
\begin{figure}[h]
\hspace{2.5cm}  \psfig{figure=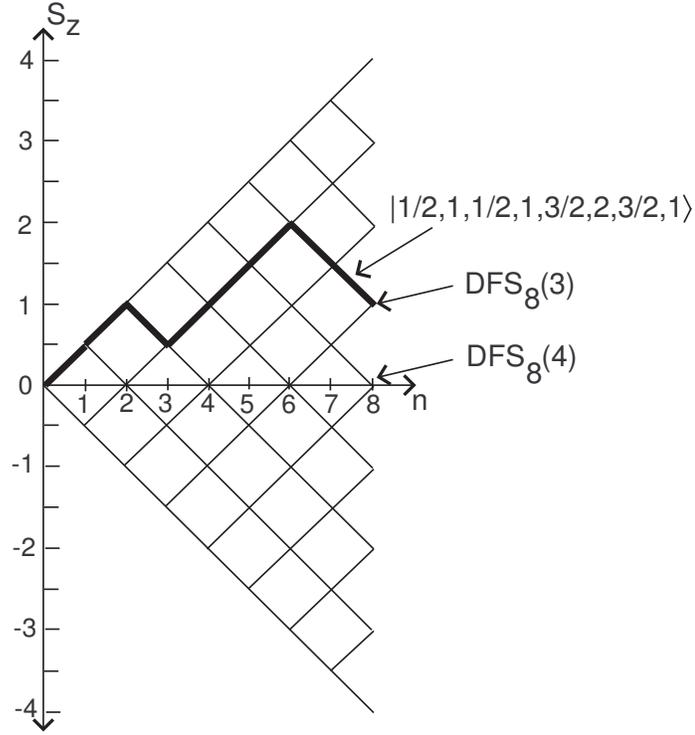,width=4in}  \caption{\em Weak
collective decoherence DFS graphical depiction} \label{fig:weakdfs}
\end{figure}
In Figure \ref{fig:weakdfs}, the horizontal axis marks the number of qubits and
the vertical axis measures the eigenvalue of ${\bf S}_z$.  Each state in the
basis $|S_z^{[1]},S_z^{[2]},\dots,S_z^{[n]}\rangle$ corresponds to a path from
the origin to the given DFS in which only connections which act from left to
right are allowed.

A simple example will help explain our notation.  For $n=3$, there are $4$ DF
subspaces.  These correspond to Hamming distances $h=0$, $h=1$, $h=2$, and
$h=3$.  The basis states for these DFSs in the standard computational basis are
\begin{eqnarray}
{\rm DFS}_3(0)&=& \left \{ |000\rangle  \right., \quad {\rm DFS}_3(1) = \left
\{
\begin{array}{c} |001\rangle \\ |010\rangle \\ |100\rangle \end{array} \right.
, \nonumber \\ {\rm DFS}_3(2)&=& \left \{
\begin{array}{c} |110\rangle \\ |101\rangle \\ |011\rangle \end{array} \right.
, \quad {\rm DFS}_3(3) = \left \{ |111\rangle \right.
\end{eqnarray}
In the weak DFS basis, these states would be denoted by
\begin{eqnarray}
|000\rangle&=&|S_z^{[1]}={1 \over 2},S_z^{[2]}=1, S_z^{[3]}={3 \over 2} \rangle
\nonumber \\
 |001\rangle&=&|S_z^{[1]}={1 \over 2},S_z^{[2]}=1, S_z^{[3]}={1
\over 2} \rangle \nonumber \\
 |010\rangle&=&|S_z^{[1]}={1 \over 2},S_z^{[2]}=0, S_z^{[3]}={1 \over 2} \rangle
\nonumber \\ |100\rangle&=&|S_z^{[1]}=-{1 \over 2},S_z^{[2]}=0, S_z^{[3]}={1
\over 2} \rangle \nonumber \\
 |011\rangle&=&|S_z^{[1]}={1 \over 2},S_z^{[2]}=0, S_z^{[3]}=-{1 \over 2} \rangle
\nonumber \\
 |101\rangle&=&|S_z^{[1]}=-{1 \over 2},S_z^{[2]}=0, S_z^{[3]}=-{1 \over 2} \rangle
\nonumber \\
 |110\rangle&=&|S_z^{[1]}=-{1 \over 2},S_z^{[2]}=-1, S_z^{[3]}=-{1 \over 2} \rangle
\nonumber \\ |111\rangle&=&|S_z^{[1]}=-{1 \over 2},S_z^{[2]}=-1, S_z^{[3]}=-{3
\over 2} \rangle.
\end{eqnarray}

A qutrit of information, for example, can encoded into the DFS$_3(1)$
\begin{eqnarray}
|\psi\rangle &=& \alpha |001\rangle + \beta |010\rangle + \beta |100\rangle
\nonumber \\ &=& \alpha |S_z^{[1]}={1 \over 2},S_z^{[2]}=1, S_z^{[3]}={1 \over
2} \rangle+\beta |S_z^{[1]}={1 \over 2},S_z^{[2]}=0, S_z^{[3]}={1 \over 2}
\rangle \nonumber \\ &&+\gamma |S_z^{[1]}=-{1 \over 2},S_z^{[2]}=0,
S_z^{[3]}={1 \over 2} \rangle,
\end{eqnarray}
and ${\bf S}_z^{[3]}$ acts on $|\psi\rangle$ as a scalar ${\bf S}_z^{[3]}
|\psi\rangle = {1 \over 2} |\psi\rangle$.

Finally in Table~\ref{tab:weakdfs} we assemble the dimension of the weak
collective decoherence DFS.  Notice that these numbers are just Pascal's
triangles.  It is easy then to see the connection between the number of paths
in Figure~\ref{fig:weakdfs} and the degeneracy in Eq.~(\ref{eq:weakdim})
\begin{table}[h]
\begin{center}
\begin{tabular}{c|cccccc}
\hline \hline  $h=6$ &  &  &  &  &  & $1$ \\
 $h=5$ &  &  &  &  & $1$ & $6$ \\
 $h=4$ & &  &  & $1$ & $5$ & $15$ \\
 $h=3$ & &  & $1$ & $4$ & $10$ & $20$ \\
 $h=2$ & & $1$ & $3$ & $6$ & $10$ & $15$ \\
 $h=1$ & $1$ & $2$ & $3$ & $4$ & $5$ & $6$ \\
 $h=0$ & $1$ & $1$ & $1$ & $1$ & $1$ & $1$ \\ \hline& $n=1$ & $n=2$ & $n=3$ & $n=4$ & $n=5$
& $n=6$ \\ \hline \hline
\end{tabular}
\end{center}
\caption{{\em Weak collective decoherence DFS dimensions, given by the
degeneracy $n_h$}} \label{tab:weakdfs}
\end{table}

\section{Strong collective decoherence DFSs} \label{sec:strongdfs}

Strong collective decoherence on $n$ qubits is characterized by the action of
the three operators ${\bf S}_x^{[n]}$, ${\bf S}_y^{[n]}$, and ${\bf
S}_z^{[n]}$.  These operators act as the Lie algebra $su(2)$ and this will help
us to characterize the DFSs arising from these operators.  In particular, the
rules of addition of angular momentum allow us to completely understand the
irreps of the ${\bf S}_\alpha^{[n]}$.  In particular we think of the
computational basis states $|0\rangle$ and $|1\rangle$ as spin-$1/2$ particles
under the mapping $|0\rangle \rightarrow |J={1 \over 2}, m={1 \over 2} \rangle$
and $|1\rangle \rightarrow |J={1 \over 2}, m=-{1 \over 2} \rangle$.

The operators ${\bf S}_\alpha^{[n]}$ do not commute with each other and thus
they cannot be simultaneously diagonalized.  Following standard addition of
angular momentum, we find that the operators
\begin{equation}
\left({\bf S}^{[n]}\right)^2 = \left({\bf S}_x^{[n]}\right)^2+\left({\bf
S}_y^{[n]}\right)^2+\left({\bf S}_z^{[n]}\right)^2
 \quad {\rm and} \quad {\bf S}_z^{[n]}
\end{equation}
do commute.  These two operators do not form a complete basis for the entire
Hilbert space.  Thus for given eigenvalues of these two operators we must
assign a degeneracy index which completes the basis.  By simultaneously
diagonalizing these two operators we have a basis $|J,\lambda,m\rangle$ which
are a representation of $su(2)$,
\begin{eqnarray}
\left( {\bf S}^{[n]} \right)^2 |J,\lambda,m\rangle &=& J(J+1)
|J,\lambda,m\rangle \nonumber \\
 {\bf S}_z^{[n]} |J,\lambda,m\rangle &=& m |J,\lambda, m \rangle.
\end{eqnarray}
Here, $0(1/2) \leq J \leq n/2$ and $-J \leq m \leq J$ and $\lambda$ labels the
degeneracy mentioned above.  In analogy with the addition of angular momentum,
we will of think of the qubits as spin-$1/2$ particles.  $J$ then represents
the total angular momentum of the particles and $m$ labels the projection of
the angular momentum along the $z$-axis.  It is important to realize that the
qubit does not necessarily correspond to a spin-$1/2$ particle in the physical
system.  However, using the language of angular momentum and addition of
spin-$1/2$ particles will simplify our nomenclature significantly.  Using these
basic observations, we can move on to study the irreps of the algebra ${\tt A}$
for strong collective decoherence.

The algebra ${\tt A}$ generated from ${\bf S}_x^{[n]}$, ${\bf S}_y^{[n]}$, and
${\bf S}_z^{[n]}$ plus identity ${\bf I}$ can be decomposed as
\begin{equation}
{\tt A} \cong \bigoplus_{J=0(1/2)}^{n/2} {\tt I}_{n_J} \otimes {\tt M}_{2J+1},
\label{eq:strongalgebra}
\end{equation}
where $J$ labels the total angular momentum of a particular irrep (and hence
the $0$ or $1/2$ depending on whether $n$ is even or odd, respectively), ${\tt
M}_{d}$ is the algebra of all linear operators on a $d$ dimensional space, and
${\tt I}_d$ is the algebra consisting only of the identity operator ${\bf I}$.
$n_J$ is then the degeneracy of the $J$th irrep and $d_J=2J+1$ is dimension of
the $J$th irrep.  The degeneracy of the $J$th irrep is given by
\cite{Mandel:95a}
\begin{equation}
n_J={(2J+1) n! \over (n/2+J+1)! (n/2-J)!}. \label{eq:strongdim}
\end{equation}
Corresponding to the decomposition Eq.~(\ref{eq:strongalgebra}) the action of
the ${\bf S}_\alpha$'s act as
\begin{equation}
{\bf S}_\alpha = \bigoplus_{J=0(1/2)}^{n/2} {\bf I}_{n_J} \otimes {\bf
S}_\alpha(2J+1),
\end{equation}
where ${\bf S}_\alpha(2J+1)$ is the $2J+1$ dimensional representation of
$su(2)$.  Corresponding to this representation is a basis $|J,\lambda,m\rangle$
which is acted upon as ${\bf S}_\alpha |J,\lambda,m\rangle = |J,\lambda\rangle
\otimes {\bf S}_\alpha(2J+1)|m\rangle$.  Notice that this action depends on
which $|J\rangle$ is acted upon, but is independent of the degeneracy index
$\lambda$.

\begin{definition}{\em (Strong collective decoherence DF subsystem DFS$_n(J)$)}
Strong collective decoherence on $n$ qubits supports DFS labeled by the integer
$0(1/2) \leq J \leq n/2$, DFS$_n(J)$.  DFS$_n(J)$ in general has a subsystem
structure.  The states in DFS$_n(J)$ are all eigenstates of $\left({\bf
S}^{[n]} \right)^2$ with eigenvalue $J(J+1)$.  The action of the collective
decoherence operators ${\bf S}_\alpha^{[n]}$ act as representations of $su(2)$
on the eigenstates of ${\bf S}_z^{[n]}$ for a particular total angular momentum
$J$ . Finally, the DFS is realized by the degeneracy of the $J$th irrep.
\end{definition}

The strong collective decoherence DFS, then has information which is encoded
into the degeneracy for a particular irrep label by the total angular momentum
$J$.  In addition of angular momentum, one takes two spin-$J$ and
spin-$J^\prime$ representations of $su(2)$ adds them together to form spin-$K$
representation of $su(2)$.  For the strong collective decoherence DFS, we
perform this addition of angular momentum with spin-$1/2$ particles.  Thus the
degeneracy for a given $J$ is given by {\em the different ways} in which $n$
qubits can be added together under the laws of angular momentum addition such
that the total angular momentum is $J$.

It is useful to present the first few DFS$_n(J)$ states in order to gain some
intuition for what is going on here.  DFS$_1(J)$ consists of only one DFS,
DFS$_1(1/2)$
\begin{eqnarray}
{\rm DFS}_1(1/2)= \left \{\begin{array}{l} |J={1 \over 2},\lambda=1, m={1 \over
2}\rangle = |0\rangle \\
 |J={1 \over 2},\lambda=1, m=-{1 \over 2}\rangle = |1\rangle
 \end{array} \right.
\end{eqnarray}
DFS$_2(J)$ now consists of two DFSs, DFS$_2(1)$ and DFS$_2(0)$,
\begin{eqnarray}
{\rm DFS}_2(1)&=& \left \{ \begin{array}{l} |J=1,\lambda=1,m=1\rangle =
|00\rangle \\ |J=1,\lambda=1,m=0\rangle={1 \over \sqrt{2}} (|01\rangle
+|10\rangle) \\ |J=1,\lambda=1,m=-1\rangle = |11\rangle
\end{array} \right. \nonumber \\
{\rm DFS}_2(0)&=& \{ |J=0,\lambda=1,m=1\rangle = {1 \over \sqrt{2}} (|01\rangle
-|10\rangle).
\end{eqnarray}
Here we see that the DFSs for $n=2$ simply correspond to the singlet and
triplet spaces.  Up to this point, however, there is no degeneracy ($\lambda=1$
for all DFSs).  For $n=3$ however, this changes. At $n=2$ we saw that we had a
singlet and a triplet. When we add a spin-$1/2$ particle to these states we can
produce a $J=1/2$ by either adding to the singlet or subtracting from the
triplet.  Thus we see that there is a degeneracy in the DFS corresponding to
$J=1/2$,
\begin{eqnarray}
{\rm DFS}_3(3/2)= \left \{ \begin{array}{l} |J={3 \over 2},\lambda=1,m={3 \over
2}\rangle = |000\rangle \\ |J={1 \over 2},\lambda =1 , m={1 \over 2} \rangle
={1 \over \sqrt{3}}(|001\rangle+|010\rangle+|100\rangle) \\ |J={1 \over
2},\lambda =1 , m=-{1 \over 2} \rangle ={1 \over
\sqrt{3}}(|110\rangle+|101\rangle+|011\rangle) \\ |J=-{3 \over
2},\lambda=1,m=-{3 \over 2} \rangle =|111\rangle
\end{array} \right. \nonumber \\
{\rm DFS}_3(1/2)=\left \{ \begin{array}{l}
 |J={1 \over 2}, \lambda =1 , m={ 1 \over 2} \rangle = {1 \over \sqrt{2}}
 (|010\rangle -|100\rangle) \\
 |J={1 \over 2}, \lambda =1 ,m=-{1 \over 2} \rangle = {1 \over \sqrt{2}}
 (|011\rangle -|101\rangle) \\
 |J={1 \over 2}, \lambda =2 , m={1 \over 2} \rangle = {1 \over \sqrt{6}}
 (-2|001\rangle +|010\rangle +|100\rangle) \\
 |J={1 \over 2}, \lambda=2, m=-{1 \over 2} \rangle = {1 \over \sqrt{6}}
 (2|110\rangle -|101\rangle -|011\rangle)
 \end{array}
 \right. \label{eq:3strongdfs}
\end{eqnarray}
The states with $\lambda=1$ were obtained by taking a singlet and adding a
single spin-$1/2$ and the states with $\lambda=2$ were obtained by taking a
triplet and subtracting a single spin-$1/2$.  Thus we see that for $n=3$, we
can encoded one qubit of information into the degeneracy index $\lambda$.

\subsection{The strong DFS basis} \label{sec:strongbasis}

The DFS corresponding to different $J$ values for a given $n$ can be computed
using standard methods for the addition of angular momentum\cite{Greiner:92a}.
This can best be illustrated by examining a full basis for the entire Hilbert
space.  The set of operators
\begin{equation}
\{ ({\bf S}^{[1]})^2, ({\bf S}^{[2]})^2, \dots, ({\bf S}^{[n]})^2, {\bf
S}_z^{[n]} \} \label{eq:cscostrong}
\end{equation}
forms a complete set of commuting observables for the Hilbert space of $n$
qubits, $\CC^{2n}$\cite{Kempe:01a}.  Corresponding to this set of observables
is a basis which we will label as
\begin{equation}
|J_1,J_2,\dots,J_{n-1},J_n,m\rangle.
\end{equation}
This basis is acted upon by the complete set of commuting observables in
Eq.~(\ref{eq:cscostrong}) as
\begin{eqnarray}
({\bf S}^{[k]})^2 |J_1,J_2,\dots,J_{n-1},J_n,m\rangle &=& J_k (J_k+1)
|J_1,J_2,\dots,J_{n-1},J_n,m\rangle \nonumber \\ {\bf S}_z^{[n]}
|J_1,J_2,\dots,J_{n-1},J_n,m\rangle&=& m |J_1,J_2,\dots,J_{n-1},J_n,m\rangle.
\end{eqnarray}
We call this basis the strong DFS basis.  We will always assume that the $J_1,
J_2,\dots, J_n$ and $m$ are consistent with the laws of the addition of angular
momentum.

One can understand this basis by thinking of the addition of angular momentum
in a piecewise fashion.  We start with a spin-$1/2$ particle.  Adding another
qubit which is just a spin-$1/2$ particle, we can then create a spin-$1$ or a
spin-$0$ particle.  If we proceed in this manner, for $k$ qubits we may have a
spin-$J$ particle and adding another qubit allows for the creation of
spin-$J+1/2$ or spin-$J-1/2$ (if $J-1/2$ is positive) particles.  This
graphical addition of angular momentum can be easily visualized as in
Figure~\ref{fig:strongdfs} below.
\begin{figure}[h]
\hspace{2.5cm}  \psfig{figure=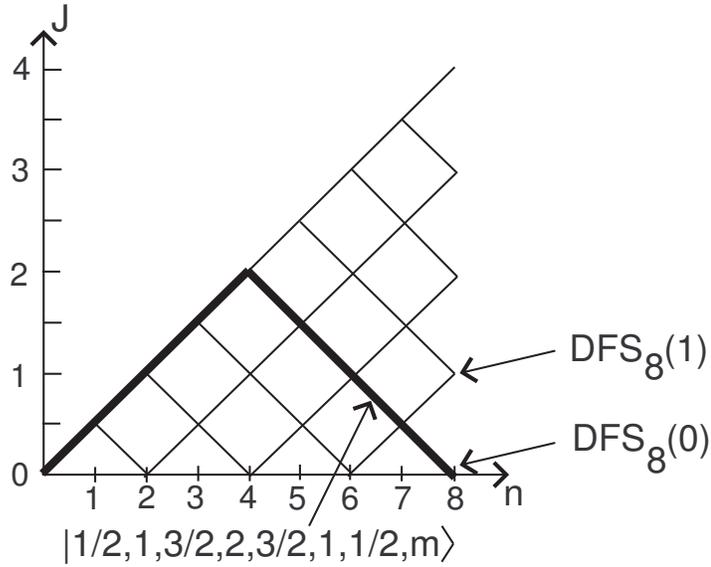,width=4in}  \caption{\em Strong
collective decoherence DFS graphical depiction} \label{fig:strongdfs}
\end{figure}
The horizontal axis of Figure~\ref{fig:strongdfs} is the number of qubits $n$
and the vertical axis is the total angular momentum $J$ obtained by summing
angular momenta of $n$ spin-$1/2$ particles.  Each state in a DFS is
represented by a pathway from the origin always moving from left to right.

Thus we find that the degeneracy $\lambda$ is labeled by the set of pathways
via which one can piecewise construct a given $J$ dimensional representation of
$su(2)$.  Symbolically we might express this as $|J,\lambda,m\rangle = |J,
\lambda=(J_1,J_2,\dots,J_{n-1}),m\rangle$.  When we are talking about a
particular $n$ qubit DFS we will often use the notation $|J_n,\lambda,m\rangle$
to mesh with the strong DFS basis.

Finally we include in Table~\ref{tab:strongdfs} the degeneracy of the $J$th
irreducible representation for $n$ qubits.  The entries of this table are
obtained just as in Pascal's triangle, except half of the triangle is missing
because negative angular momentum $J$ is not allowed.  The entries are exactly
those in Eq.~(\ref{eq:strongdim}).
\begin{table}[h]
\begin{center}
\begin{tabular}{c|cccccc}
\hline \hline  $J={3}$ &  &  &  &  &  & $1$ \\ $J={\frac{5}{2}}$ &  &  &  &  &
$1$ &
\\ $J=2$ &  &  &  & $1$ &  & $5$ \\ $J={\frac{3}{2}}$ &  &  & $1$ &  & $4$ &
\\ $J=1$ & & $1$ &  & $3$ &  & $9$ \\ $J={\frac{1}{2}}$ & $1$ &  & $2$ &  & $5$
&
\\ $J=0$ &  & $1$ &  & $2$ &  & $5$ \\ \hline& $n=1$ & $n=2$ & $n=3$ & $n=4$ & $n=5$
& $n=6$ \\ \hline \hline
\end{tabular}
\end{center}
\caption{{\em Strong collective decoherence DFS dimensions, given by the
degeneracy $n_J$}} \label{tab:strongdfs}
\end{table}

\section{Collective amplitude damping DF subspaces} \label{sec:colampdfs}

Finally let us consider the DFSs for collective amplitude
damping\cite{Duan:98b}. On $n$ qubits, collective amplitude damping consists of
a Hamiltonian evolution ${\bf S}_+^{[n]} {\bf S}_-^{[n]}$ and a collective
annihilation Lindblad operator ${\bf S}_-^{[n]}$.  Using the
$|J,\lambda,m\rangle$ basis from Section~\ref{sec:strongdfs} the action of both
of these operators can be evaluated:
\begin{eqnarray}
{\bf S}_-^{[n]} |J_n,\lambda, m\rangle &=&\sqrt{ (J_n+m)(J_n-m+1)}
|J_n,\lambda,m-1\rangle \nonumber \\ {\bf S}_+^{[n]} {\bf S}_-^{[n]}
|J_n,\lambda,m\rangle &=& (J_n+m)(J_n-m+1) |J_n,\lambda,m\rangle.
\label{eq:campaction}
\end{eqnarray}
As mentioned previously, if we extend these operators to form a
$\dagger$-closed complex associative algebra, we obtain exactly the case of
strong collective decoherence.  Thus it is clear that information encoded into
the degeneracy of the strong collective decoherence DFS can be used to store
information in the collective amplitude damping case.  However we recall that
the condition we used to show the strong collective decoherence DFS was a
sufficient but not necessary condition for the existence of a DFS.

Here, then we will examine the DF subspaces of collective amplitude damping
where we have a criteria which is both necessary and sufficient.  The DF
subspace condition is that the Lindblad operators act as identity on the states
in the DF subspace.  In the case of the collective amplitude damping the
Lindblad operator is only ${\bf S}_-^{[n]}$.  From equation
Eq.~(\ref{eq:campaction}), the only states for which this holds true are the
states $|J_n,\lambda,m =-J_n\rangle$.  In particular we see that
\begin{equation}
{\bf S}_-^{[n]} |J_n,\lambda,m=-J_n \rangle = 0.
\end{equation}
Furthermore, the Hamiltonian term ${\bf S}_+^{[n]} {\bf S}_-^{[n]}$ preserves
this subspace
\begin{equation}
{\bf S}_+^{[n]} {\bf S}_-^{[n]} |J_n,\lambda,m=-J_n\rangle = 0.
\end{equation}

Thus we find that
\begin{definition}{\em (Collective amplitude damping DF subspace DFS$_n$)}
Collective amplitude damping, in addition to supporting the DF subsystem of
strong collective decoherence, supports a DF subspace.  The elements of this
subspace are the states annihilated by the ${\bf S}_-^{[n]}$ operator.  These
states have a projection of the total angular momentum along the $z$-axis which
is negative the total angular momentum of the state.
\end{definition}

Below we list the elements of the collective amplitude damping DFS for between
$1$ and $3$ qubits
\begin{eqnarray}
{\rm DFS}_1 &=& \left \{ |1\rangle\right . \nonumber \\
 {\rm DFS}_2 &=& \left \{ \begin{array}{l} |00\rangle \\ {1 \over \sqrt{2}} (|01\rangle-|10\rangle \end{array}
 \right.\nonumber \\
 {\rm DFS}_3 &=& \left \{ \begin{array}{l}{1 \over \sqrt{6}} (-2|001\rangle +|010\rangle +|100\rangle ) \\
 {1 \over \sqrt{2}} (|011\rangle - |101\rangle) \\ |000\rangle \end{array}\right .
\end{eqnarray}
From this list we find that we can encode a single qubit of information into
two physical qubits.

The dimension of the collective amplitude damping DF subspace is given by
\begin{equation}
n_c= {n \choose \lfloor {n \over 2} \rfloor},
\end{equation}
which can be found by summing the degeneracy of the appropriate strong
collective decoherence DFSs $n_J$.

Since we will not work with universality or quantum computing structures on the
collective amplitude damping DF subspace, we will not construct a nice basis
for this DFS.

\section{Collective decoherence}

In this chapter we have seen how collective coupling of a system to a bath can
occur under reasonably generic conditions.  In latter chapters we will
encounter physical systems which explicitly realize this regime.  The value of
the collective decoherence model, of course, is limited by how realistic
collective coupling is as a source of decoherence.

\chapter{Universality on Collective Decoherence Decoherence-Free Subsystems}
\label{ch:collectiveuniv}

\begin{quote}
{\em Is it possible to compute on collective decoherence decoherence-free
subsystems or are these decoherence-free subsystems useless for quantum
computation?}
\end{quote}

In this chapter we discuss how to use the weak and strong collective
decoherence DFSs for quantum computation.  The first issue we address is
understanding how to perform universal quantum computation on the weak and
strong DFS.  We begin this task by examining the nontrivial one and two qubit
interactions which preserve the relevant DFS structure.  We then discuss
universal control on both the strong and weak collective decoherence DFSs.  A
discussion of the issue of conjoining DFSs then allow us to claim universal
unitary manipulation on the DFSs.  Preparation and measurements on the
collective DFSs is then discussed.  Finally fault-tolerant quantum computation
using concatenated collective DFSs is discussed.

\section{Nontrivial one and two qubit interactions on the collective DFSs}

It is always possible to construct a set of interactions which is universal on
an encoding corresponding to a given DFSs (see Section~\ref{sec:existuniv}.)
For physical reasons, however, we would like to limit the interactions on qubit
subsystems to be either single qubit or multiple qubit operators.

In this section we find the one and two-qubit interactions which are in the
commutant of the relevant algebra ${\tt A}$ for the weak and strong collective
decoherence DFSs.  It will turn out to be sufficient for universality to
examine only elements of the commutant ${\tt A}^\prime$.

\subsection{Weak collective decoherence DFS commutant operations}

Weak collective decoherence on $n$ qubits has an OSR or SME algebra ${\tt A}$
generated by the operations $\{{\bf I},{\bf S}_z^{[n]} \}$.

Consider the single qubit Hamiltonian acting on the $k$th qubit, ${\bf
H}_1^{(k)}=\vec{n} \cdot \bmath{\sigma}^{(k)}$.  Taking the commutator of this
operator with the nontrivial element of ${\tt A}$, we find that
\begin{equation}
\left[ {\bf S}_z^{[n]},{\bf H}_1^{(k)} \right] = 2i \left(n_x
\bmath{\sigma}_y^{(k)} - n_y \bmath{\sigma}_x^{(k)} \right).
\end{equation}
Using the trace-inner product, this implies that the only single qubit
operators which are in the commutant of ${\tt A}$ are the operators $n_z
\bmath{\sigma}_z^{(k)}$.

Consider next a two qubit Hamiltonian acting between the $k$th and $l$th qubit,
${\bf H}_2^{(kl)}= \sum_{\alpha,\beta=1}^3 h_{\alpha \beta}
\bmath{\sigma}_\alpha^{(k)} \bmath{\sigma}_\beta^{(l)}$.  Taking the commutator
of this with ${\bf S}_z^{[n]}$ we find that
\begin{equation}
\left[ {\bf S}_z^{[n]},{\bf H}_2^{(kl)} \right] = 2i \sum_{\beta=1}^3 \left(
h_{1\beta} \bmath{\sigma}_y^{(k)} \bmath{\sigma}_\beta^{(l)} - h_{2\beta}
\bmath{\sigma}_x^{(k)} \bmath{\sigma}_\beta^{(l)}+ h_{\beta1}
\bmath{\sigma}_\beta^{(k)} \bmath{\sigma}_y^{(l)}  - h_{\beta2}
\bmath{\sigma}_\beta^{(k)}
 \bmath{\sigma}_x^{(l)}  \right).
\end{equation}
The first point directly relevant is the $\bmath{\sigma}_z^{(k)}
\bmath{\sigma}_z^{(l)}$ operator commutes with ${\bf S}_z^{[n]}$.  Furthermore
if we collect like terms on the right hand side of the above commutator, and
use the trace inner product we find that we can make the commutator vanish by
setting $h_{12}=-h_{21}$ and $h_{11}=h_{22}$.  Thus the two-qubit operators
which are in ${\tt A}$ are all given by
\begin{equation}
h_{33} \bmath{\sigma}_z^{(k)} \bmath{\sigma}_z^{(l)}+ h_{12}
\left(\bmath{\sigma}_x^{(k)} \bmath{\sigma}_y^{(l)} - \bmath{\sigma}_y^{(k)}
\bmath{\sigma}_x^{(l)} \right) + h_{11} \left(\bmath{\sigma}_x^{(k)}
\bmath{\sigma}_x^{(l)} + \bmath{\sigma}_y^{(k)} \bmath{\sigma}_y^{(l)} \right).
\end{equation}
The most general Hamiltonian on two qubits $i$ and $j$ is then of the form
\begin{equation}
{\bf T}_{ij}(z_{1},z_{2},z_{3},z_{4},h)=\left(
\begin{array}{cccc}
z_{1} & 0 & 0 & 0 \\ 0 & z_{2} & h & 0 \\ 0 & h^{\ast } & z_{3} & 0 \\ 0 & 0 &
0 & z_{4}
\end{array}
\right), \label{eq:tdef}
\end{equation}
where we have expressed the operator in the standard computational
basis\cite{Kempe:01a}.

\subsection{Strong collective decoherence DFS commutant operations}

Strong collective decoherence on $n$ qubits has an OSR or SME algebra ${\tt A}$
generated by $\{{\bf I},{\bf S}_x^{[n]},{\bf S}_y^{[n]},{\bf S}_z^{[n]} \}$.

There are no single-qubit operators in the commutant of ${\tt A}$.  To see this
note that for a single qubit operator $\vec{n}\cdot\vec{\bmath{\sigma}}^{(i)}$,
one can always construct a collective operator $\vec{m} \cdot \vec{\bf
S}^{[n]}$ for some $\vec{m}$ such that $\vec{m} \cdot \vec{n} \neq 0$ and thus
$\left[ \vec{n}\cdot\vec{\bmath{\sigma}}^{(i)}, \vec{m} \cdot \vec{\bf
S}^{[n]}\right]\neq 0$.

For the two-qubit operators, we can immediately reduce the possible commuting
Hamiltonians to the two-qubit operators which are in the commutant for the weak
collective decoherence DFS,
\begin{equation}
{\bf H}^{(kl)}(h_{33},h_{12},h_{11})=h_{33} \bmath{\sigma}_z^{(k)}
\bmath{\sigma}_z^{(l)}+ h_{12} \left(\bmath{\sigma}_x^{(k)}
\bmath{\sigma}_y^{(l)} - \bmath{\sigma}_y^{(k)} \bmath{\sigma}_x^{(l)} \right)
+ h_{11} \left(\bmath{\sigma}_x^{(k)} \bmath{\sigma}_x^{(l)} +
\bmath{\sigma}_y^{(k)} \bmath{\sigma}_y^{(l)} \right).
\end{equation}
Taking the commutator of this operator with ${\bf S}_x^{[n]}$ we find that
\begin{eqnarray}
\left[{\bf S}_x^{[n]},{\bf H}^{(kl)}(h_{33},h_{12},h_{11}) \right] &=& 2i
\left[-h_{33}  \left(\bmath{\sigma}_y^{(k)} \bmath{\sigma}_z^{(l)}  +
\bmath{\sigma}_z^{(l)} \bmath{\sigma}_y^{(k)} \right) \right. \nonumber \\
&&+h_{12} \left( \bmath{\sigma}_x^{(k)} \bmath{\sigma}_z^{(l)} -
\bmath{\sigma}_z^{(k)} \bmath{\sigma}_x^{(l)} \right) \nonumber \\ &&+h_{11}
\left( \bmath{\sigma}_z^{(k)} \bmath{\sigma}_y^{(l)} + \bmath{\sigma}_y^{(k)}
\bmath{\sigma}_z^{(l)} \right) \left. \right] ,
\end{eqnarray}
which vanishes only if $h_{11}=h_{33}$ and $h_{12}=0$.  Thus we see that the
Hamiltonian
\begin{equation}
{\bf H}^{(kl)} = h \left( \bmath{\sigma}_x^{(k)} \bmath{\sigma}_x^{(l)}
+\bmath{\sigma}_y^{(k)} \bmath{\sigma}_y^{(l)} +\bmath{\sigma}_z^{(k)}
\bmath{\sigma}_z^{(l)} \right).
\end{equation}
Including a global phase ${\bf I}$ operator and scaling appropriately, this
operator is the exchange interaction between qubits $i$ and $j$
\begin{equation}
{\bf E}_{ij}= \left( \begin{array}{cccc} 1 & 0 & 0 & 0 \\ 0 & 0 & 1 & 0
\\ 0 & 1 & 0 & 0 \\ 0 & 0 & 0 & 1
\end{array}
\right)= {1 \over 2} \left({\bf I}+ \bmath{\sigma}_x^{(i)}
\bmath{\sigma}_x^{(j)} +\bmath{\sigma}_y^{(i)} \bmath{\sigma}_y^{(j)}
+\bmath{\sigma}_z^{(i)} \bmath{\sigma}_z^{(j)} \right)= {1 \over 2} \left( {\bf
I}+ \vec{\bmath{\sigma}}^{(i)} \cdot \vec{\bmath{\sigma}}^{(j)} \right),
\end{equation}
where we have expressed the exchange operator in a matrix form over the
standard computational basis over the two qubits $i$ and $j$.  The exchange
operator ${\bf E}_{ij}$ exchanges qubits $i$ and $j$: ${\bf E}_{ij}
|\psi\rangle_i |\phi\rangle_j = |\phi\rangle_i |\psi\rangle_j$.

\section{Weak collective decoherence DFS universality}

In this section we discuss universal quantum computation on the weak collective
decoherence DFSs.  Recalling Eq.~(\ref{eq:tdef}), define the operators
\begin{eqnarray}
{\bf T}_{ij}^P&=&{\bf T}_{ij}(1,0,0,0,0)={\rm diag}_{ij}(1,0,0,0) \nonumber \\
 {\bf T}_{ij}^Q&=&{\bf T}_{ij}(0,0,0,1,0)={\rm diag}_{ij}(0,0,0,1) \nonumber \\
 \bar{\bf Z}_{ij}&=&{\bf T}_{ij}(0,0,1,0,0)={\rm diag}_{ij}(0,0,1,0),
\end{eqnarray}
where ${\rm diag}_{ij}(a,b,c,d)=a|00\rangle \langle 00| +b|01\rangle\langle
01|+c|10\rangle \langle 10|+d|11\rangle \langle 11|$ represents the a matrix
with diagonal elements in the standard computational basis. Define the set
\begin{equation}
{\sf H}=\{{\bf E}_{i,i+1},{\bf T}_{i,i+1}^{P},{\bf T}_{i,i+1}^{Q},\bar{{\bf
Z}}_{i,i+1}:i=1,\dots ,n-1\}.
\end{equation}
where ${\bf E}_{i,i+1}$ is the exchange interaction between the $i$ and $i+1$th
qubit.  Notice that this set ${\sf H}$ contains nearest neighbor interactions.
All of the operators in this set are in the commutant of the algebra generated
by $\{ {\bf I}, {\bf S}_z^{[n]}$ and thus preserve the DFS structure of the
weak collective decoherence DFS.

The control afforded over the weak collective decoherence DFSs with
Hamiltonians from ${\sf H}$ is described by the following theorem:
\begin{theorem}\label{th:weakuniv} \cite{Kempe:01a}
For any $n \geq 2$ qubits undergoing weak collective decoherence, the set of
Hamiltonians ${\sf H}$ generates (in the sense of a Lie algebra) a Lie algebra
which acts independently as $su(n_h)$ on DFS$_n(h)$.  If ${\mathcal L}$ denotes
the Lie algebra generated by ${\sf H}$, then
\begin{equation}
{\mathcal L}=\bigoplus_{h=0}^n su\left( {n \choose h} \right).
\end{equation}
To say that the Lie algebra $su(n_h)$ acts independently on DFS$_n(h)$ means
that there are elements in the Lie algebra which act only on DFS$_n(h)$ and
annihilate all other DFS$_n(h^\prime)$, $h\prime \neq h$.
\end{theorem}
Proof: See Appendix~\ref{apb}.

Let us reflect on what this theorem implies.  This theorem tells us that given
control over the Hamiltonians in ${\sf H}$, any unitary action on an encoded
weak collective decoherence DFS can be enacted.   Since these operators are in
the commutant of the weak collective decoherence algebra ${\tt A}$, these
operators are in some sense maximal: they can not mix different DFSs and they
operate as full $su(n_h)$ on the DFSs.

In Chapter \ref{ch:ion} we will have the opportunity to calculate explicit
representations of the gates needed for the physically relevant case of an ion
trap quantum computer.

\subsection{Conjoining weak collective decoherence DFSs and universality}

In order to use weak collective decoherence DFSs for universal quantum
computation, there must be map from the DFSs to the quantum circuit model.  In
particular the mapping from the encoded information to the subsystem structure
of the quantum circuit model must be made.  The fact that weak collective
decoherence is most likely to occur when qubits are closely spaced puts certain
constraints on the subsystem structure.  Suppose we use a weak collective
decoherence DFS on $k$ qubits as our basic subsystem which encodes $d$ qubits
of information.  Notice that the subsystem structure of the physical qubits is
mapped to the subsystem structure of the quantum circuit model in the weak
collective decoherence DFS case.  Theorem~\ref{th:weakuniv} implies that given
the operators in ${\sf H}$ we can construct $su(d)$ operations on these encoded
subsystems.

But what about when we bring the two subsystems together to implement more
complicated gates?  When two encoded subsystems are thus {\em conjoined} we
would like to maintain the DF property of these states.  When we conjoin two
weak collective decoherence DFSs, these states inhabit a DFS of the combined
space. If $h$ is the Hamming weight of the first subsystems DFS and $h$ is also
the Hamming weight of the second subsystems DFS, then the conjoined system
inhabits the $2h$ Hamming weight DFS (more general situations where the DFSs
are of differing Hamming weights follow similar arguments.) Furthermore,
Theorem~\ref{th:weakuniv} tells us that given the operators in ${\sf H}$ we can
perform operations which preserve the $h$ Hamming weight DFS.  Among these
operations are the operations which have an input output property which
individually preserve each $h$  Hamming weight DFS. Thus we can perform
nontrivial operations between the two subsystems which always maintain the
combined ($2h$ Hamming weight) weak collective decoherence DFS.

Thus we see that Theorem~\ref{th:weakuniv} allows for universal quantum
computation on subsystems while maintaining the DF condition under the caveat
that conjoined subsystems must also be DF.  Since weak collective decoherence
is conditioned on the close spacing of the qubits, one would therefore expect
that subsystems involving the smallest number of qubits would be used in such a
conjoining scheme.

\section{Strong collective decoherence DFS universality}

The following theorem demonstrates how the exchange interaction can be used for
quantum computation on the strong collective decoherence DFS:
\begin{theorem} \label{th:stronguniv} \cite{Kempe:01a}
For any $n\geq 2$ qubits undergoing strong collective decoherence let
${\mathcal S}$ be the set of exchange Hamiltonians ${\bf E}_{ij}$ acting
between qubits $i$ and $j$.  The Lie algebra generated by ${\mathcal S}$
contains the ability to perform $su(n_J)$ independently on the degeneracy of
every irrep $J$.  If ${\mathcal L}$ is the Lie algebra generated by ${\mathcal
S}$ then
\begin{equation}
{\mathcal L} \cong \bigoplus_{J=0(1/2)}^{n/2} su(n_J) \otimes {\mathcal
I}_{d_J},
\end{equation}
where ${\mathcal I}_{d_J}$ represents an identity operator on the $d_J$
dimensional irrep space.  The ability to perform each $su(n_J)$ independently
means that the Lie algebra contains elements which act nontrivial on the $J$th
irrep but annihilate all states outside of this irrep.
\end{theorem}
Proof: See Appendix~\ref{apc}.

This theorem implies that the only interaction needed to perform computation on
the strong collective decoherence DFS is the exchange interaction.  Like the
weak collective decoherence case, this result is in some sense maximal: the
operations do not mix DFSs but act fully on the DFS encoded information.

This remarkable theorem implies that quantum computation can be performed with
only the exchange Hamiltonian between qubits.  In Chapter~\ref{ch:exchange} we
will have the opportunity to give explicit gate constructions in the context of
a solid-state exchange based quantum computer.

\subsection{Conjoining strong collective decoherence DFSs for universality}

Conjoining strong collective decoherence DFSs is slightly more complicated than
in the weak collective decoherence case because the DFSs are now subsystems and
not subspaces.  Suppose we use a strong collective decoherence DFS on $k$
qubits as our basic subsystem which encodes $d$ qubits of information.  On
these $k$ qubits suppose we encode into the subsystem with total angular
momentum $J$.  Theorem~\ref{th:stronguniv} implies that the exchange
Hamiltonian can be used to perform any encoded $su(d)$ on each of these
individual subsystems.

When the two subsystems are conjoined, the resulting states inhabit many
different irreps of the conjoined system.  This can be understood via the rules
of addition of angular momentum.  If two $J$ irreps are conjoined, then the
resulting system will have support over irreps on the conjoined system with
total angular momentum $J^\prime=0, J^\prime=1, \dots, J^\prime=2J$.
\begin{equation}
|J,\lambda_1,m_1\rangle \otimes |J,\lambda_2,m_2\rangle = \sum_{J^\prime
=0}^{2J} \sum_{m_{12}=-J^\prime}^{J^\prime} c_{J^\prime,m_{12}} |J^\prime,
\lambda_{12},m_{12}\rangle,
\end{equation}
where $\lambda_1$ and $\lambda_2$ label the degeneracies of the individual
subsystem and $\lambda_{12}$ denotes the total degeneracy when the subsystems
are conjoined. In particular $|\lambda_{12}\rangle$ contains the tensor product
of $|\lambda_1 \rangle$ and $|\lambda_2\rangle$.  If we let ${\mathcal
H}_{\lambda_{12}}$ denote the Hilbert space of this degenerate information,
then
\begin{equation}
{\mathcal H}_{\lambda_{12}} \cong  \left( {\mathcal H}_{\lambda_1} \otimes
{\mathcal H}_{\lambda_2} \right) \oplus {\mathcal H}_{\lambda_{12}^\prime},
\end{equation}
where ${\mathcal H}_{\lambda_i}$ contains the information in the $i$th
degeneracy and ${\mathcal H}_{\lambda_{12}^\prime}$ denotes all of the other
degeneracies.  Via Theorem~\ref{th:stronguniv} we can now perform any unitary
manipulation on each of the subsystems.  Thus we can perform operations which
act as operations whose final result is an operation on ${\mathcal
H}_{\lambda_1} \otimes {\mathcal H}_{\lambda_2}$.  This will represent an
encoded action between the encoded subsystem.

It is important to note that while the two subsystem are conjoined, strong
collective decoherence errors will affect the different DFSs indexed by
$J^\prime$.  This decoherence can distinguish between the different DFSs and
thus it might appear that this would lead to problems for the conjoined
information.  To see that this is not a problem, one notes that the actions
which distinguish between each of the different $J^\prime$ only act to change
the manner in which the conjoining is achieved.  During the course of an
operation on two conjoined DFSs, strong collective decoherence errors act on
the $|m_{12}\rangle$ index.  When the action on each of the $J^\prime$ irreps
on the ${\mathcal H}_{\lambda_1} \otimes {\mathcal H}_{\lambda_2}$ components
is identical, however, the effect of these errors only serve to perhaps
entangle the $|m_1\rangle$ and $|m_2\rangle$ degrees of freedom.

\section{Weak collective decoherence DFS preparation and measurement}
\label{sec:conjweak}

In order to make use of a weak collective decoherence DFS for quantum
computation, we must, in addition to the universal manipulations described
above, be able to prepare and measure the states in the DFS.

Here we would like to note that it is not necessary to prepare states that have
support exclusively within the DFS, i.e. that have no component outside of the
DFS.  This follows from the fact that in our construction, while a computation
is performed, there is no mixing of states inside and outside of the DFS.  If
an initially prepared state is ``contaminated'' (has support outside of the DFS
we want to compute on), then the result of the computation will have the same
amount of contamination, i.e. the initial error does not spread.

For example, suppose we can prepare the state $\bmath{\rho}=(1-p)|\psi\rangle
\langle \psi | + p |\psi_\perp\rangle \langle \psi_\perp|$ where $|\psi\rangle$
is a state of a particular DFS and $|\psi_\perp\rangle$ is a state outside of
the DFS.  Computation on the DFS will proceed independently on the DFS and the
states outside of the DFS.  Readout will then obtain the result of the
computation with probability $(1-p)$.  Repeated application can then be used to
magnify this computation.  Thus perfect preparation is not a strict
requirement. Preparation which is not perfect, however, will hinder the quantum
computer and thus it is desirable to be able to prepare DFS states.

For weak collective decoherence DFSs preparation of initial pure state is
rather simple.  Purse state preparation into a DFS with a Hamming weight $h$
corresponds to the preparation of a state with a specific number of $|0\rangle$
and $|1\rangle$ (eigenstates of the $\bmath{\sigma}_z$).  This can be easily
accomplished if measurements in the $\bmath{\sigma}_z^{(i)}$ basis are possible
as well as the ability to perform $\bmath{\sigma}_x^{(i)}$ gates (to ``flip''
the bits).

The second crucial ingredient for computation on a DFS (in addition to
preparation) is the decoding or readout of quantum information resulting from a
computation. Once again, there are many options for how this can be performed.
For example, in the weak collective decoherence case one can make a measurement
which distinguishes all of the DFSs and all of the states within this DFS by
simply making a measurement in the $\sigma_z$ basis on every qubit.  Further,
all measurements with a given number of distinct eigenvalues can be performed
by first rotating the observable into one corresponding to a measurement in the
computational basis (which, in turn, corresponds to a unitary operation on the
DFS) and then performing the given measurement in the $\sigma_z$ basis, and
finally rotating back. There are other situations where one would like to, say,
make a measurement of an observable over the DFS which has only two different
eigenvalues. This type of measurement can be most easily performed by a
conjoined measurement \cite{Bacon:00a}. In this scheme, one attaches another
DFS to the original DFS, forming a single larger DFS. Then, assuming universal
quantum computation over this larger DFS one can always perform operations
which allow a measurement of the first DFS by entangling it with the second
DFS, and reading out (destructively as described for the weak collective
decoherence case above) the second DFS.

For example, suppose the first DFS encodes two bits of quantum information,
$|k,l\rangle_L$, $k,l=\{0,1\}$, and the second DFS encodes a single bit of
quantum information $\{|0\rangle_L$, $|1\rangle_L\}$. Then one can make a
measurement of the observable $\sigma_z \otimes {\bf I}$ on the first DFS by
performing an encoded controlled-{\sc NOT} operation between the first and the
second DFS, and reading out the second DFS in the encoded $\sigma_z$ basis. For
the weak collective decoherence case the ability to make this destructive
measurement on the ancilla (not on the code) simply corresponds to the ability
to measure single $\sigma_z$ operations.
\begin{figure}[h]
 \psfig{figure=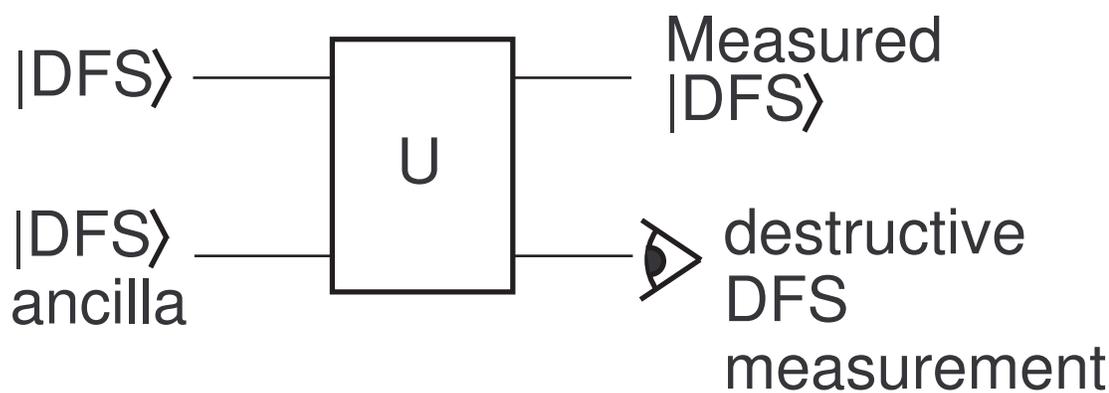,width=6in}
 \caption{\em The conjoined measurement scheme}  \label{fig:conjmeas}
\end{figure}

Finally, we note that for a weak collective decoherence DFS there is a
destructive measurement which distinguishes between different DFSs
(corresponding to a measurement of the number of $|1\rangle$'s). One can
fault-tolerantly prepare a weak collective decoherence DFS state by repeatedly
performing such a measurement to guarantee that the state is in the proper DFS.
The conjoined measurement procedures described above for any DFS are naturally
fault-tolerant in the sense that they can be repeated and are non-destructive
\cite{Gottesman:97a,Bacon:00a}. Thus fault-tolerant preparation and decoding is
available for the weak collective decoherence DFS.

\section{Strong collective decoherence DFS preparation and measurement}

At first glance it might seem difficult to prepare pure states of a strong
collective decoherence DFS, because these states are nontrivially entangled.
However, it is easy to see that every DF {\em subspace} contains a state which
is a tensor product of singlet states:
\begin{equation}
|0_D\rangle = \left({1 \over \sqrt{2}} \right)^{n/2} \otimes_{j=1}^{n/2} (|01
\rangle -|10\rangle),
\end{equation}
because these states have zero total angular momentum. Thus a supply of singlet
states is sufficient to prepare DF subspace states. Further, DF {\em
subsystems} always contain a state which is a tensor product of a DF subspace
and a pure state of the form $|1\rangle \otimes \cdots \otimes |1\rangle$. This
can be seen from Figure~(\ref{fig:strongdfs}), where the lowest path leading to
a specific DFS$_n(J)$ is composed of a segment passing through a DF subspace
(and is thus of the form $|0_D\rangle$), and a segment going straight up from
there to DFS$_n(J)$. The corresponding state is equivalent to adding a spin-$0$
(DF subspace) and a spin-$J$ DF subsystem (the $|J,m_J=J\rangle$ state of the
latter is seen to be made up entirely of $|1\rangle \otimes \cdots \otimes
|1\rangle$). In general, addition of a spin-$0$ DFS and a spin-$J$ DFS simply
corresponds to tensoring the two states. Note, however, that addition of two
arbitrary DF subsystems into a larger DFS is not nearly as simple:
concatenation of two $J\neq0$ DFSs does {\em not} correspond to tensoring.

Pure state preparation for a strong collective decoherence DFS can thus be as
simple as the ability to produce singlet states and $|1\rangle$ states (it is
also possible to use the $|J,m_J=-J\rangle=|0\rangle \otimes \cdots \otimes
|0\rangle$ or any of the other $|J,m_J\rangle$ states plus singlets). Other,
more complicated pure state preparation procedures are also conceivable, and
the decision as to which procedure to use is clearly determined by the
available resources to manipulate quantum states. The pure state preparation of
singlets and computational basis states has the distinct advantage that
verification of these states should be experimentally achievable. Such
verification is necessary for fault-tolerant preparation \cite{Gottesman:97a}.

Measurements on the strong collective decoherence DFS can be performed by using
the conjoined measurement scheme detailed in the weak collective decoherence
DFS discussion in Section~\ref{sec:conjweak}.  In particular, by attaching a
strong collective decoherence DF {\em subspace} ancilla via such conjoining,
one can construct any conjoined measurement scenario.  All that remains to be
shown is how to perform a destructive measurement on such an ancilla.

One way to perform a destructive measurement on the $n=4$ strong collective
decoherence DF subspace was presented in \cite{Bacon:00a} (for another see
\cite{DiVincenzo:99a}).  This scheme involves measuring $\bmath{\sigma}
_{z}^{(1)}$, $\bmath{\sigma}_{z}^{(2)}$, $\bmath{\sigma}_{x}^{(3)}$,
$\bmath{\sigma}_{x}^{(4)}$ on the four qubits.  To see how this measurement
works, note that the four qubit strong DF subspace is spanned by the two states
\begin{eqnarray}
|0_L\rangle &=& {1 \over 2} \left( |01\rangle -|10\rangle \right)
\left(|01\rangle -|10\rangle \right) \nonumber \\
 |1_L\rangle &=& {1 \over \sqrt{12}} \left( 2 |0011\rangle +2 |1100\rangle
 -|0101\rangle -|1010\rangle -|0110\rangle -|1001\rangle \right).
\end{eqnarray}
If measurement of $\bmath{\sigma}_z^{(1)}$ and $\bmath{\sigma}_z^{(2)}$ yields
$|00\rangle$ or $|11\rangle$, then one declares that the state must be
$|1_L\rangle$.  If however, the measurement yields, $|01\rangle$ or
$|10\rangle$, then the remaining two qubits are in the states
\begin{eqnarray}
|0_L\rangle &\rightarrow& {1 \over \sqrt{2}}( |01\rangle -|10\rangle ) = {1
\over \sqrt{2}} (|-+\rangle - |+-\rangle )\nonumber \\
 |1_L\rangle &\rightarrow& {1 \over \sqrt{2}}
 (|01\rangle +|10\rangle)= {1 \over \sqrt{2}} (|++\rangle - |--\rangle ).
\end{eqnarray}
Where we have rewritten the states in the eigenstates of $\bmath{\sigma}_x$:
$\bmath{\sigma}_x|\pm\rangle =\pm |\pm \rangle$.  Measurement of
$\bmath{\sigma}_x$ on the remaining qubits then destructively distinguishes
between $|0_L\rangle$ and $|1_L\rangle$.

Further, we note that the ability to perform a conjoined measurement scenario
by conjoining an ancilla DFS composed of a single encoded-qubit, can be used to
perform  any possible conjoined DFS measurement scenario. As mentioned in the
weak collective decoherence case, the conjoined measurement procedures are
fault-tolerant. Thus we have shown how to perform fault-tolerant preparation
and decoding on the strong collective decoherence DFS.

\section{Fault-tolerant quantum computation and collective decoherence DFSs}

So far we have shown how to implement universal computation with local
Hamiltonians on a collective DFS corresponding to a single block of qubits.
This construction assumes that the only errors are collective. This is a very
stringent symmetry requirement, which obviously becomes less realistic as the
number of particles $n$ increases significantly.  It is thus necessary to be
able to deal with perturbations that break the collective-decoherence
(permutation)\ symmetry.  To deal with these perturbations we will have to use
a quantum error correcting code (QECC).  This quantum error correcting code
will work on the encoded DFS information.  We then say that the DFSs are
concatenated into a QECC.

One particular realization of this concatenation scheme was proposed in
\cite{Lidar:99b}.  In \cite{Lidar:99b} DFS blocks of four particles (each block
constituting a single encoded qubit) into a QECC.  The QECC in the outer layer
then takes care of any single encoded-qubit errors on each of its constituent
DFS-blocks.  By choosing an appropriate QECC it is thus possible to deal with
the appropriate type of non-collective error on the encoded DFS-qubits.  More
generally any dimensional collective DFS can be concatenated into a
fault-tolerant QECC scheme.  In the previous sections we have shown how to
manipulate this information, how to fault-tolerantly measure the information,
and how to prepare the information.

One issue arising with concatenation which we have not yet addressed is the
ability to fault-tolerantly detect leakage errors on a DFS. Concatenation
resulting in unreliable leakage detection would be useless. However, this is
not a problem here, since detection can easily be performed when one has the
ability to make some fault-tolerant measurements on the DFS and also to perform
universal manipulations over any combination of DFS states.  Both of these are
valid with the DFS-QECC concatenation, as we have summarized above. In
particular, it is always possible to measure the relevant observables for
leakage by (i) attaching ancilla encoded DFS states, (ii) performing the
leakage syndrome detection routine onto the ancilla states, and (iii)
fault-tolerantly measuring this ancilla (\cite{Bacon:00a,Kempe:01a}).

We re-emphasize that the fault-tolerance in our proposed scheme is not solely a
result of properties of decoherence-free subsystems.  Decoherence-free
subsystems must be combined with quantum error correcting codes to achieve full
fault-tolerant quantum computation.

\section{Collective decoherence and quantum computation}

In this chapter we have seen how collective decoherence DFSs can be used as a
quantum computer.  Of particular importance was the discovery that one and
two-body interactions are sufficient for universal quantum computation on the
encoding corresponding to the DFS.  Furthermore, realistically implementable
preparation and measurement scenarios were put forth.  Thus we see that under
some fairly non-stringent conditions collective decoherence DFSs can be put to
use to build a quantum computer.  This being said, the actual details of the
implementations in physical systems will have many important issues of actual
execution of the tasks we have described in this chapter.  In the following two
chapters we detail some of the details of the using collective decoherence DFSs
in specific physical systems.

\chapter{The Weak Collective Decoherence Ion Trap Quantum Computer}
\label{ch:ion}

The first physical realization of a decoherence-free subspace under ambient
conditions (i.e. naturally occuring decoherence) was realized in a trapped ion
experiment performed by a group at NIST\cite{Kielpinski:01a,Kielpinski:01b} in
2001. In this chapter we discuss how to perform universal quantum computation
on the ion trap DFS of this experiment. The ion trap DFS corresponds of
\cite{Kielpinski:01a,Kielpinski:01b} is the weak collective decoherence DFS. In
this chapter we discuss how to perform universal quantum computation on
clusters of these two qubit DFSs within the context of an ion trap multi-qubit
manipulation scheme proposed by S{{\o}}rensen and
M{{\o}}lmer\cite{Sorensen:99a,Molmer:99a,Sorensen:00a}.  This is an important
concrete application of the concepts presented in previous sections for
universal quantum computation on a DFS.

\section{The ion trap quantum computer}

Ion traps are among the leading architectures for a future quantum computer. In
the ion trap quantum computing architecture multiple ions are confined strongly
in two directions ($x$ and $y$) compared to the confinement along a third
direction ($z$).  With few numbers of ions confined into an appropriate trap,
the ions form a linear chain.  The physical qubits of an ion trap quantum
computer are associated with internal quantum numbers for each ion (usually
hyperfine levels).  The internal state of the ions can be prepared using
optical pumping and highly efficient readout of the qubit state can be achieved
via electron shelving\cite{Nagourney:86a,Sauter:86a,Bergquist:86a}.

The first proposal for an ion trap quantum computer was the proposal of Cirac
and Zoller\cite{Cirac:95a}.  These authors showed how to use the collective
center of mass motion of the trapped ions as a logical bus state for enacting a
nontrivial quantum operation between the internal states of two ions.  Combined
with single qubit gates on the qubits and the preparation and readout mentioned
above, this showed that ions traps could in principle realize all of the
components needed for quantum computation.  In order to make this architecture
scalable, some method of moving ions between traps\cite{Wineland:98a} or of
coupling multiple traps together\cite{Cirac:97a,Pellizzari:97a,vanEnk:97a} must
be added onto this basic scheme.

Much progress has been made in the experimental demonstration of ion traps as
coherent manipulators of quantum information culminating with the recent
demonstration of an entangled state of four ions\cite{Sackett:00a}.  The reader
is referred to \cite{Steane:97b,Wineland:98a} for a review of some of the
basics of the ion trap quantum computer.

\section{The ion trap DFS}

Among the particular achievements of ion trap quantum computing is the recent
demonstration of a DF subspace of two ions\cite{Kielpinski:01a,Kielpinski:01b}.
In the experiment described in \cite{Kielpinski:01a,Kielpinski:01b} a single
ion was initially prepared in the state $|\psi\rangle={1 \over
\sqrt{2}}\left(|0\rangle +e^{i\phi}|1\rangle\right)$.  The physical qubits
$|0\rangle$ and $|1\rangle$ in this experiment corresponded to the $F=2$,
$m_F=-2$ and $F=1$, $m_F=-1$ sublevels of the $^2S_{1/2}$ ground state of a
$^9$Be$^+$ ion.  A two qubit interaction (of the form described in
Section~\ref{sec:sormol} below) was then applied to this single qubit state and
a prepared $|0\rangle$ state of a second ion.  This two qubit interaction has
the effect of moving the information in the single qubit to a two qubit
encoding
\begin{equation}
|0\rangle \otimes {1 \over \sqrt{2}}\left(|0\rangle+e^{i\phi}|1\rangle\right)
\rightarrow {1 \over \sqrt{2}}\left( |\psi_-\rangle+e^{i\phi}|\psi_+\rangle
\right),
\end{equation}
where $|\psi_\pm\rangle = {1 \over \sqrt{2}}\left(|01\rangle \pm i|10\rangle
\right)$.  Note that $|\psi_\pm\rangle$ span the same space as $|01\rangle$,
$|10\rangle$.  The state then has been encoded into the weak collective
decoherence DFS$_2(1)$.  In the ambient conditions experiment, this state was
then allowed exposure to the environment and then the reverse encoding
procedure was applied and the state of the qubit was read out.  A similar
experiment with no encoding and decoding but with preparation into the state
$|0\rangle \otimes {1 \over \sqrt{2}}\left(|0\rangle+e^{i\phi}|1\rangle\right)$
was also performed.  From these two experiments, the decoherence time without
encoding was $(7.9 \pm 1.5) {\rm ns}$ while the decoherence time with the DFS
encoding was $(2.2 \pm 0.3) {\rm ns}$. This, then, clearly demonstrates how DF
coding can result in protection of quantum information from decoherence.
Furthermore, the decoherence rates in this experiment were severely limited by
the fidelity of the encoding, decoding, and preparation mechanisms.  Thus it
appears that the limiting decoherence rate attained with the DFS encoding is
mostly the result of the heating of the trap.  This heating is not seen as a
fundamental obstacle to ion trap quantum computing\cite{Wineland:98a} but has
so far defied identification.

We would like to address the issue of how to use the DFS encoded states for
quantum computing in the ``quantum CCD'' model of an ion trap quantum
computer\cite{Kielpinski:01b}.  In the quantum CCD model a large trap with many
independent microtraps is envisioned.  The ions in the microtraps can perform
local quantum computations (using the microtraps vibrational modes for the
computation) and the ions from the microtraps can be moved in between
individual microtraps to realize the quantum circuit model.  A particularly
nice feature of encoding into the DFS states in the quantum CCD model is that
spatial variation of magnetic fields will not dephase the ions as they are
moved around between the microtraps.  The question which arises when using the
two qubit weak collective decoherence DFS for ion trap quantum computing is how
to perform manipulations of the information stored in the traps without leaving
the DFS.  We will call a quantum computer based around the encoded weak
collective decoherence DFS the {\em ion trap DFS quantum computer}.
\begin{figure}[h]
 \psfig{figure=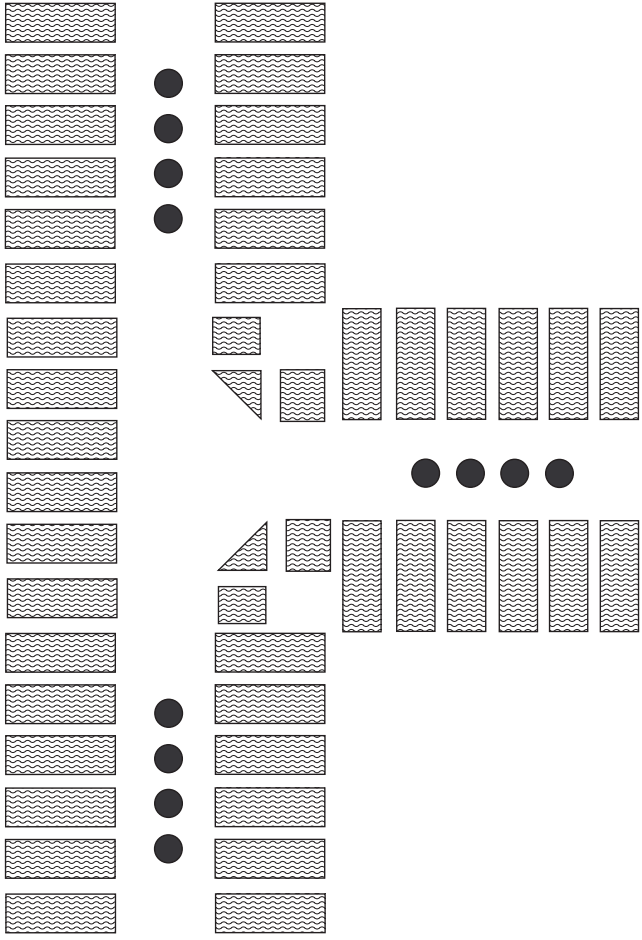,width=5in,angle=270}
\vspace{0.5cm} \caption{\em The quantum CCD ion trap quantum computer}
\label{fig:qccd}
\end{figure}

\section{The S{\o}rensen and M{\o}lmer quantum gates} \label{sec:sormol}

The four ion entanglement experiment\cite{Sackett:00a} and the two ion DFS
experiment\cite{Kielpinski:01a,Kielpinski:01b} both used a method for
manipulating trapped ions which was devised by S{\o}rensen and M{\o}lmer.  This
scheme\cite{Sorensen:99a,Molmer:99a,Sorensen:00a} is an improvement over the
proposal given by Cirac and Zoller\cite{Cirac:95a} in that it does not require
that the vibrational state of the center of mass of the ions be cooled to the
ground state.  The S{\o}rensen and M{\o}lmer scheme is fairly insensitive to
the occupations of the vibrational states of the trap.  In this section we
present an overview of the S{\o}rensen and M{\o}lmer scheme and demonstrate how
it can be used to enact two different operations which we will then use to show
how to perform universal quantum computer on the ion trap DFS quantum computer.

$n$ ions in a linear trap interacting with a laser field of frequency $\omega$
are described by the Hamiltonian
\begin{equation}
{\bf H}={\bf H}_0+{\bf V}(t),
\end{equation}
where
\begin{eqnarray}
{\bf H}_0 &=& \nu {\bf a}^\dagger {\bf a} + {\omega_0 \over 2} \sum_i
\bmath{\sigma}_z^{(i)} \nonumber \\ {\bf V}(t)&=&\sum_i {\Omega_i \over 2}
\left( \bmath{\sigma}_+^{(i)} e^{i \eta_i({\bf a}+{\bf a}^\dagger)-i \omega t
+i \phi} + \bmath{\sigma}_-^{(i)} e^{-i \eta_i({\bf a}+{\bf a}^\dagger)
+i\omega t -i \phi}  \right).
\end{eqnarray}
Here $\nu$ is the frequency of the vibrational mode, ${\bf a}^\dagger$ and
${\bf a}$ are the ladder operators for this mode, $\omega_0$ is the energy
difference between the ions internal states which are being used as qubits, and
$\Omega_i$ is the Rabi frequency of the $i$th ion.  $\eta_i$ is the Lamb-Dicke
parameter which represents the projection of the laser $k$ vector along the
direction of the string ions and rms excursion of the ionic center-of-mass
along this direction and $\phi_i$ is the phase of the laser on the $i$th ion.
We have replaced the position of the ions by the ladder operators $k {\bf x}_i
= \eta_i ({\bf a} + {\bf a}^\dagger)$ and assumed that the laser is close to a
sideband $\omega \approx \omega_0 \pm \nu$ for a single vibrational mode.  For
simplicity, we will also assume that $\eta_i = \eta$: i.e. the coupling of the
recoil to vibration is the same for all ions.  The center of mass mode is one
for which this assumption is valid.  We assume also that the mode has been
sufficiently cooled so that we are in the Lamb-Dicke regime $\eta^2 (n+1) \ll
1$ so that $e^{i \eta ({\bf a} + {\bf a}^\dagger) }\approx {\bf I}+i \eta ({\bf
a}+ {\bf a}^\dagger)$.  For simplicity of notation, we place the phase into a
new operator ${\bf s}_\pm^{(i)} = \bmath{\sigma}_\pm^{(i)}e^{\pm i \phi}$. We
will further assume ions experience identical Rabi frequencies,
$\Omega_i=\Omega$.

Notice that we have assumed that we can control the phase $\phi_i$ of the laser
on each ions.  We will only need this single ion phase control for the two
qubit case.  In this case the phase between the ions can be adjusted by
changing the oscillation frequency of the trap.  By changing the oscillation
frequency of the trap, the ion spacing can be precisely controlled and
therefore the relative phase between the two ions can be
controlled\cite{Kielpinski:01a}.

Consider two lasers acting on the string of ions and assume that these are
tuned to frequencies $\omega+\delta$ and $\omega-\delta$.  In the Lamb-Dicke
limit in the interaction picture with respect to ${\bf H}_0$, the interaction
Hamiltonian is given by
\begin{eqnarray}
\tilde{\bf V}(t)&=&2 \Omega {\bf J}_x(\vec{\phi}) \cos(\delta t) - \sqrt{2}
\eta {\bf J}_y(\vec{\phi}) \left[  {\bf x} \left(\cos(\nu-\delta)t +
\cos(\nu+\delta)t \right) \right. \nonumber \\ &&+ \left. {\bf p} \left(
\sin(\nu-\delta)t + \sin(\nu+\delta)t \right) \right],
\end{eqnarray}
where ${\bf x}={1 \over \sqrt{2}}({\bf a}+ {\bf a}^\dagger)$, ${\bf p}={i \over
\sqrt{2}} ({\bf a}^\dagger-{\bf a})$, and the we have defined the operators
\begin{eqnarray}
{\bf J}_x (\vec{\phi}) &=& {1 \over 2} \sum_i \left[{\bf s}_+^{(i)}(\phi_i) +
{\bf s}_-^{(i)}(\phi_i) \right]={1 \over 2} \sum_i \left[ e^{i \phi_i}
\bmath{\sigma}_+^{(i)} + e^{-i \phi_i} \bmath{\sigma}_-^{(i)}  \right]
\nonumber
\\
 {\bf J}_y(\vec{\phi}) &=& {i \over 2 } \sum_i \left[{\bf s}_+^{(i)}(\phi_i) -
 {\bf s}_-^{(i)}(\phi_i) \right] = {i \over 2} \sum_i \left[
 e^{i \phi_i} \bmath{\sigma}_+^{(i)} - e^{-i \phi_i} \bmath{\sigma}_-^{(i)}
 \right].
\end{eqnarray}
If the laser intensity is less than the detuning $\Omega \ll \delta$ and the
detuning is close the sidebands then the Hamiltonian becomes
\begin{eqnarray}
\tilde{\bf V}(t)&=&-\sqrt{2} \eta \Omega \left(\cos(\nu-\delta)t {\bf
J}_y(\vec{\phi}) {\bf x} + \sin(\nu-\delta)t {\bf J}_y(\vec{\phi}) {\bf p}
\right) \nonumber \\
 \tilde{\bf V}(t)&=& f(t) {\bf J}_y(\vec{\phi}) {\bf x}+g(t) {\bf
 J}_y(\vec{\phi}) {\bf p}.
\end{eqnarray}
The evolution operator for this Hamiltonian is of the form
\begin{eqnarray}
{\bf U}(t)= e^{-i A(t) {\bf J}_y^2(\vec{\phi}) }e^{-i F(t) {\bf
J}_y(\vec{\phi}) {\bf x} } e^{-i G(t) {\bf J}_y(\vec{\phi}) {\bf p}},
\end{eqnarray}
where
\begin{eqnarray}
F(t)&=&\int_0^t f(\tau) d \tau = - {\sqrt{2} \eta \Omega \over \nu - \delta}
\sin(\nu-\delta)t \nonumber
\\
 G(t)&=&\int_0^t g(\tau) d \tau=-{\sqrt{2} \eta \Omega \over \nu - \delta} \left[ 1 -\cos(\nu -\delta)t \right] \nonumber \\
 A(t)&=& \int_0^t F(\tau)g(\tau) d \tau = - {\eta^2 \Omega^2 \over \nu -\delta}
 \left[ t- {1 \over 2 (\nu -\delta)} \sin2(\nu-\delta)t \right].
\end{eqnarray}
By choosing the time $(\nu-\delta)t_K=K 2\pi$ the ion-mode entangling
components of the gate vanish $F(t_K)=0$, $G(t_K)=0$ and
\begin{equation}
A(t_K)={2 \pi \eta^2 \Omega^2 \over (\nu-\delta)^2}K ,
\end{equation}
such that the evolution is
\begin{equation}
{\bf S}(\vec{\phi},K)=\exp\left[-i A(t_K) {\bf J}_y^2(\vec{\phi}) \right].
\end{equation}
We will call the gate ${\bf S}(K,\vec{\phi}), K \in \NN^+$ S{\o}rensen and
M{\o}lmer gates\cite{Sorensen:00a}.  By adjusting $K$, $\Omega$ and $\delta$,
the S{\o}rensen and M{\o}lmer gates gives us basic Hamiltonian control over the
(effective) Hamiltonian ${\bf J}_y(\vec{\phi})$.

\section{Universal quantum computation on the ion trap DFS quantum computer}

In this section we discuss how to use the S{\o}rensen and M{\o}lmer gates to
perform quantum computation on the ion trap DFS quantum computer.

\subsection{Single qubit rotations using S{\o}rensen and M{\o}lmer gates}

Notice that the operation ${\bf J}_y^2(\vec{\phi})$ is not in the commutant of
the weak collective decoherence OSR algebra, $\left[ {\bf J}_y^2(\vec{\phi}),
{\bf S}_z \right] \neq 0$.  For the single qubit gates, however, the
S{\o}rensen and M{\o}lmer gates can still be used to perform computation
entirely within the ion trap DFS.  Note however, that during the operation of
the S{\o}rensen and M{\o}lmer gates, the states are entangled with the
vibrational modes and are also not within the DFS.  Before and after the gates
we will describe below, however, the DFS is preserved.  Thus these gates must
be executed faster than the weak collective decoherence of the system in order
to not expose the system to too much weak collective decoherence.

The single qubit gates on the ion trap DFS will be executed when $2$ ions have
been maneuvered such these two ions are the only ions in a microtrap.  Consider
the following two ion operators
\begin{eqnarray}
\bar{\bf X} &=& 2{\bf J}_y^2 \left(\phi_1=-{\pi \over 2}, \phi_2=-{\pi \over 2}
\right) = \bmath{\sigma}_x^{(1)} \bmath{\sigma}_x ^{(2)} + {\bf I} \nonumber
\\
 \bar{\bf Y} &=& 2{\bf J}_y^2 \left(\phi_1=0,\phi_2=-{\pi \over 2} \right)=
 \bmath{\sigma}_y^{(1)} \bmath{\sigma}_x^{(2)} + {\bf I}.
\end{eqnarray}
While neither of these operators in the in the commutant of the OSR algebra
form the weak collective decoherence case, the operations do preserve the two
qubit weak collective decoherence DFS.  Specifically we see that, neglecting
the global phase shift produced by the identity ${\bf I}$,
\begin{eqnarray}
\bar{\bf X}|01\rangle &=& |10\rangle, \quad \bar{\bf X}|10\rangle = |01\rangle
\nonumber \\ \bar{\bf Y}|01\rangle &=& i|10\rangle, \quad \bar{\bf Y}|10\rangle
= -i|01\rangle.
\end{eqnarray}
We thus see that $\bar{\bf X}$ and $\bar{\bf Y}$ act as encoded
$\bmath{\sigma}_x$ and $\bmath{\sigma}_y$ respectively on the $|01\rangle$,
$|10\rangle$ basis.  These are examples of operations which are not in the
commutant but which preserve a particular DFS.  Note that these operations do
not preserve all of the $2$ qubit weak collective decoherence DFSs: DFS$_2(2)$
($|00\rangle$) and DFS$_2(0)$ ($|11\rangle$) are mixed.

Using the S{\o}rensen and M{\o}lmer gates we can implement the two Hamiltonian
evolutions
\begin{eqnarray}
\exp\left[ -i \bar{\bf X} t \right] \quad {\rm and} \quad \exp \left[ -i
\bar{\bf Y} t \right].
\end{eqnarray}
Thus encoded rotations about $\bmath{\sigma}_x$ and $\bmath{\sigma}_y$ are
possible using the S{\o}rensen and M{\o}lmer gates.  These two operations in
combination then serve to generate any single qubit rotation on the encoded
states $|01\rangle$ and $|10\rangle$.

\subsection{A nontrivial two qubit gate utilizing S{\o}rensen and M{\o}lmer gates}

Having shown how to implement single qubit gates on the ion trap DFS, we now
address the question of encoded two qubit operations.  For encoded two qubit
operations two two-ion DFSs are brought together into a microtrap where the
four ions are subjected to S{\o}rensen and M{\o}lmer gates.  The conjoined
qubits are now given by the states $|0101\rangle$, $|0110\rangle$,
$|1001\rangle$, and $|1010\rangle$.

Let us show that there is a particular choice of parameters for which we can
construct a S{\o}rensen and M{\o}lmer gate which acts non-trivially on the DFS,
preserving the conjoined DFS space, but which must take the state out of the
DFS during the course of the gate operation.

Consider the four ion operator
\begin{eqnarray}
\bar{\bf YY} &=& 2{\bf J}_y^2(\phi_1=0,\phi_2=0,\phi_3=0,\phi_4=0) \nonumber \\
 &=& \sum_{i=1}^4 \sum_{j=i+1}^4 \bmath{\sigma}_y^{(i)}
 \bmath{\sigma}_y^{(j)}+2 {\bf I}.
\end{eqnarray}
Disregarding the irrelevant global phase producing ${\bf I}$, we find that
\begin{eqnarray}
\exp[-i t \bar {\bf YY}]&=& \prod_{i=1}^4 \prod_{i=j+1}^4 \left( \cos(t) {\bf
I} -i \sin(t) \bmath{\sigma}_y^{(i)} \bmath{\sigma}_y^{(j)} \right).
\end{eqnarray}
Evaluating this for $t=\pi/4$, we find that
\begin{equation}
\exp\left[-i {\pi \over 4} \bar {\bf YY}\right]={e^{i{\pi \over 4}} \over
\sqrt{2}} \left[ {\bf I} - i \bmath{\sigma}_y^{(1)}
\bmath{\sigma}_y^{(2)}\bmath{\sigma}_y^{(3)} \bmath{\sigma}_y^{(4)} \right ].
\end{equation}
This is a nontrivial gate on the two encoded DFSs:
\begin{eqnarray}
\exp\left[-i {\pi \over 4} \bar {\bf YY}\right] |0101\rangle &=& {1 \over
\sqrt{2}} \left( |0101\rangle -i |1010\rangle \right) \nonumber \\
 \exp\left[-i {\pi \over 4} \bar {\bf YY}\right] |0110\rangle &=& {1 \over \sqrt{2}} \left(
|0110\rangle -i |1001\rangle \right) \nonumber \\ \exp\left[-i {\pi \over 4}
\bar {\bf YY}\right] |1001\rangle &=& {1 \over \sqrt{2}} \left( |1001\rangle -i
|0110\rangle \right) \nonumber \\ \exp\left[-i {\pi \over 4} \bar {\bf
YY}\right] |1010\rangle &=& {1 \over \sqrt{2}} \left( |1010\rangle -i
|0101\rangle \right).
\end{eqnarray}
This is a nontrivial encoded two-qubit gate between the ion trap DFSs. Together
with single qubit rotations, this forms a universal set of gates.

The gate $\exp\left[-i{\pi \over 2} \bar{\bf YY} \right]$ is a S{\o}rensen and
M{\o}lmer gate executed with $A(t_K)=\pi$.  This condition is met if ${4 \eta^2
\Omega^2 \over (\nu - \delta)^2 }K = 1$ and the time to execute this operation
is given by\cite{Sorensen:00a}
\begin{equation}
t_K={\pi \over \eta \Omega} \sqrt{K}.
\end{equation}
If $K>1$ is required to satisfy the above condition, then the ion will
repeatedly cycle through being entangled with the system and the vibrational
mode and will only return fully to the DFS after the completion of the
operation.  The gate we described above is exactly the gate used to create four
body entanglement in \cite{Sackett:00a}.

\section{Universal quantum computation on the ion trap DFS}

In the previous section we have seen how to perform gates on the ion trap DFS
which preserve the DFS.  These gates, unlike our previous discussion of
universal gates on a DFS, do not preserve the DFS during the entire operation
of the interaction.  This is reminiscent of the universal set of operators
described by Lidar, Bacon, Kempe, and Whaley in \cite{Lidar:01b,Lidar:01c}.  If
these gates are fast on the time-scale of the weak collective decoherence
mechanism, then these gates mesh nicely with the theory of fault-tolerant
quantum error correcting codes.  The reader is referred to
$\cite{Lidar:01b,Lidar:01c}$ for more information on this topic.

The example of the ion trap DFS is a good example of how encoding can be used
to reduce decoherence in a quantum computing architecture.  Just having an
encoding which can help, however, is not in and of itself the only necessary
component of building a quantum computer.  In this chapter we have seen that
using already developed methods for manipulating trapped ions universal control
of the encoded information can also be easily achieved.

\chapter{The Exchange-Based Quantum Computer} \label{ch:exchange}

In this chapter we discuss a quantum computer based only on the exchange
interaction.  This is particularly relevant to solid-state proposals for
quantum computation due to the difficulty in supplementing the exchange
interaction with other interactions to make the architecture fully universal.
In contrast to these original proposals, in this chapter we discuss how to use
encoded universality with the exchange interaction as the basis for universal
quantum computation.  We begin by discussing some of the generic properties of
solid-state proposals including the difficulty of engineering single qubit
manipulations on these systems.  We then discuss the relevance of geometry,
parallel operations, and subsystems in an exclusively exchange-based solid
state quantum computer.  An explicit proposal using the smallest possible
encoding is then proposed.  Single qubit gates and a two-qubit gate are then
explicitly calculated.  Finally, preparation, measurement, and leakage are
discussed so as the present a complete proposal for solid state quantum
computation using only the exchange interaction.

\section{Solid-state quantum computer proposals and the exchange interaction}

Among the plethora of experimental proposals for quantum computers there has
been widespread interest in a number of solid-state
approaches\cite{Loss:98a,Kane:98a,Vrijen:00a}. In the majority of these
proposals, a genuine spin-$1/2$ particle is used as the basic qubit for the
architecture.  These approaches have proposed as their basic qubit, for
example, the spin of a single electron on quantum dots\cite{Loss:98a},
donor-atom nuclear spins\cite{Kane:98a}, and electron spins in
heterostructures\cite{Vrijen:00a}.  A common thread throughout all of these
spin-based solid-state architectures is their use of the exchange interaction
(also known as the Heisenberg interaction) in order to produce two qubit gates
between neighboring spins.  In all of these proposals, control of this exchange
interaction is then supplemented by single qubit gates in order to generate a
fully universal quantum computer.

Compared to the exchange interaction, the single qubit gates in most
solid-state proposals are considerably slower, require greater device
complexity and potentially lead to an increase in the decoherence rate of the
device.  In the Table~\ref{tab:solidstate}, we assemble estimated exchange
interactions strengths, single qubit interaction times, and the difficulty in
constructing such single qubit interactions in a few of the solid state based
quantum computers.

\begin{table}[h]
\caption{\em Solid-state quantum computer estimated parameters}
\label{tab:solidstate}
\begin{tabular}{|l|l|l|l|}
\hline Proposal & Exchange & Single qubit & Single qubit difficulties \\ & gate
& gate & \\

 \hline
 Donor-atom nuclear spins & $\approx 100$ MHz & $\approx 100$ kHz & Slow
single qubit gates.
\\ in Silicon\cite{Kane:98a} & & & Strong magnetic fields \\
 & & & at low temperature\cite{Kane:00a}. \\
 \hline Electron spins & $\approx 1$ GHz & $\approx 1$ GHz & Strong inhomogenous \\
 on quantum dots\cite{Loss:98a} & & & magnetic fields\cite{Loss:98a,Burkard:99a}.\\
 & & & ``g-factor'' engineering \\
 & & & \cite{DiVincenzo:00b}.\\
\hline
 Electron spins in & $\approx 1$ GHz & $\approx 1$ GHz & ``g-factor'' engineering \\
 Si-Ge heterostructures\cite{Vrijen:00a} & & & \cite{Vrijen:00a,DiVincenzo:00b}.\\
 \hline
\end{tabular}
\end{table}
Table~\ref{tab:solidstate} illustrates that removal of the requirement of
single qubit gates may greatly benefit these solid-state proposals.  Luckily,
we have seen in Chapter~\ref{ch:exchange} that the exchange interaction without
the single qubit gates can be used to perform encoded universal quantum
computation.  The idea, then, is to use the exchange interaction alone for
solid-state quantum computers via encoding the quantum information.  In
principle, the proof (see Appendix~\ref{apc}) of the universality of the
exchange interaction tells us that such a construction is possible. Possibility
however has little say in practicality.  In this chapter we will address some
of the details of such an solely-exchange-based quantum computer.  From
explicit gate constructions, to description of preparation and measurement
procedures, we therefore will construct the basic outline of how an
exchange-only based solid state quantum computer would function.

\section{Universality and practicality}

To get an idea of why it is important to understand the specifics of the
exchange interaction universality for practical purposes, consider the results
presented by Bacon {\em et al.} in \cite{Bacon:00a}.  This was the first work
to demonstrate that the exchange operation alone could be used to perform
quantum computation.  In this work, the four qubit strong collective
decoherence DF subspace was used as the basis of the subsystems for the quantum
computer.  After demonstrating how the exchange interaction could be used to
perform single qubit gates on this encoding, it was shown that a controlled
phase Hamiltonian could be realized on this encoding via executing a
complicated series of commutators involving exchange interactions.  In
particular defining
\begin{eqnarray}
{\bf H}_1&=&\left[ {\bf E}_{26},{\bf E}_{12} + {\bf E}_{25} \right] + \left[
{\bf E}_{15}, {\bf E}_{12} + {\bf E}_{16} \right] \nonumber \\
 {\bf H}_2&=& \sum_{j=5}^8 \left( {\bf E}_{1j} + {\bf E}_{2j} \right) \nonumber
 \\
 {\bf C}&=& {1 \over 32} \left[ {\bf H}_1, \left[  {\bf H}_2, {\bf H}_1 \right]
 \right].
\end{eqnarray}
We then find that the operator ${\bf C}$ acts a two-qubit interaction between
two four-qubit encoded DFSs.  If $|0_L\rangle$ and $|1_L\rangle$ denote the
encoded qubits in a particular basis, then ${\bf C}$ acts as $|0_L 0_L\rangle
\rightarrow 0$, $|0_L 1_L\rangle \rightarrow |0_L 1_L\rangle$, $|1_L 0_L\rangle
\rightarrow 0$, $|1_L 1_L\rangle \rightarrow 0$.  This operation can be used,
in conjunction with single qubit operations to perform universal quantum
computation.

In a similar manner, because the proof in Appendix~\ref{apc} is inductively
constructive it is always possible to exhibit such complex commutator and
linear combinations which enact any operation on the strong collective
decoherence DFS.  Via the Kitaev-Solovay theorem, we know that this gate set
will be on equivalent footing with any other gate set, yet, in a practical
sense we have not seen how to implement this interaction without resorting to
the approximate formula Eq.~(\ref{eq:liestructure}).

An example of the problem we face will help explain this problem.  Suppose we
were given the ability to perform the Hamiltonians ${\bf H}_1=\bmath{\sigma}_x$
and ${\bf H}_2=\bmath{\sigma}_y$ and we wished to implement the Hamiltonian
${\bf H}_3=\bmath{\sigma}_x +\bmath{\sigma}_y$ for a time $T$. Using a standard
Euler angle construction we could perform a series of evolution with ${\bf
H}_1$ and ${\bf H}_2$ which would result in this evolution.  Suppose, however,
instead of this Euler angle construction we decided to use the Trotter
approximation formula
\begin{equation}
\left( \exp \left[ -{i {\bf H}_1 T \over N} \right] \exp \left[-{i {\bf H}_2 T
\over N} \right] \right)^N = \exp \left[-i ({\bf H}_1 + {\bf H}_2)T \right]
+O\left({1 \over N^2} \right),
\end{equation}
to execute this gate.  Recalling the definition of the error between two
unitary operators from Section~\ref{sec:approxu} we can explicitly calculate
this error for our simple example.  The results of this calculation are plotted
in Figure~\ref{fig:errorplot}
\begin{figure}[h]
\psfig{figure=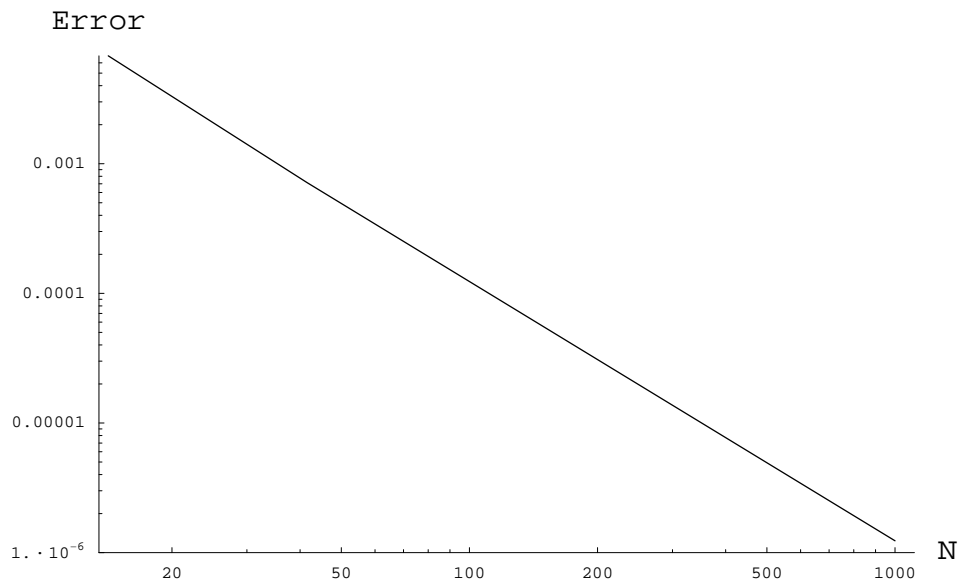,width=5in}
 \caption{\em Plot of the error in the example using the Trotter approximation} \label{fig:errorplot}
\end{figure}
Since the error scales line $1/N^2$, in order to obtain an accuracy, say, that
is sufficient for the threshold for quantum computation which is currently
estimated at $\approx 10^{-6}$, we see that $N^2$ must be of order $10^3$.
Given gates with an interaction strength $g$, this implies that these
interactions must we switched on an off at a rate of $g/N^2$ in order to obtain
a reasonable approximation.  For almost all proposals, however, such rapid
control of the system will not be achievable.  Without knowing about the Euler
angle construction, then, the real world functioning of the universality is
unclear.

We are faced with the problem of knowing that the exchange gates are universal
but not knowing the explicit methods for explicit construction of the gates in
this set.  Of course, one can always resort to the Kitaev-Solovay theorem,
which is constructive, to determine such gate sequences.  For situations larger
than a few qubits, however, this is an extremely daunting task to approach by
brute force.

\section{Subsystems and geometric layout}

Before we discuss universality on the solely-exchange-based quantum computer,
we must first discuss the subsystem structure of such a quantum computer.  In
this chapter we will focus on the smallest encoding which supports universality
using the exchange interaction.  This is the subsystems encoding of one logical
qubit into three physical qubits.  In this chapter we will not be concerned
with the decoherence-free properties of these states and will instead just
focus on their use in a quantum computer.  Specifically, we will focus on the
encoding
\begin{eqnarray}
|0_L\rangle &=& {1 \over \sqrt{2}} \left(|01\rangle -|10\rangle \right)
|1\rangle \nonumber
\\
|1_L\rangle &=& {2 \over \sqrt{3}} |001\rangle -{1 \over \sqrt{3}}\left(
|01\rangle + |10\rangle \right) |0\rangle. \label{eq:logicalbasis}
\end{eqnarray}

An important component of any quantum computer is the geometric layout and
connectivity of the physical qubits.  In an encoded universality construction
it is especially important to consider the geometry of the encoded qubits. We
will consider three different geometries which will probably best represent
future solid-state device layouts.  Other arrangements are, of course,
possible, but these layouts should be representative of real world constraints
imposed on most solid-solid state systems.

In the first layout, which will call the {\em one-dimensional layout}, the
physical qubits are assumed to lie in linear succession.  Only nearest neighbor
exchange interactions are allowed such that within an encoded qubit only two of
three possible exchanger interactions can be implemented.  This model is
sketched in Figure~\ref{fig:1d}.
\begin{figure}[h]
\quad \quad \psfig{figure=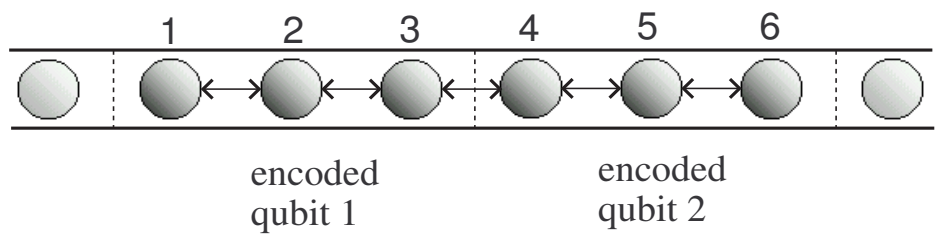,width=5in}
 \caption{\em The one-dimensional layout} \label{fig:1d}
\end{figure}

In the second layout, which we will call the {\em triangular layout}, the
physical qubits are assumed to be arranged in a linear succession of triangular
encoded qubits.  Each triangle represents an encoded qubit and successive
triangles are only coupled by one exchange interaction.  This model is sketched
in Figure~\ref{fig:trid}.
\begin{figure}[h]
\quad \quad \psfig{figure=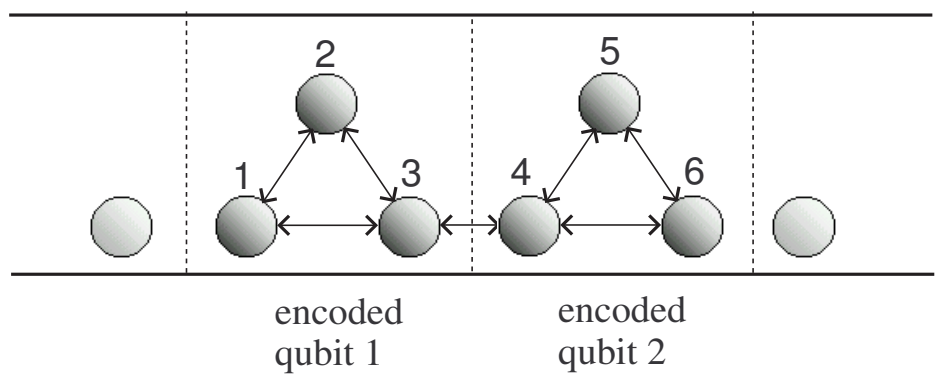,width=5in}
 \caption{\em The triangular layout} \label{fig:trid}
\end{figure}

Finally the third layout, which we call the {\em two-dimensional layout},
consists of a square grid layout of physical qubits.  The encoded qubits are
then grouped into triplets of physical qubits which can couple only with other
physical qubits which are nearest neighbors.  This model is sketched in
Figure~\ref{fig:twod}.
\begin{figure}[h]
\quad \quad \psfig{figure=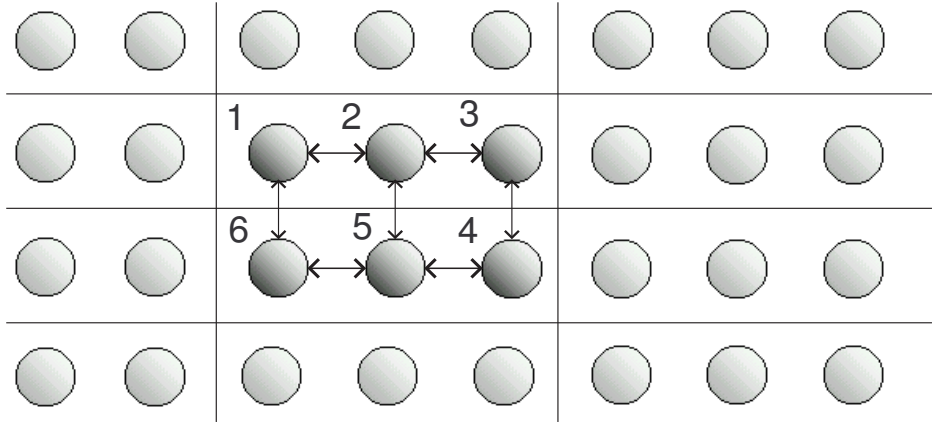,width=5in}
 \caption{\em The two-dimensional layout} \label{fig:twod}
\end{figure}

Finally we will also have the opportunity to consider serial and parallel
operation of the device.  In {\em serial} operation it is assumed that only one
exchange interaction between qubits can be turned on for a single time period.
In {\em parallel} operation, multiple exchange interactions, perhaps with
varying strengths, can be turned on for a single time period.  Of course, for
full quantum error correction some amount of parallel operation is
necessary\cite{Aharonov:96a}, however for early implementations of the
exchange-only solid-state proposals, this will not be an issue and considerable
experimental simplification is expected when operations are not enacted in
parallel.

\section{Single encoded qubit gates using the \\ exchange interaction}

For single encoded qubit gates, the geometries described above motivate two
different scenarios.  In the first scenario, only the exchange interaction
between qubits $1,2$ and $2,3$ can be enacted and in the other scenario,
interaction between all qubits can be enacted.  We call these situations the
constrained and unconstrained geometries respectively.  Furthermore we must
also consider the case where parallel or serial operation is allowed.  Thus we
have four scenarios: parallel constrained, parallel unconstrained, serial
constrained and serial unconstrained.

First, it is easy to calculate the explicit action of the exchange gates on the
logical basis defined in Eq.~(\ref{eq:logicalbasis}):
\begin{eqnarray}
{\bf E}_{12}= \left( \begin{array}{cc} -1 & 0 \\ 0 & 1 \end{array}\right),\quad
 {\bf E}_{23}=\left( \begin{array}{cc} {1 \over 2} & -{\sqrt{3} \over 2} \\  -{\sqrt{3} \over 2} & -{1 \over
 2}\end{array}\right),\quad
 {\bf E}_{23}=\left( \begin{array}{cc} {1 \over 2} & {\sqrt{3} \over 2} \\  {\sqrt{3} \over 2} & -{1 \over
 2}\end{array}\right),
\end{eqnarray}
where we have used the $|0_L\rangle$, $|1_L\rangle$ basis.  We can also define
the encoded $\bmath{\sigma}_\alpha$ matrices in the obvious manner such that
\begin{eqnarray}
{\bf E}_{12}=-\bmath{\sigma}_z, \quad {\bf E}_{23}= {1 \over 2}
\bmath{\sigma}_z - {\sqrt{3} \over 2} \bmath{\sigma}_x, \quad {\bf E}_{13}= {1
\over 2} \bmath{\sigma}_z + {\sqrt{3} \over 2} \bmath{\sigma}_x.
\end{eqnarray}

Let us deal with two of the four scenarios, the parallel unconstrained scenario
and the parallel constrained scenario.  In particular we can use the fact that
\begin{eqnarray}
\bmath{\sigma}_x=-{2 \over \sqrt{3}} \left( {\bf E}_{23} + {1 \over 2}{\bf
E}_{12} \right).
\end{eqnarray}
Therefore if we allow parallel operations, then in both the unconstrained and
constrained geometries, we have the ability to enact the Hamiltonians
$\bmath{\sigma}_z$ ($-{\bf E}_{12}$) and $\bmath{\sigma}_x$ (from above). Using
an Euler angle construction, we therefore have a method for constructing every
possible single qubit gate on this encoding.  In particular every single qubit
gate can be constructed via a sequence like
\begin{equation}
\exp[-i \alpha {\bf E}_{12}] \exp \left[-i \beta \left({1 \over 2} {\bf
E}_{12}+ {\bf E}_{23} \right) \right] \exp \left[-i\gamma {\bf E}_{12}\right],
\end{equation}
for some combination of $\alpha,\beta,\gamma$.  The circuit for this procedure
is given in Figure~\ref{fig:parallel1} where the arrows indicate an exchange
interaction between the connected qubits for the duration specified beside the
arrow.
\begin{figure}[h]
\quad \quad \psfig{figure=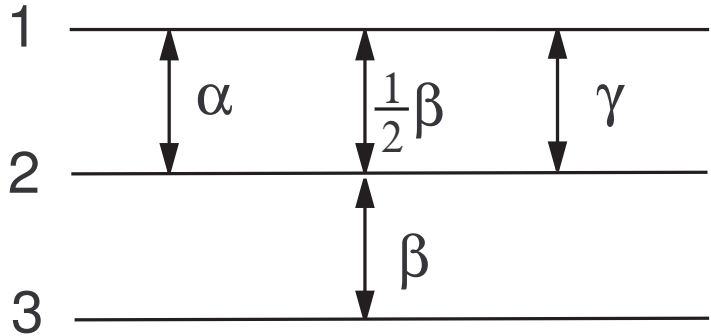,width=4in}
 \caption{\em Single qubit encoded Euler angle construction with parallel operations} \label{fig:parallel1}
\end{figure}

When serial operations are required and a constrained geometry is used it is
impossible to construct certain rotations with only three applications of the
exchange gates.  Notice that in this case there are two possible orders for the
application of the exchange interactions
\begin{eqnarray}
&&\exp[-i{\bf E}_{12}\alpha ] \exp[-i{\bf E}_{23} \beta] \exp[-i{\bf E}_{12}
\gamma] \nonumber \\ && \exp[-i{\bf E}_{23}\alpha ] \exp[-i{\bf E}_{12} \beta]
\exp[-i{\bf E}_{23} \gamma]. \label{eq:sequence}
\end{eqnarray}
In order to understand why it is not possible to perform all single qubit gates
with these two orders of rotations it is useful to work in the Bloch sphere
description of single qubit rotations (see Chapter 4 of \cite{Nielsen:00a}),
i.e. mapping the $SU(2)$ rotations onto $SO(3)$.

Suppose we are given two vectors on the block sphere.  If we can perform any
single qubit rotation, then we can manipulate these two vectors such that they
point in any direction consistent with the inner product between the vectors
unchanged.  But now consider the first sequence in Eq.~(\ref{eq:sequence}).  If
we start the state in $|0_L\rangle$, then the first rotation does nothing, the
second rotation can reach a ring on the bloch sphere which is not a great
circle and the third rotation will finally be able to rotate this state to
everywhere on the Bloch sphere except a small cap.  This is illustrated in
Figure~\ref{fig:whynot}.
\begin{figure}[h]
\psfig{figure=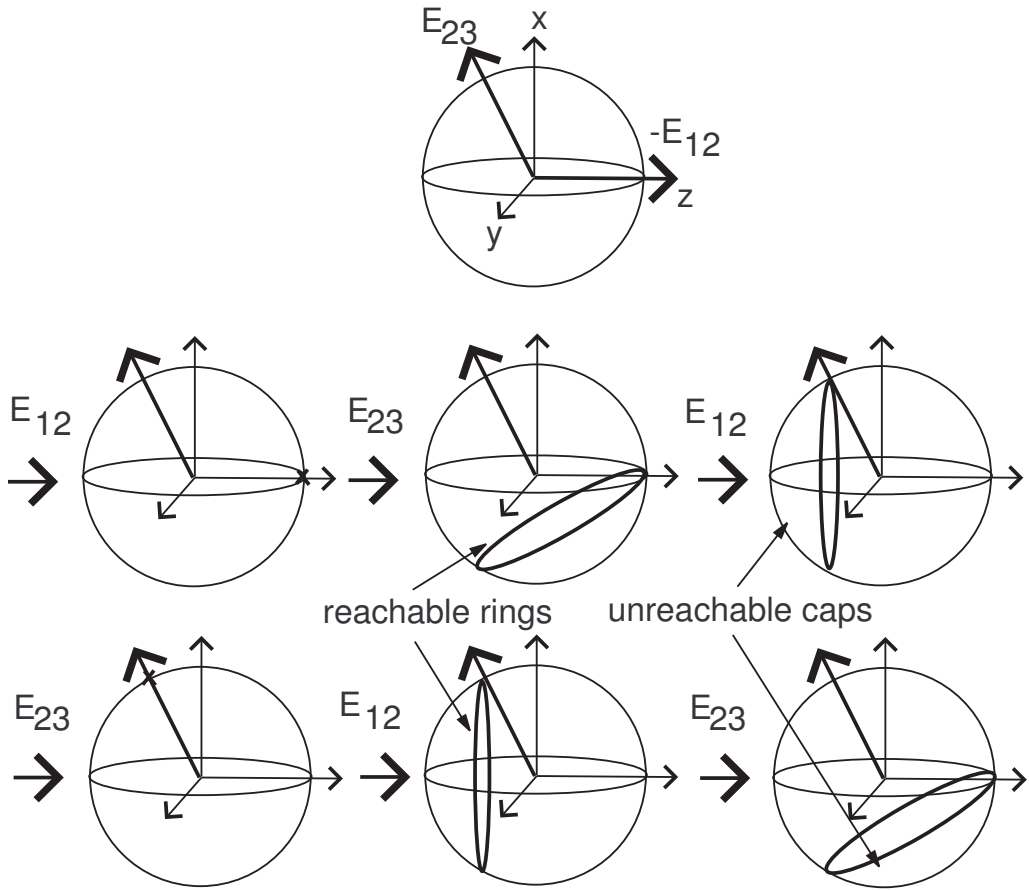,width=5.5in}
 \caption{\em Bloch sphere picture of reachable operations in serial mode} \label{fig:whynot}
\end{figure}
Similarly using the second sequence in Eq.~(\ref{eq:sequence}) one can start
with the $+1$ eigenstate of ${\bf E}_{23}$ and there is an unreachable cap for
this sequence of operations.  See the second sequence in
Figure~\ref{fig:whynot}.

Thus if we take these two vectors (the $+1$ eigenstates of ${\bf E}_{12}$ and
${\bf E}_{23}$) for each sequence there is a region which the vector cannot be
rotated to.  Thus there are rotations which cannot be achieved by the sequences
in Eq.~(\ref{eq:sequence}).

In order to be able to implement any single qubit gate on the serial
constrained scenario we must, in fact, use four exchange interactions.  In
fact, with four operations there is an equivalence between constrained exchange
interactions and the three interaction unconstrained case.  One possible order
for the four interactions is given by
\begin{equation}
\exp[-i \alpha {\bf E}_{12} ]\exp[-i \beta {\bf E}_{23} ] \exp[-i \gamma {\bf
E}_{12} ]\exp[-i \delta {\bf E}_{23} ].
\end{equation}
Choosing $\alpha=-{\pi \over 2}$ and $\gamma=\gamma^\prime + {\pi \over 2}$,
this becomes
\begin{eqnarray}
&&{\bf E}_{12} \exp [-i \beta {\bf E}_{23}] {\bf E}_{12} \exp[-i \gamma^\prime
{\bf E}_{12}] \exp[-i \delta {\bf E}_{23}] \nonumber \\ &&=\exp[-i\beta{\bf
E}_{13}] \exp[-i\gamma^\prime {\bf E}_{12} ] \exp[-i\delta {\bf E}_{23}].
\end{eqnarray}
Different orderings of the four interactions allow for different three
interaction orderings.  We therefore see that the four interaction constrained
model can be mapped onto the three interaction unconstrained model.

To show that the three interaction unconstrained model is sufficient to perform
any single qubit operation one simply follows the standard argument of an Euler
angle constructions.

\section{Explicit encoded controlled-not using a sequence of exchange
interactions}\label{sec:only}

Having shown how to explicitly construct the encoded single qubit rotations the
question now arises as to how to coupled together different encoded qubits.
This is a challenging question.  The first item to note is that there is no
Hamiltonian which by itself will directly enact a coupling which preserves the
two logical qubits.  To see this, examine the two angular momenta on two
conjoined encoded qubits
\begin{eqnarray}
\left({\bf S}_1\right) &=& \sum_{\alpha=1}^3 (\sum_{i=1}^3 {\bf
s}_\alpha^{(i)})^2 \nonumber \\
 \left( {\bf S}_2\right)&=&\sum_{\alpha=1}^3 (\sum_{i=4}^6 {\bf
 s}_\alpha^{(i)})^2,
\end{eqnarray}
where $s_\alpha^{(i)}={1 \over 2} \bmath{\sigma}_\alpha^{(i)}$. It is easy to
see that no linear combination of exchanges commutes with both of these
operators unless the exchanges in the linear combination act exclusively
between the first three qubits or exclusively between the final three qubits.
But these are just the encoded single qubit operators.  Therefore there is no
linear combination of exchange operations which preserves the original DFSs.

In Appendix~\ref{ape} we present a gate sequence for enacting a
controlled-phase between two four qubit strong collective decoherence DFSs.
This sequence uses parallel operations and was analytically derived using
insights gained from using the strong DFS basis.  Working with the three qubit
strong collective decoherence DFSs is not as amenable to such analysis because
conjoining two strong collective decoherence DF subsystems is not as
straightforward as in the subspace case.

In order to deal with deriving some nontrivial gate on our encoded qubit, it is
therefore necessary to resort to numerical searches.  Much of the difficulty of
these searches arises from the fact that while the four basis states $|0_L
0_L\rangle$, $|0_L 1_L\rangle$, $|1_L 0_L\rangle$, and $|1_L 1_L\rangle$ have
totals spin $S=1$, the complete space with these quantum numbers for six spins
has nine states and exchanges perform rotations on this nine dimensional space.
The numerical search algorithm then must search for a series of exchanges
confined to this $9$ dimensional space which performs a nontrivial gate $G$ on
the encoded qubits and any unitary matrix on the five dimensional component of
the space perpendicular to this encoded space, i.e. ${\bf U}={\bf G} \oplus
{\bf A}_5$ where ${\bf A}_5$ is any unitary matrix on the $5$ dimensional
perpendicular space.  A numerical search for optimal gates on the encoded
states was performed in \cite{DiVincenzo:00a}.  In this work search for a
controlled-not gate was performed with the aid of two invariants identified by
Makhlin\label{Makhlin:00a}.  Explicitly, a controlled-not on the basis states
acts as
\begin{equation}
\left( \begin{array}{cccc} $1$ & $0$ & $0$ & $0$ \\ $0$ & $1$ & $0$ & $0$ \\
$0$ & $0$ & $0$ & $1$
\\ $0$ & $0$ & $1$ & $0$
\end{array} \right) \begin{array}{c} |0_L 0_L \rangle \\ |0_L 1_L \rangle \\ |1_L 0_L \rangle \\ |1_L 1_L \rangle
 \end{array}.
\end{equation}
Figure~\ref{fig:cnot} presents the optimal (in the sense of fewest exchange
interaction gates) serial operation solution for the one-dimensional layout.
In this figure, the $t_i$ values represent the duration of the exchange
interaction as in $\exp[i \pi t_i {\bf E}_{ij}]$.  In Figure~\ref{fig:cnot} the
serial operation has been compressed where gates commute.  The uncertainty of
the final digits is indicated in parenthesis and the accuracy of the gate is to
$6\times 10^{-5}$.

\begin{figure}[h]
\quad \quad \psfig{figure=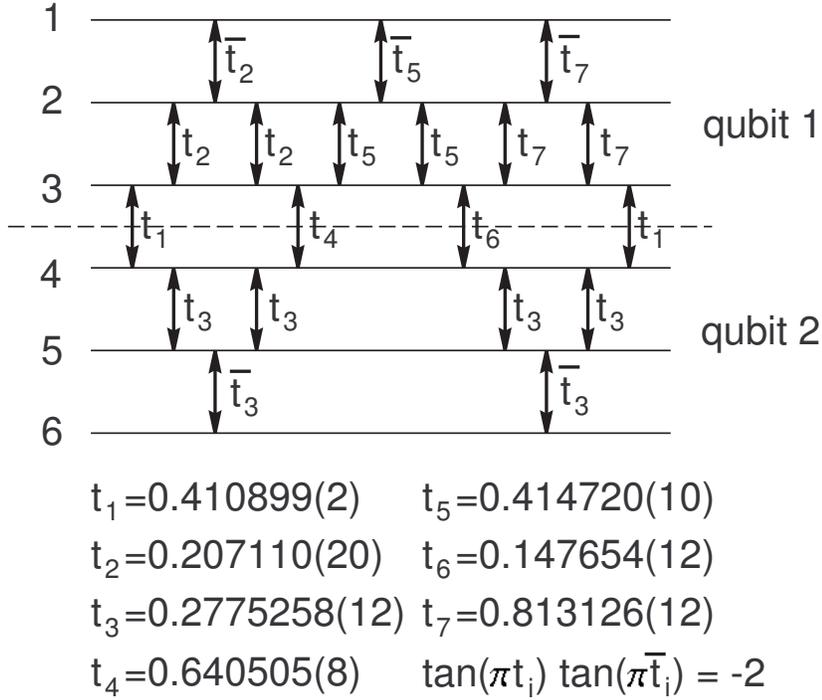,width=4.5in}
 \caption{\em Encoded controlled-not} \label{fig:cnot}
\end{figure}
\begin{table}[h]
\caption{\em Number of exchange gates in different scenarios} \label{tab:times}
\begin{tabular}{|l|l|l|l|}
\hline Gate size & Operation mode & Geometry & Gates \\ \hline
 single qubit & serial & one-dimensional, two-dimensional & 4 \\
 single qubit & serial & triangular & 3 \\
 single qubit & parallel & one-dimensional, two-dimensional & 3 \\
  & & triangular & \\
  \hline
  two qubit & serial & one-dimensional & 19 \\
  two qubit & parallel & one-dimensional & 8 \\
  two qubit & parallel & two-dimensional & 7 \\
  \hline
\end{tabular}
\end{table}

In Table~\ref{tab:times} we assemble a list of the different optimal (in the
sense of the best found by the search algorithm) solutions for different
operation modes of described above.  One shortcoming of our construction is
that it does not make use of the subsystem nature of collective decoherence. A
true controlled-not which preserves the subsystem structure was searched for
and none was found for less than $26$ serial exchange interactions, although in
parallel operation, sequences with $8$ exchange interactions where found. It
would be worthwhile to obtain optimal gate sequences not only for the three
qubit subsystem example, but also to examine the four qubit subspace example.
There are certain simplifications which seem to imply that these gate sequences
for the subspace case might be simple enough even for analytical treatment (as
in the parallel operation of Appendix~\ref{ape}).

Together with the single qubit rotations described above the controlled-not
forms a universal gate set.  The tradeoffs inherit in the exchange-only based
techniques are thus clear.  For a factor of $3$ in space and $\approx 10$ in
clock cycles universality can be achieved using only the exchange interaction.

\section{Preparation, measurement, and leakage}

Finally let us describe preparation, measurement, and leakage detection on the
exchange-based quantum computer.

Preparation can be achieved by preparing the state $|0_L\rangle ={1 \over
\sqrt{2}} (|01\rangle -|10\rangle) |0\rangle$.  This state can be prepared by
turning on an exchange interaction of strength $J$ between the first two qubits
and a moderately strong magnetic field $B$ pointing along the $z$ direction is
applied such that $k_B T \ll g \mu_B B < J$.  With these physical parameters,
the ground state of the system is $|0_L\rangle$ and a gap to the next excited
state is of energy $g\mu_B B$. Thus at low temperature in comparison to this
gap the encoded register will be initialized to the $|0_L\rangle$ state.

Measurement on the encoded qubit can be achieved in a variety of manners.  In
particular if the singlet states of the first two qubits can be distinguished
from the triplet states of the first two qubits then the encoded qubits can be
distinguished.  A method which meshes nicely with our current scheme is the
a.c. capacitance scheme proposed by Kane\cite{Kane:98a}. When two electrons
occupy a common potential well, in the absence of a magnetic field, the Pauli
exclusion principle mandates that the singlet state of the two electrons lies
at a lower energy than the triplet state.  Therefore an electrometer capable of
detecting the number of electrons occupying a bound state can be used to
determine whether the singlet or triplet is occupied.  The a.c. capacitance
scheme of Kane is directly analogous to this procedure for the solid-state
quantum computing proposals.

Finally we can briefly address the problem of leakage in the exchange-only
setup.  While encoded universality allows for an interaction which was
previously not fully universal to be used in a universal manner, one of the
tradeoffs is that there is a particular nasty type of error which can occur in
which the information in leaks out of the encoded subspace.  One particularly
simple manner of dealing with leakage errors is to engineer a system such that
the subspace upon which one is working on is the ground state of the system. If
this is the case, then at low enough temperature, the leakage errors will be
self-corrected by energy exchange with the environment.  One way to achieve
this in our case is to apply exchange interactions with equal strength between
all three qubits and a moderate magnetic field along the $z$ direction.  In
this scenario the $|0_L\rangle$ and $|1_L\rangle$ states are the degenerate
ground state of the system.  Further, a.c. capacitance probing of this system
can be used to determine if states have leaked outside of the subspace.
Therefore leakage detection can be achieved via this fairly straightforward
methodology.

\section{Exchange-based quantum computation}

In this chapter we have seen how to achieve quantum computation using only the
exchange interaction.  This allows for a considerable device simplification as
well as fundamental speed increases for certain solid-state quantum computation
proposals.  An important open question is whether other quantum computing
architectures would benefit from a similar encoded universality.

\chapter{Decoherence-Free Subspaces in Multilevel Atomic Systems}
\label{ch:atom}

\begin{quote}
{\em Decoherence-free subspaces in single atomic systems?}
\end{quote}

In this chapter we discuss the application of the theory of decoherence-free
subspaces to multilevel atomic systems.  It is interesting to question whether
decoherence-free conditions can exist in single atomic systems.  We begin by
developing a sort of no-go theorem for DF subspaces with nondegenerate energy
spectra. This leads to a discussion of coherent population trapping and we
demonstrate how coherent population trapping can be thought of as a
semi-classical DF subspace.

\section{OSR DF subspaces in multilevel systems}

Suppose we are given an atomic multilevel system with $N$ levels labeled by the
states $|i\rangle$, $1 \leq i \leq N$, with energies $\omega_i$, which is
coupled to a free space electromagnetic field with modes labeled by $\vec{k}$
and polarizations $\epsilon \in \{1,2\}$.  The Hamiltonian for this system is
given by:
\begin{eqnarray}
{\bf H}&=&{\bf H}_0 + {\bf V} \nonumber \\ {\bf H}_0&=& \sum_i \omega_i
|i\rangle \langle i| + \sum_{\vec{k},\epsilon} \omega_k {\bf
a}_{\vec{k},\epsilon}^\dagger {\bf a}_{\vec{k},\epsilon} \nonumber \\
 {\bf V}&=& \sum_{\vec{k},\epsilon} \sum_{i>j=1}^N \bmath{\sigma}_{ij} \left( g_{\vec{k},\epsilon}(i,j)
  {\bf a}_{\vec{k},\epsilon} + g_{\vec{k},\epsilon}^*(i,j) {\bf a}_{\vec{k},\epsilon}^\dagger
  \right),\label{eq:hint}
\end{eqnarray}
where $\bmath{\sigma}_{ij}=|i\rangle \langle j| + |j\rangle \langle i|$.  In
general the coefficient $g_{\vec{k},\epsilon}(i,j)$ will separate out into
functions of $\vec{k},\epsilon$ and of $i,j$: $g_{\vec{k},\epsilon}(i,j)=
f_{\vec{k},\epsilon} g_{ij}$.  This decomposition implies
\begin{eqnarray}
{\bf V}= \sum_{\vec{k},\epsilon} \sum_{i>j=1}^N \bmath{\sigma}_{ij}
\left(g_{ij} f_{\vec{k},\epsilon}  {\bf a}_{\vec{k},\epsilon} + g_{ij}^*
f_{\vec{k},\epsilon}^* {\bf a}_{\vec{k},\epsilon}^\dagger \right).
\end{eqnarray}
The creation and annihilation operators ${\bf a}_{\vec{k},\epsilon}$ and ${\bf
a}_{\vec{k},\epsilon}^\dagger$ combined as we have done form a single operator
and therefore the OSR algebra from this interaction is generated by
$\sum_{i>j=1}^N g_{ij} \bmath{\sigma}_{ij}$ and $\sum_{i>j=1}^N g_{ij}^*
\bmath{\sigma}_{ij}$. The DF subspace condition then becomes
\begin{equation}
\sum_{i>j=1}^N g_{ij} \bmath{\sigma}_{ij} |\psi\rangle = \alpha_1 |\psi\rangle
\quad \sum_{i>j=1}^N g_{ij}^* \bmath{\sigma}_{ij} |\psi\rangle = \alpha_2
|\psi\rangle,
\end{equation}
for $|\psi\rangle$ in a DF subspace defined by $\alpha_1$ and $\alpha_2$.  In
order for both of these conditions to be met, $\alpha_1=\alpha_2^*\equiv
\alpha$ and the condition is really one condition
\begin{equation}
\sum_{i>j=1}^N g_{ij} \bmath{\sigma}_{ij} |\psi\rangle = \alpha|\psi\rangle
\quad {\rm or} \quad\sum_{i,j=1}^n G_{ij} |i\rangle \langle j |\psi\rangle =
\alpha |\psi\rangle,
\end{equation}
where $G_{ij}=g_{ij}, i>j=1\dots n$, $G_{ii}=0$, and $G_{ij}=g_{ji}, i<j=1
\dots N$.  Thus the spectrum of $G_{ij}$ essentially determines the states for
which the DF subspace condition is fulfilled.  However, we also desire that
${\bf H}_0$ not take the state outside of the DFS.  We recall that this will be
true if the diagonal form of system Hamiltonian can be written so that it
contains on states from a particular DF subspace.

The first result we will prove along these lines is that degeneracy of the
system energy spectrum is necessary for a perfect DF subspace under the
condition that there are no completely isolated levels of the multilevel
system.
\begin{lemma}
A multilevel system with a non-degenerate energy spectrum does not support a DF
subspace (in the strict sense of not evolving) with respect to interaction with
an electromagnetic field if every energy level has at least one non-vanishing
transition from the state to another state ($G_{ij} \neq 0$ for every fixed $i$
for at least one $1\leq j\leq N$).
\end{lemma} \label{lem:nogo}
Proof: Suppose that such a system did support a DF subspace.  The diagonalized
form of the system Hamiltonian ${\bf H}_0$ is unique ${\bf H}_0=\sum_i \omega_i
|i\rangle \langle i|$ because the energy spectrum is non-degenerate by
assumption.  Via arguments in Section~\ref{sec:syshamdfs}, the state
$|i\rangle$ must be a DF state for the supposed DF subspace in order that ${\bf
H}_0$ preserve the DF subspace.  However $\sum_{i,j=1}^N G_{jk} |j\rangle
\langle k|$ acting on $|i\rangle$ does not satisfy the DF condition because
there is at least one transition from $|i\rangle$ to another state. Therefore
$|i\rangle$ cannot be DF and there can be no DF subspace for this setup.

This lemma implies that there can be no perfect DF subspace for a multilevel
atomic system unless there is a degeneracy in the energy spectrum of the
multi-level system.  Furthermore it follows from the proof of this lemma that
the only way a DF subspace can exist in a multilevel atomic system is if the DF
subspace has support over degenerate states of the system.

\section{Coherent population trapping and DF subspaces}

Consider now an example from quantum optics which appears to be a DF subspace
but which violates Lemma~\ref{lem:nogo}.  This is the case of coherent
population trapping.  Consider a three level system $|a\rangle$, $|b\rangle$,
and $|c\rangle$ with nondegenerate energies $\omega_a$, $\omega_b$ and
$\omega_c$, respectively.  The levels are in the so-called $\Lambda$
configuration in which the lower two levels $|a\rangle$ and $|b\rangle$ are
coupled to a single higher level $|c\rangle$.  The transition between the
$|a\rangle$ and $|b\rangle$ levels is assumed to be strongly forbidden.  We
suppose that this atom is being driven by two lasers of frequency $\nu_1$ and
$\nu_2$.  See Figure~\ref{fig:cohtrap}.

\begin{figure}[ht]
\quad \quad  \quad\psfig{figure=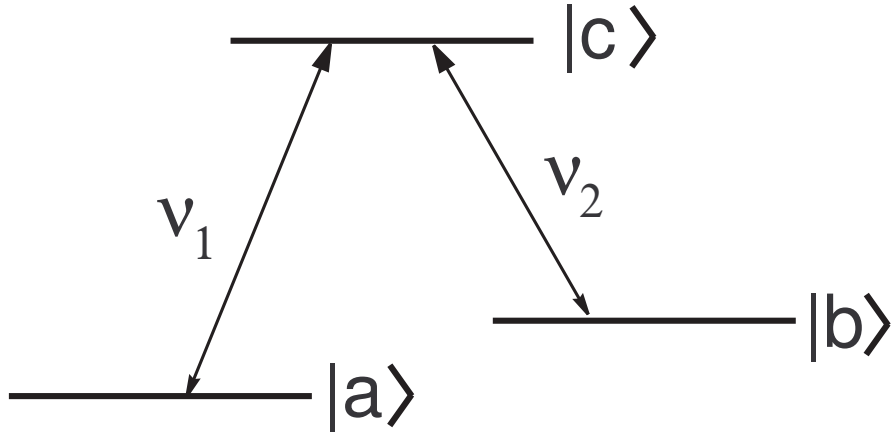,width=5in}
 \caption{\em Coherent population trapping} \label{fig:cohtrap}
\end{figure}

The semi-classical description of this problem is given by the Hamiltonian
\begin{eqnarray}
{\bf H}&=&{\bf H}_0+{\bf H}_1 \nonumber \\
 {\bf H}_0&=& \omega_a |a\rangle \langle a| + \omega_b |b\rangle \langle b| +
 \omega_c |c\rangle \langle c| \nonumber \\
 {\bf H}_1&=& {\Omega_1 \over 2} \left( e^{-i\phi_1-i \nu_1 t} |c\rangle
 \langle a| + e^{i\phi+i\nu_1 t} |a\rangle \langle c | \right)
 +{\Omega_2 \over 2} \left(e^{-i \phi_2 -i \nu_2 t} |c \rangle \langle b| + e^{i \phi_2 + i \nu_2 t} |b\rangle \langle
 c|\right),\nonumber \\
\end{eqnarray}
where $\Omega_i$ is the Rabi frequency associated with the laser with frequency
$\nu_i$.  In the interaction picture with respect to ${\bf H}_0$, this
Hamiltonian becomes
\begin{eqnarray}
{\bf V}(t)&=&{\Omega_1 \over 2} \left( e^{-i \phi_1 - i (\nu_1 -\omega_c +
\omega_a)t} |c\rangle \langle a|+e^{i \phi_1+i (\nu_1-\omega_c+\omega_a)t }
|a\rangle \langle c|  \right) \nonumber \\ &&+ {\Omega_2 \over 2} \left( e^{-i
\phi_2 -i (\nu_2-\omega_c + \omega_b)t} |c \rangle \langle b| + e^{i \phi_2 +i
(\nu_2 -\omega_c + \omega_b)t} |b\rangle \langle c| \right).
\end{eqnarray}
At resonance this becomes
\begin{equation} {\bf V}_r(t)={\Omega_1 \over 2}
\left( e^{-i \phi_1} |c \rangle \langle a | +e^{i \phi_1} |a\rangle \langle c|
\right) + {\Omega_2 \over 2} \left( e^{-i \phi_2} |c\rangle \langle b| +e^{i
\phi_2} |b\rangle \langle c| \right).
\end{equation}
This Hamiltonian has three eigenstates, two of which contain components along
the $|c\rangle$ state and one of which does not.  The eigenstate which does not
have a component along the $|c\rangle$ state is given by the coherent
population trapped state
\begin{equation}
|\psi\rangle= {1 \over \sqrt{ \Omega_1^2+\Omega_2^2}} \left( \Omega_2e^{-i
\phi_2}  |a\rangle + \Omega_1 e^{-i \phi_1} |b\rangle \right).
\label{eq:trapped}
\end{equation}
This state shares some of the characteristics of a state in a DF subspace.  The
state does not couple to resonant radiation field even though it is a state
which is a superposition of states which individually couple to the resonant
radiation field.

Notice however that if the resonance condition is not met or the phases
$\phi_i$ fluctuate the trapped state is different than that given in
Eq.~(\ref{eq:trapped}).  In particular, this state is not robust to interaction
with {\em any} environmental electromagnetic field mode.  However, in the
semiclassical picture given above, the trapped state is indeed isolated from
the driving lasers.

\section{DF subspaces with respect to spontaneous emission}

Despite the fact that coherent trapped states are not examples of DF subspaces,
we can still use the idea of this trapped states within the context of DF
subspace for a multilevel atomic system subject to an approximation of which
processes are most likely to cause decoherence.

Consider again the $\Lambda$ configuration but now in a fully quantum
treatment.  In this system, spontaneous emission from $|c\rangle$ into the
$|a\rangle$ and $|b\rangle$ state is clearly possible.  This is due to a ${\bf
a}^\dagger_k |a \rangle \langle c|$ or ${\bf a}^\dagger_k |b \rangle \langle
c|$ term in the coupling Hamiltonian where ${\bf a}_k^\dagger$ is the creation
operator for the photon mode $k$.  Suppose $\omega_b>\omega_a$. Spontaneous
emission from $|b\rangle$ to $|a\rangle$ can still occur, but now it must
transverse virtually through the $|c\rangle$ mode.  This occurs from higher
order interactions like $({\bf a}_k^\dagger |a\rangle \langle c| ) ({\bf a}_k
|c\rangle \langle b|)={\bf a}_k^\dagger {\bf a} |a\rangle \langle b|$ and will
in most cases be much weaker that the spontaneous emission from $|c\rangle$.

If one derives a master equation for the $\Lambda$ configuration, unless the
interaction is taken to high enough order, the only decohering terms when
interacting with the electromagnetic vacuum are the spontaneous emission terms.
In such a treatment  the Lindblad operators for the $\Lambda$ configuration are
given by
\begin{equation}
{\bf L}_1= |a\rangle \langle c| \quad {\rm and } \quad {\bf L_2}=|b\rangle
\langle c|.
\end{equation}
With respect to these Lindblad operators it is clear that there is a DF
subspace is given by the two ground states $|a\rangle$ and $|b\rangle$ (${\bf
L}_i$ annihilates both of these states).

\begin{figure}[ht]
\quad  \quad\psfig{figure=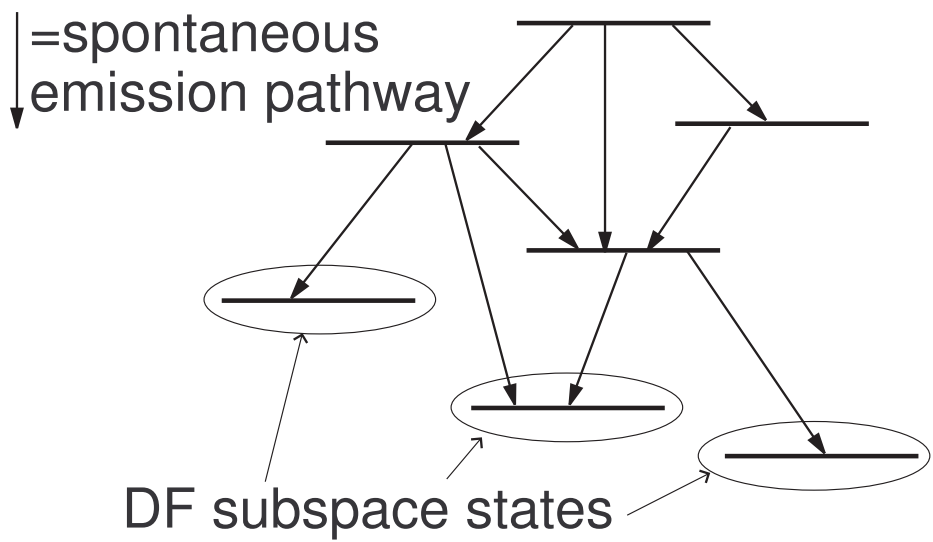,width=5in}
 \caption{\em Multilevel spontaneous emission DF subspace} \label{fig:spon}
\end{figure}

In general, multilevel atomic systems then can have DF subspaces with respect
to the spontaneous emission given by the states which do not spontaneously
emit.  This observation is not very profound.  Given, however, that such states
exist we can however ask the more interesting question of how one can
manipulate the information in these states.  Of significance here is that the
multiple states which do not spontaneously emit do not have strong transitions
between themselves because if they did this would be a spontaneous emission
pathway.

\section{Manipulation of information in spontaneous emission DF subspaces}

In order to understand how it might be possible to manipulate the information
stored in states which do not spontaneously emit, consider the scenario
diagramed in Figure~\ref{fig:fourlevel} below.

\begin{figure}[ht]
\quad  \quad\psfig{figure=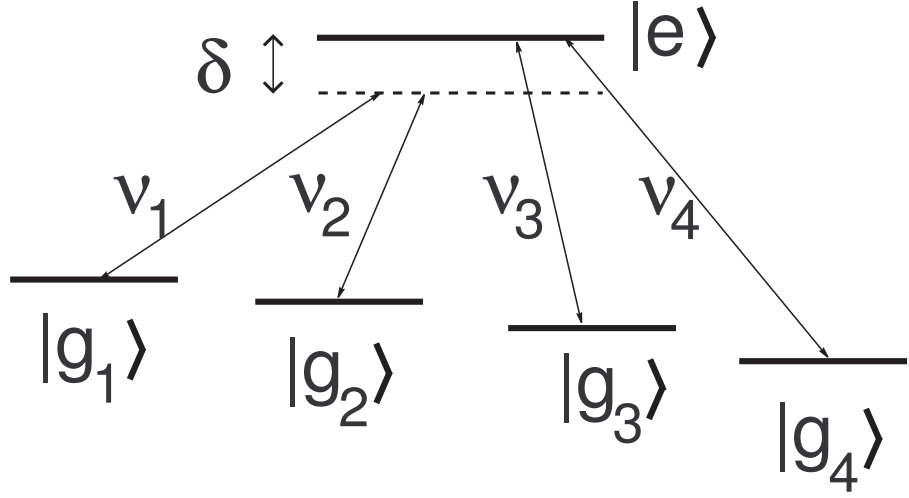,width=5in}
 \caption{\em Four level DF subspace scheme} \label{fig:fourlevel}
\end{figure}

Here four laser with frequencies $\nu_i$ illuminate a five level system, four
states $|g_i\rangle$ which do not spontaneously emit and a fifth level
$|e\rangle$ which is subject to spontaneous emission.  Two of these transitions
are driven on resonance and the other two are driven at a detuning of $\delta$
as shown in Figure~\ref{fig:fourlevel}.  The semiclassical Hamiltonian for this
system in the interaction picture is given by
\begin{eqnarray}
{\bf V}(t)&=&{1 \over 2}\left(\Omega_1 e^{-i\delta t} |g_1\rangle + \Omega_2
e^{-i\delta t}|g_2\rangle + \Omega_3 |g_3\rangle + \Omega_4 |g_4\rangle \right)
\langle e | \nonumber \\
  &&+{1 \over 2} |e\rangle \left(\Omega_1 e^{i \delta t} |g_1\rangle + \Omega_2 e^{i
  \delta t}  |g_2 \rangle + \Omega_3 |g_3\rangle + \Omega_4 |g_4\rangle
  \right),
\end{eqnarray}
where we have assumed a fixed phase for all of the incident light and
$\Omega_i$ are the Rabi frequencies of the transitions.  Moving into a frame
rotating with $|g_1\rangle$ and $|g_2\rangle$, this becomes
\begin{equation}
\tilde{\bf V}(t)= \left({1 \over 2} \sum_{i=1}^4 \Omega_i \left[ |g_i\rangle
\langle e | + |e \rangle \langle g_i| \right]  \right)+ \delta |g_1\rangle
\langle g_1| + \delta |g_2\rangle \langle g_1|.
\end{equation}
For simplicity we will assume that $\Omega_i=\Omega$ for all $i$.  There are
then two eigenstates of $\tilde{\bf V}(t)$ which have no support on the
spontaneously emitting state $|e\rangle$.  These states are
\begin{equation}
|\psi_1\rangle = {1 \over \sqrt{2}} (|g_1\rangle - |g_2\rangle) \quad {\rm and}
\quad |\psi_2\rangle ={1 \over \sqrt{2}} (|g_3\rangle - |g_4\rangle),
\end{equation}
which are eigenstates of $\tilde{\bf V}(t)$ with eigenvalue $e$ and $0$
respectively.  Thus we could imagine encoding a qubit of information into the
states $|\psi_1\rangle$ and $|\psi_2\rangle$.  It is then possible to use the
above transitions so that a gate which acts as $\bmath{\sigma}_z$ over these
two states is achieved.  Thus we see that it is possible to achieve differing
phase evolutions via the application of resonant coherent population trapping
beams and detuned coherent population trapping beams.

However, the question which now remains is how to perform other single qubit
operations other that the encoded $\bmath{\sigma}_z$?  We can show, in fact,
that it is not possible to first order in time for such transitions to occur.

To see this, we note that the operations which we wish to enact are of the form
\begin{equation}
c |\psi_1\rangle \langle \psi_2| +c^*|\psi_2\rangle \langle \psi_1| + {\rm
any~operator~on~the~non-DFS~states}. \label{eq:wanted}
\end{equation}
The general coupling Hamiltonian of these five levels to laser fields is,
assuming only the allowed transitions,
\begin{eqnarray}
{\bf V}(t)=\sum_{i=1}^4 \sum_j e^{-i \phi_{ij} -i ( \nu_j- \omega_e+\omega_i
)t} |e\rangle \langle g_i| + e^{i\phi_{ij} +i(\nu_j -\omega_e + \omega_i) t}
|g_i\rangle \langle e|,
\end{eqnarray}
where $j$ labels the laser mode of frequency $\nu_j$, $\omega_e$ is the energy
of $|e\rangle$, $\omega_i$ is the energy of $|g_i\rangle$ and $\phi_{ij}$ is
the phase of the $j$th mode on the $i$th level $|g_i\rangle$.  Examining this
Hamiltonian, it is apparent that there is no way to get an operator like that
in Eq.~(\ref{eq:wanted}).  In particular, no terms like $|g_1\rangle \langle
g_3|$ appear in this formula.

\section{Outlook}

In this chapter we have addressed the issue of DF subspaces in multi-level
atomic systems.  Interestingly we have seen that such subspaces can exist under
ideal assumptions only when a level is degenerate.  When levels are not
degenerate, it is still possible that ground states may be decoherence-free
under the assumption of a vacuum environment.  Unfortunately, there are no
first or processes which preserve the DF subspace in such multilevel-level
atomic systems for the same reason that the DF subspace exists.  An interesting
question is the existence of DF subsystems in multi-level atomic systems.

\chapter{Decoherence-free Subsystems for Quantum Computation}

In part II of this thesis we have had the opportunity to examine a particular
method for avoiding the detrimental process of decoherence.  The experimental
demonstration of a DF subspace in the ion trap quantum computing
architecture\cite{Kielpinski:01a,Kielpinski:01b} (as well as proof-of-principle
demonstrations with engineered decoherence done by Kwiat, {\em et
al.}\cite{Kwiat:00a}) lends credit to the notion that the notion of DFSs will
be an important of future quantum computers.  Theoretical arguments for DFSs in
other physical systems (of particular note are the solid state proposals of
Zanardi and Rossi\cite{Zanardi:98b,Zanardi:99a} and the use of a
decoherence-free subspaces for creating Schr\"{o}dinger cat states of a
Bose-Einstein condensate\cite{Dalvit:00a}) also lend credit to the notion that
decoherence-free systems will play an important part in overcoming decoherence
in constructing a quantum computer.

An important lesson to be taken from DFSs is the notion that just because a
system has a high decoherence rate that does not mean that the system cannot be
used for robust quantum computation.  Symmetries of the system-environment
coupling allow for system which might otherwise be discarded as having ``too
high a decoherence rate'' to be managed into a realm where quantum computation
may become possible.

DFSs are a good example of a small subsystems technique for dealing with
decoherence.  They cannot and are not the end-all solution for quantum
computation for reasons which we have detailed in the previous chapters.  That
being said, they can represent a large step towards making such a solution
technologically feasible.

\part{Natural Fault Tolerance}

\chapter{The Road Ahead}

\begin{quote}
{\em How difficult is it to build a quantum computer?}
\end{quote}

The discovery that fault-tolerant quantum computation can be used to solve the
decoherence problem was among one of the greatest theoretical achievements of
the end of the twentieth century.  With enough control, decoherence can be
reversed!  While this discovery is heartening to the prospects of building a
quantum computer, the road towards the eventual construction of a quantum
computer is far from paved and it is certainly unknown if this pavement is made
of gold or rather, as some pessimists believe, mere asphalt.

Towards this end, there is much to be said for thinking deeper about error
correction, fault-tolerance, and the ultimate use of physical systems to
achieve the goal of quantum computation.  In Part III of this thesis, we make
first steps towards the idea that building a quantum computer may not be as
difficult as early experiments and theoretical understanding indicates.  In
particular we will eschew the notion that quantum computers must be build from
the single quantum system up in lieu of the idea that there may be many-body
quantum systems which are {\em naturally fault-tolerant}.

One way to look at this is from the perspective of why classical computers have
achieved such success.  When Turing, von Neumann and others began thinking
about constructing a physical device which carried out the theoretical concepts
of computer science, it was certainly unclear that classical computers would
eventually attain today's amazing speeds and versatility.  Having achieved so
much with the modern day silicon revolution, it is important to realize,
however, that {\em strict physical principles are responsible for the
robustness of classical computers}.  The discovery that decoherence is not a
fundamental roadblock towards building a quantum computer leads us to question
whether there are similar physical principles which can lead to the robustness
of a naturally fault-tolerant quantum computer.

\chapter{Supercoherence} \label{ch:sup}

\begin{quote}
{\em No, decoherence, you cannot forever walk uphill!}
\end{quote}

One of the physical principles which helps make classical computers robust is
their use of energetics.  In particular, a classical current in a transistor is
robust because the energetics of the device directs the flow of information
through the transistor.  Classical conservation of the energy helps transistors
from not making errors.  In this chapter we take the first steps towards
developing quantum systems which similarly harness the power of the transfer of
energy to reduce the destructive effect of decoherence.  We begin with a
discussion of the relationship between (near) energy conservation and
decoherence pathways.  We then introduce a simple example of a supercoherent
system using a Pauli stabilizer error detecting code.  A supercoherent system
is a multi-qubit system which has a ground state in which degenerate quantum
information is encoded.  The degeneracy of this ground state is broken by
single qubit error and these single qubits errors take the ground state to a
state of higher energy.  At low environment temperatures, decoherence is then
ineffective in destroying the coherence of the degenerate supercoherent
information.  We then discuss the difficulties of manipulating the information
in the particular Pauli stabilizer example.  We then present an example of a
supercoherent qubit for which universal manipulation of the quantum information
can be obtained while still retaining supercoherence.  Solid state
implementation of the supercoherent qubit is then discussed.  A bath of
harmonic oscillator  coupling to a supercoherent system is then analyzed and
the supercoherence is directly demonstrated.  Finally it is demonstrated the
Cooper pairs are quantum error detecting codes for resistance causing processes
and the relationship of this to supercoherence is discussed.

\section{Energetics and decoherence}

\begin{quote}
{\em It is typical of modern physicists that they erect skyscrapers of theory
upon the slender foundations of outrageously simplified models}
\begin{flushright} --J.M. Ziman\cite{Ziman:62a}
\end{flushright}
\end{quote}

In the absence of coupling, a system and its environments have separate
dynamics governed by separate energy spectra.  If a perturbing interaction
between the system and environment is then introduced, the dynamics of the
system and environment is {\em dominated} by mechanisms which conserve the
energies of the unperturbed system and environment energy spectra.  This is the
essence of the rotating wave approximation in quantum optics (see, for example,
\cite{Allen:75a,Scully:97a}).  Thus the flow (or lack of flow) of energy
between a system and the environment is, in the perturbing regime commonly
encountered, essential to determining the effective mechanisms of decoherence.
Let us examine a simple analytical example which we can use to gain an
understanding of this principle.

Consider a system consisting of a qubit and its environment also made of a
qubit. In the absence of coupling, we suppose the Hamiltonian of the system is
given by
\begin{equation}
{\bf H}_0= \omega_0 \bmath{\sigma}_z \otimes {\bf I} + (\omega_o+\Delta) {\bf
I} \otimes \bmath{\sigma}_z,
\end{equation}
where $2\omega_0$ is the energy of the system and $2\Delta$ is the difference
between the energy of the system and the environment qubit.  We set
$\Delta>-\omega_0$ and $\omega_0>0$ so as to set the positivity of the energies
on a solid footing.  Now suppose that a perturbing interaction of strength $g$
is introduced between these two qubits and is of the form
\begin{equation}
{\bf V}=g \bmath{\sigma}_x \otimes \bmath{\sigma}_x,
\end{equation}
where $g$ is the interaction energy.  The evolution operator for this system
can be exactly calculated and found to be
\begin{eqnarray}
{\bf U}(t)&=&\left[ \cos(\Omega_1 t) (|01\rangle \langle 01| +|10\rangle
\langle10|)-i \sin(\Omega_1 t)\left( {\Delta \over \Omega_1}(|10\rangle \langle
10|- |01 \rangle \langle 01|)\right. \right. \nonumber \\ &&\left. \left. + {g
\over \Omega_1} (|01\rangle \langle 10| + |10\rangle \langle 01| )\right)
\right] \nonumber
\\
&&+\left[ \cos(\Omega_2 t) (|00\rangle \langle 00| +|11\rangle \langle11|)-i
\sin(\Omega_2 t)\left( {2 \omega_0 + \Delta \over \Omega_2}(|00\rangle \langle
00|- |11 \rangle \langle 11|) \right. \right. \nonumber \\ &&\left. \left. + {g
\over \Omega_2} (|00\rangle \langle 11| + |11\rangle \langle 00| )  \right)
\right],
\end{eqnarray}
where $\Omega_1=\sqrt{\Delta^2+g^2}$ and $\Omega_2=\sqrt{(2 \omega_0 +
\Delta)^2 + g^2 }$.  In the limit of $g \ll \omega_0,\omega_0+\Delta$ (the
perturbing interaction limit), the coupling between the $|00\rangle$ and
$|11\rangle$ states reduces to the unperturbed evolution
\begin{eqnarray}
&&\left[ \cos(\Omega_2 t) (|00\rangle \langle 00| +|11\rangle \langle11|)-i
\sin(\Omega_2 t)\left( {2 \omega_0 + \Delta \over \Omega_2}(|00\rangle \langle
00|- |11 \rangle \langle 11|) \right. \right. \nonumber \\ &&\left. \left. + {g
\over \Omega_2} (|00\rangle \langle 11| + |11\rangle \langle 00| )  \right)
\right] \underbrace{\rightarrow}_{g \ll \omega_0,\omega_0+\Delta} e^{-i (2
\omega_0+\Delta) t }|00\rangle \langle 00|+e^{i(2 \omega_0 +\Delta)t}|11\rangle
\langle 11|. \nonumber \\
\end{eqnarray}
Thus the evolution of the $|00\rangle$ and $|11\rangle$ subspace in the
perturbing limit is nearly identical to the unperturbed dynamics.  The dynamics
of the $|01\rangle$ and $|10\rangle$ subspace, does not escape so easily and is
more drastically affected by the perturbing interaction.  In the absence of
perturbation, the system has energies $-\omega_0$ ($|0\rangle$) and $+\omega_0$
($|1\rangle$) and the environment has energies $-\omega_0-\Delta$ ($|0\rangle$)
and $\omega_0+\Delta$ ($|1\rangle$). When the perturbation is now turned on,
the dynamics is dominated by the action on the $|01\rangle$, $|10\rangle$
pathways. These are exactly the pathways which most closely conserve the
original energies of the system and environment.  Furthermore the pathways
which least conserve energy and act on the $|00\rangle$ and $|11\rangle$ states
contribute little dynamics different from the normal evolution of these states.
This simple example then demonstrates how decoherence is dominated by pathways
which most nearly preserve the unperturbed energies of the system and
environment.

Under the assumption of such a perturbative interaction, energetics play a key
role in determining the rate of decoherence processes.  The notion that
energetics plays a key role in determining decoherence rates is often confused
with the statement that ``the most damaging decoherence is that in which energy
is not exchanged between the system and the environment''.  We emphasize here
that the fact that the most damaging decoherence is often of a form where no
energy is exchanged between the system and the environment is different from
the fact that energetics play a key role in determining the dynamics of
decoherence.  Certainly it is true that the fact that energetics determines the
decoherence pathways allows decoherence which does not exchange energy to act,
but the reason why such decoherence is typically more destructive is not
related to the fact that decoherence is dominated by nearly energy conserving
dynamics.

Having emphasized that energetics is key in determining decoherence dynamics,
it is useful to place decoherence in three different categories. Specifically,
energy conserving decoherence has three possible forms: energy is supplied from
the system to the environment ({\em cooling}), energy is supplied from the
environment to the system ({\em heating}), or no energy is exchanged at all
({\em non-dissipative}).  Thus even when the environment is a heat bath at zero
temperature, cooling and especially non-dissipative decoherence processes
occur.  A schematic of these process is presented in
Figure~\ref{fig:threekings}.

\begin{figure}[h]
\quad  \quad\psfig{figure=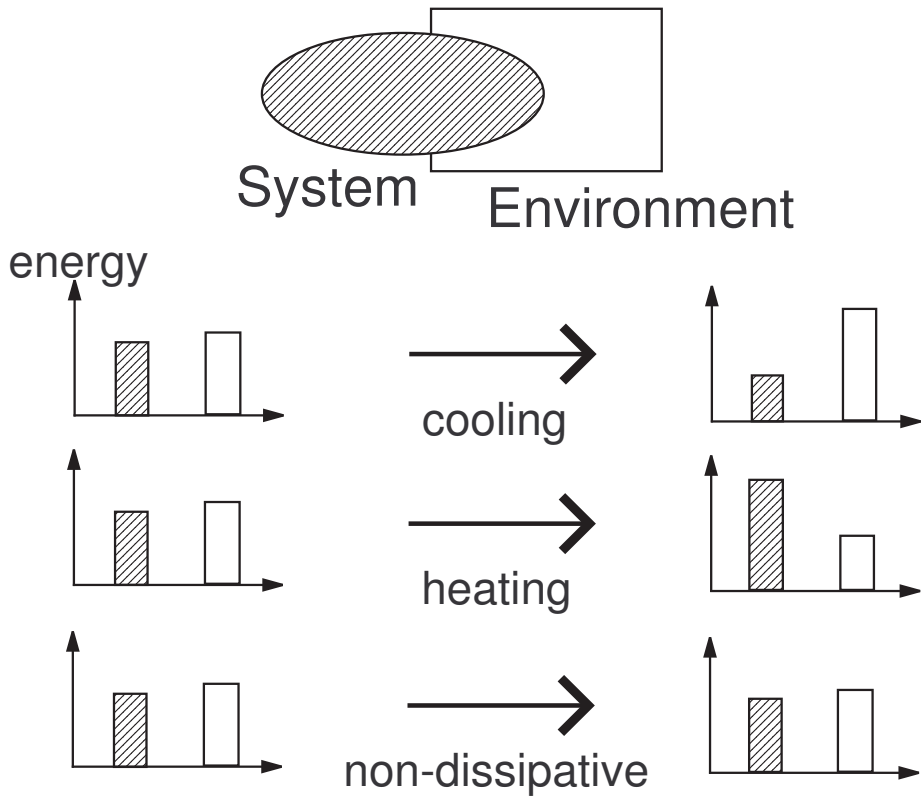,width=5in}
 \caption{\em Heating, cooling, and non-dissipative decoherence dynamics} \label{fig:threekings}
\end{figure}

Among the possible decoherence energetics, non-dissipative decoherence is
often, but not always, the most damaging source of decoherence.  Of the
possible energetically favored pathways, the easiest to eliminate is heating
where energy is transferred from the environment to the system.  By cooling
down the environment, heating can often be nearly completely eliminated as a
decoherence pathway.  Cooling and non-dissipative dynamics, on the other hand,
cannot be eliminated by simply cooling the environment.

\section{A simple Pauli stabilizer supercoherent quantum bit}

Having shown that energetics dominates the allowed dynamics of decoherence, we
now present an approach to reducing decoherence which relies on this
observation.

Consider the smallest possible additive quantum error detecting code which
detects single qubit errors, the $[4,2,2]$ code (see Appendix~\ref{apa:pauli}
for information on this nomenclature and stabilizer codes). The stabilizer of
this code is generated by the two Pauli operators
\begin{equation}
{\bf S}_1=\bmath{\sigma}_x\otimes \bmath{\sigma}_x\otimes
\bmath{\sigma}_x\otimes \bmath{\sigma}_x , \quad {\bf
S}_2=\bmath{\sigma}_z\otimes \bmath{\sigma}_z\otimes \bmath{\sigma}_z\otimes
\bmath{\sigma}_z,
\end{equation}
and encodes two qubits of information. The logical (informational) operators
for this code (modulo the stabilizer) are given by
\begin{eqnarray}
\bar{\bf X}_1=\bmath{\sigma}_x\otimes \bmath{\sigma}_x\otimes {\bf I} \otimes
{\bf I}, \quad \bar{\bf Z}_1={\bf I}\otimes \bmath{\sigma}_z \otimes
\bmath{\sigma}_z \otimes {\bf I} \nonumber
\\ \bar{\bf X}_2={\bf I} \otimes \bmath{\sigma}_x\otimes \bmath{\sigma}_x \otimes {\bf I},
\quad \bar{\bf Z}_2=\bmath{\sigma}_z \otimes \bmath{\sigma}_z \otimes {\bf I}
\otimes {\bf I},
\end{eqnarray}
 ($\bar{\bf Y}_i={1 \over 2i}
[\bar{\bf Z}_i,\bar{\bf X}_i]$, for each code, of course).  A complete set of
commuting operators for this code is thus given by (for example) ${\bf S}_1,
{\bf S}_2, \bar{\bf Z}_1, \bar{\bf Z}_2$.  We will denote the basis
corresponding to this complete set of commuting operators by
$|S_1,S_2,Z_1,Z_2\rangle$.

Now consider the Hamiltonian
\begin{equation}
{\bf H}= \bmath{\sigma}_x \otimes \bmath{\sigma}_x \otimes {\bf I} \otimes {\bf
I} + {\bf I} \otimes {\bf I} \otimes \bmath{\sigma}_x \otimes \bmath{\sigma}_x
 + {\bf I} \otimes \bmath{\sigma}_z \otimes \bmath{\sigma}_z \otimes {\bf I}+
 \bmath{\sigma}_z\otimes {\bf I}\otimes {\bf I}\otimes \bmath{\sigma}_z.
\end{equation}
The stabilizer of the $[4,2,2]$ code commutes with this Hamiltonian.  $\bar{\bf
Z}_2$ and $\bar{\bf X}_2$ also both commute with this Hamiltonian.  This
implies that the action of the Hamiltonian acts only on the first encoded
qubit. Indeed we see that this Hamiltonian can be written in terms of this code
as
\begin{eqnarray}
{\bf H}&=&\bar{\bf X}_1 + \bar{\bf X}_1 {\bf S}_1 + \bar{\bf Z}_1 + \bar{\bf
Z}_1 {\bf S}_2 \nonumber \\ &=& \bar{\bf X}_1 ({\bf I}+{\bf S}_1) + \bar{\bf
Z}_1 ({\bf I}+{\bf S}_2).
\end{eqnarray}
Notice how this Hamiltonian does not depend on the second qubit.  Therefore
this Hamiltonian will have a spectrum which is two-fold degenerate: this
degeneracy corresponding to the second encoded qubit.

We recall that the codespace of the $[4,2,2]$ code is labeled by the
eigenvalues, $S_1$ and $S_2$, of the stabilizer generators, ${\bf S}_1$ and
${\bf S}_2$ respectively. For each set of eigenvalues for the stabilizer
generators the action of ${\bf H}$ on the corresponding subspace is different.
In fact we see that
\begin{eqnarray}
&&S_1=+1, S_2=+1 \Rightarrow {\bf H}= 2(\bar{\bf X}_1 +\bar{\bf Z}_1) \nonumber
\\ &&S_1=+1, S_2=-1 \Rightarrow {\bf H}= 2\bar{\bf X}_1 \nonumber \\&& S_1=-1,
S_2=+1 \Rightarrow {\bf H}= 2\bar{\bf Z}_1 \nonumber \\&& S_1=-1, S_2=-1
\Rightarrow {\bf H}= {\bf 0}.
\end{eqnarray}
The eigenvalues of ${\bf H}$ for these four cases are thus $\pm 2 \sqrt{2}, \pm
2, \pm 2,$ and $0$ respectively.  We therefore see that the ground state of
${\bf H}$ will be within the $S_1=+1, S_2=+1$ subspace of the $[4,2,2]$ code
and will be the $-2 \sqrt{2}$ eigenvalue of ${\bf H}$ over the first encoded
qubit.  But what about the second encoded qubit?  Here we see that the ground
state of the code is actually two-fold degenerate: corresponding directly to
the second encoded qubit.  The spectrum of this Hamiltonian is given in
Figure~\ref{fig:paulisuper}.

\begin{figure}[h]
\quad  \quad \quad \psfig{figure=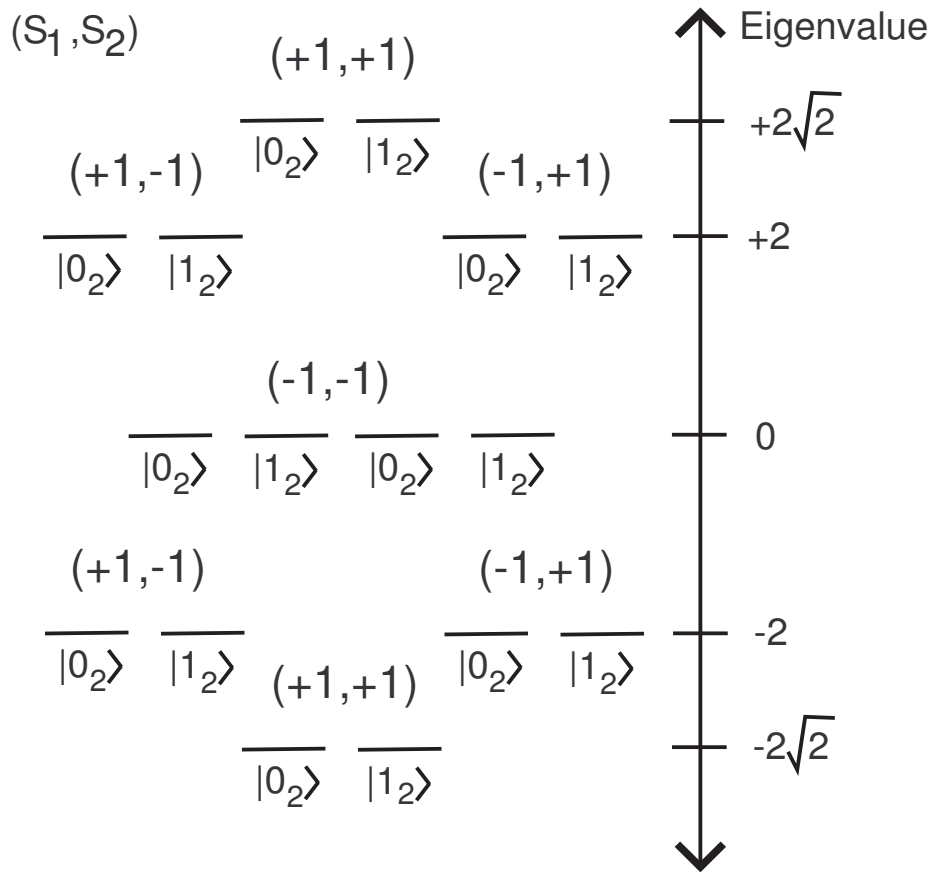,width=5in}
 \caption{\em Spectrum of the Pauli supercoherent qubit} \label{fig:paulisuper}
\end{figure}

The ground state of this Hamiltonian has a spectacular property with respect to
single qubit operators on this four qubit system.  First recall that the code
stabilized by ${\bf S}_1$ and ${\bf S}_2$ is a single qubit error detecting
code.  Thus every single qubit operator $\bmath{\sigma}_\alpha^{(i)}$
anticommutes with at least one of ${\bf S}_1$ and ${\bf S}_2$.  This in turn
implies that every single qubit error flips the value of the eigenvalue of
${\bf S}_1$ or ${\bf S}_2$ for every basis state $|S_1,S_2,Z_1,Z_2\rangle$.

For example, suppose we are in the state labeled by $(S_1=+1,S_2=+1)$, with the
Hamiltonian eigenvalue of $-2\sqrt{2}$, and logical basis $|0_2\rangle$ for the
second encoded qubit.  This state is one of the ground states of ${\bf H}$, the
other being the logical basis $|1_2\rangle$.  The action of a single qubit
operation will act to change the value of $S_1$ or $S_2$ or possibly both. In
any case, this implies that the action of the single qubit operator is to take
the system from this ground state to one of the higher energy eigenvalue states
of ${\bf H}$.  Diagrammed in Figure~\ref{fig:pauliz} is the action of
$\bmath{\sigma}_z^{(1)}$ on the energy levels of ${\bf H}$

\begin{figure}[h]
\quad  \quad \quad \psfig{figure=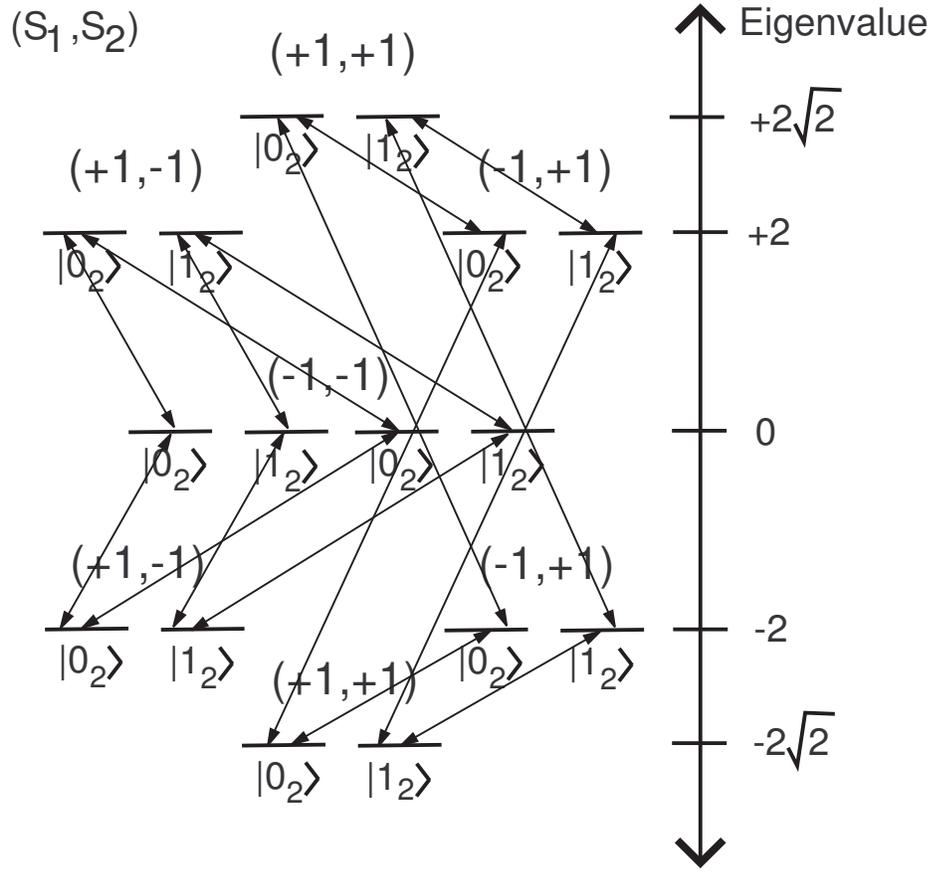,width=5in}
 \caption{\em Effect of $\bmath{\sigma}_z^{(1)}$ on the Pauli supercoherent qubit} \label{fig:pauliz}
\end{figure}

In fact we see that this is generically true for single qubit operators acting
on a system with this Hamiltonian.  Every single qubit operator changes a value
of $S_1$ and $S_2$ and therefore takes the system from the ground state to a
higher energy state.  We term such a Hamiltonian a supercoherent Hamiltonian
and make the following general definition:
\begin{definition} {\em (Supercoherence)}
A system of qubits with a system Hamiltonian ${\bf H}$ which has a degenerate
ground state and for which every single qubit operator takes the system out
states in this degenerate ground state is called a {\em supercoherent
Hamiltonian}. In more generality, we may allow the subsystems which make up the
system to be take any desired subsystem structure.  The criteria used for
supercoherence is then that every operator on an individual subsystem must take
the system out of the degenerate ground state.  If $|i\rangle$ denotes the
degenerate ground state of the Hamiltonian ${\bf H}$, then the condition for
supercoherence is
\begin{equation}
\langle j|{\bf o}^{(k)}|i\rangle=0, \quad \forall {\bf o}^{(k)},
\label{eq:superdef}
\end{equation}
for all $|i\rangle$, $|j\rangle$ in the degenerate ground state and ${\bf
o}^{(k)}$ is the single subsystem operator ${\bf o}$ acting on the $k$th
subsystem.  Information which has been encoded into the degenerate ground state
of a supercoherent Hamiltonian is referred to as a supercoherent qubit or
supercoherent qudit, depending on the dimension of the degeneracy.
\end{definition}

Note that the supercoherence condition Eq.~(\ref{eq:superdef}) implies the
ground states are an error detecting code for the operators single qubit
(subsystem) operators.

Why do we label such degenerate ground states supercoherent?  The main reason
for this lies in the fact that the Hamiltonian of such a system has been
constructed so that the only single qubit operations which can destroy the
coherence of the system are interactions which heat the system.  As we
mentioned previously, it is often possible to substantially decrease such
decoherence mechanisms by simply cooling the system's environment.  We will
have a chance to analytically demonstrate this effect in
Section~\ref{sec:hbathex}.  In general, we expect that the condition for
supercoherence to hold will occur when the temperature $T$ (we set $k_B=1$) is
much less than the energy gap of from the degenerate ground state to the lowest
state excited by the single qubit operators.  What kind of robustness should we
expect for the supercoherent qubit?  If the individual baths have a temperature
$T$, then we expect the decoherence rate of the supercoherent qubit to scale at
low temperatures as $\approx e^{-\beta \Delta}$, where $\beta=(kT)^{-1}$.  At
low temperatures there should thus be an exponential suppression of the
decoherence.

It is helpful to compare a supercoherent qubit to a single qubit with two
different Hamiltonians.

First compare a supercoherent qubit to a single qubit with non-degenerate
energy levels ${\bf H}=\epsilon \bmath{\sigma}_z$.  Now the single qubit error
$|0\rangle \langle 1|$ is a error which takes the system from a state of higher
energy to one of lower energy.  This error, then, will be involved in a cooling
type decoherence.   The single qubit error $|1\rangle \langle 0|$ takes the
system from a state of lower energy to one of higher energy.  This is a heating
type decoherence.  Finally the single qubit error $\bmath{\sigma}_z$ does not
change the energy of the system but acts to dephase the system.  Therefore this
error is of the non-dissipative form.  In contrast to this single qubit
example, all single qubit errors acting on a supercoherent ground state must
take the system from the ground state to a state of higher energy.  Thus all of
the above error $|0\rangle \langle 1|$, $|1\rangle \langle 0|$ and
$\bmath{\sigma}_z$ are errors on the supercoherent Hamiltonian of the heating
form.

Second it is useful to compare the supercoherent qubit to a single qubit with a
degenerate Hamiltonian.  In this case all $\bmath{\sigma}_\alpha$ errors act on
this qubit are of the non-dissipative form because the two qubits have the same
energy.  This is in direct contrast to the supercoherent qubit for which all
errors are of the heating form.  Degeneracy alone is not enough for
supercoherence.

\subsection{Encoded operations on the Pauli supercoherent qubit}

Returning now the specific Pauli stabilizer supercoherent example we began
with, we can now ask the question of how to manipulate the information encoded
into the degeneracy.  Nothing is worse than a quantum memory which one cannot
manipulate!

In this case, we already know how to manipulate the quantum information in the
degeneracy because this degeneracy is simply the second encoded qubit.  In
particular the two-qubit operations $\bar{\bf X}_2={\bf I} \otimes
\bmath{\sigma}_x \otimes \bmath{\sigma}_x \otimes {\bf I}$ and $\bar{\bf
Z}_2=\bmath{\sigma}_z \otimes \bmath{\sigma}_z \otimes {\bf I} \otimes {\bf I}$
are encoded single qubit rotations on this qubit.  By turning on an off these
interactions it is therefore possible to perform computation any $SU(2)$
rotation on the supercoherent qubit.

The question which is immediately raised, however, is how turning on an off the
interactions affects the supercoherent property of the system.  Suppose that we
have a supercoherent system with a gap from the ground state to the single
qubit excited states of energy $\Delta$.  When we turn on $\bar{\bf X}_2$ or
$\bar{\bf Z}_2$, we want to make sure that this gap is still preserved and the
supercoherent property that any single qubit gate will take the supercoherent
qubit to a state of higher energy is maintained.

In order for the gap to be still maintained the interaction strength must be
weak in comparison with the temperature of the bath.  To see this, we note that
$\bar{\bf X}_2$ and $\bar{\bf Z}_2$ break the degeneracy of the system.  If
these interactions are of strength $\epsilon$, then this degeneracy is split by
$\epsilon$.  The gap of size $\Delta$ is therefore shrunk by $c \epsilon$ where
$c$ is a prefactor which may be state dependent.  In order to maintain the
supercoherent condition, this shrinking must not remove the condition that the
gap is large compared to the temperature of the bath.  Thus if $\epsilon \ll
T$, the gap is maintained while the computation is being performed.

Also notice that the manipulation of the degenerate information does not change
the fact that single qubit interactions take the ground states (which are no
longer degenerate due to the interaction) to states of higher energy.  To see
this is sufficient to notice that the interactions we implement act {\em only}
on the degeneracy.  If, for example, we tried to implement a $\bmath{\sigma}_x$
on the codespace with $\bar{\bf X}_2 \bar{\bf X}_1= \bmath{\sigma}_x \otimes
{\bf I} \otimes \bmath{\sigma}_x \otimes {\bf I}$, we would act not only on the
degenerate supercoherent qubit, but also on the space of the Hamiltonian ${\bf
H}$.

Having shown how to perform single qubit rotations on the information in the
supercoherent qubit, we can now ask the question of whether it is possible to
perform universal quantum computation to perform universal quantum computation
on conjoined sets of such supercoherent qubits.  The difficulty here is that
any two qubit interaction which acts between two conjoined supercoherent qubits
act as single qubit operations on each individual supercoherent systems.
Therefore any two qubit interaction necessarily excites the information in the
supercoherent ground state to an excited energy level.  Of course, we could
resort to more than two qubit interactions.  This method, however, is highly
unlikely to be practical and we will therefore disregard this technique.

One method for overcoming this problem is, when an interaction between the
encoded qubits is desired, another set of Hamiltonians is turned on which
maintains the ground state condition of the individual supercoherent qubits but
also creates other ground states which are degenerate with these states.
Manipulation of the quantum computation is then performed over this larger
supercoherent ground state.  We know of no such method for the Pauli stabilizer
supercoherent example we have just presented, however, in the next section we
will present a more complicated supercoherent system which can be used to
perform universal quantum computation in a manner similar to what we have just
described.

\section{Exchange-based supercoherent quantum bit}

Amazingly there is a supercoherent system which is intimately related to the
strong collective decoherence DFS states which we have so thoroughly studied in
Part II of this thesis.  Consider the Hamiltonian
\begin{equation}
{\bf H}_0^{[n]}={\Delta \over 2}(\vec{\bf S}^{[n]})^2, \label{eq:superh}
\end{equation}
where we recall that ${\bf S}_\alpha^{[n]}={1 \over 2} \sum_{i=1}^n
\bmath{\sigma}_\alpha^{(i)}$.  This Hamiltonian has eigenvalues ${\Delta \over
2} J_n (J_n+1)$, with corresponding eigenstates given by the strong DFS basis
$|\lambda,J_n,m_\alpha \rangle$.

Let us briefly recall the definitions of the strong DFS basis for completeness.
Let ${\mathcal H}_n=(\CC^2)^{\otimes n}$ be a Hilbert space of $n$ qubits, and
let ${\bf s}_\alpha^{(i)}$ be the $\alpha$th Pauli spin operator acting on the
$i$th qubit tensored with identity on all other qubits. The ${\bf
s}_\alpha^{(i)}$ satisfy the commutation and anticommutation rules, $[{\bf
s}_\alpha^{(j)},{\bf s}_\beta^{(k)}]=i\delta_{jk} \epsilon_{\alpha \beta
\gamma} {\bf s}_\gamma^{(j)}$ and $ \{ {\bf s}_\alpha^{(j)}, {\bf
s}_\beta^{(k)}  \} = {1 \over 2} \delta_{jk} \delta_{\alpha \beta} {\bf I} +
2(1-\delta_{jk}){\bf s}_\alpha^{(j)} {\bf s}_\beta^{(k)}$.  We define the $k$th
partial collective spin operators on the $n$ qubits, ${\bf S}_\alpha^{[k]} =
\sum_{i=1}^k {\bf s}_\alpha^{(i)}$. The total collective spin operators acting
on all $n$ qubits, ${\bf S}_{\alpha}^{[n]}$, form a Lie algebra ${\mathcal L}$
which provides a representation of the Lie algebra $su(2)$: $[{\bf
S}_{\alpha}^{[n]},{\bf S}_\beta^{[n]}]=i \epsilon_{\alpha \beta \gamma} {\bf
S}_{\gamma}^{[n]}$.  Thus ${\mathcal L}$ can be decomposed in a direct product
of irreducible representations (irreps) of $su(2)$, ${\mathcal L} \simeq
\bigoplus_{J=0,1/2}^{n/2} \bigoplus_{k=1}^{n_J} {\mathcal L}_{2J+1}$, where
${\mathcal L}_{2J+1}$ is the $2J+1$ dimensional irrep of $su(2)$ which appears
with a multiplicity $n_J$.  If we let $({\bf J}_d)_\alpha$ be the operators of
the $d$ dimensional irrep of $su(2)$, then there exists a basis for the total
collective spin operators such that ${\bf S}_{\alpha}^{[n]}=
\bigoplus_{J=0,1/2}^{n/2} {\bf I}_{n_J} \otimes ({\bf J}_{2J+1})_{\alpha}$.
Corresponding to this decomposition of ${\bf S}_{\alpha}^{[n]}$, the Hilbert
space ${\mathcal H}$ can be decomposed into states $|\lambda,J_n,m\rangle$
classified by quantum numbers labeling the irrep, $J_n$, the degeneracy index
of the irrep, $\lambda$, and an additional internal degree of freedom, $m$. A
complete set of commuting operators consistent with this decomposition and
providing explicit values for these labels is given by $B_\alpha= \{ (\vec{\bf
S}^{[1]})^2, (\vec{\bf S}^{[2]})^2,\dots, (\vec{\bf S}^{[n-1]})^2, (\vec{\bf
S}^{[n]})^2, {\bf S}^{[n]}_\alpha \}$\cite{Kempe:01a}. Therefore a basis for
the entire Hilbert space is given by $|J_1,J_2,\dots,J_{n-1},J_n,m_\alpha
\rangle$, where $(\vec{\bf S}^{[k]})^2|J_1,\dots,J_n,m_\alpha\rangle = J_k
(J_k+1) |J_1,\dots,J_n,m_\alpha\rangle$ and ${\bf
S}_\alpha^{[n]}|J_1,\dots,J_n,m_\alpha\rangle = m_\alpha
|J_1,\dots,J_n,m_\alpha\rangle$.  The degeneracy index $\lambda$ of a
particular irrep having total collective spin $J_n$ is completely specified by
the set of partial collective spin eigenvalues $J_k$, $k<n$: $\lambda \equiv \{
J_1,\dots, J_{n-1} \}$. This degeneracy is simply due to the ($n_J$) different
possible ways of constructing a spin-$J_n$ out of $n$ qubits.

Thus the (possibly degenerate) ground state of the Hamiltonian in
Eq.~(\ref{eq:superh}) is given by the lowest $J_n$ states for a particular $n$.
For $n$ even, these states have $J_n=0$, and for $n$ odd they have $J_n=1/2$.
Furthermore, ${\bf H}_0^{[n]}$ can be constructed from two-qubit interactions
alone:
\begin{equation}
{\bf H}_0^{[n]} = {\Delta \over 2} \left(\sum_{i \neq j=1}^n \vec{\bf s}^{(i)}
\cdot \vec{\bf s}^{(j)} + {3 n \over 4} {\bf I} \right).
\end{equation}
Thus we see that ${\bf H}_0^{[n]}$ is nothing more than the Heisenberg coupling
$\vec{\bf s}^{(i)} \cdot \vec{\bf s}^{(j)}$ acting with equal magnitude between
every pair of qubits.  The identity component of ${\bf H}_0$ produces only a
trivial global phase on the system and is not relevant to our discussion.

${\bf H}_0^{[n]}$ has a highly degenerate spectrum, with energies determined by
$J_n$.  To determine the effect of single qubit operations on these states we
first examine the effect of a single qubit operation on the $n$th qubit, ${\bf
s}_\alpha^{(n)}$.  Since $[{\bf s}_\alpha^{(n)},(\vec{\bf S}^{[k]})^2]=0$ for
$k<n$, we see that ${\bf s}_\alpha^{(n)}$ can not change the degeneracy index
$\lambda$ of a state $|\lambda,J_n,m_\alpha\rangle$.  Let ${\bf O}_n=-{1 \over
4} {\bf I}+(\vec{\bf S}^{[n]})^2-(\vec{\bf S}^{[n-1]})^2$ (defined for $n>1$).
${\bf O}_n$ determines which final step is taken in making the addition from
qubit $n-1$ to qubit $n$ (see Figure~\ref{fig:supercoherent}). If the final
step from $J_{n-1}$ to $J_n$ was taken by adding $1/2$, then the eigenvalue of
${\bf O}_n$ will be $O_n=J_{n-1}+{1 \over 2}$, while if it was taken by
subtracting $1/2$, then $O_n=-(J_{n-1}+{1 \over 2})$.  It is convenient to
replace $(\vec{\bf S}^{[n]})^2$ by ${\bf O}_n$ in our set of commuting
operators, which can clearly be done while still maintaining a complete set of
commuting operators. We can then replace the quantum number $J_n$ by $O_n$, to
obtain the basis $|\lambda,O_n,m_\alpha\rangle$.  It is easy to verify that $
\{{\bf O}_n,{\bf s}_\alpha^{(n)} \} = {\bf S}_\alpha^{[n]}$. If we examine the
effect of ${\bf s}_\alpha^{[n]}$ on the basis $|\lambda,O_n,m_\alpha \rangle$
(where we have defined $m_\alpha$ as the orientation corresponding to ${\bf
S}_\alpha^{(n)}$), then we find that
\begin{eqnarray}
(O_n^\prime+O_n)\langle \lambda, O_n,m_\alpha| {\bf s}_\alpha^{(n)}
|\lambda^\prime, O_n^\prime ,m_\alpha^\prime \rangle \nonumber \\ = m_\alpha
\delta_{\lambda, \lambda^\prime } \delta_{O_n,O_n^\prime} \delta_{m_\alpha
,m_\alpha^\prime}. \label{eq:1}
\end{eqnarray}
Thus we see that the only non-zero matrix elements occur when $O_n^\prime=O_n$
or $O_n^\prime=-O_n$.  From this it follows that the final step in the paths of
Figure~\ref{fig:supercoherent} can either flip sign ({\it e.g.}, $1 \rightarrow
-1$) or else must remain the same. Using the relation between $O_n$ and $J_n$
above, shows that this results in the selection rules $\Delta {\bf J}_n = \pm
1,0$ for ${\bf s}_\alpha^{(n)}$ acting on states in the $|\lambda, J_n,
m_\alpha \rangle$ basis. Note further that if we had chosen a basis with
$m_\beta$ instead of $m_\alpha$ in Eq.~(\ref{eq:1}) ($\beta \neq \alpha$), we
would have obtained the same selection rule but now the $m_\alpha$ components
could be mixed by the operation of ${\bf s}_\beta^{(n)}$. We recall (see
Appendix~\ref{apc}) that the exchange operation ${\bf E}_{ij}={1 \over 2} {\bf
I}+2 \vec{\bf s}^{(i)} \vec {\bf s}^{(j)}$ which exchanges qubits $i$ and $j$
modifies only the degeneracy index $\lambda$ of the
$|\lambda,J_n,m_\alpha\rangle$ basis.  Because ${\bf s}_\alpha^{(j)}= {\bf
E}_{jn} {\bf s}_\alpha^{(n)} {\bf E}_{jn}$, this implies that any single qubit
operator ${\bf s}_\beta^{(i)}$ can therefore give rise to mixing of both the
spin projections $m_\alpha$, and of the degeneracy indices $\lambda$.

These selection rules must be modified for the $J_n=0$ states. $O_n=-1$ and
$m_\alpha=0$ for all $J_n=0$ states and any transitions between these states
will therefore have zero matrix element, {\it i.e.}, $\langle
\lambda,J_n=0,m_\alpha| {\bf s}_\alpha^{(n)}|\lambda^\prime, J_n^\prime=0
,m_\alpha^\prime \rangle =0$. Thus the transitions $\Delta J = 0$ are forbidden
for $J_n = 0$, and ${\bf s}_\alpha^{(n)}$ {\em must} take $J_n=0$ states to
$J_n=1$ states. Furthermore, since $\langle \lambda, J_n=0,0| {\bf
s}_\alpha^{(n)} |\lambda^\prime,J_n'=0,0 \rangle =0$, the degeneracy index
$\lambda$ for $J_n=0$ states is not affected by any single qubit operation.

To summarize, we have shown that any single qubit operation ${\bf
s}_\alpha^{(i)}$ enforces the selection rules $\Delta J_n=\pm 1,0$ with {\em
the important exception of $J_n=0$ which must have $\Delta J_n=+1$}.  The
degenerate $J_n=0$ states are therefore a quantum error detecting code for
single qubit errors\cite{Bacon:99a,Kempe:01a}, with the special property that
they are also the ground state of a realistically implementable Hamiltonian.

\begin{figure}[h]
\psfig{figure=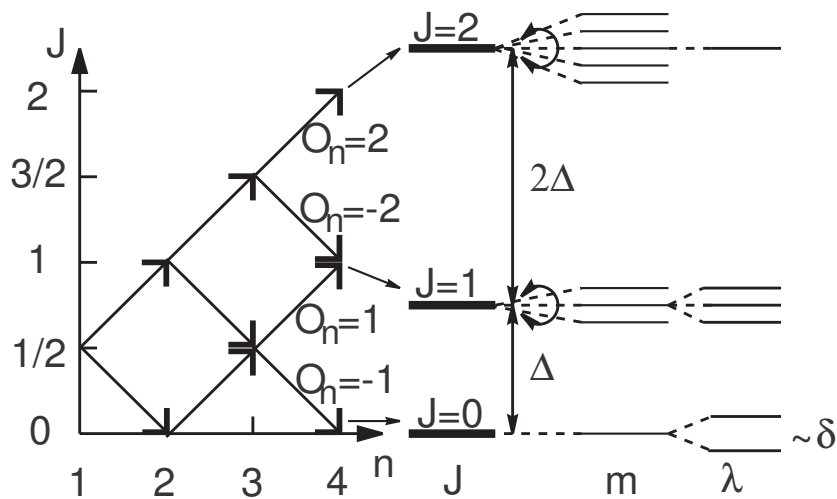,width=5in} \caption{\em Diagram showing formation of
the $|\lambda,J_n,m\rangle$ states}  \label{fig:supercoherent}
\end{figure}

Figure \ref{fig:supercoherent} shows that for an even number of qubits the
$J_n=0$ ground state of ${\bf H}_0^{[n]}$ is degenerate.  For $n=4$ physical
qubits the ground state is two-fold degenerate \cite{Zanardi:97a}.  This
degeneracy cannot be broken by any single qubit operator and single qubit
operations must take the $J_n=0$ states to $J_n=1$ states as described above.
This ground state is therefore a supercoherent qubit.  If each qubit couples to
its own individual environment, we expect that the major source of decoherence
for this ground states will indeed be the processes which take the system from
$J_n=0$ to $J_n=1$.

\subsection{Encoded operations on the exchange-based supercoherent qubit}

We now turn to the question which we could not solve for the Pauli stabilized
supercoherent qubits. In order to be useful for quantum computation, the
supercoherent qubits should allow for universal quantum computation. Extensive
discussion of fault-tolerant universal quantum computation on qubits encoded in
decoherence-free subsystems has been given in Chapter~\ref{ch:collectiveuniv}
where it was shown that computation on these encoded states can be achieved by
turning on Heisenberg couplings between neighboring physical qubits.  This
means that we need to add extra Heisenberg couplings to the supercoherent
Hamiltonian ${\bf H}_0^{[4]}$.  For a single supercoherent qubit these
additional Heisenberg couplings can be used to perform any SU$(2)$ rotation,
{\it i.e.}, an encoded one-qubit operation. In the present scheme one would
like this additional coupling to avoid destroying the energy gap which
suppresses decoherence. This can be achieved if the strength of the additional
couplings, $\delta$, is much less than the energy gap, {\it i.e.}, $\delta \ll
\Delta$. The trade-off between the decoherence rate and the speed of the one
qubit operations can be quantified by calculating the gate fidelity $F \propto
{\delta} e^{\beta (\Delta - \delta)}$. $F$ quantifies the number of operations
which can be done within a typical decoherence time of the system. For small
$\delta$ the gates are slower while for larger $\delta$ the gap is smaller
resulting in a tradeoff.  $F$ is minimized for $\delta_0=k T$.  At this minimum
$F$ is still exponentially suppressed for lower temperatures, in particular, $F
\mid_{\delta=\delta_0} \propto \beta^{-1} e^{\beta \Delta}$.

Of more concern for the present scheme is how to perform computation between
two encoded supercoherent qubits, the question which perplexed us in the
previous section.  In Section~\ref{sec:only} we saw that using only Heisenberg
couplings, a nontrivial two encoded qubit gate cannot be done without breaking
the degeneracy of the ${\bf H}_0^{[4]}$ Hamiltonian on the two sets of four
qubits. This can be circumvented by considering a joint Hamiltonian of the
eight qubits, ${\bf H}_0^{[8]}$.  This Hamiltonian has a ground state which is
$14$-fold degenerate, including the tensor product states of the degenerate
ground state of the ${\bf H}_0^{[4]}$ Hamiltonian.  The universality
constructions previously presented in Chapter~\ref{ch:collectiveuniv} and
explicitly in Appendix~\ref{ape} can then easily be shown to never leave the
ground state of this combined system.  Thus we see that we can circumvent the
problem of the previous section.  When a gate between two conjoined
exchange-based supercoherent qubits is needed, additional exchange interactions
must be turned on to obtain the Hamiltonian ${\bf H}_0^{[8]}$ and additional
exchanges must be used to produce a nontrivial gate between the two conjoined
supercoherent qubits.

\subsection{Implementation of the exchange supercoherent qubit in quantum dot
arrays}

The technological difficulties in building a supercoherent qubit are daunting
but we believe within the reach of present experiments.  In particular the
supercoherent qubit states appear perfect for solid state implementations of a
quantum computer using quantum
dots\cite{Loss:98a,DiVincenzo:99a,DiVincenzo:00b}. Related encodings on
$3$-qubit states were recently shown to permit universal computation with the
exchange interaction alone in \cite{DiVincenzo:00a}.  The main new requirement
for the supercoherent encoding, which allows the additional exponential
suppression of decoherence, is the construction of ${\bf H}_0^{[4]}$ and ${\bf
H}_0^{[8]}$. ${\bf H}_0^{[4]}$ can be implemented by a two dimensional array
with Heisenberg couplings between all four qubits. ${\bf H}_0^{[8]}$ poses a
more severe challenge, since the most natural geometry for implementing this
Hamiltonian is eight qubits on a cube with couplings between all qubits.  Such
structures should be possible in quantum dots by combining lateral and vertical
coupling scheme. Finally, estimates of the strength of the Heisenberg coupling
in the quantum dot implementations are expected to be on the order of $0.1$
meV~\cite{Loss:98a,DiVincenzo:99a,DiVincenzo:00b}. Thus we expect that at
temperatures below $0.1$ meV $\approx 1$ K, decoherence should be suppressed
for such coupled dots by encoding into the supercoherent states proposed here.

\section{Harmonic baths coupled to a supercoherent qubit}
\label{sec:hbathex}

As an example of the expected supercoherence we consider a quite general model
of $4$ qubits coupling to $4$ independent harmonic baths for the exchange-based
supercoherent qubit.  The unperturbed Hamiltonian of the system and bath is
${\bf H}_0 ^{[4]}\otimes {\bf I} + {\bf I} \otimes \sum_{i=1}^4 \sum_{k_i}
\hbar \omega_{k_i} {\bf a}_{k_i}^\dagger {\bf a}_{k_i}$ where ${\bf
a}_{k_i}^\dagger$ is the creation operator for the $i$th bath mode with energy
$\hbar \omega_{k_i}$. The most general linear coupling between each system
qubit and its individual bath is $\sum_{i=1}^4 \sum_{k_i} \sum_{\alpha}
 {\bf s}_\alpha^{(i)} \otimes (g_{i,\alpha} {\bf a}_{k_i}+ g_{i,\alpha}^*{\bf a}_{k_i}^\dagger
)$.  According to the selection rules described above we can write ${\bf
s}_\alpha^{(i)}=\sum_{(m,n) \in {\mathcal S}} {\bf A}_{i,\alpha}^{(m,n)} +{\rm
h.c.}$, where ${\bf A}_{i,\alpha}^{m,n \dagger}$ takes states $J_n=m$ to
$J_n=n$ (and acts on $\lambda$ and $m_\alpha$ in some possibly nontrivial
manner) and ${\mathcal S}$ is the set of allowed transitions ${\mathcal S}=\{
(0,1), (1,2), (1,1), (2,2) \}$. In the interaction picture, after making the
rotating-wave approximation \cite{Scully:97a}, this becomes
\begin{eqnarray}
{\bf V}(t)= \sum_{i,\alpha,k_i,(m,n) \in {\mathcal S}} g_{i,\alpha}^* e^{-{i
({\Delta \over \hbar}  f(m,n)- \omega_{k_i}) t}} {\bf A}_{i,\alpha}^{(m,n)}
{\bf a}_{k_i}^\dagger \nonumber \\ + g_{i,\alpha} e^{{i ({\Delta \over \hbar}
f(m,n)- \omega_{k_i}) t}} {\bf A}_{i,\alpha}^{(m,n) \dagger} {\bf a}_{k_i},
\end{eqnarray}
where $f(m,n)=n(n+1)-m(m+1)$.  Coupling to thermal environments of the same
temperature, under quite general circumstances (Markovian dynamics, smooth
spectral density of the field modes) we are led to a master equation (see for
example \cite{Scully:97a})
\begin{equation}
{ \partial \rho \over \partial t} = \sum_{i,\alpha,(m,n) \in {\mathcal S}}
\gamma_{i,\alpha}^{(m,n)} {\mathcal L}_{i,\alpha}^{(m,n)}
[\rho]+\gamma_{i,\alpha}^{(n,m)} {\mathcal L}_{i,\alpha}^{(n,m)} [\rho],
\end{equation}
with ${\mathcal L}_{i,\alpha}^{(m,n)}[\rho]=([{\bf A}_{i,\alpha}^{(m,n)} \rho ,
{\bf A}_{i,\alpha}^{(m,n)\dagger }] + [ {\bf A}_{i,\alpha}^{(m,n)},\rho {\bf
A}_{i,\alpha}^{(m,n)\dagger}])$.  The only operators which act directly on the
supercoherent qubit are ${\bf A}_{i,\alpha}^{(0,1)}$.  The relative decoherence
rates satisfy $\gamma_{i,\alpha}^{(0,1)} \propto n(T)$ where $n(T)$ is the
thermal average Bose occupation number $n(T)=\left[ \exp( \beta \Delta) -1
\right]^{-1}$. Thus we see, as predicted that the supercoherent qubit decoheres
at a rate which decreases exponentially as $kT$ decreases below $\Delta$.

\section{Cooper pairs as error detecting codes and supercoherence}

Finally let us mention an interesting connection between our supercoherent
constructions and Cooper pairs in superconductivity.  In the standard
derivation of superconductivity as initially put forth by Bardeen, Cooper, and
Schrieffer \cite{Bardeen:57a} electrons with energies near the Fermi energy of
a metal interact via the exchange of a phonon producing an attractive effective
potential between the electrons.  The Hamiltonian which describes the system is
well described by \cite{Bardeen:57a}
\begin{equation}
{\bf H} = \sum_k E(k) {\bf c}_{k,\uparrow}^\dagger {\bf c}_{k,\uparrow} -
\sum_{k,k^\prime} V_{kk^\prime} {\bf c}_{k^\prime,\downarrow}^\dagger {\bf
c}_{-k^\prime,\downarrow}^\dagger {\bf c}_{-k,\uparrow} {\bf c}_{k,\uparrow},
\label{eq:supercondH}
\end{equation}
where ${\bf c}_{k,s}$ is the single electron annihilation operator for an
electron with wavenumber $k$ and spin $s$, $E(k)$ is the energy of an electron
with wavenumber $k$, and $V_{kk^\prime}$ represents the attractive coupling.
The ground state of the superconductor to a good approximation (in the
thermodynamic limit) is \cite{Bardeen:57a}
\begin{equation}
\prod_k (\alpha_k  + \beta_k {\bf c}_{k,\uparrow}^\dagger {\bf
c}_{-k,\downarrow}^\dagger) |0\rangle, \label{eq:supercondg}
\end{equation}
where $\alpha_k$ and $\beta_k$ are real coefficients and $|0\rangle$ is the
vacuum state.  The ground state is composed of Cooper pairs of electrons with
opposite momentum $|k,\uparrow\rangle |-k,\downarrow \rangle$.  If we work in
the frame of reference which is drifting with the superconducting current, then
the types of effects which normally establish resistivity in a conductor are
those that change the momentum of a single electron (via scattering from
impurities, phonons, etc).  We will now show that Cooper pairs are a form of
quantum error detecting code for these single electron scattering processes.

We recall that the open system evolution of a system which is initially
decoupled from its environment is described in the operator-sum
representation\cite{Kraus:83a} as $\bmath{\rho}(t) = \sum_i {\bf A}_i(t) \bmath
{ \rho}(0) {\bf A}_i^\dagger(t)$ where $\sum_i {\bf A}_i^\dagger(t) {\bf
A}_i(t)={\bf I}$.  Quantum error correction and detection begin by expanding
${\bf A}_i(t)$ in terms of a suitable basis ${\bf E}_a$ of possibly non-unitary
``error'' operators.   A sufficient condition for the detection of such
processes on a code with states $|i\rangle$ representing the encoded quantum
information is given by \cite{Knill:97a}
\begin{equation}
\langle j | {\bf E}_a |i \rangle= c_a \delta_{ij}. \label{eq:qed}
\end{equation}
Consider now a single Cooper pair with different wavenumbers $k$ and
$k^\prime$: $|k,\uparrow,-k,\downarrow\rangle$ and
$|k^\prime,\uparrow,-k^\prime,\downarrow\rangle$.  Any error operator ${\bf E}$
which acts on only one of the electrons and changes the momentum of the
electron, the operators which would normally cause resistance, can easily be
seen to satisfy $ \langle k,\uparrow,-k,\downarrow| {\bf E} | k^\prime,\uparrow
,-k^\prime ,\downarrow \rangle = 0$ because of the orthogonality of states on
the electron which is not operated on.  Further because ${\bf E}$ changes the
momentum of the single electron,
\begin{equation}
\langle l,\uparrow,-l,\downarrow| {\bf E} | l,\uparrow ,-l ,\downarrow \rangle
= 0,
\end{equation}
for both $l=k$ and $l=k^\prime$.  We therefore see that Cooper pairs satisfy
Eq.~(\ref{eq:qed}) for all resistance causing interactions.  Cooper pairs,
then, are quantum error detecting codes for resistance cause scattering.  If we
could store quantum information in the wavenumber of a Cooper pair then we
could use these Cooper pairs as a supercoherent system.  We note here that the
fact that Cooper pairs are single electron error detecting codes which exhibit
supercoherence does {\em not} however explain the zero electrical resistance of
superconductors.  It is interesting to note, however, the connections between
Cooper pairs and supercoherence.

\section{Supercoherence and the importance of energetics}

In this chapter we have introduced the notion of supercoherence.  When the
interaction between a system and its environment is perturbing (which is the
case in most systems of interest) decoherence follows pathways which preserve
the unperturbed system and environment energies.  This allows us to construct a
method for avoiding decoherence by engineering the system Hamiltonian such that
all single qubit decoherence processes are processes which heat the system.
Thus by cooling the environment decoherence in a supercoherent system can be
minimized.  We have therefore harnessed the power of energetics to help
strengthen the resistance of quantum information to decoherence.  This
represents a small step towards constructing a system which has resistance to
decoherence built into the natural evolution of the system.

\chapter{A Supercoherent Spin Ladder with Error Correcting Properties}
\label{ch:ladder}

\begin{quote}
{\em To be an Error and to be Cast out is part of God's Design} \\
\begin{flushright} -- William Blake \end{flushright}
\end{quote}

In this chapter we study a spin ladder which has both supercoherent and error
correcting properties.  We begin by presenting a stabilizer encoding which maps
this model into clusters of Ising models with transverse fields.  We then
explicitly calculate the spectrum of this model and show that there is a unique
two-fold degenerate ground state for this spin ladder.  It is then shown that
this ground state detects single bit flips and corrects multiple phase errors.
The ground state is therefore supercoherent with the added benefit of being
quantum error correcting.  We then discuss the role of encoded operations on
this state and conclude with some discussion of the shortcomings of this spin
ladder for quantum computation.

\section{Description of the spin ladder}

Suppose we are given a spin ladder of $2n$ qubits.  We label these qubits via
the indices $(i,j)$ where $1 \leq i \leq n$ and $j \in \{1,2\}$ with the
operator ${\bf O}$ acting on the $(i,j)$th qubit tensored with identity on all
other qubits as ${\bf O}^{(i,j)}$.  Define the two operators
\begin{eqnarray}
{\bf H}_Z &=& \sum_{i=1}^n \bmath{\sigma}_z^{(i,1)} \bmath{\sigma}_z^{(i,2)}
\nonumber \\
 {\bf H}_X &=& \sum_{i=1}^{n-1} \bmath{\sigma}_x^{(i,1)} \bmath{\sigma}_x^{(i+1,1)}
 + \bmath{\sigma}_x^{(i,2)} \bmath{\sigma}_x^{(i+1,2)}.
\end{eqnarray}
The spin ladder we consider is the sum of these two Hamiltonians with equal
negative strengths
\begin{equation}
{\bf H}_n= -\omega_0({\bf H}_Z+ {\bf H}_X),
\end{equation}
where $\omega_0>0$.  This spin ladder is sketched in
Figure~\ref{fig:spinladder}

\begin{figure}[h]
\quad \quad \psfig{figure=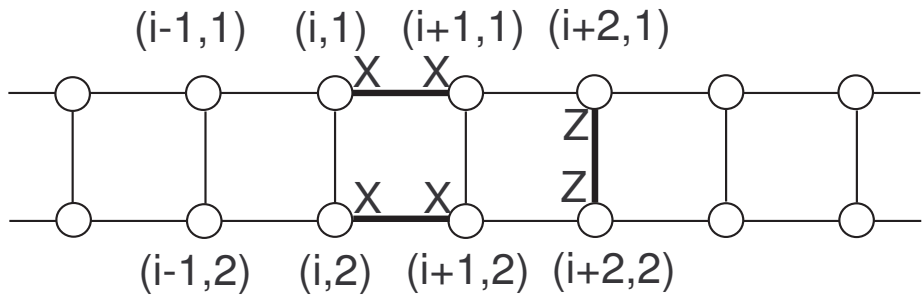,width=5in} \caption{\em A
supercoherent spin ladder.} \label{fig:spinladder}
\end{figure}

Let us begin by understanding the intuition behind why this spin ladder system
may have interesting supercoherent properties.  The ground state of this spin
ladder system will attempt to minimize the energy of the total Hamiltonian. Any
given qubit is acted upon by an interaction which acts as $\bmath{\sigma}_x
\otimes \bmath{\sigma}_x$ or $\bmath{\sigma}_z \otimes \bmath{\sigma}_z$ where
the first qubit is the qubit of concern and the other qubit is one of the
qubits neighbors.  We call such couplings between the qubits bonds.  If we
individually diagonalize the interactions corresponding to the bonds, single
qubit interactions $\bmath{\sigma}_\alpha$ act to change the eigenvalue of each
of these operators.  In particular, because we are dealing with Pauli
operators, the eigenvalue will flip sign and therefore increase in energy.  The
real ground state, of the system, of course cannot be analyzed in such a manner
because all of the bond operators do not commute. However, it is not
unreasonable that the ground state will maintain some of this intuition, that
single qubit interactions increase the energy and indeed we will see that our
intuition does pay off and this is exactly what happens.  Such spin ladders are
known as frustrated spin ladders\cite{Toulouse:77a} due to the competition of
the different bonds in establishing a ground state.

\subsection{Stabilizer encoding}

There are two transformations which make exact calculation of the spectrum of
this spin ladder possible.  The first of these is a Pauli stabilizer encoding
(see Appendix~\ref{apa:pauli}).  Instead of the Pauli basis
$\bmath{\sigma}_\alpha^{(i,j)}$, consider instead the following set of Pauli
operators
\begin{eqnarray}
{\bf X}_1^{(i)}&=& \bmath{\sigma}_x^{(i,1)} \bmath{\sigma}_x^{(i,2)}, \quad
{\bf Z}_1^{(i)} = \bmath{\sigma}_z^{(i,1)}, \nonumber \\
 {\bf X}_2^{(i)}&=& \bmath{\sigma}_x^{(i,2)},  \quad {\bf Z}_2^{(i)} =
 \bmath{\sigma}_z^{(i,1)} \bmath{\sigma}_z^{(i,2)}. \label{eq:newpb}
\end{eqnarray}
In particular, there is an encoding of information such that ${\bf X}_1, {\bf
Z}_1$ act as corresponding Pauli operators on the first qubit and ${\bf X}_2,
{\bf Z}_2$ act as corresponding Pauli operators on the second qubit.  In fact
this encoding is simply the controlled-not basis change on the qubits from
adjacent qubits connected by the rungs of the ladder
\begin{equation}
|00\rangle \rightarrow |00\rangle, \quad |01\rangle \rightarrow |01\rangle,
\quad |10\rangle \rightarrow |11\rangle, \quad |11\rangle \rightarrow
|10\rangle.
\end{equation}
Under this basis, we find that the spin ladder Hamiltonian becomes
\begin{eqnarray}
{\bf H}&=&-\omega_0 \left( \sum_{i=1}^n {\bf Z}_2^{(i)} + \sum_{i=1}^{n-1} {\bf
X}_1^{(i)} {\bf X}_2^{(i)} {\bf X}_1^{(i+1)} {\bf X}_2^{(i+1)} + {\bf
X}_2^{(i)} {\bf X}_2^{(i+1)} \right) \nonumber \\
 &=& -\omega_0 \left( \sum_{i=1}^n {\bf Z}_2^{(i)} + \sum_{i=1}^{n-1} {\bf X}_2^{(i)} {\bf
 X}_2^{(i+1)}\left( {\bf I} + {\bf X}_1^{(i)} {\bf X}_1^{(i+1)} \right)
 \right).
 \label{eq:spinlad}
\end{eqnarray}

At this point it is useful to introduce a basis corresponding to the operators
Eq.~(\ref{eq:newpb}).  A complete set of commuting operators corresponding to
this basis is given by the operator ${\bf Z}_2^{(i)}$ and ${\bf X}_1^{(i)}$. We
label this basis by the $\pm 1$ eigenvalues of these operators as
$|z_2^{(1)},z_2^{(2)},\dots,z_2^{(n)},x_1^{(1)},x_2^{(2)},\dots,x_2^{(n)}
\rangle$.  We will sometimes abbreviate this as $|\vec{z}_2,\vec{x}_1\rangle$
under the obvious correspondence.

Under this basis
\begin{equation}
\left( {\bf I}+{\bf X}_1^{(i)} {\bf X}_1^{(i+1)} \right)
|\vec{z}_2,\vec{x}_1\rangle =\left(1+x_1^{(i)}x_1^{(i+1)} \right)
|\vec{z}_2,\vec{x}_1\rangle.
\end{equation}
Now $1+x_1^{(i)}x_1^{(i+1)}$ is either $0$ if the signs of $x_1^{(i)}$ and
$x_1^{(i+1)}$ differ or $2$ if the signs of $x_1^{(i)}$ and $x_1^{(i+1)}$ are
identical.  This identification allows us to see that the spin ladder
Hamiltonian Eq.~(\ref{eq:spinlad}) acts as a different Hamiltonian depending
only the value of $\vec{x}_1$.  In particular we see that
\begin{equation}
{\bf H} = -\omega_0 \bigoplus_{x_1^{(1)},\dots,x_1^{(n)}=\{-1,+1\}}
\sum_{i=1}^n \left( {\bf Z}_2^{(i)} + 2 c_i(\vec{x}_1) {\bf X}_2^{(i)} {\bf
X}_2^{(i+1)} \right) \otimes |\vec{x}_1 \rangle \langle \vec{x}_1 |,
\end{equation}
where
\begin{equation}
c_i(\vec{x}_1)= {1 \over 2} \left(1+ x_1^{(i)} x_1^{(i+1)}\right).
\end{equation}
Notice $c_i(\vec{x})$ is a list over $i$ of either $+1$ or $0$.  We therefore
see that the spin ladder Hamiltonian has been brought to a block diagonal form
where each of the blocks corresponds to a given $\vec{x}_1$.  Given a
particular block with a $\vec{x}_1$ the values of the $n-1$ $c_i(\vec{x}_1)$
then specify the exact form of the Hamiltonian in this block.

We will now focus on these block diagonal Hamiltonians for a fixed $\vec{x}_1$.
We see that a Hamiltonian for a particular $c_i(\vec{x}_1)$ corresponds to
multiple Ising chains in a transverse field.  Define the Ising chain with a
transverse field Hamiltonian as
\begin{equation}
{\bf H}_I^{j,k} = - \omega_0 \left( \sum_{i=j}^{k-1} 2 {\bf X}_2^{(i)} {\bf
X}_2^{(i+1)} + \sum_{i=j}^{k} {\bf Z}_2^{(i)} \right).
\end{equation}
and the transverse field only Hamiltonian as
\begin{equation}
{\bf H}_T^{j,k}= - \omega_0 \sum_{i=j}^k {\bf Z}_2^{(i)}.
\end{equation}
The Hamiltonian over the $\vec{z}_2$ qubits for a fixed $\vec{x}_1$ is given by
a sum of such chains with transverse fields and transverse fields only:
\begin{equation}
{\bf H}(c_i(\vec{x}_1))= {\bf H}_I^{i_1,i_2} + {\bf H}_I^{i_3,i_4} + \cdots +
{\bf H}_I^{i_{2k-1},i_{2k}} + {\bf H}_T^{j_1,j_2} + {\bf H}_T^{j_3,j_4} +
\cdots + {\bf H}_I^{j_{2l-1},i_{2l}}.
\end{equation}
Since each of the ${\bf H}_I^{i_j,i_{j+1}}$ ${\bf H}_T^{i_j,i_{j+1}}$ act on
different $|x_2^{(i)}\rangle$ qubits they can each be individually diagonalized
and the total energy added up.  For the systems with simply a transverse field
this is trivially achieved.  The eigenstates are simply the single qubit
configurations of the qubits pointing with or anti to the transverse field.
Luckily, also, we can analyze the Ising chains with a transverse field and find
analytical expressions for the energy and eigenstates of these chains up to a
small correction.

\subsection{The one dimensional Ising chain in a transverse magnetic field}

We need to consider an Ising chain of length $k$ in a transverse field of the
form
\begin{equation}
{\bf H}_I=-\sum_{i=1}^k {\bf Z}_i - 2\sum_{i=1}^{k-1} {\bf X}_{i} {\bf
X}_{i+1},
\end{equation}
where we have relabeled our qubit operators in an obvious notion for simplicity
in this calculation.  We follow the calculation in \cite{Chakrabarti:96a}.
Define the raising and lowering operations
\begin{equation}
{\bf S}_i^\pm = {1 \over 2} \left[ {\bf X}_i \pm i {\bf Y}_i \right],
\end{equation}
such that
\begin{eqnarray}
{\bf H}_I&=&\sum_{i=1}^k ({\bf I}-2{\bf S}_i^+ {\bf S}_i^-) - 2\sum_{i=1}^{k-1}
\left[ {\bf S}_i^+ + {\bf S}_i^- \right] \left[ {\bf S}_{i+1}^+ + {\bf
S}_{i+1}^-\right] \nonumber \\ &=&k {\bf I}-2 \left(\sum_{i=1}^k {\bf S}_i^+
{\bf S}_i^- + \sum_{i=1}^{k-1} \left[ {\bf S}_i^+ + {\bf S}_i^- \right] \left[
{\bf S}_{i+1}^+ + {\bf S}_{i+1}^-\right] \right).
\end{eqnarray}
Henceforth, we will drop the identity constant $k{\bf I}$ and recover such
constant terms at the end of our calculation (they will turn out to be
important!).

Next we can use the Jordan-Wigner transformation to take this model, which is
that of hard-core bosons, from spin operators to fermions.  In particular if we
define
\begin{equation}
{\bf c}_i= \prod_{j=1}^{i-1} (-{\bf Z}_j) {\bf S}_i^-, \quad {\bf c}_i^\dagger
= \prod_{j=1}^{i-1} (-{\bf Z}_j) {\bf S}_i^+,
\end{equation}
then we see that the operators ${\bf c}_i, {\bf c}_i^\dagger$ are fermionic
operators satisfying
\begin{equation}
\{ {\bf c}_i, {\bf c}_j \} =0, \quad \{ {\bf c}_i, {\bf c}_j^\dagger \} =
\delta_{ij}.
\end{equation}
Expressing our model in terms of the fermionic operators we find
\begin{equation}
{\bf H}_I = -2 \left( \sum_{i=1}^k {\bf c}_i^\dagger {\bf c}_i + \sum_{i=1}^{k}
\left[ {\bf c}_i^\dagger - {\bf c}_i \right] \left[ {\bf c}_{i+1}^\dagger +
{\bf c}_{i+1} \right] \right)+{\bf C}, \label{eq:fermham}
\end{equation}
where addition is done modulo $k$ and ${\bf C}$ is a correction
\begin{equation}
{\bf C}=+2 ({\bf c}_1^\dagger + {\bf c}_1)({\bf c}_k^\dagger-{\bf c}_k).
\end{equation}
We will ignore the correction term ${\bf C}$ for now and return to the effect
of this term later.  Notice also that this correction term only appears when
$k>2$.

To diagonalize this Hamiltonian it is useful to work first with fermions in
momentum space
\begin{eqnarray}
{\bf c}_q &=& {1 \over \sqrt{k}} \sum_{j=1}^k {\bf c}_j \exp \left( i q
j\right) \nonumber \\
 {\bf c}_q^\dagger &=& {1 \over \sqrt{k}} \sum_{j=1}^k {\bf c}_j^\dagger \exp
 \left( - i q j \right),
\end{eqnarray}
where $q={2 \pi m \over k}$ with $m=-{k \over 2}, \dots ,{k \over 2}$ for $k$
even and $m=-{k-1 \over 2}, \dots, {k-1 \over 2}$ for $k$ odd.  Check that
these still obey the fermion rules:
\begin{eqnarray}
\{ {\bf c}_q , {\bf c}_{q^\prime} \}&=& \{ {\bf c}_q^\dagger, {\bf
c}_{q^\prime}^\dagger \} =0 \\
 \{ {\bf c}_q , {\bf c}_{q^\prime}^\dagger \} &=& {1 \over k} \sum_{a,b=1}^k \{ {\bf c}_a,
 {\bf c}_b^\dagger \} \exp(i (qa-q^\prime b) )= {1 \over k} \sum_{a=1}^k \exp(i a
 (q-q^\prime)\nonumber
 )=\delta_{q,q^\prime}.
\end{eqnarray}
We can compute that
\begin{equation}
{\bf c}_q^\dagger {\bf c}_q = {1 \over k} \sum_{a,b=1}^k {\bf c}_a^\dagger {\bf
c}_b \exp( i q (b-a)).
\end{equation}
So that
\begin{eqnarray}
\sum_q {\bf c}_q^\dagger {\bf c}_q &=& \sum_{a=1}^k {\bf c}_a^\dagger {\bf
c}_a, \nonumber \\ 2 \sum_q \cos(q) {\bf c}_q ^\dagger {\bf c}_q &=&\sum_q
\left(
 \exp(iq) + \exp(-iq) \right) \sum_{a,b=1}^k {\bf c}_a^\dagger {\bf c}_b
 \exp(iq (b-a))  \nonumber \\ &=& \sum_{a=1}^{k-1} \left( {\bf c}_a^\dagger {\bf c}_{a+1} - {\bf c}_a {\bf
 c}_{a+1}^\dagger \right) \nonumber \\
 \sum_q \exp(-iq) {\bf c}_q^\dagger {\bf c}_{-q}^\dagger &=&{1 \over k}\sum_q
 \sum_{a,b=1}^k {\bf c}_a^\dagger {\bf c}_b^\dagger \exp(iq (b-a-1) ) =
 \sum_{a=1}^{k-1} {\bf c}_a^\dagger {\bf c}_{a+1}^\dagger \nonumber \\
 \sum_q \exp(iq) {\bf c}_q {\bf c}_{-q} &=& {1 \over k} \sum_q \sum_{a,b=1}^k
 {\bf c}_a {\bf c}_b \exp(iq(a-b+1) = -\sum_{a=1}^{k-1} {\bf
 c}_a {\bf c}_{a+1}.
\end{eqnarray}
Thus we find that
\begin{eqnarray}
{\bf H}_I&=&-2 \left( \sum_q (1+2 \cos(q)) {\bf c}_q^\dagger {\bf c}_q - \sum_q
(\exp(-iq) {\bf c}_q^\dagger {\bf c}_{-q}^\dagger + \exp(iq) {\bf c}_q {\bf
c}_{-q}) \right) \\ &=&- 2 \left( \sum_{q>0} (1 +2\cos(q)) ( {\bf c}_q^\dagger
{\bf c}_q +{\bf c}_{-q}^\dagger {\bf c}_{-q}) + 2i \sum_{q>0} \sin(q) \left(
{\bf c}_q^\dagger {\bf c}_{-q}^\dagger + {\bf c}_q {\bf c}_{-q}) \right)
\right). \nonumber
\end{eqnarray}
To diagonalize this Hamiltonian we apply a Bogoliubov transformation
\begin{eqnarray}
\bmath{\eta}_q &=& u_q {\bf c}_q + i v_q {\bf c}_{-q}^\dagger, \quad \bmath{
\eta}_{-q} = u_q {\bf c}_{-q} - i v_q {\bf c}_q^\dagger \nonumber \\ \bmath{
\eta}_q^\dagger &=& u_q {\bf c}_q^\dagger - i v_q {\bf c}_{-q}, \quad \bmath{
\eta}_{-q}^\dagger = u_q {\bf c}_{-q}^\dagger + i v_q {\bf c}_q,
\end{eqnarray}
where $q>0$ everywhere and $u_q, v_q$ are both real.  We require that the
$\bmath{\eta}_q, \bmath{\eta}_q^\dagger$ are fermionic operators:
\begin{equation}
\{ \bmath{\eta}_{q^\prime}, \bmath{\eta}_q \} = 0,\quad \{ \bmath{
\eta}_{q^\prime}^\dagger,\bmath{\eta}_{q} \}=\delta_{q^\prime,q} \quad
\Rightarrow \quad u_q^2 +v_q^2 =1.
\end{equation}
Thus we parameterize $u_q$ and $v_q$ via $u_q=\sin(\theta_q),
v_q=\cos(\theta_q)$.  The inverse transformation to the Bogoliubov fermions is
given by
\begin{eqnarray}
{\bf c}_q&=& u_q \bmath{\eta}_q -i v_q \bmath{\eta}_{-q}^\dagger, \quad {\bf
c}_{-q}=u_q \bmath{\eta}_{-q} + i v_q \bmath{\eta}_q^\dagger \nonumber \\
 {\bf c}_q^\dagger&=&u_q \bmath{\eta}_q^\dagger +i v_q \bmath{\eta}_{-q} ,
 \quad{\bf c}_{-q}^\dagger=u_q \bmath{\eta}_{-q}^\dagger - i v_q
 \bmath{\eta}_q.
\end{eqnarray}
Which can be used to rewrite the Hamiltonian in terms of the Bogoliubov
fermions:
\begin{eqnarray}
{\bf H}_I&=&- 2\left(   \sum_{q>0}(1 +2\cos(q)) \left( (u_q
\bmath{\eta}_q^\dagger +i v_q \bmath{\eta}_{-q})(u_q \bmath{\eta}_q -i v_q
\bmath{\eta}_{-q}^\dagger)  \right. \right. \nonumber \\ && \left. +(u_q
\bmath{\eta}_{-q}^\dagger - i v_q \bmath{\eta}_q)(u_q \bmath{\eta}_{-q} + i v_q
\bmath{\eta}_q^\dagger)) \right) \nonumber \\  && \left. +  \sum_{q>0} 2i
\sin(q) \left( (u_q \bmath{\eta}_q^\dagger +i v_q \bmath{\eta}_{-q}) (u_q
\bmath{\eta}_{-q}^\dagger - i v_q \bmath{\eta}_q) \right. \right. \nonumber \\
&& \left. \left. +(u_q \bmath{\eta}_q -i v_q \bmath{\eta}_{-q}^\dagger)(u_q
\bmath{\eta}_{-q} + i v_q \bmath{\eta}_q^\dagger)) \right) \right)
\\&=& -2\left( \sum_{q>0} \left[ (1+2 \cos(q) )(u_q^2-v_q^2) - 4 \sin(q) u_q
v_q \right](\bmath{ \eta}_q^\dagger \bmath{\eta}_q + \bmath{\eta}_{-q}^\dagger
\bmath{\eta}_{-q} ) \right. \nonumber \\ &&+ \left. \sum_{q>0} \left[ 4i (1+ 2
\cos(q) ) u_q v_q + 4i \sin(q) (u_q^2-v_q^2) \right] (\bmath{\eta}_q^\dagger
\bmath{ \eta}_{-q}^\dagger + \bmath{\eta}_q \bmath{\eta}_{-q} )\right),
\nonumber
\end{eqnarray}
up to a constant vacuum energy.  We can make the off-diagonal terms vanish if
\begin{equation}
(1+2 \cos(q)) u_q v_q +\sin(q) (u_q^2-v_q^2)=0 \rightarrow \tan(2\theta_q)= -
{2 \sin(q) \over (1+2 \cos(q))}.
\end{equation}
Thus
\begin{eqnarray}
{\bf H}_I= -2 \sum_{q>0} \left[ (1+2\cos(q)) - 2 \sin(2\theta_q) \sin(q)
\right] (\bmath{\eta}_q^\dagger \bmath{\eta}_q + \bmath{\eta}_{-q}^\dagger
\bmath{\eta}_{-q} ),
\end{eqnarray}
or
\begin{equation}
{\bf H}_I=2 \sum_q \sqrt{ 5 + 4 \cos(q) } \bmath{\eta}_q^\dagger
\bmath{\eta}_q.
\end{equation}
We can recover the constant vacuum energy by noting that the original
Hamiltonian was traceless and the trace should be preserved under the canonical
transformations we have performed.  Since ${\rm Tr} \left[ {\bf
H}_I\right]=2k\sum_q \sqrt{5 +4 \cos{q}}$ we find that
\begin{equation}
{\bf H}_I= 2 \sum_q \sqrt{5+4 \cos(q)} \left(\bmath{\eta}_q^\dagger
\bmath{\eta}_q -{1 \over 2} \right). \label{eq:isk}
\end{equation}
The vacuum (ground) state of this system has no Bogoliubov fermions occupying
any sites.  Note that there is a gap between this state and excited states.
Further note that the energy of this vacuum state is really dependent on $k$:
\begin{equation}
E_g(k)=-\sum_{q} \sqrt{5+4\cos(q)}=-\sum_{m=-{k \over 2} \left(-{k-1 \over 2}
\right)}^{{ k \over 2} \left({{k-1} \over 2} \right)} \sqrt{5+4 \cos \left( {2
\pi m \over k} \right) }.
\end{equation}
We note that $E_g(k)<-k$ because each term in the sum is greater than unity.
Thus if we compare a Ising chain in a transverse field to one which is simply
in a transverse field the Ising chain in the transverse field always has a
lower energy ground state.  Further we note that $E_g(k)>E_g(j)$ if $k>j$.

Let us return to the correction term ${\bf C}$ in Eq.~(\ref{eq:fermham}).  When
we express this term in momentum space we find that
\begin{equation}
{\bf C}=  {2 \over k} \sum_{q,q^\prime} \left( e^{i q} {\bf c}_q^\dagger +
e^{-iq} {\bf c}_q^\dagger \right) \left( e^{ikq^\prime} {\bf
c}_{q^\prime}^\dagger- e^{-ikq^\prime} {\bf c}_{q^\prime} \right).
\end{equation}
The first observation is that for large $k$, this term becomes a small
correction to the energy derived above.  Furthermore, each given term has an
eigenvalue $\lambda$ which has a value between $-{2 \over k} \leq \lambda \leq
{2 \over k}$.  This implies that the correction to our expression
Eq.~(\ref{eq:isk}) will be bound from above by ${4 \over k}$.  Thus we have
found that
\begin{equation}
{\bf H}_I= 2 \sum_q \sqrt{5+4 \cos(q)} \left(\bmath{\eta}_q^\dagger
\bmath{\eta}_q -{1 \over 2} \right)+\bmath{\Omega}_k,~{\rm where}~ {\rm Tr}
\bmath|\bmath{\Omega}_k|<{4 \over k}.
\end{equation}

\section{Clusters, clusters, everywhere} \label{sec:cluster}

Having nearly exactly calculated the spectra of the Ising with transverse field
Hamiltonian of length $k$ we understand the spectrum of the total spin ladder
Hamiltonian.  For each subspace corresponding to a specification of
$c_i(\vec{x}_1)$ we can construct the binary string
$\vec{c}(\vec{x}_1)=(c_1(\vec{x}_1),c_2(\vec{x}_1),\dots,c_{n-1}(\vec{x}_1))$
labeling the structure of the Hamiltonian on the $\vec{x}_1$ specified
subspace. Each such string can further be specified by the values where the
elements take the value $+1$ and, in particular, we wish to simply label such a
subspace by the structure of such $+1$ clusters.  A $+1$ cluster from the $i$th
to the $j$ qubit will be denoted by $(i,j)$.  Thus every string will correspond
to some cluster structure $\widehat{c}(\vec{x}_1)= (i_1,i_2) (i_3,i_4) \dots
(i_{2r-1},i_{2r})$ where $r$ is the number of $+1$ clusters in a string. Thus,
for example $\vec{c}=(+1,+1,-1,-1,+1,-1,+1,+1,-1) \Rightarrow \widehat{c}=
(1,2)(5,5)(7,8)$ which has $3$ $+1$ clusters, two of length $2$ and one of
length $1$.  For each cluster labeling the spectrum of the Hamiltonian ${\bf
H}$ has a structure related only to the number and size of the clusters.  In
particular if two subspaces have identical cluster structure (number of cluster
of a given length is identical) then they have an identical spectrum (with
different eigenstates however). This is because, for a given cluster of length
$l$, the Hamiltonian takes on the structure of an Ising chain with a transverse
magnetic field which we have analyzed above.  Let us label the cluster
structure by $[m_2,\dots,m_{n-1}]$ where $m_i$ is the number of clusters of
length $i$. Clearly $\sum_{i=2}^{n-1} m_i i \leq n-1$.  Let $E_g(k)$ label the
ground state energy of a cluster of length $k$ physical qubits which
corresponds to a cluster of $k$ in the bit string $\vec{c}$.  In particular
from the previous section we know that
\begin{equation}
E_g(k)=-\omega_0\sum_{m=-{k \over 2} \left(-{k-1 \over 2} \right)}^{{ k \over
2} \left({{k-1} \over 2} \right)} \sqrt{5+4 \cos \left( {2 \pi m \over k}
\right) }+\Omega_k,~{\rm where}~ {\rm Tr} \bmath|\bmath{\Omega}_k|<{4 \over k}.
\end{equation}
For elements which are not members acted upon by a cluster, only the ${\bf
H}_T$ Hamiltonian contributes to the spectrum of these sights. For a given
cluster structure $[m_2,m_3,\dots,m_{n-1}]$ the vacuum state of the Hamiltonian
has an energy
\begin{equation}
E([m_2,m_3,\dots,m_{n-1}])=  \sum_{i=2}^{n-1} m_i E_g(i)  - \omega_0
(n-\sum_{i=2}^{n-1} m_i i )).
\end{equation}
It is easy to then verify that the global ground state corresponds to the
subspace where $[m_2=0,m_3=0,\dots,m_{n-2}=0,m_{n-1}=1]$, i.e. the full cluster
situation.

Further we note that the for every cluster configuration $\vec{c}(\vec{x}_1)$
corresponds to two different $\vec{x}_1$ configurations and thus all of the
levels of our Hamiltonian are two-fold degenerate.  To see this note that
$\vec{c}(\vec{x}_1)$ is unchanged if the value of every element in $\vec{x}_1$
flip signs.  Every element of the spin ladder Hamiltonian, therefore, is
two-fold degenerate.  The ground state of the system then corresponds to the
Hamiltonian over the subspace defined by $\vec{x}_1=(+1,+1,\cdots,+1)$ and also
by $\vec{x}_1=(-1,-1,\cdots,-1)$.

\section{Quantum error correcting properties}

Let us now examine the error detecting and correcting properties of the spin
ladders ground state.  We will examine the error properties of information
encoded into the degeneracy of the ground state of the Hamiltonian.

Instead of using the basis $|\vec{z}_2,\vec{x}_1\rangle$ it is convenient to
work with the basis $|\vec{z}_2,\vec{c},x_1^{(1)}\rangle$ where
$\vec{c}(\vec{x}_1)=(c_1(\vec{x}_1),c_2(\vec{x}_1),\dots,c_{n-1}(\vec{x}_1))$
where we recall that $c_i(\vec{x}_1) = {1 \over 2} (1+ x_1^{(i)} x_1^{(i+1)})$.
The ground state is therefore labeled by $|\vec{z}_2=g\rangle \otimes
|1,1,\dots,1\rangle \otimes |x_1^{(1)}\rangle$ where $|\vec{z}_2=g\rangle$ is
the ground state of the full cluster Hamiltonian, $|1,1,\dots,1\rangle$
represents $c_1=1,c_2=1,\dots,c_{n-1}=1$ and $x_1^{(1)}$ now labels the
degeneracy of this ground state.

The first thing to notice is that any operator which acts as identity on the
degeneracy is a detectable error.  In other words
\begin{eqnarray}
&&\langle \vec{z}_2=g,\vec{c}=(1,1,\dots,1), x_1^{(1)} | {\bf E} |\vec{z}_2=
g,\vec{c} =(1,1,\dots,1),y_1^{(1)} \rangle  \\ && = \langle
\vec{z}_2=g,\vec{c}=(1,1,\dots,1)| {\bf E} | \vec{z}_2=g,\vec{c}=(1,1,\dots,1)
\rangle \langle x_1^{(1)}| y_1^{(1)} \rangle = c \delta_{x_1^{(1)},y_1^{(1)}},
\nonumber
\end{eqnarray}
where $c$ is the matrix element $\langle g,\vec{c}=(1,1,\dots,1)| {\bf E} |
g,\vec{c}=(1,1,\dots,1) \rangle$.

We will now show that
\begin{equation}
\langle \vec{z}_2=g,\vec{c}=(1,1,\dots,1), x_1^{(1)} | {\bf E}_\alpha
|\vec{z}_2= g,\vec{c} =(1,1,\dots,1),y_1^{(1)} \rangle= C_\alpha \delta_{ij},
\label{eq:erro}
\end{equation}
where ${\bf E}_\alpha$ is any product of a single $\bmath{\sigma}_x^{(i,j)}$
operator and up to $n-1$ $\bmath{\sigma}_z^{(i,1)}$ or
$\bmath{\sigma}_z^{(i,2)}$ operators.

First note that any product of up to $n-1$ $\bmath{\sigma}_z^{(i,1)}$ or
$\bmath{\sigma}_z^{(i,2)}$ operators is a product of up to $n-1$ ${\bf
Z}_1^{(i)}$ operators and $n-1$ ${\bf Z}_2^{(i)}$ operators.  Let us examine
the case where ${\bf E}_\alpha$ contains a $\bmath{\sigma}_x^{(i,j)}$ operator
and then we will examine the case where $\bmath{\sigma}_x^{(i,j)}$ does not
appear in ${\bf E}_\alpha$.  A single $\bmath{\sigma}_x^{(i,j)}$ is either
${\bf X}_2^{(i)}$ or ${\bf X}_2^{(i)} {\bf X}_1^{(i)}$.  Under the
Jordan-Wigner transformation, ${\bf X}_2^{(i)}$ is a sum of a product of an odd
number of Bogoliubov fermions.  Furthermore any product of ${\bf Z}_2^{(i)}$'s
is given by a sum of an even number of Bogoliubov fermions.  Multiplying
together a single ${\bf X}_2^{(i)}$ and any number of ${\bf Z}_2^{(i)}$
operators, we thus create an operator with a sum over an odd number of
Bogoliubov fermions.  It is an elementary result of fermion operators, then,
that an error ${\bf E}$ constructed from ${\bf X}_2^{(i)}$ and any number of
${\bf Z}_2^{(i)}$ has a vanishing matrix element over the ground state
\begin{equation}
\langle \vec{z}_2=g,\vec{c}=(1,1,\dots,1), x_1^{(1)} | {\bf E} |\vec{z}_2=
g,\vec{c} =(1,1,\dots,1),y_1^{(1)} \rangle= 0.
\end{equation}
Furthermore, multiplying such an error ${\bf E}$ any product of ${\bf
Z}_1^{(i)}$ and ${\bf X}_1^{(i)}$ operators does not change this result because
these operators act on a different tensor product subsystem.  We therefore see
that any error which contains a single $\bmath{\sigma}_x^{(i,j)}$ any
combination of $\bmath{\sigma}_z^{(i,j)}$'s satisfies the error detection
criteria, Eq.~(\ref{eq:erro}).

Next let us examine the case where $\bmath{\sigma}_x^{(i,j)}$ does not appear
in the error ${\bf E}_\alpha$, but the product of $n-1$
$\bmath{\sigma}_z^{(i,j)}$ operators do appear in error ${\bf E}_\alpha$. Every
such error will be a product of up to $n-1$ ${\bf Z}_1^{(i)}$ operators and
$n-1$ ${\bf Z}_2^{(i)}$ operators.  $n-1$ ${\bf Z}_1^{(i)}$ operators acting on
the ground state $|\vec{z}_2=g,\vec{c}=(1,1,\dots,1),x_1\rangle$ changes at
lest one value of $\vec{c}$.  Therefore if ${\bf E}$ is the product of $n-1$
${\bf Z}_1^{(i)}$ operators and operators ${\bf Z}_2^{(i)}$,
\begin{equation}
\vec{z}_2=g,\vec{c}=(1,1,\dots,1), x_1^{(1)} | {\bf E} |\vec{z}_2= g,\vec{c}
=(1,1,\dots,1),y_1^{(1)} \rangle= 0.
\end{equation}
as long as ${\bf E}$ contains at least one ${\bf Z}_1^{(i)}$.  If, on the other
hand ${\bf E}$ contains only ${\bf Z}_2^{(i)}$, then
\begin{equation}
\vec{z}_2=g,\vec{c}=(1,1,\dots,1), x_1^{(1)} | {\bf E} |\vec{z}_2= g,\vec{c}
=(1,1,\dots,1),y_1^{(1)} \rangle= E \delta_{x_1^{(1)},y_1^{(1)}}.
\end{equation}
because ${\bf Z}_2^{(i)}$ acts only on the first tensor product of the ground
state.  $E$ is some constant independent of $x_1^{(1)}$ and $y_1^{(1)}$.

We have therefore shown that the ground state of the spin ladder is an error
detecting code for any product of a single $\bmath{\sigma}_x^{(i,j)}$ and $n-1$
products of $\bmath{\sigma}_z^{(i,j)}$.  This result is equivalent to saying
that the code is an error {\em detecting} code for single
$\bmath{\sigma}_x^{(i,j)}$ operators and is an error {\em correcting} code for
$\left \lfloor {n-1 \over 2} \right \rfloor$ $\bmath{\sigma}_z^{(i,j)}$
operators.

\section{Supercoherent properties of the spin ladder ground state}

We have now shown that the ground state of the spin ladder is an error
detecting code for single $\bmath{\sigma}_z^{(i,j)}$ errors and error detecting
for $\left \lfloor {n-1 \over 2} \right \rfloor$ $\bmath{\sigma}_z^{(i,j)}$
errors.  In order for this to qualify as a supercoherent spin ladder, there
must be a gap between the ground state energy and higher energy levels.  We
know that this is true because we have found that a unique ground state.
However, it is useful to qualify the size of this gap.

As we calculated in Section~\ref{sec:cluster} the ground state of the spin
ladder occupies a specific subspace assignment of $\vec{c}$.  Specifically the
ground state corresponded to $\vec{c}=(+1,+1,\dots,+1)$ which is the ``full
cluster'' subspace.  There are two types of excitations which can occur on this
ground state. The first type of excitation is where operators maintain this
subspace. These operators will act as Bogoliubov excitations on the ground
state.  Recall that the Hamiltonian for this full cluster is given by
\begin{equation}
{\bf H}_I= 2 \omega_0 \sum_{m=-{n \over 2}\left( -{n-1 \over 2}\right)}^{{n
\over 2}\left( {n-1 \over 2}\right)} \sqrt{5+4 \cos\left( {2 \pi m \over n
}\right)} \left(\bmath{\eta}_q^\dagger \bmath{\eta}_q -{1 \over 2}
\right)+\bmath{\Omega}_k,~{\rm where}~ {\rm Tr} \bmath|\bmath{\Omega}_k|<{4
\over n}.
\end{equation}
Now $\sqrt{5+4\cos\left({2 \pi m \over n} \right)}$ varies from $1$ to $3$.
Thus there is always an energy gap in between the vacuum of this subspace and
any Bogoliubov excitations of this vacuum.  The size of such a gap is
$2\omega_0$.  Note that this is true for any Bogoliubov excitations in any of
the subspaces corresponding to a particular $\vec{c}$.

The second type of excitation which can occur is from the ground state to a
state with a different $\vec{c}$.  This type of excitation has a gap which is
the difference in the vacuum energies of the ground state and the new state.
The smallest such gap occurs when only one element of $\vec{c}$ is flipped.
This will then divide the system into two clusters.  One of length $l$ and the
other of length $n-l$.  The energy of the vacuum for this configuration is
given by
\begin{equation}
E_g(l)=- \omega_0 \sum_{m=-{l \over 2} \left(-{l-1 \over 2} \right)}^{{ l \over
2} \left({{l-1} \over 2} \right)} \sqrt{5+4 \cos \left( {2 \pi m \over l}
\right)} - \omega_0 \sum_{m=-{n-l \over 2} \left(-{n-l-1 \over 2} \right)}^{{
n-l \over 2} \left({{n-l-1} \over 2} \right)} \sqrt{5+4 \cos \left( {2 \pi m
\over n-l} \right) },
\end{equation}
if the element of $\vec{c}$ which was flipped was not $c_1$ or $c_{n-1}$.  If
the element which was flipped was at the end, then there is a single qubit
which just feels a transverse field.  The difference between the ground state
vacuum and all of the other vacuums can easily be estimated to be approximately
$\omega_0$.  Figure~\ref{fig:laddergap} shows this gap for $n$ even.
\begin{figure}[h]
\quad \quad \quad \psfig{figure=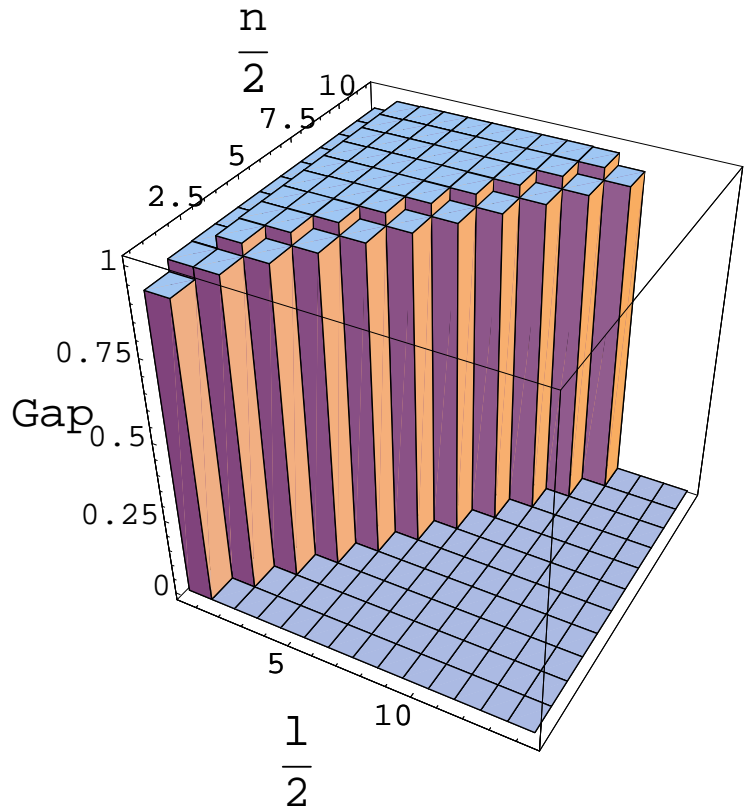,width=4in} \caption{\em The
energy gap in units of $\omega_0$ between clusters of length $l$ and $n-l$ for
the $n$ even spin ladder.} \label{fig:laddergap}
\end{figure}

We have therefore seen that the ground state of the spin ladder is separated
from all excitations by $\approx \omega_0$.  Thus, for the excitations which
are error detecting all of these errors take the state up in energy.

An interesting property of this spin ladder was the fact that not only was the
ground state error detecting for the $\bmath{\sigma}_x^{(i,j)}$ errors, the
spin ladder is also error correcting for a limited number of
$\bmath{\sigma}_z^{(i,j)}$ errors.  When such Pauli phase error will be
suppressed at low temperatures as in the supercoherent case, but it is also
possible to now correct these errors.  To see how this is done, we note that a
$\bmath{\sigma}_z^{(i,j)}$ error takes the ground state causes only an
excitation which changes the subspace labeled by $\vec{c}$.  Therefore
correcting these error corresponds to making a measurement of the operators
corresponding to $\vec{c}$.  These are the operators
\begin{equation}
{\bf e}^{(i)}=\bmath{\sigma}_x^{(i,1)} \bmath{\sigma}_x^{(i,2)}
\bmath{\sigma}_x^{(i+1,1)} \bmath{\sigma}_x^{(i+1,2)}.
\end{equation}
Measurement of these observable diagnoses the flipped $\vec{c}$ elements and
this can be used to flip these elements back and hence correct to the error.

We have thus seen that the spin ladder we have constructed has some amazing
properties.  The ground state of the spin ladder is doubly degenerate and
separated from all other states by an energy of $\approx \omega_0$.  All single
$\bmath{\sigma}_x^{(i,j)}$ error act to take the state from its ground state to
a state of higher energy and at low temperatures these errors should be
suppressed.  Similarly multiple $\bmath{\sigma}_z^{(i,j)}$ errors do not break
the degeneracy of this ground state.  Furthermore if $\left \lfloor  {n-1 \over
2} \right \rfloor$  of these $\bmath{\sigma}_z^{(i,j)}$ errors occur,
measurements can be performed such that all of these errors can be corrected.
Thus this spin ladder is a hybrid with both supercoherent and error correcting
properties.  Since $\bmath{\sigma}_z$ errors are generally more damaging to the
coherence of a system, the remarkable error correcting property should make
this spin chain extremely useful for protecting quantum information.

\section{Encoded operations}

Having shown that the ground state of the spin ladder supports a supercoherent
qubit with the extra property that it can error correct certain errors we now
ask the question of what are the encoded operations on this degeneracy.

In fact, we could have begun our discussion of this spin ladder by examining
the degeneracy of the spin ladder.  The operators
\begin{equation}
\bar{\bf Z}=\prod_{i=1}^n \bmath{\sigma}_z^{(i,1)} \quad \bar{\bf
X}=\bmath{\sigma}_x^{(1,1)} \bmath{\sigma}_x^{(1,2)}
\end{equation}
commute with the spin-ladder Hamiltonian ${\bf H}$.  Since each of these
operators square to identity we know that these operators generate
$2$-dimensional representations of the Pauli group on one qubit.  In other
words they act like $2$ dimensional single qubit operations.  Since these
operators commute with ${\bf H}$ we therefore know that these operators act on
the two-fold degeneracy of ${\bf H}$.

A further important point is necessary here.  Every state in ${\bf H}$ is
two-fold degenerate.  Not only the $\bar{\bf Z}$ and $\bar{\bf X}$ given above
commute with this Hamiltonian, but also $\prod_{i=1}^n
\bmath{\sigma}_z^{(i,j)}$ for $j=1,2$ and $\bmath{\sigma}_x^{(i,1)}
\bmath{\sigma}_x^{(2,1)}$ for $i=1,\dots,n$.  All of these operators enact an
encoded $\bmath{\sigma}_z$ or $\bmath{\sigma}_x$ on the degeneracy, however,
each may enact a different representation of these operators on the different
levels of ${\bf H}$.  For any given level, however, the action of all of these
operators is identical. Therefore one can enact an encoded $\bmath{\sigma}_z$
via either the operator $\prod_{i=1}^n \bmath{\sigma}_z^{(i,1)}$ or the
operator $\prod_{i=1}^n \bmath{\sigma}_z^{(i,2)}$.  The action of this operator
on the ground state is identical.

Unfortunately, while we could easily implement $\bar{\bf X}$ as a Hamiltonian
on the code, the operator $\bar{\bf Z}$ is not so easily to implement as a
Hamiltonian on this code (see \cite{Lidar:01b,Lidar:01c} for possible methods).
Thus, like our earlier Pauli stabilizer code example, we are left we a very
good quantum memory without the ability to manipulate the information.

Another interesting problem with this spin chain is that while $\left \lfloor
{n-1 \over 2} \right \rfloor$ $\bmath{\sigma}_z^{(i,j)}$ errors can be
corrected, it is possible for the environment to enact and error which cannot
be corrected by using only $\omega_0$ energy.  We will not delve into the
derivation of this result now as this point will be taken up in
Chapter~\ref{ch:nft} where we discuss natural fault-tolerant quantum
computation.

\section{The supercoherent spin ladder}

In this chapter we studied an interesting spin ladder.  This spin ladder has a
supercoherent ground state, and this ground state also has additional error
correcting properties.  This is an important first step towards incorporating
more than just the error detection properties of supercoherence but also error
correction. Unfortunately this spin ladder's information is not useful for
quantum computation because encoded actions cannot be enacted on the ground
state. Further it is unfortunate that only phase errors can be corrected.  Bit
flip errors are only detectable and robustness to these errors must come from
the low temperature of the environment.  In the next chapter we will see how it
is possible to encode a full single qubit quantum error correcting code into
the degenerate ground state of a system.

\chapter{A Naturally Quantum Error Correcting Ground State} \label{ch:lattice}

\begin{quote}
{\em To err is human; to forget, divine} \\
\begin{flushright} --J. H. Goldfuss \end{flushright}
\end{quote}

In this chapter we demonstrate a spin lattice system whose ground state is a
single qubit quantum error correcting code.  This is the first example of a
fully quantum error correcting ground state constructed with only two-qubit
interaction in the Hamiltonian.  The spectrum of the Hamiltonian is presented
using a simplified Stabilizer encoding and it is shown that the ground state of
this system is indeed a quantum error correcting code for single qubit errors.
We discuss the natural error correcting properties of this spin lattice ground
state and encoded operations on this code.  Finally we discuss how adiabatic
passage can be used to prepare the ground state of the spin lattice.

\section{The three-by-three quantum error correcting ground state}

Consider a three-by-three square lattice with qubits on the vertices of the
lattice (nine qubits total).  We label the elements by the row and column
indices $(i,j)$ respectively and an operator ${\bf O}$ which acts on this qubit
tensored with identity on all other qubits is ${\bf O}^{(i,j)}$.  The
Hamiltonian we are interested is given by ${\bf H}$
\begin{eqnarray}
{\bf G}&=& \bmath{\sigma}_x^{(1,1)}
\bmath{\sigma}_x^{(1,2)}+\bmath{\sigma}_x^{(1,2)}
\bmath{\sigma}_x^{(1,3)}+\bmath{\sigma}_x^{(2,1)}
\bmath{\sigma}_x^{(2,2)}+\bmath{\sigma}_x^{(2,2)}
\bmath{\sigma}_x^{(2,3)}+\bmath{\sigma}_x^{(3,1)}
\bmath{\sigma}_x^{(3,2)}+\bmath{\sigma}_x^{(3,2)} \bmath{\sigma}_x^{(3,3)}
\nonumber \\ &+&\bmath{\sigma}_z^{(1,1)}
\bmath{\sigma}_z^{(2,1)}+\bmath{\sigma}_z^{(2,1)}
\bmath{\sigma}_z^{(3,1)}+\bmath{\sigma}_z^{(1,2)}
\bmath{\sigma}_z^{(2,2)}+\bmath{\sigma}_z^{(2,2)}
\bmath{\sigma}_z^{(3,2)}+\bmath{\sigma}_z^{(1,3)}
\bmath{\sigma}_z^{(2,3)}+\bmath{\sigma}_z^{(2,3)} \bmath{\sigma}_z^{(3,3)}
\nonumber \\ {\bf H}&=&- \omega_0 {\bf G}.
\end{eqnarray}
This spin-lattice system is sketched in Figure~\ref{fig:qecc}.
\begin{figure}[h]
 \quad \quad \quad \quad \quad \psfig{figure=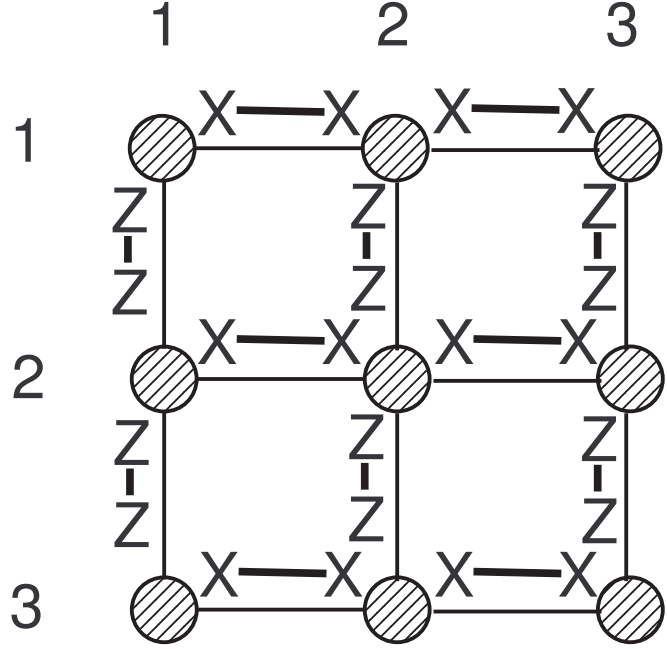,width=3.8in}
 \caption{\em The spin-lattice with a quantum error correcting ground state}
 \label{fig:qecc}
\end{figure}

\subsection{Stabilizer encoding}

Once again, in order to understand this Hamiltonian it is useful to work in a
different basis.  Particularly useful in this case is a Pauli stabilized
quantum error correction code (see Appendix~\ref{apa:pauli}).  In particular
consider the stabilizer code with stabilizer elements generated by the
operators
\begin{eqnarray}
{\bf S}_1&=& \bmath{\sigma}_z^{(1,1)} \bmath{\sigma}_z^{(1,2)}
\bmath{\sigma}_z^{(1,3)} \bmath{\sigma}_z^{(2,1)} \bmath{\sigma}_z^{(2,2)}
\bmath{\sigma}_z^{(2,3)}, \nonumber \\
 {\bf S}_2&=& \bmath{\sigma}_z^{(2,1)}
\bmath{\sigma}_z^{(2,2)} \bmath{\sigma}_z^{(2,3)} \bmath{\sigma}_z^{(3,1)}
\bmath{\sigma}_z^{(3,2)} \bmath{\sigma}_z^{(3,3)}, \nonumber \\
 {\bf S}_3&=&\bmath{\sigma}_x^{(1,1)} \bmath{\sigma}_x^{(2,1)}
 \bmath{\sigma}_x^{(3,1)} \bmath{\sigma}_x^{(1,2)} \bmath{\sigma}_x^{(2,2)}
 \bmath{\sigma}_x^{(3,2)}, \nonumber \\
  {\bf S}_4&=&\bmath{\sigma}_x^{(1,2)} \bmath{\sigma}_x^{(2,2)}
 \bmath{\sigma}_x^{(3,2)} \bmath{\sigma}_x^{(1,3)} \bmath{\sigma}_x^{(2,3)}
 \bmath{\sigma}_x^{(3,3)},
\end{eqnarray}
and corresponding to this code are the five logical operators
\begin{eqnarray}
\bar{\bf X}_1&=& \bmath{\sigma}_x^{(1,1)} \bmath{\sigma}_x^{(1,2)}, \quad
\bar{\bf Z}_1= \bmath{\sigma}_z^{(1,1)} \bmath{\sigma}_z^{(2,1)} \nonumber \\
 \bar{\bf X}_2&=& \bmath{\sigma}_x^{(1,2)} \bmath{\sigma}_x^{(1,3)}, \quad
\bar{\bf Z}_2= \bmath{\sigma}_z^{(1,3)} \bmath{\sigma}_z^{(2,3)} \nonumber \\
 \bar{\bf X}_3&=& \bmath{\sigma}_x^{(3,1)} \bmath{\sigma}_x^{(3,2)}, \quad
\bar{\bf Z}_3= \bmath{\sigma}_z^{(2,1)} \bmath{\sigma}_z^{(3,1)} \nonumber \\
 \bar{\bf X}_4&=& \bmath{\sigma}_x^{(3,2)} \bmath{\sigma}_x^{(3,3)}, \quad
\bar{\bf Z}_4= \bmath{\sigma}_z^{(2,3)} \bmath{\sigma}_z^{(3,3)} \nonumber \\
 \bar{\bf X}_5&=& \bmath{\sigma}_x^{(1,1)} \bmath{\sigma}_x^{(1,2)} \bmath{\sigma}_x^{(1,3)}, \quad
\bar{\bf Z}_5= \bmath{\sigma}_z^{(1,1)} \bmath{\sigma}_z^{(2,1)}
\bmath{\sigma}_z^{(3,1)}.
\end{eqnarray}
Using this code, we can express that Hamiltonian as
\begin{eqnarray}
{\bf H}&=&- \omega_0 \left(\bar{\bf X}_1+\bar{\bf X}_2+\bar{\bf X}_3+\bar{\bf
X}_4+\bar{\bf X}_1 \bar{\bf X}_3 {\bf S}_3 + \bar{\bf X}_2 \bar{\bf X}_4 {\bf
S}_4 \right. \nonumber
\\ && \left. +\bar{\bf Z}_1 + \bar{\bf Z}_2+\bar{\bf Z}_3+\bar{\bf Z}_4+ \bar{\bf Z}_1
\bar{\bf Z}_2 {\bf S}_1 +\bar{\bf Z}_3 \bar{\bf Z}_4 {\bf S}_2 \right).
\label{eq:qeccham}
\end{eqnarray}
Notice, as in the original supercoherent Pauli example, the fifth encoded qubit
does not appear in this Hamiltonian.  This will be degenerate codespace we will
use to store the quantum information.

Unfortunately, even after the reduction to four encoded qubits, we have not
found the exact eigenvalues and eigenstates of this Hamiltonian.  Instead we
resort to the mathematical package Mathematica to calculate the spectrum.

\section{The spin-lattice spectrum}

Corresponding to the eigenvalues of ${\bf S}_1$, ${\bf S}_2$, ${\bf S}_3$, and
${\bf S}_4$, Eq.~(\ref{eq:qeccham}) has a specific form.  Moreover, ${\bf
S}_1$, ${\bf S}_2$, ${\bf S}_3$, and ${\bf S}_4$ can be simultaneously
diagonalized. We label each of the subspaces defined by these operators via
their eigenvalues $S_1, S_2, S_3, S_4$.  For an assignment of $S_1$, $S_2$,
$S_3$, $S_4$, the four qubit Hamiltonian in Eq.~(\ref{eq:qeccham}) The spectrum
of the four qubit Hamiltonian Eq.~(\ref{eq:qeccham}) was calculated using the
program Mathematica.  These energies are assembled in Table~\ref{tab:spin}.

\begin{table}[h]
\caption{\em Energy levels of the spin-lattice} \label{tab:spin}
\begin{tabular}{|l|l|l|l|l|}
\hline
 $S_1$ & $S_2$ & $S_3$ & $S_4$ & Sorted energies in units of $\omega_0$ (degeneracy) rounded to $10^{-2}$ \\ \hline
 $+1$&$+1$&$+1$&$+1$& $-7.79, -4.69, -3.46(2), -2, -0.94, -0.79, 0(2), 2(2),
 2.58,$ \\
 & & & & \quad \quad $3.46(2), 3.62,4(2)$ \\
  \hline
 $+1$&$+1$&$+1$&$-1$& $-7.27, -4.75, -3.69, -3.09, -2, -1.20, -0.85, -0.23, \
1.19,$ \\
 $+1$& $+1$ &$-1$ &$+1$ & \quad \quad $1.31, 2(3), 4.13, 4.75, 5.69$ \\
 $+1$& $-1$ &$+1$ &$+1$ & \\
 $-1$& $+1$ &$+1$ &$+1$ & \\ \hline
 $+1$ & $+1$ &$-1$ & $-1$ & $-6.85, -4, -3.46, -3.23, -2(2), -1.62, 0(2), 1.62, 2(2),
3.23,$ \\
 $-1$ & $-1$ &$+1$ & $+1$ & \quad \quad $3.46, 4, 6.85$ \\ \hline
 $+1$ & $-1$ & $+1$ & $-1$ & $-6.46, -5.18, -3.46(2), -1.52, -1.09, 0(4), 1.09, 1.52,
 3.46(2),$ \\
 $+1$ & $-1$ & $-1$ & $+1$ & \quad \quad $5.18,
 6.46$\\
 $-1$ & $+1$ & $+1$ & $-1$ & \\
  $-1$ & $+1$ & $-1$ & $+1$ & \\ \hline
  $+1$ & $-1$ & $-1$ & $-1$ & $-5.69, -4.75, -4.13, -2(3), -1.32, -1.19, 0.23,
0.85, 1.19, 2,$ \\
  $-1$ & $+1$ & $-1$ & $-1$ & \quad \quad $3.09, 3.69, 4.75, 7.27$ \\
 $-1$ & $-1$ & $+1$ & $-1$ & \\
 $-1$ & $-1$ & $-1$ & $+1$ & \\ \hline
 $-1$ & $-1$ & $-1$ & $-1$ & $-4(2), -3.63, -3.46(2), -2.58, -2, 0(2), 0.79, 0.94, 2, 3.46(2),$ \\
 & & & & \quad \quad $4.69, 7.79$ \\ \hline
\end{tabular}
\end{table}

We see from Table~\ref{tab:spin} that the ground state of ${\bf H}$ over the
four encoded qubits is unique and inhabits the $S_1=+1, S_2=+1, S_3=+1, S_4=+1$
subspace.  As mentioned above, the fifth encoded qubit is not involved in ${\bf
H}$ and therefore all of the states in Table~\ref{tab:spin} will be two-fold
degenerate corresponding to this encoded qubit.

In order to label the states of the spin-lattice we use the basis
$|S_1,S_2,S_3,S_4,j,z_5\rangle$ where $S_i$ are the $\pm 1$ eigenvalues of
${\bf S}_i$, $j$ labels the energy levels sorted from $j=0$ the lowest energy
to $j=15$ the highest energy (and picking some arbitrary ordering and basis for
the degenerate states), and $z_5$ is the $\pm 1$ eigenvalue of ${\bf Z}_5$.

\section{Ground state error correcting properties }

The two-fold degenerate ground state of the spin-lattice system is given by the
state $|+1,+1,+1,+1,0,z_5\rangle$.  We will now show that this state is an
error correcting code for all single qubit errors.  The condition that the
ground state is an error correcting code for all single qubit errors is given
by
\begin{equation}
\langle +1,+1,+1,+1,0,z_5 | \bmath{\sigma}_\alpha^{(i,j)}
\bmath{\sigma}_\beta^{(j,l)} | +1,+1,+1,+1,0,z_5^\prime \rangle =
C_{\alpha,\beta,i,j,k,l} \delta_{z_5,z_5^\prime}, \label{eq:qecon}
\end{equation}
where $\alpha,\beta=\{0,1,2,3\}$.  Notice we allow the identity operators in
this expression.

Every operator of the form $\bmath{\sigma}_\alpha^{(i,j)}
\bmath{\sigma}_\beta^{(i,j)}$ where $\alpha \neq \beta$ anticommutes with at
least one element of the stabilizer generators ${\bf S}_i$.  This implies that
\begin{equation}
\langle +1,+1,+1,+1,0,z_5 | \bmath{\sigma}_\alpha^{(i,j)}
\bmath{\sigma}_\beta^{(k,l)} | +1,+1,+1,+1,0,z_5^\prime \rangle = 0 \quad
\alpha \neq \beta.
\end{equation}
This follows from the standard reasoning about stabilizer codes.  If the error
element anticommutes with one of the stabilizer elements, the action of this
error is to flip the value of the corresponding $S_i$ eigenvalue.  Therefore
the matrix element vanishes.

Thus we need only concern ourselves with the $\bmath{\sigma}_\alpha^{(i,j)}
\bmath{\sigma}_\alpha^{(k,l)}$ elements.  The identity case, $\alpha=0$ is
trivially filled.  Some of these elements anticommute with a generator of the
stabilizer ${\bf S}_i$ and therefore, via the argument of the previous
paragraph satisfy Eq~.(\ref{eq:qecon}).  It is easy to check that all of the
elements which do not anticommute with a generator of the stabilizer ${\bf
S}_i$ can be written as a product of the first four encoded qubit operators and
the stabilizer elements
\begin{equation}
\bmath{\sigma}_\alpha^{(i,j)} \bmath{\sigma}_\alpha^{(j,l)}= p {\bf X}_1^{c_1}
{\bf X}_2^{c_2} {\bf X}_3^{c_3} {\bf X}_4^{c_4} {\bf Z}_1^{d_1} {\bf Z}_2^{d_2}
{\bf Z}_3^{d_3} {\bf Z}_4^{d_4} {\bf S}_1^{s_1} {\bf S}_2^{s_2} {\bf S}_3^{s_3}
{\bf S}_4^{s_4},
\end{equation}
where $p=\pm 1$ or $p=\pm i$ and $c_i,d_i,s_i \in \{0,1\}$.  Therefore this
operator only acts on the first four encoded qubits and not the encoded qubit.
Therefore we see that for these elements
\begin{eqnarray}
\langle +1,+1,+1,+1,0,z_5 | \bmath{\sigma}_\alpha^{(i,j)}
\bmath{\sigma}_\alpha^{(k,l)} | +1,+1,+1,+1,0,z_5^\prime \rangle =
E_{\alpha,i,j,k,l} \delta_{z_5,z_5^\prime}.
\end{eqnarray}
where $E_{\alpha,i,j,k,l}$ does not depend on $z_5$ or $z_5^\prime$ and
therefore satisfies the error correcting requirement.

We have thus seen that the ground state of the spin-lattice system is a error
correcting code for single qubit errors.

How does one perform the error correction procedure for this ground state?  One
manner is as follows.  There are two kinds of errors.  The first type of error
takes the state from the $S_i=+1,\forall i$ subspace to another $S_i$ labeled
subspace. By measuring the stabilizer elements, these errors can be detected
and corrected.  The second type of error preserves ${\bf S}_i=1, \forall i$
subspace but acts as an excitation on the four qubit encoded Hamiltonian.  One
way to  determine if there error has occurred is to measure the Hamiltonian
itself ${\bf H}$.  If the value is not that of the ground state, then
appropriate manipulations can be applied to restore the system to the ground
state.  This method of error correction is, however, appears very difficult to
implement on the ground state.  However, in the next section we shall argue
that the system will apply much of the error correction procedure through the
{\em naturally} evolution of the system plus environment.

\section{Natural error correction}

Inspection of Table~\ref{tab:spin} shows that the ground state is separated
from states reached by an error  $\approx 0.52 \omega_0$.  All single qubit
errors, as in supercoherence, take the system from the global ground state to a
state of higher energy.  There is an important consequence, however, of the
fact that the ground state of the spin-lattice is a quantum error correcting
code.

Consider supercoherence first.  Suppose a single qubit error occurs on the
supercoherent ground state.  The state will then be excited to higher energy
levels.  Since the supercoherent ground state is only error detecting, it is in
general impossible to restore the system to the ground state without destroying
the quantum information stored into the degeneracy of the supercoherent system.
In supercoherence then, a single qubit error will occur and any relaxation of
the system back to the ground state will occur in such a way that the
degeneracy is acted upon nontrivially.  Once an error has happened on the
supercoherent ground state, the supercoherent information is in trouble of
being decohered.

Now consider the spin lattice we have described above.  As in the supercoherent
case if a single qubit error occurs on the ground state of the spin-lattice the
state will be excited to higher energy levels.  Now, however, because the
ground state is error correcting there is the possibility of restoring the
information to the ground state without decohering the information stored in
the degeneracy.  In fact, one of the relaxation pathways open to a system which
decoheres back to the ground state will be exactly the error correction
procedure necessary to restore the system to the ground state without
destroying the degeneracy of the system.  The fact that one of the open system
evolution pathways open to the system is the error correcting procedure follows
directly from the Hermiticity of the Hamiltonian.  It is important to note that
in our spin-lattice case the relaxation back to the ground state is not always
error correcting.  It is possible for the state to take a relaxation pathway
which goes through other energy levels and thus destroys the degeneracy of the
ground state.  This corresponds to a two qubit error which our single qubit
error correcting code is not designed to correct.  However, the Hermiticity of
the Hamiltonian implies that the relaxation pathway which fixed the error is
open an thus evolution of the spin-lattice has a non-negligible component along
the error correction pathways.

\begin{figure}[h]
 \quad \psfig{figure=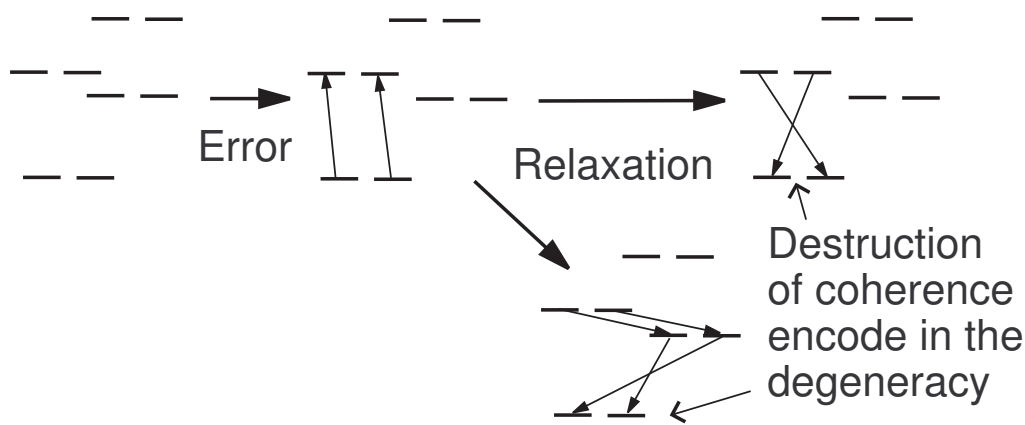,width=5.5in}
 \caption{\em Supercoherence evolution pathways}
 \label{fig:supervs}
\end{figure}

\begin{figure}[h]
 \quad \psfig{figure=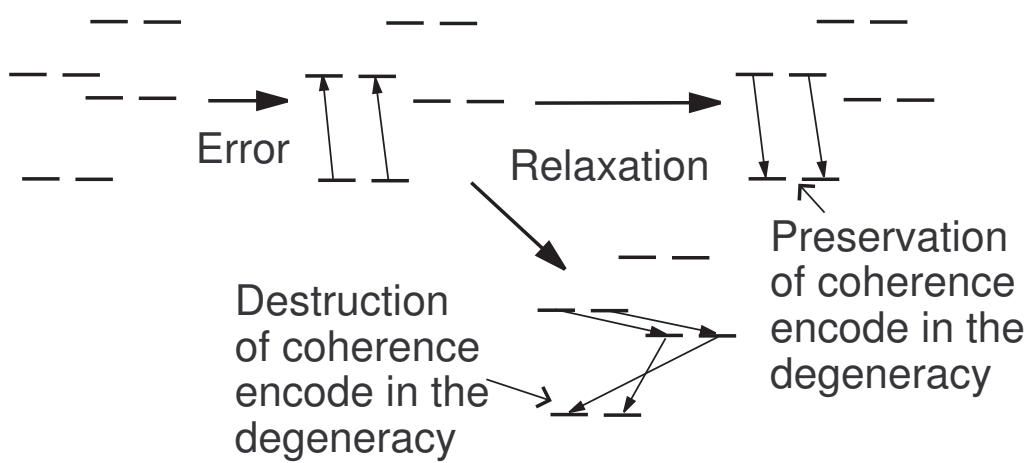,width=5.5in}
 \caption{\em Quantum error correcting ground state evolution pathways}
 \label{fig:qeccvs}
\end{figure}

In Figures~\ref{fig:supervs} and \ref{fig:qeccvs} we show a schematic of the
difference between supercoherence and the quantum error correcting
spin-lattice.

The ability of a system to self-correct decoherence processes is an interesting
property of our spin-lattice system.  In fact, our spin-lattice system is the
first example of such automatic or natural error correction which uses only
two-body interactions between qubits.  There are two precedents for such
automatic error correction, one by Barnes and Warren\cite{Barnes:00a} and the
other from Kitaev and
coworkers\cite{Kitaev:97a,Kitaev:97c,Bravyi:98a,Freedman:01a}.

Barnes and Warren\cite{Barnes:00a} present a scheme where errors are
automatically corrected.  These authors present an NMR implementation whose
ground state is an error correcting code. We note, however, that this
implementation only corrects limited types of errors. In particular their
system does not correct single qubit phase errors.  In fact, as we will discuss
in Chapter~\ref{ch:nft}, the system presented by Barnes and Warren is not any
more special than a two-dimensional Ising system.  In contrast to the proposal
of Barnes and Warren, the spin lattice we present can correct all single qubit
errors.  On the other hand, our system has the shortcoming that correction does
not always succeed.

The second precedent for our spin lattice is the work of Kitaev and
coworkers\cite{Kitaev:97a,Kitaev:97c,Bravyi:98a,Preskill:98a,Ogburn:99a,Freedman:01a}.
In this work, codes are constructed which have a ground state which is quantum
error correcting.  However, in these systems the interactions needed in order
to make this system naturally error correcting require either interactions
between greater than three subsystems or require two-body interactions between
subsystems with greater than $60$ levels for each subsystem!  The benefit of
our spin-lattice system should be obvious in this respect as it removes this
many-body or many-level restriction.

\section{Encoded operations}

The encoded operation on the degeneracy are easy to find.  The encoded
$\bmath{\sigma}_x$ and $\bmath{\sigma}_z$ are simply the $\bar{\bf X}_5$ and
$\bar{\bf Z}_5$ operators.  As in the spin ladder system, we see that there is
a difficulty in implementing the operators on this spin lattice.  We shall not
delve into method for fixing this problem here.  Needless to say, it {\em is}
possible to construct lattices which are error correct but for which can also
be manipulated.

However, let us note two interesting properties of the encoded operators.
Suppose we wanted to perform the {\em gate} $\bmath{\sigma}_x$ on the
degeneracy.  This corresponds to the operator $\bar{\bf
X}_5=\bmath{\sigma}_x^{(1,1)} \bmath{\sigma}_x^{(1,2)}
\bmath{\sigma}_x^{(1,3)}$ (or such an operator times a stabilizer element.  One
way in which this gate can be enacted is by performing single qubit rotations
on each of the listed qubits.  Suppose that one of these single qubit rotations
was over rotated.  Such an over rotation now becomes an error on the ground
state.  But this error will be correctable (either naturally or by our error
correction procedure).  This then is a form of {\em fault-tolerance}. The gate
we use to implement the rotation can be faulty and we still will obtain the
correct operation.  We will have a chance to considerably extend this notion in
Chapter~\ref{ch:nft}.

Second we note that systems much like the spin-lattice we have constructed here
can most easily be constructed from the encoded operations backwards.  In
particular, the encoded operations will be the operations which are errors on
the system.  Thus given encoded Pauli operators, constructing operations which
are products of the remaining Pauli operators guarantees the error correcting
properties of the ground state.  This is a powerful tool for constructing such
codes: {\em work with the operators on the code first!}

\section{Preparation via adiabatic passage}

An interesting question which arises in the context of using the information
the ground state of the spin lattice is the question how to prepare the
information.  A method for doing this can be achieved using the adiabatic
theorem.  Suppose, for instance that we could completely turn off the
$\bmath{\sigma}_x^{(i,j)}$ operators in ${\bf H}$ and then slowly turn these
operators back on.  In particular consider the ability to enact the time
dependent Hamiltonian
\begin{eqnarray}
{\bf H}(t)&=& -{t \over T} \omega_0 \left( \bmath{\sigma}_x^{(1,1)}
\bmath{\sigma}_x^{(1,2)}+\bmath{\sigma}_x^{(1,2)}
\bmath{\sigma}_x^{(1,3)}+\bmath{\sigma}_x^{(2,1)} \bmath{\sigma}_x^{(2,2)}
 +\bmath{\sigma}_x^{(2,2)}
\bmath{\sigma}_x^{(2,3)}+\bmath{\sigma}_x^{(3,1)} \bmath{\sigma}_x^{(3,2)}
\right. \nonumber \\ && \left. +\bmath{\sigma}_x^{(3,2)}
\bmath{\sigma}_x^{(3,3)} \right)  - \omega_0 \left(\bmath{\sigma}_z^{(1,1)}
\bmath{\sigma}_z^{(2,1)}+\bmath{\sigma}_z^{(2,1)}
\bmath{\sigma}_z^{(3,1)}+\bmath{\sigma}_z^{(1,2)}
\bmath{\sigma}_z^{(2,2)}+\bmath{\sigma}_z^{(2,2)} \bmath{\sigma}_z^{(3,2)}
\right. \nonumber \\ &&\left. +\bmath{\sigma}_z^{(1,3)}
\bmath{\sigma}_z^{(2,3)}+\bmath{\sigma}_z^{(2,3)} \bmath{\sigma}_z^{(3,3)}
\right),
\end{eqnarray}
where $T$ is constant with units of time.

\begin{figure}[h]
 \quad \psfig{figure=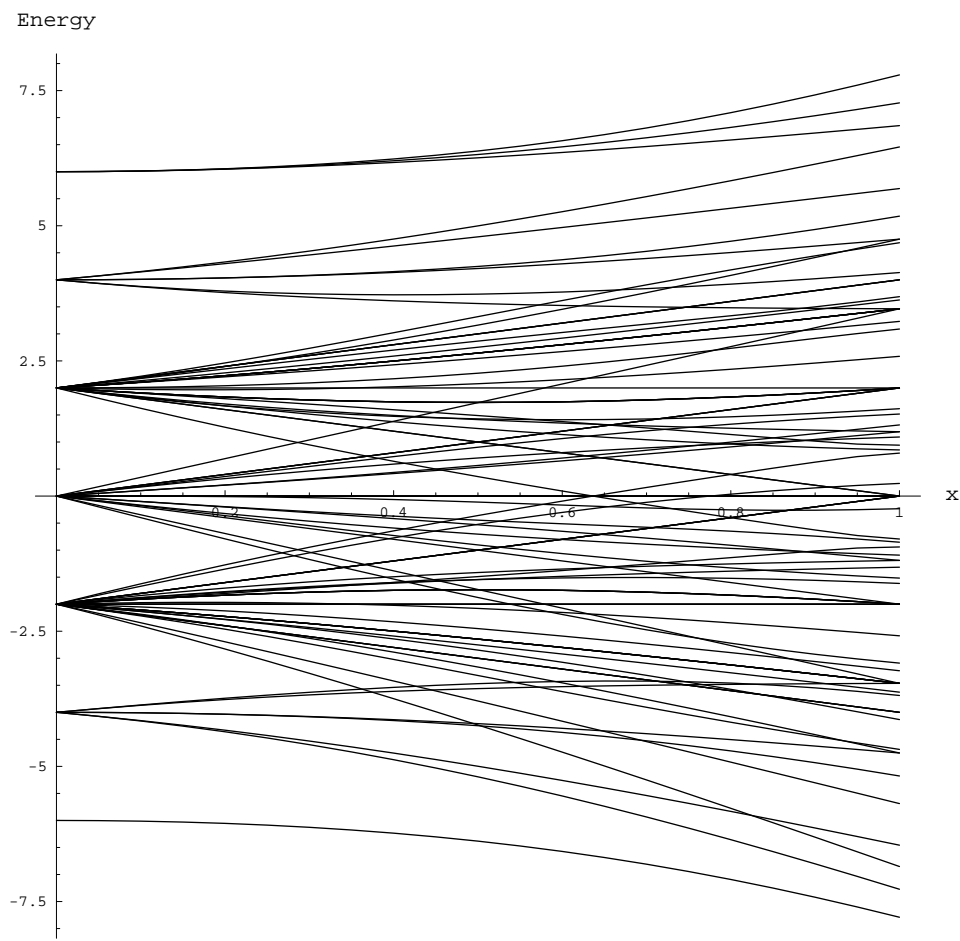,width=5in}
 \caption{\em Energy levels in units of $\omega_0$ as a function of the x component ($x=t/T$) of the spin lattice Hamiltonian}
 \label{fig:adiab}
\end{figure}

The adiabatic theorem\cite{Messiah:76a} states that a system which is in an
eigenstate of a time dependent Hamiltonian will remain the instantaneous
eigenstate of the system if variation of this Hamiltonian is slow enough and
the energies of the Hamiltonian do not cross.  In Figure~\ref{fig:adiab} we see
that the ground state does not cross any other state.   Thus for sufficiently
long $T$, if we can prepare the state into the ground state of ${\bf H}(0)$ we
can then guarantee that we end up in the ground of the spin lattice Hamiltonian
${\bf H}(T)$.  Furthermore, the degeneracy of the system will remain intact
throughout this evolution.  But preparation into the ground state of ${\bf
H}(0)$ with a given degeneracy is easy.  In particular the state where every
qubit is $|0\rangle$ is such a ground state.  How slow do we have to ramp up
the field?  From the adiabatic theorem\cite{Messiah:76a} we find that we
require $T > {1 \over \omega_0}$.  Thus there is an easy method for preparing
the state via the adiabatic theorem.

\section{Natural quantum error correction}

In this chapter we have presented the first example of a ground state which is
a full single qubit quantum error correcting code.  This ground state has the
intriguing property that all single qubit error excite the system to an energy
level of higher energy and there is a non-vanishing probability that the system
will then decay back to the ground state in such a way as the correct the
error.  Clearly the next step along these lines is to demonstrate how one can
obtain perfect automatic error correction where each single error is always
corrected unless the system is excited to a higher energy.  Furthermore the
issue of how to robustly perform operations on the spin lattice ground state
was not satisfactorily addressed.  In the next chapter we will have the chance
to address these issues from the context of a more distanced perspective.

\chapter{Towards Naturally Fault-tolerant Systems} \label{ch:nft}

\begin{quote}
{\em What passes for optimism is most often the effect of an intellectual
error} \\
\begin{flushright} --Raymond Aron, {\em The Opium of the Intellectuals}\cite{Aron:01a}\end{flushright}
\end{quote}

Before the discovery of quantum error correction and fault-tolerant quantum
computation, there was much reason to be
pessimistic\cite{Unruh:95a,Landauer:96a} about the future technological
prospects of the construction of a quantum computer.  Having discovered quantum
error correction followed by the penning of the threshold theorem for
indefinite fault-tolerant quantum computation, the prospects for building a
quantum computer has brightened considerably.  The influence of the discovery
of quantum error correction, however, has not had much of an impact on the
experimental proposals for quantum computation.  True, many proposals now
mention the explicit requirement that parallel operations are necessary for
fault-tolerant quantum computation\cite{Aharonov:96a}, but the notions of
fault-tolerance are mostly viewed as an {\em eventual} goal of a given physical
proposal.  Calculate your error rate, demonstrate you have universal control,
and you have a quantum computer!  To proceed in this manner calls on the
argument of technological inevitability, but is it not possible that there are
systems which are naturally fault-tolerant for quantum computation just as such
system exist for classical computers?  In this chapter we lay out the
schematics for such a naturally fault-tolerant quantum system built not from
the single qubit up but from large numbers of qubits whose collective
properties are used for quantum information manipulation.

\section{The classical stability of information}

Why is it that classical computers are, to date, so overly robust to
interaction with their environment?  In fact, it is a mistake to say that all
classical computers are robust to interactions with their environment.  One
need only take a standard household personal computer out into the hard
radiation of space to see that classical computers are only robust in certain
environments. Further it is also obvious that the actual physical
implementation of the classical computer is essential to the robustness of the
classical computer: building a classical computer out of billiard balls is
possible, but then substantial error correction is needed to make the system
robust to small deviations in the trajectories of the billiard balls.  So we
should really ask, why are today's silicon-based computers with magnetic
recording devices so robust to sufficiently non-harsh environments?

We will begin by examining the question of what make the robust long-term
storage of classical information possible.

\subsection{Classical memory, security in numbers, and the lesson of dimensionality}

Today's classical memory devices come in two forms, read-only and read-write
memories.  In a read-only memory, information is imprinted once and can be read
out but not changed without substantial technical prowess.  We will focus on
the read-write memories where classical information can both be imprinted and
easily manipulated.  In particular we will focus on the use of magnetic media
which is the media used for storage in most hard drives.

Information on a hard drive is stored in  spatially distinct grains of a
ferromagnetic substance.  The grains of the ferromagnetic substance consist of
complete magnetic domains and are magnetized in one of two possible directions
which are the logical $0$ and $1$ of the classical information. This
information is read and written using a device known as a read-write head. This
head can read the information by sensing the direction of the magnetized
domains and writes information by applying a magnetic field magnetizes the
domain.

The question of the stability of such a memory is therefore a question of the
stability of the magnetization of a ferromagnetic substance.  The properties of
a ferromagnetic substance are dominated by an exchange energy between the
electron spins of the substance
\begin{equation}
{\bf H}=-J \vec{\bf s}_i \cdot \vec{\bf s}_j.
\end{equation}
In addition to this energy, most materials have a magnetocrystalline energy (or
anisotropy energy) in which different directions of magnetization have a
preferential (lower) energy.  The exchange energy, however, dominates the
ferromagnetic properties of the substance.  The exchange energy is minimized
when all of the spins are completely aligned.  This is the origin of domains in
magnetization of a ferromagnetic material: the electrons would rather align
with each other. The magnetocrystalline energy is essential in determining
which directions within the solid are preferred.

Let us now show how such ferromagnetic substances can be understood to be a
classical automatic error correcting code.  In order to see this, consider the
simplified model of a ferromagnet given by the Ising model\cite{Ising:25a}.  In
this Ising model, spins on a lattice are assumed to point in one of two
directions $s_i=\pm 1$ and the energy of system is given by nearest neighbor
interactions of the form
\begin{equation}
E=-J\sum_{<ij>} s_i s_j + B\sum_i s_i,
\end{equation}
where $J$ is the energy of the nearest neighbor bonds $<ij>$ and $B$ is an
applied field.  In the absence of an applied field $B=0$, the ground state of
this system is a two-fold degenerate with all of the spins parallel.  In the
presence of a magnetic field, the ground state is one of the two configuration
with all of the spins parallel determined by sign of the applied field $B$. Let
us ignore $B$ for now but we will return to nonzero $B$ later.

The degenerate ground state of the Ising model with no magnetic field can be
considered an encoding of classical information.  Let us define logical $0$ as
the case where $s_i=+1$ for all lattice sites $i$ and logical $1$ as the case
where $s_i=-1$ for all lattice sites $i$.  An important consideration enters
into the stability of this encoded information: the dimensionality of the
lattice of spins.  For now we will assume that this dimension is greater than
or equal to two.  At $T=0$ (i.e. completely isolated from any environment), the
ground states will just stay where they are.  But as the temperature is turned
up $T>0$, it is possible for the environment to excite the spins in the system
and destroy the information encoded into this degeneracy.  What is it that
protects the information encoded into this degeneracy from such errors?

Define the spontaneous magnetization as the expectation value of all of the
lattice spins $m={1 \over N} \sum_i\langle s_i \rangle$ where $N$ is the number
of spins in the lattice.  For the two-dimensional Ising system on a square
lattice, for example, it is possible to exactly solve for the spontaneous
magnetization\cite{Plischke:94a} which is given by
\begin{eqnarray}
m(T)&=&\pm \left[ 1- {(1 - \tanh^2(\beta J))^4 \over16 \tanh^4 (\beta J)  }
\right]^{1/8} \quad T<T_c \nonumber
\\
 &=& 0 \quad T>T_c,
\end{eqnarray}
where $\beta={1 \over T}$.  At low enough temperatures, $T \ll T_c$, the
spontaneous magnetization of the system persists.  Even though there the
environment can heat the system, the information stored in the total
magnetization is unaffected by these fluctuations.  In ferromagnetic materials,
the temperature $T_c$ is the Currie temperature of the material: usually on the
order of a thousand Kelvin.  So the mystery has become why does this
magnetization persist at non-zero temperature.

Consider taking a two-dimensional Ising model with all of the spins parallel
and flipping a single one of the spins.  This will result in a change of energy
of the system by $2J r$ where $r$ is the number of neighbors to which the spin
is attached.  This is the lowest energy excitation which can occur on the
system without flipping $N-1$ out of $N$ spins.  The second of these options,
flipping $N-1$ out of $N$ spins requires an extraordinary amount of energy
which just is not available.  Suppose after we flip the single spin, we flip
another spin.  The energy of this new configuration must now be even higher
than the system with just one spin flipped.  To see this, first note that if
the next spin flipped is not a neighbor of the first spin flipped, then there
is certainly an increase in energy.  If the second spin is a neighbor of the
first spin, because the dimension of the spin lattice is greater than one the
total number of violated Ising energies must increase.  See
Figure~\ref{fig:ising2d}.

\begin{figure}[h]
 \quad \psfig{figure=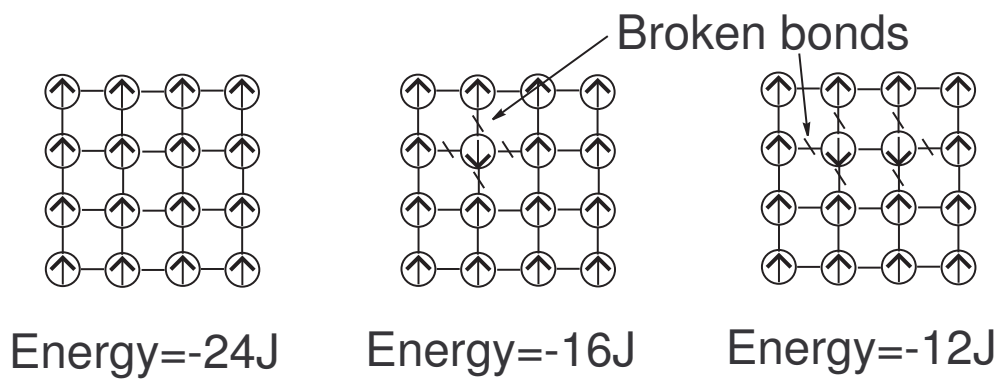,width=5.5in}
 \caption{\em Ising model in two dimensions showing the energy proportional to domain perimeter effect}
 \label{fig:ising2d}
\end{figure}

We can now see how the Ising model in greater than two dimensions can be viewed
as an automatic error correcting code.  The codewords are the majority labeled
states with all of the spins aligned in parallel.  Each bit flip error that
occurs on the system, until over $N/2$ spins have been flipped, causes an
increase in the energy of the system.  Therefore the tendency of the system is
to self correct the errors which have occurred.  There is security in numbers
here and one sees the most trivial error correcting code, the majority code, at
work.   When one performs thermodynamical calculations of the magnetization of
this system, the unlikelyhood of spin flips which flip between the $0$ and $1$
states is reflected in the persistence of spontaneous magnetization.  At low
enough temperature, this magnetization thus persists. Long rang off diagonal
order\cite{Mattis:85a} is therefore an indication of the ability of the system
to self-correct errors on the system. We also see how an applied field can
change the state of affairs.  If the applied field is strong enough, then it
can overcome the Ising bonds and flip the information encoded into the
degeneracy of the ground state.  Note also that even the states which have up
to $N/2$ qubits flip still maintain the information about the classical
information.  It is only when the sign of the total magnetization flips sign
does the classical information get destroyed.

Of central importance in the argument for the stability of the information in
the degeneracy of the Ising model was the dimension of the system.  When a spin
is flipped in the two dimensional Ising model from the ground state, the energy
of the system is proportional to the number of bonds with nearest neighbors
broken.  More generally domains of flipped spins have an energy greater than
the ground state energy which is proportional to the area of the perimeter of
the domain.  In $d$ dimensions, the energy of a domain is proportional to the
$d-1$ dimensional surface area of a domain.  For one dimension, we therefore
see that this energy is a constant.  This implies that it is possible to
exchange a minimal amount of energy between the environment and the system
while destroying the information stored in the degeneracy.  Consider the one
dimensional Ising model.  One can flip a single spin which requires only the
bond energy $2J$, and then proceed to flip neighboring spins without expending
any energy.  Thus it is possible to use only $2J$ energy in destroying the
degeneracy of the ground state.  This is shown in Figure~\ref{fig:ising1d}. The
condensed matter theorist would say, ``there is no long range order at nonzero
temperature for the one dimensional Ising model'' which we see is equivalent to
the statement that the system will not automatically correct its own errors. We
will return to this question when we consider quantum models, but we note here
that all of the examples we have demonstrated in Part III of this thesis are
analogous to the one dimensional Ising case in that errors (now quantum) can
occur which only exert a minimal amount of energy in order to decohere the
degenerate quantum information.

\begin{figure}[h]
 \quad \psfig{figure=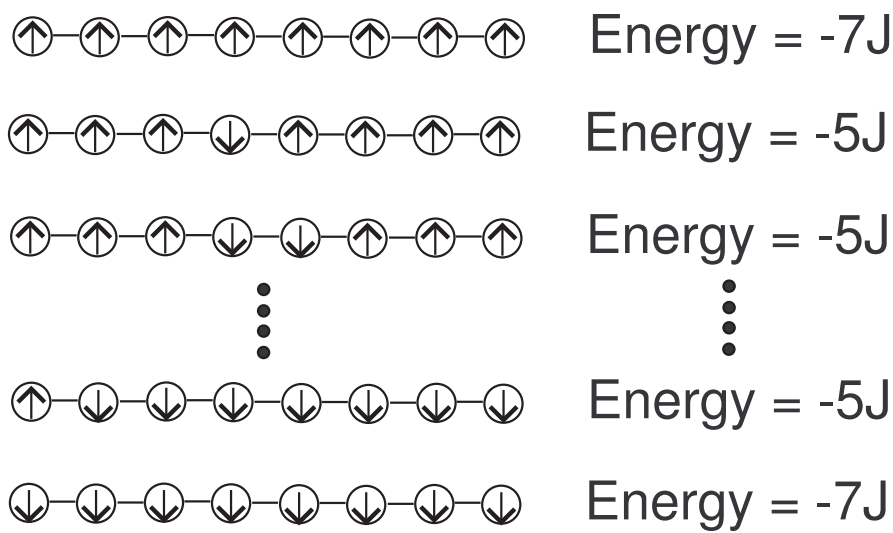,width=5.5in}
 \caption{\em One dimension Ising model showing how the degeneracy can be
 adversely changed by flipping spins with only a minimal amount of work}
 \label{fig:ising1d}
\end{figure}

We have thus seen that classical information stored in a magnetic media is
robust due to a robust automatic error correcting code.  This information is
robust because errors which occur to the information are robustly fixed.
Further the information can be written on by the application of a field which
breaks the degeneracy of the information and further changed the energetics of
the system which allowed for the degeneracy to be protected.

\subsection{Classical gates}

Next we ask the question of what makes the classical manipulation of
information so robust to error.  Classical computation occurs on the
manipulation of current in an integrated circuit and the important manipulation
of the current is that of a simple switch.  The prototype of such manipulation
is the transistor.  Consider, for example a standard bipolar junction
transistor.  In such a transistor, a small voltage bias between the emitter and
the base can lead to a large change in the current running from (for a npn
transistor) the collector to the emitter.

There are two lessons to be learned from the manipulation of information by a
transistor.  The first lesson comes from the use of current to represent
information.  It is important to realize that current represents a majority
voting correcting code in a method similar to that encountered in magnetic
medium above.  Surely current can run the wrong way in a circuit, but only at
the expense of energy conservation.  If insufficient energy is provided by an
environment, then the majority of electrons will flow in the correct direction.
The second lesson taken from the transistor is that a small change in one
information carrier can make a large change in the information of another
information carrier.  In particular this change must be digital in the sense
that the system is robust to small variations in the controlling mechanism. The
transistor either does or does not allow current flow between the collector and
emitter.  This is a very important property of a fault-tolerant system:
applying the gate and not applying the gate are two macroscopically separate
actions.

Classical gates therefore rely on a form of natural error correction by
majority manipulation of the information as well as a completely digital
manipulation of the information.

\section{Natural quantum error correction}

Having briskly described the reasons why classical computers are so robust we
now seek to extend these notions to quantum systems.  The lesson of quantum
error correction is that quantum information is a bit more complicated, but
very similar to classical information.  Quantum error correction works because
it deals not just with the bit flips of classical error correction but also
because it deals with phase errors.  It would not be surprising then to find
that, just as there are natural classical error correcting codes, there may be
natural quantum error correcting codes.

For simplicity we will assume a collection of qubits which we desire to be
naturally endowned with error correction.  Obvious generalizations are possible
to other subsystems and we shall leave these generalization implicit.

A natural quantum error correcting code is a collection of $N$ qubits and a
system Hamiltonian ${\bf H}(N)$ which satisfies
\begin{enumerate}
    \item The ground state of the Hamiltonian ${\bf H}(N)$ is degenerate.
    \item The ground state of the Hamiltonian ${\bf H}(N)$ is a quantum error correcting
    code for $l=O(N)$ qubit errors.
    \item Let $E_i$ denote the energy of the lowest level reachable by $i$
    single qubit errors on the ground state.  For $i$ less than the number of
    correctable errors, $E_i<E_{i+1}$ and $\sum_{i=1}^l E_i = O(N).$
\end{enumerate}
Item 1 and 2 insure that information can be encoded into the ground state and
is a quantum error correcting code.  Item 3 insures that each error must supply
energy from the environment to the system and that the total energy needed to
induce an error on the ground state is an extensive variable.  This last
requirement is extremely important for natural error correction as it requires
that the amount of energy needed to break the degeneracy is a macroscopic
amount of energy in the sense that it depends on the size of the system.
Finally we note that due to the Hermiticity of the system-environment
Hamiltonian, error pathways are always accompanied by correction pathways.  The
basic idea is there that of the automatic error correction of Barnes and
Warren\cite{Barnes:00a} where errors are automatically fixed as they move up in
energy.  We must stress again, however, that the example of Barnes and Warren
is only a classical error correcting code (just because a system is quantum
this does not make the dynamics uniquely quantum).  Here we would like to fully
extend the notion to quantum error correction.  Furthermore we would also like
to stress the requirement that breaking the degeneracy of the system require a
macroscopic expenditure of energy.  This requirement protects the quantum
information from all but the most energetic environmental fluctuations.

One of the best ways to examine an error correcting code is to examine the
encoded operations which manipulate the code.  Encoded operations must consist
of operators which act on greater than the number of qubits which the error
correcting code can correct.  This requirement implies that all encoded
operators must be of size $O(N)$ on a natural error correcting code.  By
examining the smallest encoded operations on the code it is possible to
determine whether the code can satisfy the natural quantum error correcting
criteria.

Consider, for example, an extension of the Pauli supercoherent example of
Chapter~\ref{ch:sup} and the quantum error correcting ground state of
Chapter~\ref{ch:lattice}.  Given a $l \times l$ square lattice with sites
$(i,j)$ acted upon by the Hamiltonian
\begin{equation}
{\bf H}=- \omega_0 \left( \sum_{i=1}^{l-1} \sum_{j=1}^l
\bmath{\sigma}_x^{(i,j)} \bmath{\sigma}_x^{(i+1,j)} \sum_{i=1}^l
\sum_{j=1}^{l-1}  \bmath{\sigma}_z^{(i,j)}\bmath{\sigma}_z^{(i,j+1)} \right).
\end{equation}
This Hamiltonian has a degenerate ground state for the following reason.  The
smallest Pauli operators which commute with ${\bf H}$ are the column
$\bmath{\sigma}_x$ operators: ${\bf X}_j =\prod_{i=1}^l
\bmath{\sigma}_x^{(i,j)}$ and the row $\bmath{\sigma}_z$ operators: ${\bf
Z}_i=\prod_{j=1}^l \bmath{\sigma}_z^{(i,j)}$.  Take any two of these operators.
Since they square to identity and form a group isomorphic to the single qubit
Pauli group, these operators act as $2^{(l^2)-1}$ $2$-dimensional irreps of the
Pauli group.  Therefore the Hamiltonian ${\bf H}$ must have a degenerate ground
state which is at least two-fold degenerate.  The degeneracy of this ground
state may, in fact, be more than this two-fold degeneracy.  But notice that we
have shown that there are operations on the degenerate ground state which
involve $l$ single qubit operators on a system with $l^2$ qubits.  This system,
then, does not satisfy the requirement that the system is an error correcting
code for $O(N)$

It is perhaps best, then, when constructing a naturally quantum error
correcting code to start from the operators which manipulate the information
and work backwards. Further we note that naturally quantum error correcting
codes do in fact exist\cite{Preskill:00a}.  However, the known constructions
involve four dimensional spatial configurations as well as unreasonably
complicated many-body or many-level interactions.  The challenge of natural
quantum error correction is to achieve a naturally error correcting code
without these unphysical assumptions.

Finally we would like to note some physical properties of a natural quantum
error correcting code.  The quantum error correcting code condition implies
that for a naturally quantum error correcting code
\begin{equation}
\langle j| \sum_i \bmath{\sigma}_\alpha^{(i)} |k\rangle = C_\alpha \delta_{ij},
\end{equation}
where $|j\rangle$ and $|k\rangle$ are the ground state codewords and the sum is
over all lattice sites.  If, for example, each qubit is a spin, this would
imply that the codewords all have the same net magnetization.  It is therefore
impossible to measure the information encoded into the degeneracy by simply
measuring the bulk magnetization of such a naturally error correcting code. The
question of the readout of information will be addressed in the next section
where we discuss fault-tolerant quantum computation.

\section{Natural fault-tolerant quantum computation}

The notion of natural quantum error correction is not enough for quantum
computation.  Natural quantum error correcting codes will protect the quantum
information, but this says nothing about preparing the information,
manipulating the information, and reading out the information in a robust
manner.

In classical information manipulation, we saw that there were two requirements
for fault-tolerance: the digital nature of manipulations as well as the self
correcting energetics like standard quantum error correction.

Suppose we are given a natural quantum error correcting code and wish to
perform a manipulation of the information stored in the code.  This will be
achieved by turning on some Hamiltonian which manipulates the information.  Due
to the natural error correcting criteria specified above the only operators
which can affect the degeneracy of the code states are those which involve
qubit operators which are of size $O(n)$.  This gate will enact and operation
on $O(n)$ qubits.  A set of gates ${\mathcal G}$ is said to be a fault-tolerant
gate set if
\begin{enumerate}
\item Every gate can be implemented in a manner that faulty gates correspond to
errors which can be corrected by the natural quantum error correcting code.
\item Any two gates are separated from each other by $O(N)$ qubit operators.
\end{enumerate}
When item 1 is fulfilled, faulty gate creation corresponds to errors which can
and will be naturally corrected by the code.  This is the requirement of
energetics: a macroscopic expenditure of energy enacts the operation and
fluctuations in this enacting cannot destroy the quantum information unless
these fluctuations convey a macroscopic amount of energy to the system.  The
second requirement is the requirement of the digital nature of the gate set. It
must not be possible for different encoded actions to be enacted which are
close together in the space of errors.

Consider, as an example, a code on $N$ qubits in which an encoded action is
performed by enacting the ${\bf G}$ operator which consists of a ${\bf
G}^{(i)}$ on every qubit $\prod_i {\bf G}^{(i)}$.  Now suppose that these
operators ${\bf G}^{(i)}$ are created using a Hamiltonian ${\bf H}_G^{(i)}$:
${\bf G}^{(i)}=\exp \left[-i {\bf H}_G^{(i)} T \right]$ where $T$ is a fixed
constant.  Now suppose that an over rotation in the enacting of this gate
occurs.  Instead of ${\bf G}^{(i)}$ on each qubit, the gates enacted are $\exp
\left[-i {\bf H}_G^{(i)} (T+\Delta T) \right]$. For small $\Delta T$ this can
be expressed as ${\bf G}^{(i)} \left[ {\bf I}-i {\bf H}_G^{(i)} \Delta T
\right]$.  The full evolution if each gate is overrotated by $\Delta T$ is
given by
\begin{equation}
\prod_i {\bf G}^{(i)} \left[ {\bf I} -i {\bf H}_G^{(i)} \Delta T  \right]= {\bf
G} \left[ {\bf I} - i \sum_i {\bf H}_G^{(i)} \Delta T - \Delta T^2 \sum_{ij}
{\bf H}_G^{(i)} {\bf H}_G^{(j)} + \cdots \right].
\end{equation}
The major corrections to the evolution are therefore single, two, etc. qubit
errors.  If the error correcting code can naturally correct $O(N)$ errors, then
the only evolution which escapes detection is a correction $\Delta T^{O(N)}$
which is exponentially small.  Thus microscopically faulty errors will not be
able to destroy the quantum information.

Two issues remain to be addressed for a naturally fault-tolerant quantum
computer.  The first is preparation and the second is measurement.  An
important realization for a natural quantum error correcting code is that
preparation does not mean perfect preparation of the degenerate ground state,
but instead means perfect preparation of the degenerate ground state plus a
minimal amount of errors occurring on to this ground state.  The important
point of fault-tolerant preparation is that preparation should prepare a state
which may have errors but none of these errors are macroscopic errors  which
act nontrivially on the degeneracy.  One method which appears to be extremely
useful for preparation and measurement of information in a naturally
fault-tolerant code is the use adiabatic continuity.  In order to prepare a
state, the degeneracy of the ground state should be macroscopically broken.
This then will correspond to a macroscopic breaking of the degeneracy of the
ground state.  By adiabatically changing the system Hamiltonian it should be
possible to move from a state were this degeneracy is broken to the state where
this degeneracy is not broken while maintaining a robust error correcting
criteria. Finally reversing this process adiabatically the degeneracy can again
be restored and a measurement of a macroscopic variable can be used to read out
the quantum information.

\section{The road ahead}

\begin{quote}
{\em We can't solve problems by using the same kind of thinking we used when we
created them.} \\
\begin{flushright}
--Albert Einstein
\end{flushright}
\end{quote}

In this chapter we have sketched out a road map for the possibility of natural
quantum computing systems.  There is much work to be done!  Besides the obvious
physical consequences of natural fault-tolerant quantum computation, the
uniformity of such natural fault-tolerant system can serve as a good test bed
for a rigorous proof of the quantum computing threshold.  There are also
interesting connections between the idea of natural fault-tolerant quantum
computation and non-abelian gauge fields\cite{Kitaev:97c,Ogburn:99a}.  Of
particular interest are theories of high-temperature
superconductivity\cite{Senthill:00a,Senthill:01a,Senthill:01b} and nonabelian
effects in the fractional quantum Hall effect\cite{Freedman:01a,Moore:91a}
which support discrete gauge groups.

The path towards building a quantum computer will by no means be an easy
journey.  Certainly the technological revolution of modern classical computers
was an amazingly complex and difficult revolution.  However, it is unclear that
all of the present experimental proposals for a quantum computer, which build
the quantum computer from the qubit up, will be the ultimate manner in which a
quantum computer will be built.  In part III of this thesis we have given
simple examples of systems which begin to exhibit many of the conditions
necessary for natural fault-tolerant quantum computation.  The natural
assumption that a quantum computer must be an entirely different type of device
than a classical computer is, we believe, a fallacy born of the mystery
attributed to quantum theory.

\chapter{Conclusion}

\begin{quote}
{\em We will either find a way or make one.} \\
\begin{flushright}
--Hannibal
\end{flushright}
\end{quote}

In the beginning there was Alan Turing, poking away at the foundations of
computer science, dreaming that machines could perform amazing feats of
calculation.  Among Turing's other interests were the foundations of quantum
mechanics\cite{Hodges:00a}\footnote{Alan Turing in fact rediscovered the Zeno
paradox in quantum mechanics\cite{Hodges:00a}.}.  Today we stand in the middle
of a computer revolution far outstripping anything possibly imagined by Turing.
There is a hint, however, that Alan's other interest, the quantum theory of
nature may hold even more revolutionary computational power than his basic
insights into classical computer science.  In this thesis we have, hopefully,
provided helpful steps towards the construction of a quantum computer. Someday,
we may even dream, we may even be as lucky as Alan Turing: the pokings of this
thesis may turn into the revolutionary technologies of tomorrow.

\bibliographystyle{plain}
\bibliography{thesisb3.0}

\appendix

\chapter{The quantum information language} \label{apa}

One strength of studying quantum information is that it provides a language for
understanding generic informational properties of quantum systems.  In this
section we introduce some of the basic language and machinery which we use
freely in this thesis.

\section{Basic quantum computation notion}

In classical information the basic unit of information is a bit which is
conventionally described by the two possible states $0$ and $1$.  In quantum
information the most basic unit of information is the {\em
qubit}\cite{Schumacher:95a}. The state of a qubit inhabits a two dimensional
Hilbert space $\CC^2$.  Whereas the classical bit has just two possible states
$0$ and $1$, the state of qubit is a unit vector in $\CC^2$:
$|\psi\rangle=\alpha |0 \rangle + \beta |1 \rangle$ where $\alpha,\beta \in
\CC$, $|\alpha|^2+|\beta|^2=1$, and we have picked some convenient orthogonal
basis $|0\rangle,|1\rangle$ for this state. More precisely, because the global
phase of a quantum state has no physical relevance, the state of a qubit is
identified with a ray in the Hilbert space $|\psi\rangle=e^{i\theta} \left(
\cos(\omega) |0\rangle + \sin(\omega) e^{i\phi}|1 \rangle \right), \forall
\theta \in \RR$.

In quantum computation, a particular basis $|0\rangle$, $|1\rangle$ for a qubit
is usually singled out as the {\em computational basis}.  In physical systems,
this basis is usually determined by some physically motivated definition (i.e.
the eigenstates of the system Hamiltonian).  Given the fixed basis $|0\rangle$,
$|1\rangle$ we can define the Pauli matrices
\begin{eqnarray}
\bmath{\sigma}_0&=&{\bf I}=\left[\begin{array}{cc} 1 & 0 \\ 0 & 1
\end{array}\right], \quad
 \bmath{\sigma}_1=\bmath{\sigma}_x=\left[\begin{array}{cc} 0 & 1 \\ 1 & 0
\end{array}\right], \nonumber \\
\bmath{\sigma}_2&=&\bmath{\sigma}_y=\left[\begin{array}{cc} 0 & -i \\ i & 0
\end{array}\right], \quad
 \bmath{\sigma}_3=\bmath{\sigma}_z=\left[\begin{array}{cc} 1 & 0 \\ 0 & -1
\end{array}\right].
\end{eqnarray}
These matrices for a basis for linear operators on the qubit.  Real
combinations of these matrices are a basis for Hermitian operators on a qubit.
It is also convenient to define the Pauli spin matrices ${\bf s}_\alpha={1
\over 2} \bmath{\sigma}_\alpha$.

More generally the state of a single qubit is described by a density matrix
$\bmath{\rho}$ which has a particularly useful parametrization as a vector in
the ``Bloch'' sphere: $\bmath{\rho}={1 \over 2} {\bf I} + \vec{n} \cdot
\vec{\bmath{\sigma}}$ where $\vec{\bmath{\sigma}} = (\bmath{\sigma}_1,
\bmath{\sigma}_2, \bmath{\sigma}_3)$.

\section{Entangled and separable} \label{apa:sepdef}

Suppose one is given a bipartite Hilbert space ${\mathcal H}={\mathcal H}_A
\otimes {\mathcal H}_B$.  A density matrix, $\bmath{\rho}$ on this Hilbert
space is defined as {\em separable} if it can be written as
\begin{equation}
\bmath{\rho}=\sum_i p_i |\psi_i \rangle \langle \psi_i | \otimes |\phi_i
\rangle \langle \phi_i |,
\end{equation}
where $0<p_i\leq1$, $\sum_i p_i=1$, $|\psi_i\rangle \in {\mathcal H}_A$, and
$|\phi_i\rangle \in {\mathcal H}_B$.  A density matrix which cannot be written
in this form is called {\em entangled}.

\section{Fixed basis formalism} \label{apa:fixedbasis}

Suppose one is given a $d$ dimensional Hilbert space ${\mathcal H}$.  It is
convenient when examining linear operators on this space to work with a fixed
basis.  The space of linear operators on ${\mathcal H}$ is spanned by a fixed
basis of hermitian traceless operators ${\bf F}_\alpha$ with
$\alpha=1\dots2d^2-1$ and a scaled identity operator ${\bf F}_0={1 \over
\sqrt{d}} {\bf I}$.  These operators can always be chosen to be trace
orthogonal:
\begin{equation}
{\rm Tr}\left[{\bf F}_\alpha^\dagger {\bf F}_\beta \right]= \delta_{\alpha
\beta}.
\end{equation}
Linear combinations of the ${\bf F}_\alpha$'s over $\CC$ span all operators on
${\mathcal H}$ while linear combinations of the ${\bf F}_\alpha$'s over $\RR$
span all Hermitian operators.

\section{Positive operator valued measurements} \label{apa:povm}

The most general notion of a measurement on a quantum system is given by the
concept of a positive operator valued measurement (POVM).  A POVM is specified
by a set of positive operators ${\bf E}_\alpha$. where $\alpha$ labels the
measurement outcome, which satisfy $\sum_\alpha {\bf E}_\alpha= {\bf I}$.  The
result of a POVM on the quantum state $\bmath{\rho}$ is the result $\alpha$
with probability $p_\alpha={\rm Tr} \left[ \bmath{\rho} {\bf E}_\alpha
\right]$.

\section{Distance measures on density matrices} \label{apa:tracenorm}

The trace distance between two density matrices is defined as
\begin{equation}
D\left(\bmath{\rho},\bmath{\sigma} \right)= {1 \over 2} {\rm Tr} |\bmath{\rho}
- \bmath{\sigma} |,
\end{equation}
where $|{\bf A}|=\sqrt{{\bf A}^\dagger {\bf A}}$.  Suppose that one was
attempting to perform a measurement which distinguished the two density
matrices $\bmath{\rho}$ and $\bmath{\sigma}$.  Performing a POVM with elements
$\{ {\bf M}_\alpha \}$ one obtain result $\alpha$ on the density matrices
$\bmath{\rho}$ and $\bmath{\sigma}$ with probabilities $p_\alpha={\rm Tr}[ {\bf
M}_\alpha \bmath{\rho}]$ and $q_\alpha={\rm Tr}[{\bf M}_\alpha \bmath{\sigma}]$
respectively.  A basic theorem of distance measures on density matrices (see,
for example \cite{Fuchs:96a}) tells us that
\begin{equation}
D\left(\bmath{\rho},\bmath{\sigma} \right)= \max_{\{{\bf M}_\alpha \}} {1 \over
2} \sum_\alpha |p_\alpha - q_\alpha|,
\end{equation}
where the maximization is taken over all possible POVMs.  Density matrices
which are close in trace distance are therefore hard to distinguish by a
maximally distinguishing measurement.

\section{Decoherence rates under the trace inner product} \label{apa:rates}

The trace inner product $\langle {\bf A}, {\bf B} \rangle={\rm Tr}[{\bf
A}^\dagger {\bf B}]$ is an easy to use metric for examining how a density
matrix strays from its initial state.  If we have a state $\bmath{\rho}(0)$ at
$t=0$, we define the {\em mixed-state memory fidelity} of this state at time
$t$ later as
\begin{equation}
F_m(t)={\rm Tr}[\bmath{\rho}(0) \bmath{\rho}(t)].
\end{equation}
While this fidelity has no intrinsic relation to, say, how distinguishable
$\bmath{\rho}(t)$ has become from $\bmath{\rho}(0)$, the Taylor expansion of
the memory fidelity can give us a good idea about the rate of change of the
density matrix from its initial state.  Performing this Taylor series
expansion,
\begin{equation}
F_m(t)=\sum_{n=0}^\infty {1 \over n!} \left({t \over \tau_n} \right)^n,
\end{equation}
where we have defined the {\em decoherence rates} \cite{Duan:97a,Duan:97b}
\begin{equation}
{1 \over \tau_n}= \left\{{\rm Tr}\left[\bmath{\rho}(0) \bmath{\rho}^{(n)}(0)
\right] \right\}^{1 \over n},
\end{equation}
where $\bmath{\rho}^{(n)}(0) = \left. {\partial ^n \bmath{\rho}(t) \over
\partial t^n} \right|_{t=0}$.

\section{The Pauli group and Pauli stabilizer codes} \label{apa:pauli}

The Pauli group ${\mathcal P}$ on $n$ qubits is the group made up of all
possible tensor products of the Pauli operators $\bmath{\sigma}_\alpha, (\alpha
\in \{0,1,2,3\})$ together with possible global phase factors $\{\pm 1, \pm
i\}$. Elements of the Pauli group either commute or anticommute with each
other.

An abelian subgroup of the Pauli group is call a Pauli stabilizer group
$\mathcal S$. An example of a Pauli stabilizer group consists of the elements
$\bmath{\sigma}_x \otimes \bmath{\sigma}_x \otimes \bmath{\sigma}_x$,
$\bmath{\sigma}_y \otimes \bmath{\sigma}_x \otimes \bmath{\sigma}_y$,
$-\bmath{\sigma}_z \otimes {\bf I} \otimes \bmath{\sigma}_z$, and ${\bf I}$.
Every Pauli stabilizer group $\mathcal S$ can be generated by a set of
generating elements which are independent in the sense that none can be
generated from the others. If a Pauli stabilizer group has $k$ generators, then
there are $2^k$ elements in the stabilizer (i.e. the order of the Pauli
stabilizer group is $2^k$). Since the Pauli stabilizer groups are abelian, they
can be simultaneously diagonalized.

A Pauli stabilizer code is the subspace which has a $+1$ eigenvalue for all of
the elements of the stabilizer ${\bf S}|i\rangle = |i\rangle$.  We then say
that such a state is stabilized by the element.  In particular the $+1$ common
eigenspace of the stabilzer elements defines a $2^{n-k}$ dimensional subspace
which is called the stabilizer code space.

Elements of the Pauli group which anticommute with an element of the stabilizer
group act to take codewords from the stabilizer code space to the space
perpendicular to the stabilizer code space.

The set of all Pauli group members which commute with a Pauli stabilizer group
$\mathcal S$ is called the centralizer of the group $C({\mathcal S})$.
Properties of the Pauli group imply that the stabilizer is also the normalizer
of the group $N({\mathcal S})$ which is defined as the set of operators which
fix ${\mathcal S}$ under conjugation.  The operators which are in the
normalizer but not the stabilizer of ${\mathcal S}$ are the {\em logical
operators} on the stabilizer  These operators preserve the stabilizer code but
act nontrivially on the $2^{n-k}$ dimensional stabilizer space.  In fact these
operators act like a $n-k$ qubit Pauli group.

A stabilizer code can detect all errors which anticommute with at least one
stabilizer elements.  The number of elements upon which a Pauli group element
act nontrivially is called the weight of the Pauli group element.  The standard
nomenclature for a code is given by $[n,k,d]$ where $n$ is the number of qubits
for the code, $k$ is the number of encoded qubits, and $d$ is the distance of
the code.  The smallest weight of the Pauli group which does not anticommute
with any stabilizer element is called the distance of the code.  A code with
can correct $l$ single qubit errors must have a distance of at least $2l+1$.

The reader is referred to \cite{Gottesman:97a} for more detailed information on
stabilizer codes.

Some of the elements of the normalizer of the single qubit Pauli group are
often denoted as follows
\begin{eqnarray}
{\bf H}&=&{1 \over \sqrt{2}} \left( \begin{array}{cc} 1 & 1 \\ 1 & -1
\end{array} \right), \quad {\bf P}=\left( \begin{array}{cc} 1 & 0 \\ 0 & i
\end{array} \right), \nonumber \\
 {\bf Q}&=&{1 \over \sqrt{2}}\left( \begin{array}{cc} 1 & i \\ -i & -1
\end{array} \right), \quad {\bf T}={1 \over \sqrt{2}}\left( \begin{array}{cc} 1
& -i \\ 1 & i \end{array} \right).
\end{eqnarray}
All of these gates map $\bmath{\sigma}_\alpha$ operators to themselves under
conjugation by one of these elements.

\chapter{Proof of universality on the weak collective decoherence DFS}
\label{apb}

Let DFS$_{n}$($K$) denote the decoherence-free subsystem on $n$ physical qubits
with Hamming weight $h={n-K \over 2}$. We show here that
\begin{equation}
{\sf H}=\{{\bf E}_{i,i+1},{\bf T}_{i,i+1}^{P},{\bf T}_{i,i+1}^{Q},\bar{{\bf
Z}}_{i,i+1}:i=1,\ldots ,n-1\}, \label{eq:weakugs}
\end{equation}
is a universal generating set of Hamiltonian for any of the DFSs occurring in a
system of $n$ physical qubits.  It is convenient to work directly with the
Hamiltonians, and to show that ${\sf H}$ gives rise to the Lie-algebra
$su(d_{K})$ on each DFS$_{n}$($K$) (via scalar multiplication, addition, and
Lie-commutator; i.e. the allowed composition operations for a Lie algebra).
Exponentiation then gives the group $SU(d_{K})$ on the DFS. We will proceed by
induction on $n$, the number of physical qubits, building the DFS-states of $n$
qubits out of DFS-states for $n-1$ qubits. A graphical representation of this
construction is useful (and will also generalize to the strong case presented
in the following section \ref{apc}): see Figure~(\ref{figure1}) at the end of
this Appendix.

We have seen that in the weak collective decoherence case the DFS states are
simply bitstrings of $n$ qubits in either $|0\rangle $ or $|1\rangle $. The
different $n$-qubit DFSs are labeled by their eigenvalue
\begin{equation}
\lambda _{K} = ({\rm number\ of}\ 0{\rm 's} - {\rm number\ of}\ 1{\rm 's})
\equiv K_{n}.
\end{equation}
To obtain a DFS-state of $n$ qubits out of a DFS-state of $n-1$ qubits
corresponding to $K_{n-1}$ we can either add the $n^{{\rm th}}$ qubit as $
|0\rangle $ ($K_{n}=K_{n-1}+1$) or as $|1\rangle $ ($K_{n}=K_{n-1}-1$). Each
DFS-state can be built sequentially from the first qubit onward by adding
successively $|0\rangle $ or $|1\rangle $, and is uniquely defined by a
sequence $K_{1},\ldots ,K_{n}$ of eigenvalues. In the graphical representation
of Fig.~(\ref{figure1}) the horizontal axis marks $n$, the number of qubits up
to which the state is already built, and the vertical axis shows $K_{n}$, the
difference (number of 0's - number of 1's) up to the $n^{{\rm th}}$ qubit.
Adding a $|0\rangle $ at the $n+1^{{\rm th}}$ step will correspond to a line
pointing upwards, adding a $|1\rangle $ to a line pointing down. {\em Each
DFS-state of }$n${\em qubits having eigenvalue }$\lambda _{K}=K_{n}$,{\em \ is
thus in one-to-one correspondence with a path on the lattice from the origin
to} $(n,K_{n})$.

Consider the first non-trivial case, $n=2$, which gives rise to one DFS-qubit:
DFS$_{2}$($0$). This corresponds to the two states $|0_{L}\rangle =|01\rangle $
[path 2 in Fig.~(\ref{figure1})] and $|1_{L}\rangle =|10\rangle $ (path 3) with
$ K_{2}=0$. The remaining Hilbert space is spanned by the one-dimensional DFS$
_{2}$($2$) $|00\rangle $ (path 1) corresponding to $K_{2}=2$, and DFS$_{2}$($
-2$) $|11\rangle $ (path 4) corresponding to $K_{2}=-2$. The exchange ${\bf E
}_{12}$ flips $|0_{L}\rangle $ and $|1_{L}\rangle $ (path 2 and 3), and leaves
the other two paths unchanged. The interaction ${\bf A}_{12}={\rm diag}
(0,0,1,0) $ induces a phase on $|1_{L}\rangle =|10\rangle $ (path 3). Their
commutator forms an encoded $\sigma_y$ {\em acting entirely within the}
DFS$_{2}$($0$) {\em subspace}. Its commutator with ${\bf E }_{12}$ in turn
forms an encoded $\sigma_z$ with the same property. Together they form the
(encoded) Lie algebra $su(2)$ acting entirely within this DFS. The Lie algebra
is completed by forming the commutator between these $\bar{{\bf Y}}$ and
$\bar{{\bf Z}}$ operations. To summarize:
\begin{eqnarray}
\bar{{\bf Y}}_{12} &=& i[\bar{{\bf A}},{\bf E}_{12}]=\left(
\begin{array}{cccc}
0 & 0 & 0 & 0 \\ 0 & 0 & -i & 0 \\ 0 & i & 0 & 0 \\ 0 & 0 & 0 & 0
\end{array}
\right) \\  \nonumber \bar{{\bf Z}}_{12} &\equiv& i[{\bf E}_{12},\bar{{\bf
Y}}_{12}]  \nonumber \\ \bar{{\bf X}}_{12} &\equiv& i[\bar{{\bf
Y}}_{12},\bar{{\bf Z}}_{12}].
\end{eqnarray}
We call the property of acting entirely within the specified DFS {\em
independence}, meaning that the corresponding Hamiltonian has zero entries in
the rows and columns corresponding to the other DFSs
[DFS$_{2}$($2$)=$|00\rangle $ and DFS$_{2}$($-2$)=$|11\rangle $ in this case].
When the Hamiltonian is exponentiated, the corresponding gate will act as
identity on all DFSs except DFS$_{2}$(0).

To summarize these considerations, the Lie-algebra formed by ${\sf H}
_{0}^{2}=\{\bar{{\bf X}},\bar{{\bf Z}}\}$ is $su(2)$, and generates $SU(2)$ on
DFS$_{2}$(0) by exponentiation. In addition, this is an independent $ SU(2) $,
namely, these operations act as identity on the other DFSs: when written as
matrices over the basis of DFS-states, their generators in ${\sf H }_{0}^{2}$
have zeroes in the rows and columns corresponding to all other DFSs.

In the following we show how this construction generalizes to $n>2$ qubits, by
proving the following theorem:
\begin{theorem}
For any $n\geq 2$ qubits undergoing weak collective decoherence, there exist
sets of Hamiltonians ${\sf H}_{K_{n}}^{n}$ [obtained from ${\sf H}$ of
Eq.~(\ref{eq:weakugs}) via scalar multiplication, addition, and Lie-commutator]
acting as $ su(d_{K_{n}})$ on the DFS corresponding to the eigenvalue $K_{n}$.
Furthermore each set acts {\em independently} on this DFS only (i.e., with
zeroes in the matrix representation corresponding to their action on the other
DFSs).
\end{theorem}

Before proving this theorem, we first explain in detail the steps taken in
order to go from the $n=2$ to the $n=3$ case, so as to make the general
induction procedure more transparent.

The structure of the DFSs for $n=2$ and $3$ qubits is:
\begin{eqnarray}
{\rm DFS}_{2}(2)&=&\{|00\rangle \},\quad {\rm DFS}_{2}(0)=\left\{
\begin{array}{c}
|01\rangle \\ |10\rangle
\end{array}
\right. ,\quad {\rm DFS}_{2}(-2)=\{|11\rangle \}  \nonumber \\ {\rm
DFS}_{3}(3)&=&\{|000\rangle \},\quad {\rm DFS}_{3}(1)=\left\{
\begin{array}{c}
|001\rangle \\ |010\rangle \\ |100\rangle
\end{array}
\right. ,\quad {\rm DFS}_{3}(-1)=\left\{
\begin{array}{c}
|011\rangle \\ |101\rangle \\ |110\rangle
\end{array}
\right. \nonumber \\ {\rm DFS}_{3}(-3)&=&\{|111\rangle \}.
\end{eqnarray}
DFS$_{3}$($3$) is obtained by appending a $|0\rangle $ to DFS$_{2}$($2$).
Similarly DFS$_{3}$($-3$) is obtained by appending a $|1\rangle $ to DFS$ _{2}
$($-2$). Graphically, this corresponds to moving along the only allowed pathway
from DFS$_{2}$($2$) [DFS$_{2}$($-2$)] to DFS$_{3}$($3$) [DFS$_{3}$($ -3$)], as
shown in Fig.~(\ref{figure1}). The lowest and highest $\lambda _{K}$ for $n$
qubits will always be made up of the single pathway connecting the lowest and
highest $\lambda _{K}$ for $n-1$ qubits. The structure of DFS$ _{3} $($\pm 1$)
is only slightly more complicated. DFS$_{3}$($1$) is made up of one state,
$|001\rangle $, which comes from appending a $|1\rangle $ (moving down) to
DFS$_{2}$($2$). We call $|001\rangle $ a ``Top-state''\ in DFS$_{3}$($1$). The
two other states, $|010\rangle $ and $|100\rangle $, come from appending
$|0\rangle $ (moving up) to DFS$_{2}$($0$). Similarly, we call $|010\rangle $
and $|100\rangle $ ``Bottom-states'' in\ DFS$_{3}$($1$). DFS$_{3}$($-1$) is
constructed in an analogous manner (Fig.~\ref{figure1}).

We showed above that it is possible to perform independent $su(2)$ operations
on DFS$_{2}$($0$). DFS$_{2}$($\pm 2$) are also both acted upon independently,
but because they are one-dimensional subspaces, independence implies that
$su(2)$ operations annihilate them. Since the states $ \{|010\rangle
,|100\rangle \}\in $ DFS$_{3}$($1$) and the states $ \{|011\rangle ,|101\rangle
\}\in $ DFS$_{3}$($-1$) both have $\{|01\rangle ,|10\rangle \}\in $
DFS$_{2}$($0$) as their first two qubits, one immediate consequence of the
independent action on DFS$_{2}$($0$) is that one can {\em simultaneously}
perform $su(2)$ operations on the corresponding daughter subspaces created by
expanding DFS$_{2}$($0$) into DFS$_{3}$($\pm 1$). The first step in the general
inductive proof is to eliminate this simultaneous action, and to act
independently on each of these subspaces (the ``{\em independence step}''). To
see how this is achieved, it is convenient to represent the operators acting on
the $8$-dimensional Hilbert space of $3$ qubits in the basis of the $4$ DFSs:

\begin{center}
\begin{tabular}{cccccccc}
$000$ & \multicolumn{1}{|c}{$001$} & $010$ & $100$ & \multicolumn{1}{|c}{$ 011
$} & $101$ & $110$ & \multicolumn{1}{|c}{$111$} \\ \hline\hline
\multicolumn{1}{|c}{$M_{3}$} & \multicolumn{1}{|c}{} &  &  &  &  &  &  \\
\cline{1-4} & \multicolumn{1}{|c}{} &  &  & \multicolumn{1}{|c}{} &  &  &  \\ &
\multicolumn{1}{|c}{} & $M_{1}$ &  & \multicolumn{1}{|c}{} &  &  &  \\ &
\multicolumn{1}{|c}{} &  &  & \multicolumn{1}{|c}{} &  &  &  \\ \cline{2-7} &
&  &  & \multicolumn{1}{|c}{} &  &  & \multicolumn{1}{|c}{} \\ &  &  &  &
\multicolumn{1}{|c}{} & $M_{-1}$ &  & \multicolumn{1}{|c}{} \\ &  &  &  &
\multicolumn{1}{|c}{} &  &  & \multicolumn{1}{|c}{} \\ \cline{5-8} &  &  &  &
&  &  & \multicolumn{1}{|c|}{$M_{-3}$} \\ \cline{8-8}
\end{tabular}
\end{center}

The simultaneous action on DFS$_{3}$($\pm 1$) can now be visualized in terms of
both $M_{\pm 1}$ being non-zero. Let us show how to obtain an action where,
say, just $M_{1}$ is non-zero. This can be achieved by applying the commutator
of two operators with the property that their intersection has non-vanishing
action just on $M_{1}$. This is true for the ${\bf T}_{23}^{P}$ and $\bar{{\bf
X}}_{12}$ Hamiltonians: ${\bf T}_{23}^{P}$ annihilates every state except those
that are $|00\rangle $ over qubits $2$ and $3$, namely $ |100\rangle \in $
DFS$_{3}$($1$) and $|000\rangle \in $DFS$_{3}$($3$). This implies that the only
non-zero blocks in its matrix are
\begin{equation}
M_{3}({\bf T}_{23}^{P})=1,\quad M_{1}({\bf T}_{23}^{P})=\left(
\begin{array}{ccc}
0 & 0 &  \\ 0 & 0 &  \\ &  & 1
\end{array}
\right) .
\end{equation}
On the other hand, $\bar{{\bf X}}_{12}$ is non-zero only on those states that
are $|01\rangle $ or $|10\rangle $ on qubits $1$ and $2$. Therefore it will be
non-zero on all $3$-qubit states that have $|01\rangle $ or $ |10\rangle $ as
``parents''. This means that in its matrix representation $ M_{\pm 3}=0$ and
\begin{equation}
M_{1}(\bar{{\bf X}}_{12})=\left(
\begin{array}{ccc}
0 &  &  \\ & 0 & 1 \\ & 1 & 0
\end{array}
\right) ,\quad M_{-1}(\bar{{\bf X}}_{12})=\left(
\begin{array}{ccc}
0 & 1 &  \\ 1 & 0 &  \\ &  & 0
\end{array}
\right) .
\end{equation}
Clearly, taking the product of ${\bf T}_{23}^{P}$ and $\bar{{\bf X}}_{12}$
leaves non-zero just the lower $2\times 2$ block of $M_{1}$, and this is the
crucial point:\ it shows that an independent action on DFS$_{3}$($1$) can be
obtained by forming their commutator. Specifically, since the lower $2\times 2$
block of $M_{1}({\bf T}_{23}^{P})$ is just $\frac{1}{2}\left( {\bf I} -\sigma
_{z}\right) $:
\begin{equation}
i[{\bf T}_{23}^{P},\bar{{\bf X}}_{12}]=\bar{{\bf Y}}_{\{|100\rangle
,|010\rangle \}},
\end{equation}
i.e., this commutator acts as an encoded $\sigma _{y}$ inside the $
\{|100\rangle ,|010\rangle \}$ subspace of DFS$_{3}$($1$). Similarly, ${\bf
\bar{Z}}_{\{|100\rangle ,|010\rangle \}}=\frac{i}{2}[\bar{{\bf Y}}
_{\{|100\rangle ,|010\rangle \}},\bar{{\bf X}}_{12}]$. Together the two
operators $\{\bar{{\bf Y}}_{\{|100\rangle ,|010\rangle \}},\bar{{\bf
Z}}_{\{|100\rangle ,|010\rangle \}}\}$ generate $su(2)$ acting {\em
independently} on the $ \{|100\rangle ,|010\rangle \}$ subspace of
DFS$_{3}$($1$), which we achieved by subtracting out the action on
DFS$_{3}$($-1$).

In an analogous manner, an independent $su(2)$ can be enacted on the $
\{|011\rangle ,|101\rangle \}$ subspace of DFS$_{3}$($-1$) by using the
Hamiltonians acting on DFS$_{2}$($0$) in conjunction with ${\bf T}_{23}^{Q}$ to
subtract out the $su(2)$ action on DFS$_{3}$($1$).\footnote{ Since ${\bf
T}_{23}^{Q}$ annihilates every state except those that are $ |11\rangle $ over
qubits $2$ and $3$, namely $|011\rangle \in $ DFS$_{3}$($ -1 $) and
$|111\rangle \in $DFS$_{3}$($-3$), the only non-zero blocks in its matrix are
\[
M_{-3}({\bf T}_{23}^{Q})=1,\quad M_{-1}({\bf T}_{23}^{Q})=\left(
\begin{array}{ccc}
1 &  &  \\ & 0 & 0 \\ & 0 & 0
\end{array}
\right) .
\]
} Thus we can obtain independent action for each of the daughters of DFS$ _{2}
$($0$), i.e., separate actions on the subspace spanned by $ \{|010\rangle
,|100\rangle \}$ and $\{|011\rangle ,|101\rangle \}$.

Having established independent action on the two {\em subspaces} of
DFS$_{3}$($1$) and DFS$_{3}$($-1$) arising from DFS$_{2}$($0$), we need only
show that we can obtain the full action on DFS$_{3}$($1$) and DFS$_{3}$($-1$).
For DFS$_{3}$($1$) we need to {\em mix} the subspace $\{|010\rangle
,|100\rangle \}$ over which we can already perform independent $su(2)$, with
the $|001\rangle $ state. To do so, note that the effect of the exchange
operation ${\bf E}_{23}$ is to flip $|001\rangle $ and $|010\rangle $, and
leave $|100\rangle $ invariant. Thus the matrix representation of ${\bf E}
_{23}$ is
\begin{equation}
M_{1}({\bf E}_{23})=\left(
\begin{array}{ccc}
0 & 1 &  \\ 1 & 0 &  \\ &  & 1
\end{array}
\right) .
\end{equation}
Unfortunately, ${\bf E}_{23}$ has a simultaneous action on DFS$_{3}$($-1$).
This, however, is not a problem, since we have already constructed an
independent $su(2)$ on DFS$_{3}$($1$) elements. Thus we can eliminate the
simultaneous action by simply forming commutators with these $su(2)$ elements.
The Lie algebra generated by these commutators will act independently on {\em
all} of DFS$_{3}$($1$). In fact we claim this Lie algebra to be all of $su(3)$
(see Appendix~\ref{apd} for a general proof). In other words, the Lie algebra
spanned by the $su(2)$ elements $\{\sigma _{x},\sigma _{y},\sigma _{z}\}$
acting on the subspace $\{|100\rangle ,|010\rangle \}$, together with the
exchange operation ${\bf E} _{23}$, generate all of $su(3)$ independently on
DFS($1$). A similar argument holds for DFS$_{3}$($-1$). This construction
illustrates the induction step: we have shown that it is possible to perform
independent $su(d_{K})$ actions on all four of the DFS$_{3}$($K$) ($K=\pm 3,\pm
1$), given that we can perform independent action on the three DFS$_{2}$($K$)
($K=\pm 1,0$). In Fig.~(\ref{figure2}) we have further illustrated these
considerations by depicting the action of exchange on two of the $4$-qubit
DFSs. Let us now proceed to the general proof.

{\it Proof---} By induction.

The case $n=2$ already treated above will serve to initialize the induction.
Assume now that the theorem is true for $n-1$ qubits and let us show that it is
then true for $n$ qubits as well.

First note that each DFS$_{n}$($K$) is constructed either from the DFS$
_{n-1}$($K-1$) (to its lower left) by adding a $|0\rangle $ for the $n^{{\rm
th}}$ qubit, or from DFS$_{n-1}$($K+1$) (to its upper left) by adding a $
|1\rangle $: the states in DFS$_{n}$($K$) correspond to all paths ending in $
(n,K)$ that either come from below (B) or from the top (T). See
Fig.~(\ref{figure3}).

If we apply a certain gate ${\bf U}=\exp (i{\bf H}t)$ to DFS$_{n-1}$($K+1$),
then this operation will induce the same ${\bf U}$ on DFS$_{n}$($K$), by acting
on all paths (states) entering DFS$_{n}$($K$) from above. At the same time
${\bf U}$ is induced on DFS$_{n}$($K+2$) by acting on all paths entering this
DFS from below. So, ${\bf U}$ affects two DFSs {\em simultaneously}. In other
words, the set of valid Hamiltonians ${\sf H} _{K+1}^{n-1}$ [acting on $n-1$
qubits and generating $su(d_{K+1})$] on DFS$ _{n-1}$($K+1$), that we are given
by the induction hypothesis, induces a {\em simultaneous} action of
$su(d_{K+1})$ on DFS$_{n}$($K$) (on the paths coming from above only) and
DFS$_{n}$($K+2$) (on the paths coming from below only). Additionally, it does
not affect any other $n$-qubit DFS, since we assumed that the action on
DFS$_{n-1}$($K+1$) was {\em independent}, and the only $n$-qubit DFSs built
from DFS$_{n-1}$($K+1$) are DFS$_{n}$($K$) and DFS$ _{n}$($K+2$). These
considerations are depicted schematically in Fig.~(\ref{figure3}).

We now show how to annihilate, for a given non-trivial (i.e., dimension $>1$)
DFS$_{n}$($K$), the unwanted simultaneous action on other DFSs (the ``{\em
independence step''}). We then proceed to obtain the entire $su(d_{K})$, by
using the $su(d_{K\pm 1})$ on DFS$_{n-1}$($K\pm 1$) that are given by the
induction hypothesis (the ``{\em mixing step''}).

\subsubsection{\noindent Independence}

Let us call all the $t_{K}$ paths converging on DFS$_{n}$($K$) from above
``Top-states'', or T-states for short, and the $b_{K}$ paths converging from
below ``Bottom- (or B) states'' (recall that there is a 1-to-1 correspondence
between paths and states). The total number of paths converging on a given DFS
is exactly its dimension, so $d_{K}=t_{K}+b_{K}$. By using the induction
hypothesis on DFS$_{n-1}$($K+1$) we can obtain $ su(t_{K})$ (generated by ${\sf
H}_{K+1}^{n-1}$) on the T-states of DFS$_{n}$($K$), which will simultaneously
affect the B-states in the higher lying DFS$_{n}(K+2)$ as $su(b_{K+2}$) (note
that $t_{K}=b_{K+2}$). The set ${\sf H} _{K+1}^{n-1}$ is non-empty only if
$n-3\geq K+1\geq -(n-3)$ [because the ``highest'' and ``lowest'' DFS are always
one-dimensional and $su(1)=0$]. If this holds then DFS$_{n}$($K+2$) ``above''
DFS$_{n}$($K$) is non-trivial (dimension $>1$), and there are paths in
DFS$_{n}$($K$) ending in $ |11\rangle $ (``down, down''). This is exactly the
situation in which we can use ${\bf T}_{n-1,n}^{Q}$ to wipe out the unwanted
action on DFS$_{n}$($K+2$): recall that ${\bf T}_{n-1,n}^{Q}$ annihilates all
states except those {\em ending} in $|11\rangle $, and therefore affects
non-trivially only these special T-states in each DFS. Since the operations in
${\sf H} _{K+1}^{n-1}$ affect only B-states on DFS$_{n}$($K+2$), ${\bf
T}_{n-1,n}^{Q}$ commutes with ${\sf H}_{K+1}^{n-1}$ on DFS$_{n}$($K+2$).
Therefore the commutator of ${\bf T}_{n-1,n}^{Q}$ with elements in ${\sf
H}_{K+1}^{n-1}$ annihilates all states not in DFS$_n$($K$).\footnote{ The
argument thus far closely parallels the discussion above showing how to
generate an independent $su(2)$ on the $\{|011\rangle ,|101\rangle \}$ subspace
of DFS$_{3}$($-1$), starting from the $su(2)$ on DFS$_{2}$($0$) and ${\bf
T}_{23}^{Q}$.} To show that commuting ${\bf T}_{n-1,n}^{Q}$ with ${\sf
H}_{K+1}^{n-1}$ generates $su(t_{K})$ on the T-states of DFS$_n$($K$) we need
the following lemma, which shows how to form $su(d)$ from an overlapping
$su(d-1)$ and $su(2)$:

{\it Enlarging Lemma---} Let ${\mathcal H}$ be a Hilbert space of dimension $d$
and let $|i\rangle \in {\mathcal H}$. Assume we are given a set of Hamiltonians
$ {\sf H}_{1}$ that generates $su(d-1)$ on the subspace of ${\mathcal H}$ that
does not contain $|i\rangle $ and another set ${\sf H}_{2}$ that generates $
su(2)$ on the subspace of ${\mathcal H}$ spanned by $\{|i\rangle ,|j\rangle
\}$, where $|j\rangle $ is another state in ${\mathcal \ H}$. Then $[{\sf
H}_{1}, {\sf H}_{2}]$ (all commutators) generates $su(d)$ on ${\mathcal H}$
under closure as a Lie-algebra (i.e., via scalar multiplication, addition and
Lie-commutator).

{\it Proof---} See Appendix~\ref{apd}

Now consider two states $|i\rangle ,|j\rangle \in $DFS$_{n}$($K$) such that $
|i\rangle $ ends in $|11\rangle $ and $|j\rangle $ is a\ T-state, but does not
end in $|11\rangle .$ Then we can generate $su(2)$ on the subspace spanned by
$\{|i\rangle ,|j\rangle \}$ as follows: (i) We use the exchange interaction
$\bar{{\bf X}}_{ij}=|i^{\prime }\rangle \langle j^{\prime }|+|j^{\prime
}\rangle \langle i^{\prime }|$ [a prime indicates the bitstring with the last
bit (a $1$ in this case)\ dropped] in $su(t_{K})\in {\sf H}_{K+1}^{n-1}$ to
generate a simultaneous action on DFS$_{n}$($K$) and DFS$_{n}$($K+2$). This
interaction is represented by a $2\times 2$ $\sigma _{x}$-matrix in the
subspace spanned by $\{|i\rangle ,|j\rangle \}$. (ii) $ {\bf T}_{n-1,n}^{Q}$ is
represented by the $2\times 2$ matrix ${\rm diag} (1,0)=\frac{1}{2}\left( {\bf
I}+\sigma _{z}\right) $ in the same subspace, and commutes with $\bar{{\bf
X}}_{ij}$ on DFS$_{n}$($K+2$) (since $\bar{{\bf X}}_{ij}$ affects only B-states
in DFS$_{n}$($K+2$)$,$ and ${\bf T} _{n-1,n}^{Q}$ is non-zero only on states
ending in $|11\rangle $). Thus we can use it to create an independent action on
DFS$_{n}$($K$) alone: ${\bf \bar{Y}}_{ij}=i[{\bf T}_{n-1,n}^{Q},\bar{{\bf
X}}_{ij}]$, $\bar{{\bf Z}}_{ij}=\frac{i}{ 2}[\bar{{\bf Y}}_{ij},\bar{{\bf
X}}_{ij}]$.

Together $\{\bar{{\bf Y}}_{ij},\bar{{\bf Z}}_{ij}\}$ generate $su(2)$
independently on $\{|i\rangle ,|j\rangle \}\in$ DFS$_{n}$($K$). Since these
operators vanish everywhere except on DFS$_{n}$($K$), their commutators with
elements in $ {\sf H}_{K+1}^{n-1}$ [acting as $su(t_{K})$] will annihilate all
other DFSs. Therefore, using the Enlarging Lemma, in this way all operations in
$su(t_{K})$ acting on DFS$_n$($K$) {\em only} can be generated.

So far we have shown how to obtain an independent $su(t_{K})$ on the T-states
of DFS$_{n}$($K$) using ${\sf H}_{K+1}^{n-1}$ (for $K\leq n-4$). To obtain an
independent $su(b_{K})$ on the {\em B-states} of DFS$_{n}$($K)$ we use
Hamiltonians in ${\sf H}_{K-1}^{n-1}$ (acting on DFS$_{n-1}$($K-1)$ -- the DFS
from below). This will generate a simultaneous $su(b_{K})$ in DFS$ _{n}$($K$)
and $su(t_{K-2})$ in DFS$_{n}$($K-2$). To eliminate the unwanted action on
DFS$_{n}$($K-2$) we apply the previous arguments almost identically, except
that now we use ${\bf T}_{n-1,n}^{P}$ to wipe out the action on all states
except those ending in $|00\rangle $. We thus get an independent $su(b_{K})$ on
DFS$_{n}$($K$). Together, the ``above'' and ``below'' constructions
respectively provide independent $su(t_{K})$ and $ su(b_{K})$ on
DFS$_{n}$($K$). Finally, note that we did not really need both ${\bf
T}_{ij}^{P}$ and ${\bf T}_{ij}^{Q}$, since once we established independent
action on the T-states, we could have just subtracted out this action when
considering the B-states. Also, the specific choice of ${\bf T}_{ij}^{P,Q}$ was
rather arbitrary (though convenient): in fact almost any other diagonal
interaction would do just as well.

\subsubsection{Mixing} In order to induce operations between the two sets
of paths (from ``above'' and from ``below'') that make up DFS$_{n}$($K$)
consider the effect of ${\bf E}_{n-1,n}$. This gate does not affect any paths
that ``ascend'' two steps to $(n,K)$ (corresponding to bitstrings ending in
$|00\rangle $) and paths that ``descend'' two steps (ending in $ |11\rangle $),
but it flips the paths that pass from $(n-2,K)$ via $ (n-1,K+1) $ with the
paths from $(n-2,K)$ via $(n-1,K-1)$ [see Fig.~(\ref{figure3})]. It does this
for all DFSs simultaneously.

In order to get a full $su(d_{K})$ on DFS$_{n}$($K$) we need to ``mix'' $
su(t_{K})$ (on the T-states) and $su(b_{K})$ (on the B-states) which we already
have. We show how to obtain an {\em independent} $su(2)$ between a T-state and
a B-state. By the Enlarging Lemma this generates $su(d_{K})$.

Since $n\geq 3$ DFS$_{n}$($K$) contains states terminating in $|00\rangle $
and/or $|11\rangle $. Let us assume,  without loss of generality, that states
terminating in $ |00\rangle $ are present, and let $|i\rangle $ be such a state
(B-state). Let $|j\rangle $ be a B-state not terminating in $|00\rangle $, and
let $ |k\rangle ={\bf E}_{n-1,n}|j\rangle $ ($|k\rangle $ is a T-state). Let
${\bf \bar{Z}}_{ij}=|i\rangle \langle i|-|j\rangle \langle j|\in su(b_{K})$,
and recall that we have {\em independent} $su(b_{K})$. Then as is easily
checked, $i[{\bf E}_{n-1,n},\bar{{\bf Z}}_{ij}]\equiv \bar{{\bf Y}}_{jk}$
yields $ \sigma _{y}$ between $|j\rangle $ and $|k\rangle $ {\em
only}.\footnote{ Since ${\bf E}_{n-1,n}=|i\rangle \langle i| + |k\rangle
\langle j| + |j\rangle \langle k| + O$, where $O$ is some action on an
orthogonal subspace.} In addition, $ \bar{{\bf Z}}_{jk} \equiv \frac{i}{2}[{\bf
E}_{n-1,n}, \bar{{\bf Y}}_{jk}] $ gives $\sigma _{z}$ between $|j\rangle $ and
$|k\rangle $, thus completing a generating set for $su(2)$ on the B-state
$|j\rangle $ and the T-state $|k\rangle $, that affects these two states only
and annihilates all other states. This completes the proof.

To summarize, we have shown {\em constructively} that it is possible to
generate the entire Lie algebra $su(d_{K})$ on a given weak
collective-decoherence DFS$_{n}$($K$) of dimension $d_{K}$, from the elementary
composition of the operations of scalar multiplication, addition,
Lie-commutators (conjugation by unitaries was not necessary in the weak
collective decoherence case). Moreover, this $su(d_{K})$ can be generated
independently on each DFS, implying that universal quantum computation can be
performed inside {\em each} DFS$_{n}$($K$). Naturally, one would like to do
this on the largest DFS. Since given the number of qubits $n$ the dimensions of
the DFSs are $d_{K}= { {n \choose K} } $, the largest DFS is the
decoherence-free sub{\em space} $K=0$. In principle it is possible, by virtue
of the independence result, to universally quantum compute {\em in parallel} on
all DFSs.

\newpage

\begin{figure}[ht]
 \psfig{figure=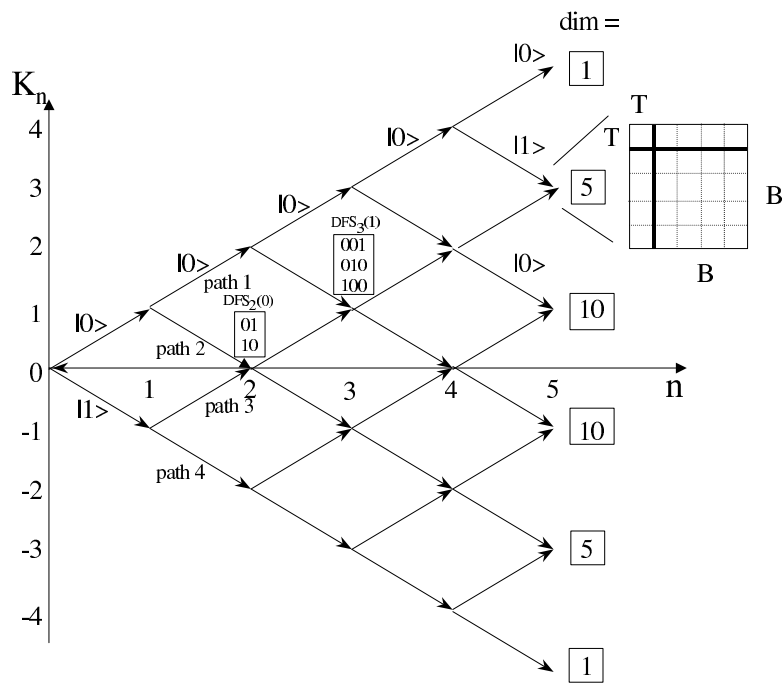,width=5.3in,angle=270}
\vspace{0.5cm} \caption{\em Graphical representation for visualizing the weak
collective decoherence universality proof}  \label{figure1} The horizontal axis
marks the number of qubits. The vertical axis shows (number of $0$'s - number
of $1$'s) in each state ($K_n$). Each state in the standard basis thus
corresponds to a path from the origin which follows the indicated arrows. The
dimension of a DFS corresponds to the multiple pathways through which one can
arrive at the same $J_n$. The DFSs are labeled by their values of $n$ and
$K_n$, as DFS$_n$($K_n)$.  The insert shows the matrix structure of operators
acting on DFS$_5(3)$, in terms of Top (T) and Bottom (B) states (see text for
definition of these). Note that there is only one T-state entering DFS$_5(3)$,
whence the action of exchange is represented by a $1\times 1$ block.
\end{figure}

\begin{figure}[ht]
 \psfig{figure=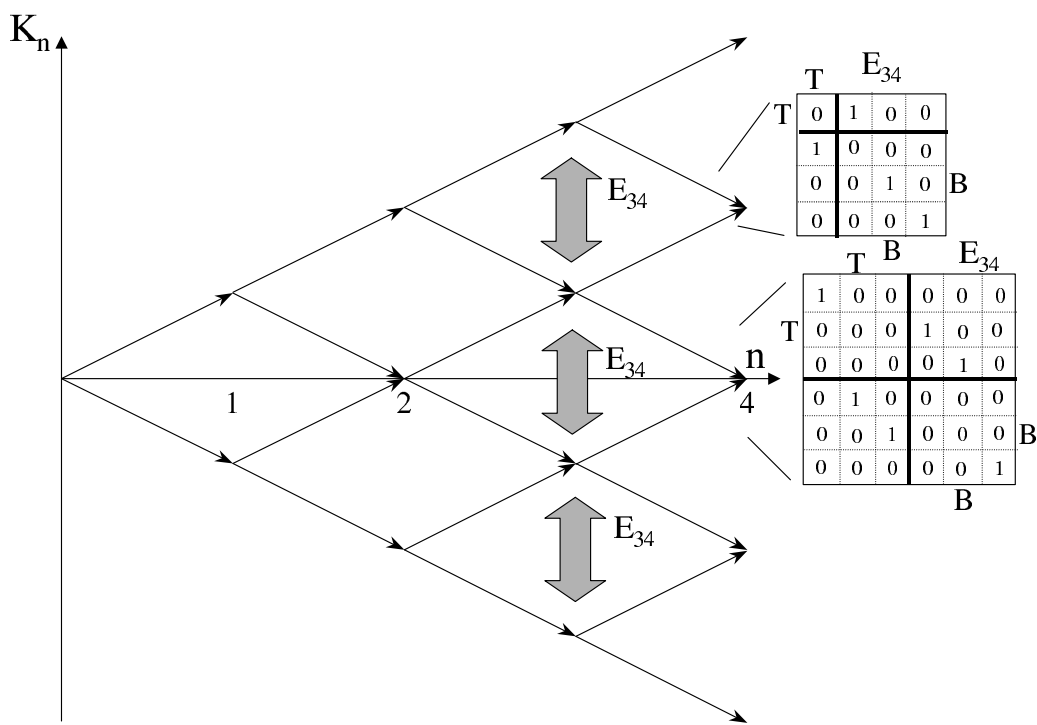,width=6in,angle=270}
 \caption{\em Graphical representation of the action of exchange on DFS states
for weak collective decoherence}  \label{figure2}  Exchange acts to
simultaneously flip different paths to a given DFS$_n$($K_n$). Axes and labels
are as defined in Figure 1.  $E_{ij}$ denotes the exchange of the $i$-th and
$j$-th qubits. The matrices displayed at right are the representations of ${\bf
E}_{34}$ on DFS$_4(0)$ (lower) and DFS$_4(2)$ (upper).
\end{figure}

\begin{figure}[ht]
\psfig{figure=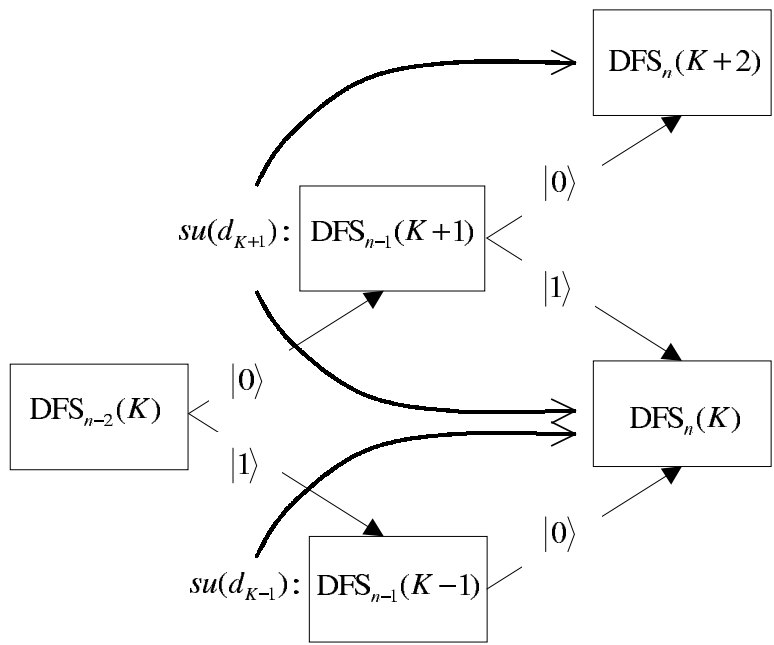,width=4in}  \caption{\em Detailed structure of the
pathways connecting adjacent DFSs in the weak collective decoherence case}
\label{figure3}  The action of the different $su$ Lie algebras is indicated by
the superposed heavy arrows. DFS$_n$($K$) denotes the DFS arising from $n$
qubits and having eigenvalue $K$.
\end{figure}

\chapter{Proof of universality on the strong collective decoherence DFS}
\label{apc}

\begin{quote}
{\em What is now proved was once only imagined}
\begin{flushright} --William Blake \end{flushright}
\end{quote}

We begin by examining the action of the exchange interaction on the three and
four qubit strong collective decoherence DFS.

\section{Quantum Computation on the $n=3$ and $n=4$ qubit strong collective
decoherence DFS} \label{n=3:exchange}

We begin our discussion of universal quantum computation on strong collective
decoherence DFSs by examining the simplest strong collective decoherence DFS
which supports encoding of quantum information: the $n=3$ decoherence-free
subsystem. We label these states by $|J,\lambda ,\mu \rangle $. Recall that the
$J=3/2$ irrep is not degenerate and the $J=1/2$ irrep has degeneracy $2$. The
$J=3/2$ states can be written as $|\frac{3}{2},0,\mu \rangle $, with $\mu = m
=\pm 3/2, \pm 1/2$. Since the action of exchange does not depend on $\mu$
(recall that it affects paths, i.e., the $\lambda$ component only) it suffices
to consider the action on the representative $\mu = 3/2$ only: $|111\rangle$.
Let us then explicitly calculate the action of exchanging the first two
physical qubits on this state and the four $J=1/2$ states. Using
Eq.~(\ref{eq:3strongdfs}):
\begin{eqnarray}
{\bf E}_{12}|\frac{3}{2},0,\frac{3}{2} \rangle &=&{\bf E}_{12}|111\rangle
=|\frac{3}{ 2},0,\frac{3}{2} \rangle  \nonumber \\ {\bf
E}_{12}|{\frac{1}{2}},0,0\rangle &=&{\bf E}_{12}{\frac{1}{\sqrt{2}}} \left(
|010\rangle -|100\rangle \right) ={\frac{1}{\sqrt{2}}}\left( |100\rangle
-|010\rangle \right) =-|{\frac{1}{2}},0,0\rangle  \nonumber \\ {\bf
E}_{12}|{\frac{1}{2}},0,1\rangle &=&{\bf E}_{12}{\frac{1}{\sqrt{2}}} \left(
|011\rangle -|101\rangle \right) ={\frac{1}{\sqrt{2}}}\left( |101\rangle
-|011\rangle \right) =-|{\frac{1}{2}},0,1\rangle  \nonumber \\ {\bf
E}_{12}|{\frac{1}{2}},1,0\rangle &=&{\bf E}_{12}{\frac{1}{\sqrt{6}}} \left(
-2|001\rangle +|010\rangle +|100\rangle \right) =|{\frac{1}{2}} ,1,0\rangle
\nonumber \\ {\bf E}_{12}|{\frac{1}{2}},1,1\rangle &=&{\bf
E}_{12}{\frac{1}{\sqrt{6}}} \left( 2|110\rangle -|101\rangle -|011\rangle
\right) =|{\frac{1}{2}} ,1,1\rangle .
\end{eqnarray}
Focusing just on the $J=1/2$ states, the exchange action on $|\lambda \rangle
\otimes |\mu \rangle $ can thus be written as:
\begin{equation}
{\bf E}_{12}=-\sigma _{z}\otimes {\bf I}.
\end{equation}
Since the action of the ${\bf S}_{\alpha }$ operators on the $J=1/2$ states is
${\bf I}_{n_{1/2}}\otimes gl(2)$, this explicit form for ${\bf E}_{12}$
confirms that is has the expected structure of operators in the commutant of
the algebra spanned by the ${\bf S}_{\alpha }$. It can also be seen that
quantum information should be encoded in the $|\lambda \rangle $ component.

Using similar algebra it is straightforward to verify that the effect of the
three possible exchanges on the $n=3$ DFS states are given by:
\begin{equation}
{\bf E}_{12}=\left(
\begin{array}{ccc}
1 & 0 & 0 \\ 0 & -1 & 0 \\ 0 & 0 & 1
\end{array}
\right) \quad {\bf E}_{23}=\left(
\begin{array}{ccc}
1 & 0 & 0 \\ 0 & {\frac{1}{2}} & -{\frac{\sqrt{3}}{2}} \\ 0 &
-{\frac{\sqrt{3}}{2}} & -{\frac{1}{2}}
\end{array}
\right) \quad {\bf E}_{13}=\left(
\begin{array}{ccc}
1 & 0 & 0 \\ 0 & {\frac{1}{2}} & {\frac{\sqrt{3}}{2}} \\ 0 &
{\frac{\sqrt{3}}{2}} & -{\frac{1}{2}}
\end{array}
\right) ,
\end{equation}
where the rows and columns of these matrices are labeled by the basis elements
$\{|J=3/2,\lambda =0\rangle ,|J=1/2,\lambda =0\rangle ,|J=1/2,\lambda =1\rangle
\}$. As expected from general properties of the commutant, the exchange
operators do not mix the different $J$ irreps. Now,
\begin{eqnarray}
{\frac{1}{3}}({\bf E}_{12}+{\bf E}_{13}+{\bf E}_{23}) &=&\left(
\begin{array}{ccc}
1 & 0 & 0 \\ 0 & 0 & 0 \\ 0 & 0 & 0
\end{array}
\right)  \nonumber \\ {\frac{1}{2}}(-{\bf E}_{12}+{\bf E}_{13}+{\bf E}_{23})
&=&\left(
\begin{array}{ccc}
0 & 0 & 0 \\ 0 & 1 & 0 \\ 0 & 0 & -1
\end{array}
\right)  \nonumber \\ {\frac{1}{\sqrt{3}}}({\bf E}_{13}-{\bf E}_{23}) &=&\left(
\begin{array}{ccc}
0 & 0 & 0 \\ 0 & 0 & 1 \\ 0 & 1 & 0
\end{array}
\right) ,
\end{eqnarray}
showing that the last two linear combinations of exchanges look like the Pauli
$\sigma _{z}$ and $\sigma _{x}$ on DFS$_{3}(1/2)$. Using a standard Euler angle
construction it is thus possible to perform any $SU(2)$ gate on this DFS.
Moreover, it is possible to act independently on DFS$_{3}(3/2)$ and
DFS$_{3}(1/2)$. In other words, we can perform $U(1)$ on DFS$_{3}(3/2)$ alone,
and $SU(2)$ on DFS$_{3}(1/2)$ alone. Note, however, that at this point we
cannot yet claim universal quantum computation on a register composed of
clusters of DFS$_{3}(J)$'s ($J$ constant) because we have not shown how to
couple such clusters.

For $n=4$ the Hilbert space splits up into one $J=2$-irrep [DFS$_{4}(2)$],
three $J=1$-irreps [DFS$_{4}(1)$], and two $J=0$-irreps [DFS$_{4}(0)$] -- see
Table (\ref{tab:strongdfs}). Direct calculation of the effect of exchange on
these DFSs shows that we can independently perform $su(1)$ (i.e. zero), $
su(3)$, and $su(2)$. In particular, we find that\cite{Bacon:00a}:
\begin{equation}
{\bf X}={\frac{1}{\sqrt{3}}}({\bf E}_{23}-{\bf E}_{13})\quad {\bf Y}={\frac{ i
}{2\sqrt{3}}}[{\bf E}_{23}-{\bf E}_{13},{\bf E}_{34}]\quad {\bf Z}={\frac{i }{
2}}[{\bf Y},{\bf X}]=-{\bf E}_{12}
\end{equation}
act as the corresponding $su(2)$ Pauli operators on DFS$_{4}(0)$ only. Further,
the following operators act independently on the $J=1$-irreps (rows and columns
are labeled by $\lambda =0,1,2$. The action occurs simultaneously on all three
$\mu $ components corresponding to a given $ \lambda $):
\begin{eqnarray}
{\bf Y}_{13} &=&{\frac{3i}{2\sqrt{2}}}[{\bf E}_{12},{\bf E}_{34}]=\left(
\begin{array}{ccc}
0 & 0 & -i \\ 0 & 0 & 0 \\ i & 0 & 0
\end{array}
\right) ,\quad {\bf X}_{13}={\frac{i}{2}}[{\bf E}_{12},{\bf Y}_{13}]=\left(
\begin{array}{ccc}
0 & 0 & 1 \\ 0 & 0 & 0 \\ 1 & 0 & 0
\end{array}
\right) ,  \nonumber \\ \quad {\bf Z}_{13} &=&{\frac{i}{2}}[{\bf Y}_{13},{\bf
X}_{13}]=\left(
\begin{array}{ccc}
1 & 0 & 0 \\ 0 & 0 & 0 \\ 0 & 0 & -1
\end{array}
\right) ,\quad {\bf Y}_{23}={\frac{2i}{\sqrt{3}}}[{\bf E}_{23},{\bf Z}
_{13}]=\left(
\begin{array}{ccc}
0 & 0 & 0 \\ 0 & 0 & -i \\ 0 & i & 0
\end{array}
\right) . \nonumber \\
\end{eqnarray}
These operators clearly generate $su(3)$, and hence we have an independent $
SU(3)$ action on DFS$_{4}(1)$.

\section{Universal Quantum Computation on the $n\geq 5$ qubit strong
collective decoherence DFSs} \label{SCDinduction}

We are now ready to prove our central result: that {\em using only the two-body
exchange Hamiltonians} every unitary operation can be performed on a strong
collective decoherence DFS. More specifically:

\begin{theorem}
 For any $n\geq 2$ qubits undergoing strong collective decoherence, there
exist sets of Hamiltonians ${\sf H}_{J}^{n}$ obtained from exchange
interactions only via scalar multiplication, addition, Lie-commutator and
unitary conjugation, acting as $su(d_{J})$ on the DFS corresponding to the
eigenvalue $J$. Furthermore each set acts {\em independently} on this DFS only
(i.e., with zeroes in the matrix representation corresponding to their action
on the other DFSs).
\end{theorem}

In preparation for the proof of this result let us note several useful facts:

(i) The exchange operators do not change the value of $m$, because they are in
the commutant of ${\tt A}=\{S_{\alpha }\}$.  Therefore in order to evaluate the
action of the exchange operators on the different DFS$_{n}(J)$ ($n$ given) it
is convenient to fix $ m$, and in particular to work in the basis given by the
maximal $m$ value ($m=J$). Expressions for these ``maximal'' states in terms of
$ |J_{1},J_{2},\dots ,J_{n-2};m\rangle $ and the single qubit states of the
last two qubits are given in Appendix~\ref{apd}.

(ii) Every $({\bf S}^{[k]})^{2}$ can be written as a sum of exchange operators
and the identity operation. This follows from noting that the exchange operator
can be expanded as
\begin{equation}
{\bf E}_{ij}={\frac{1}{2}}\left( {\bf I}+\sigma _{x}^{i}\sigma _{x}^{j}+\sigma
_{y}^{i}\sigma _{y}^{j}+\sigma _{z}^{i}\sigma _{z}^{j}\right) ,
\end{equation}
so that:
\begin{equation}
({\bf S}^{[k]})^{2}=k\left(1- {\frac{k}{4}}\right) {\bf I}+{\frac{1}{2}}
\sum_{i\neq j=1}^{k}{\bf E}_{ij}.
\end{equation}
Thus $({\bf S}^{[k]})^{2}$ is a Hamiltonian which is at our disposal.

We are now ready to present our proof by induction. Recall the DFS
dimensionality formula for $n_{J}$, Eq.~(\ref{eq:strongdim}). We assume that it
is possible to perform $su(n_{J})$ on each of the different DFS$_{n-1}(J)$ {\em
independently} using only exchange operators and the identity Hamiltonian. Our
construction above proves that this is true for $3$ and $4$ qubits. The
assumption that the actions we can perform can be performed independently
translates into the ability to construct Hamiltonians which annihilate all of
the DFSs except a desired one on which they act as $su(n_{J})$.

As in the weak collective decoherence case a specific DFS$_{n}(J)$ of dimension
$n_{J}$ splits into states which are constructed by the subtraction of angular
momentum from DFS$ _{n-1}(J+1/2)$ (T-states), or by the addition of angular
momentum to DFS$ _{n-1}(J-1/2)$ (B-states) [see Fig.~(\ref{figure5})].
Performing $su(n_{J+1/2})$ on DFS$_{n-1}(J+1/2)$ will simultaneously act on
DFS$_{n}(J)$ and DFS$_{n}(J+1)$. In other words, $ su(n_{J+1/2})$ on
DFS$_{n-1}(J+1/2)$ acts on both the B-states of DFS$ _{n}(J+1)$ and on the
T-states of DFS$_{n}(J)$. We split the proof into three steps. In the first
step we obtain an $su(2)$ set of operators which acts only on DFS$_{n}(J)$ and
mixes particular B- and T-states. In the second step we expand the set of
operators which mix B- and T-states to cover all possible $su(2)$ algebras
between any two B- and T-states. Finally, in the third step we apply a Mixing
Lemma which shows that we can obtain the full $ su(n_{J})$ (i.e., also mix
B-states and mix T-states).

\subsubsection{T- and B-Mixing}
There are two simple instances where there is no need to show independent
action in our proof: (i) The (upper) $J=n/2$ -irrep is always $1$-dimensional,
so the action on it is always trivial (i.e., the Hamiltonian vanishes and hence
the action is independent by definition); (ii) For odd $n$ the ``lowest''
DFS$_{n}(1/2)$ is acted upon independently by the $su(n_{0})$ from
DFS$_{n-1}(0)$ [i.e., $su(n_{0})$ cannot act ``downward'']. In order to
facilitate our construction we extend the notion of T and B-states one step
further in the construction of the DFS. TB-states are those states which are
constructed from T-states on ($n-1$)-qubits and from the B-states on $n$-qubit
states [see Fig.~(\ref{figure5})]. Similarly we can define the BT, TT, and
BB-states:
\begin{eqnarray}
|{\rm TT}\rangle  &\equiv &|J_{1},\dots ,J_{n-3},J_{n}+1,J_{n}+{\frac{1}{2}}
,J_{n};m=J_{n}\rangle =
\begin{array}{rl}
\searrow  &  \\ & \searrow
\end{array}
\\
|{\rm BT}\rangle  &\equiv &|J_{1},\dots ,J_{n-3},J_{n},J_{n}+{\frac{1}{2}}
,J_{n};m=J_{n}\rangle =\nearrow \searrow   \nonumber \\ |{\rm TB}\rangle
&\equiv &|J_{1},\dots ,J_{n-3},J_{n},J_{n}-{\frac{1}{2}} ,J_{n};m=J_{n}\rangle
=\searrow \nearrow   \nonumber \\ |{\rm BB}\rangle &\equiv &|J_{1},\dots
,J_{n-3},J_{n}-1,J_{n}-{\frac{1}{2}} ,J_{n};m=J_{n}\rangle =
\begin{array}{rl}
& \nearrow  \\ \nearrow  &
\end{array}
.
\end{eqnarray}
Every DFS$_{n}(J)$ can be broken down into a direct sum of TT, BT, TB, and
BB-states; e.g., as seen in Fig.~(\ref{figure4}), in DFS$_{6}(1)$ there are 1
TT, 3 TB, 3 BT and 2 BB states. Note that for $J=n/2-1$ there are no TT-states,
for $J=0$ there are no BB and BT-states, for $J=1/2$ there are no BB-states,
and otherwise there are as many TB as there are BT states,

At this point it is useful to explicitly give the action of exchange on the
last two qubits of a strong collective decoherence DFS. Using
Eq.~(\ref{eq:twodeep}) we find (assuming the existence of the given states,
i.e., $n$ large enough and $J$ not too large) the representation
\begin{equation}
{\bf E}_{n,n-1}=\left(
\begin{array}{cccc}
1 & 0 & 0 & 0 \\ 0 & -\cos (\theta _{J+1}) & \sin (\theta _{J+1}) & 0 \\ 0 &
\sin (\theta _{J+1}) & \cos (\theta _{J+1}) & 0 \\ 0 & 0 & 0 & 1
\end{array}
\right)
\begin{array}{c}
{\rm TT} \\ {\rm BT} \\ {\rm TB} \\ {\rm BB}
\end{array}
\label{eq:lastex}
\end{equation}
where $\tan (\theta _{J})=2\sqrt{J(J+1)}$. Thus exchange acts to transform the
BT and TB states entering a given DFS into linear combinations of one another,
while leaving invariant the BB and TT states.

Let us now consider the action of $su(n_{J-1/2})$ from DFS$_{n-1}(J-1/2)$ [see
Fig.~(\ref{figure5})]. It acts on DFS$_{n}(J-1)$ and DFS$_{n}(J)$
simultaneously. However, since the T-states of DFS$_{n}(J-1)$ and the B-states
of DFS$_{n}(J)$ share the same set of quantum numbers $\{J_{1},...,J_{n-1}\}$,
the action of the $su(n_{J-1/2})$ operators is identical on these two sets of
states.

We first deal with the case where the number of BT-states of DFS$_{n}(J)$ is
greater than 1. As can be inferred from Fig.~(\ref{figure4}), this condition
corresponds to $J<n/2-1$ and $n>4$. We will separately deal with the $ J=n/2-1$
case at the end of the proof. Let $|a\rangle $ and $|b\rangle $ be any two
orthogonal BT-states of DFS$_{n}(J)$ (i.e., states differing only by the paths
on the first $n-2$ qubits). Corresponding to these are $ \{|a^{\prime }\rangle
,|b^{\prime }\rangle \}$: a pair of orthogonal BT-states of DFS$_{n}(J)$. One
of the elements in $su(n_{J-1/2})$ is the traceless operator ${\bf C}=|a\rangle
\langle a|-|b\rangle \langle b|$, which we have at our disposal by the
induction hypothesis. Consider $i[{\bf E }_{n,n-1},{\bf C}]$: since ${\bf
E}_{n,n-1}$ acts as identity on BB states, even though ${\bf C}$ has an action
on DFS$_{n}(J-1)$ the commutator acting on the BB states of DFS$_{n}(J-1)$
vanishes. The action of $i[{\bf E} _{n,n-1},{\bf C}]$ on the BT and TB states
can be calculated by observing, using Eq.~(\ref{eq:lastex}), that the matrix
representations of ${\bf C}$ and ${\bf E}_{n,n-1}$ are, in the ordered
$\{|a^{\prime }\rangle ,|b^{\prime }\rangle ,|a\rangle ,|b\rangle \}$ basis:
\begin{eqnarray}
{\bf C}&=&{\rm diag}(0,0,1,-1)=\frac{1}{2}\left( {\bf I}\otimes {\sigma }_{z} -
{\sigma }_{z}\otimes {\sigma }_{z}\right)   \nonumber \\ {\bf
E}_{n,n-1}&=&\left(
\begin{array}{cccc}
-\cos (\theta _{J}) & 0 & \sin (\theta _{J}) & 0 \\ 0 & -\cos (\theta _{J}) & 0
& \sin (\theta _{J}) \\ \sin (\theta _{J}) & 0 & \cos (\theta _{J}) & 0 \\ 0 &
\sin (\theta _{J}) & 0 & \cos (\theta _{J})
\end{array}
\right)  \nonumber \\&=&-\cos (\theta _{J}){\sigma }_{z}\otimes {\bf I}+\sin
(\theta _{J}){\sigma }_{x}\otimes {\bf I}.
\end{eqnarray}
This yields:
\begin{equation}
i[{\bf E}_{n,n-1},{\bf C}]=-\sin (\theta _{J}){\sigma }_{y}\otimes {\bf \sigma
}_{z}=i\sin (\theta _{J})\left( -|a\rangle \langle a^{\prime }|+|a^{\prime
}\rangle \langle a|+|b\rangle \langle b^{\prime }|-|b^{\prime }\rangle \langle
b|\right) .
\end{equation}
Now let $|c\rangle $ be a TT-state of DFS$_{n}(J)$. Such a state always exists
unless $J=n/2-1$, which is covered at the end of the proof. Then there is an
operator ${\bf D}=|a^{\prime }\rangle \langle a^{\prime }|-|c\rangle \langle
c|$ in $ su(n_{J+1/2})$.\footnote{ We need to subtract $|c\rangle \langle c|$
in order to obtain a traceless operator.} It follows that:
\begin{equation}
{\bf X}_{aa^{\prime }}\equiv {\frac{1}{\sin (\theta _{J})}}i[i[{\bf E}
_{n,n-1},{\bf C}],{\bf D}]=|a\rangle \langle a^{\prime }|+|a^{\prime }\rangle
\langle a|,
\end{equation}
acts like an encoded ${\sigma }_{x}$ on $|a\rangle $ and $|a^{\prime }\rangle $
and annihilates all other states. Further, one can implement the commutator
\begin{equation}
{\bf Y}_{aa^{\prime }}=i[{\bf X}_{aa^{\prime }},{\bf D}]=i\left( |a\rangle
\langle a^{\prime }|-|a^{\prime }\rangle \langle a|\right) ,
\end{equation}
which acts like an encoded ${\sigma }_{y}$ on $|a\rangle $ and $|a^{\prime
}\rangle $. Finally, one can construct ${\bf Z}_{aa^{\prime }}=i[{\bf X}
_{aa^{\prime }},{\bf Y}_{aa^{\prime }}]=|a\rangle \langle a|-|a^{\prime
}\rangle \langle a^{\prime }|$. Thus we have shown that for $J<n/2-1$ we can
validly (using only exchange Hamiltonians) perform $su(2)$ operations between
$|a\rangle $, a specific B-state and $|a^{\prime }\rangle $, its corresponding
T-state, on DFS$_n(J)$ only.

\subsubsection{Extending the $su(2)$'s}
We now show that by using the operation of conjugation by a unitary we can
construct $su(2)$ between {\em any} two B and T-states. To see this recall
Eq.~(\ref{eq:Heff}), which allows one to take a Hamiltonian ${\bf H}$ and turn
it via conjugation by a unitary gate into the new Hamiltonian ${\bf H} _{{\rm
eff}}={\bf U}{\bf H}{\bf U}^{\dagger }$. By the induction hypothesis we have at
our disposal every $SU$ gate which acts on the T-states of DFS$_{n}(J)$ [and
simultaneously acts on the B-states of DFS$_{n}(J+1)$] and also every $SU$ gate
which acts on the B-states of DFS$_{n}(J)$ [and simultaneously acts on the
T-states of DFS$_{n}(J-1)$]. Above we have shown how to construct ${\bf X}$,
${\bf Y}$, and ${\bf Z}$ operators between specific T- and B-states:
$|a^{\prime }\rangle $ and $|a\rangle $. Let $ |i^{\prime }\rangle $ and
$|i\rangle $ be some other T- and B-states of DFS$_{n}(J)$, respectively. Then
we have at our disposal the gate ${\bf P}_{i^{\prime }i}=|a^{\prime }\rangle
\langle i^{\prime }|+|i^{\prime }\rangle \langle a^{\prime }|+|a\rangle \langle
i|+|i\rangle \langle a|+{\bf O}$ where ${\bf O}$ is an operator which acts on a
DFS other than DFS$_{n}(J)$ (included to make ${\bf P}_{i^{\prime }i}$ an $SU$
operator). It is simple to verify that
\begin{equation}
{\bf X}_{i^{\prime }i}={\bf P}_{i^{\prime }i}{\bf X}_{aa^{\prime }}{\bf
P}_{i^{\prime }i}^{\dagger }=|i^{\prime }\rangle \langle i|+|i\rangle \langle
i^{\prime }|,
\end{equation}
which acts as an encoded $\sigma _{x}$ between $|i^{\prime }\rangle $ and
$|i\rangle $. Note that because ${\bf X}_{aa^{\prime }}$ only acts on
DFS$_{n}(J)$, $ {\bf X}_{i^{\prime }i}$ will also only act on the same DFS.
Similarly one can construct ${\bf Y}_{i^{\prime }i}={\bf P}_{i^{\prime }i}{\bf
Y}_{aa^{\prime }}{\bf P} _{i^{\prime }i}^{\dagger }$ and ${\bf Z}_{i^{\prime
}i}={\bf P}_{i^{\prime }i}{\bf Z}_{aa^{\prime }}{\bf P} _{i^{\prime
}i}^{\dagger }$ which act, respectively, as encoded $\sigma _{y}$ and $ \sigma
_{z}$ on $|i^{\prime }\rangle $ and $|i\rangle $. Thus we have shown that one
can implement every $su(2)$ between any two T- and B-states in DFS$_{n}(J)$.
Each of these $su(2)$ operations is performed {\em independently} on
DFS$_{n}(J)$.

\subsubsection{Mixing T- and B-States}
Next we use a Lemma proved in Appendix~\ref{apd}:

{\it Mixing Lemma}: Given is a Hilbert space ${\mathcal H}$ $={\mathcal
H}_{1}\oplus {\mathcal H}_{2}$ where $\dim {\mathcal H}_{j}=$ $n_{j}$. Let
$\{|i_{1}\rangle \}$ and $\{|i_{2}\rangle \}$ be orthonormal bases for
${\mathcal H}_{1}$ and ${\mathcal H }_{2}$ respectively. If one can implement
the operators ${\bf X} _{i_1 i_2}=|i_{1}\rangle \langle i_{2}|+|i_{2}\rangle
\langle i_{1}|$, ${\bf Y} _{i_1 i_2}=i|i_{1}\rangle \langle
i_{2}|-i|i_{2}\rangle \langle i_{1}|$, and $ {\bf Z}_{i_1 i_2}=|i_{1}\rangle
\langle i_{1}|-|i_{2}\rangle \langle i_{2}|$, then one can implement
$su(n_{1}+n_{2})$ on ${\mathcal H}$.

Above we have explicitly shown that we can obtain every ${\bf X}_{i_1 i_2}$, $
{\bf Y}_{i_1 i_2}$, and ${\bf Z}_{i_1 i_2}$ acting independently on
DFS$_{n}(J)$. Thus direct application of the Mixing Lemma tells us that we can
perform $ su(n_{J})$ {\em independently} on this DFS.

{\it Special case of }$J=n/2-1$: We have neglected DFS$_{n}(n/2-1)$ because it
did not contain two different BT-states (nor a TT) state. The dimension of this
DFS is $n-1$. We now show how to perform $su(n-1)$ on this DFS using the fact
that we have already established $su(n_{J=n/2-2})$ on DFS$_{n}(n/2-2 $). First,
note that by the induction hypothesis we can perform $ su(n_{J=n/2-3/2})$
independently on DFS$_{n-1}(n/2-3/2)$. As above, this action simultaneously
affects DFS$_{n}(n/2-1)$ and DFS$_{n}(n/2-2)$. However, since we can perform
$su(n_{J=n/2-2})$ on DFS$_{n}(n/2-2)$, we can subtract out the action of
$su(n_{J=n/2-3/2})$ on DFS$_{n}(n/2-2)$. Thus we can obtain $su(n_{J=n/2-3/2})$
on all of the B-states of DFS$_{n}(n/2-1)$. But the exchange operator ${\bf
E}_{n,n-1}$ acts to mix the B-states with the single T-state of
DFS$_{n}(n/2-1)$. Thus we can construct an $su(2)$ algebra between that
single-T state and a single B-state in a manner directly analogous to the above
proof for $J<n/2-1$. Finally, by the Enlarging Lemma it follows that we can
obtain $su(n-1)$ on DFS$_{n}(n/2-1)$.

This concludes the proof that the exchange interaction is independently
universal on each of the different strong-collective-decoherence DFSs.

\begin{figure}[ht]
 \psfig{figure=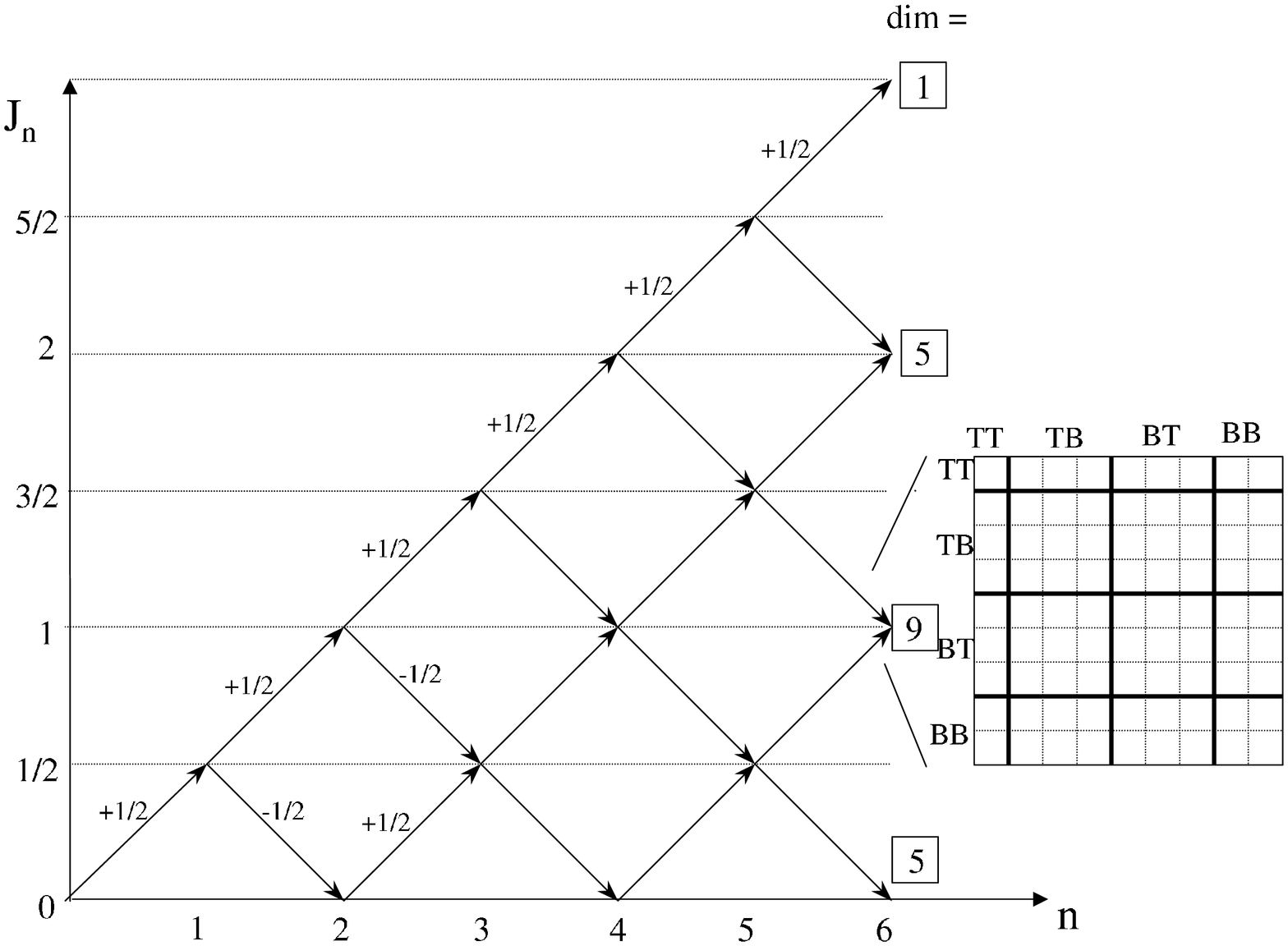,width=6in,angle=270} \vspace{0.5cm}
\caption{\em Graphical representation of DFS states for strong collective
decoherence} \label{figure4} The horizontal axis is the number of qubits, $n$,
just as in Figure 1 for weak collective decoherence.  The vertical axis is now
the total angular momentum ${\bf J}$ obtained by summing angular momenta of $n$
spin 1/2 particles representing the $n$ qubit, rather than just the
$z$-component of this. The DFSs are denoted by DFS$_n$($J$) as before.  Each
state in the DFS is represented by a pathway from the origin along the arrows
as indicated. The insert shows the matrix structure  of operators acting on
DFS$_6(1)$, given in terms of TT, TB, BT, and BB-states.
\end{figure}

\begin{figure}[ht]
\psfig{figure=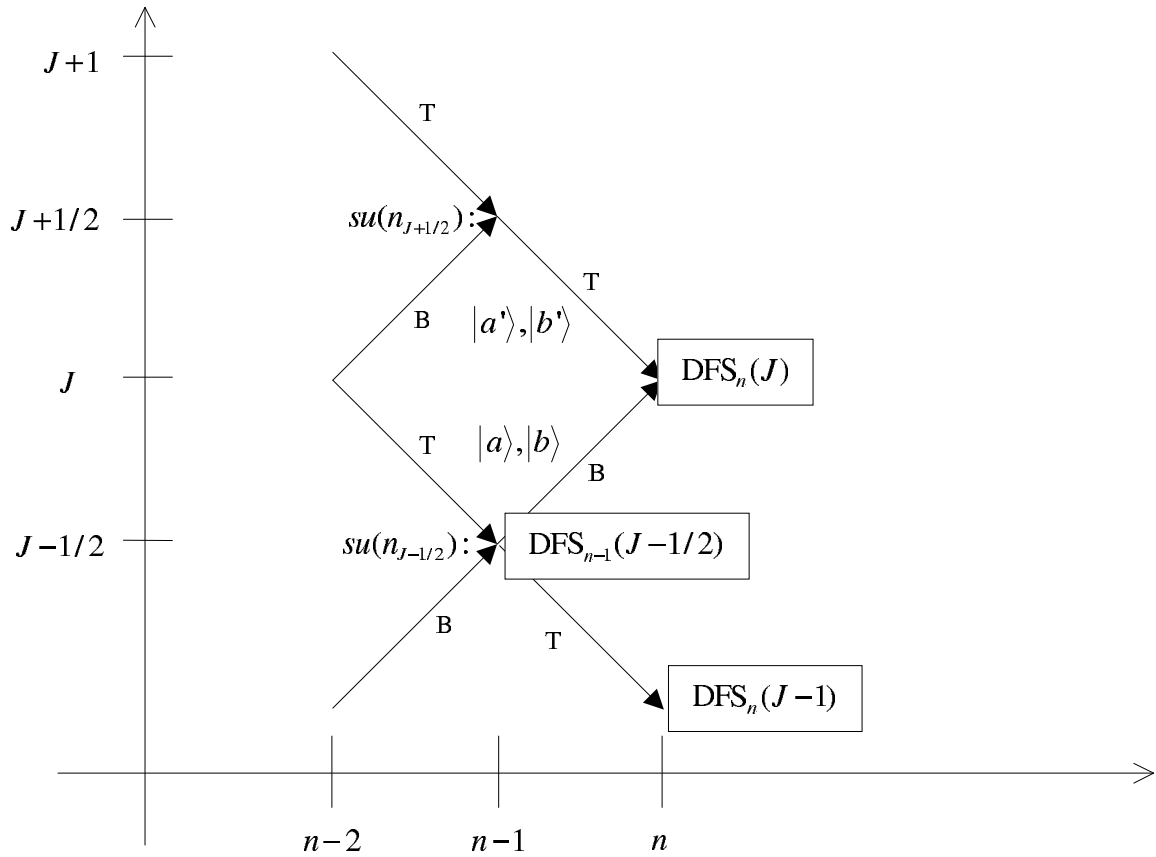,width=6in} \vspace{0.5cm} \caption{\em Scheme for
visualizing the inductive proof of universal computation using only the
exchange Hamiltonian} \label{figure5} TB- and BT-states of DFS$_n(J)$ are
indicated. $su(n_{J-1/2})$ acts on DFS$_n(J-1)$ and on DFS$_n(J)$ via
DFS$_{n-1}(J-1/2)$. See text in Section~\ref{SCDinduction} for details.
\end{figure}

\chapter{Lemmas and results for the collective universality proofs} \label{apd}

Here we collect some lemmas and results which are used in the universality
proofs for collective decoherence DFSs.

\section{Maximal-$\lowercase{m}_{J}$ States of the Strong Collective Decoherence DFS}

We show how to recursively express the $n$-particle total spin-$J$ states in
terms of ($n-1$)-particle states. Let us focus on DFS$_{n}(J)$ and in
particular on the maximal-$m_{J}$ state in it:
\begin{equation}
|\psi \rangle =|J_{1},\dots ,J_{n-1},J;m_{J}=J\rangle .
\end{equation}
In general ($J\neq 0,{n/2}$) there are two kinds of states: bottom ($|\psi
\rangle _{{\rm B}}$) and top ($|\psi \rangle _{{\rm T}}$) ones. The angular
momentum addition rule that must be satisfied for adding a single
spin-$\frac{1}{2}$ particle is that
\[
m_{J_{n-1}}\pm \frac{1}{2}=m_{J}.
\]
The B-state comes from adding a particle to the maximal $m_{J}$ state in DFS$
_{n-1}(J-1/2)$, which is:
\begin{equation}
|{\rm B}\rangle =|J_{1},\dots ,J_{n-2},J{-{\frac{1}{2}}};m_{J_{n-1}}=J-{{
\frac{1}{2}}}\rangle . \label{eq:B}
\end{equation}
There is only one way to go from $|{\rm B}\rangle $ to $|\psi \rangle _{{\rm
B}}$, namely to add $1/2$ to $m_{J_{n-1}}=J-{{\frac{1}{2}}}$ in order to obtain
$m_{J}=J$. Thus
\begin{equation}
|\psi \rangle _{{\rm B}}=|{\rm B}\rangle |{\frac{1}{2}},{\frac{1}{2}}\rangle
\label{eq:psiB},
\end{equation}
where $|{\frac{1}{2}},{\frac{1}{2}}\rangle $ is the single-particle spin-up
state. The situation is different for the T-state, which is constructed by
adding a particle to
\begin{equation}
|{\rm T}_{\pm }\rangle =|J_{1},\dots ,J_{n-2},J+{{\frac{1}{2}}}
;m_{J_{n-1}}=J\pm {{\frac{1}{2}}}\rangle . \label{eq:Tpm}
\end{equation}
These two possibilities give:
\begin{equation}
|\psi \rangle _{{\rm T}}=\alpha |{\rm T}_{+}\rangle |{\frac{1}{2}},-{\frac{1
}{2}}\rangle +\beta |{\rm T}_{-}\rangle |{\frac{1}{2}},{\frac{1}{2}}\rangle .
\label{eq:psiT}
\end{equation}
To find the coefficients $\alpha $ and $\beta $, we use the collective raising
operator ${\bf S}_{+}={\bf S}_{x}+i{\bf S}_{y}$, where we recall that ${\bf
s}_{\alpha }^{(k)}=\frac{1}{2}\sum_{i=1}^{k}\sigma _{\alpha }^{i}$. Since
$|\psi \rangle $ is a maximal-$m_{J}$ state it is annihilated by $ {\bf
s}_{+}\equiv {\bf S}_{\alpha }^{(n)}$. Similarly, $|{\rm T}_{+}\rangle $ is
annihilated by ${\bf S}_{+}^{(n-1)}$. Therefore, since ${\bf S}_{+}={\bf S
}_{+}^{(n-1)}+\frac{1}{2}\sigma _{+}^{n}$:
\begin{eqnarray*}
{\bf S}_{+}|{\rm T}_{+}\rangle |{\frac{1}{2}},-{\frac{1}{2}}\rangle &=&|{\rm
T}_{+}\rangle |{\frac{1}{2}},{\frac{1}{2}}\rangle  \\ {\bf S}_{+}|{\rm
T}_{-}\rangle |{\frac{1}{2}},{\frac{1}{2}}\rangle &=&\sqrt{ 2J+1}|{\rm
T}_{+}\rangle |{\frac{1}{2}},{\frac{1}{2}}\rangle ,
\end{eqnarray*}
where in the second line we used the elementary raising operator formula $ {\bf
J}_{+}|j,m\rangle =\left[ j(j+1)-m(m+1)\right] ^{1/2}|j,m+1\rangle $ with
$j=J+{{\frac{1}{2}}}$ and $m=J-{{\frac{1}{2}}}$. Application of ${\bf S} _{+}$
to Eq.~(\ref{eq:psiT}) thus yields:
\begin{equation}
\alpha +\sqrt{2J+1}\beta =0.
\end{equation}
Hence, up to an arbitrary phase choice, we find that
\begin{equation}
\alpha =-\sqrt{\frac{2J+1}{2J+2}}\quad \beta ={\frac{1}{\sqrt{ 2J+2}}.}
\end{equation}
The special cases of $J=0,{n/2}$ differ only in that the corresponding DFSs
support just T- and B-states, respectively. The calculation of the
coefficients, therefore, remains the same.

In a similar manner one can carry the calculation one particle deeper. Doing
this we find for the maximal-$m_{J}$ states (provided they exist):
\begin{eqnarray}
|{\rm TT}\rangle &\equiv& |J_{1}, \dots ,J_{n-3},J+1,J+{\frac{1}{2}}
,J;m_{J}=J\rangle \nonumber \\&=&\sqrt{\frac{2J+1}{2J+3}}|J_{1},\dots
,J_{n-3},J+1;m_{J_{n-2}}=J+1\rangle |{\frac{1}{2}},-{\frac{1}{2}}\rangle |{
\frac{1}{2}},-{\frac{1}{2}}\rangle   \nonumber \\
&&-\sqrt{\frac{2J+1}{(2J+2)(2J+3)}}|J_{1},\dots
,J_{n-3},J+1;m_{J_{n-2}}=J\rangle \nonumber \\ && \times \left(
|{\frac{1}{2}},{\frac{1}{2}}\rangle |{\frac{1}{2}},-{\frac{1}{2}}\rangle
+|{\frac{1}{2}},-{\frac{1}{2}}\rangle |{ \frac{1}{2}},{\frac{1}{2}}\rangle
\right)   \nonumber \\ &&+\sqrt{\frac{2}{(2J+2)(2J+3)}}|J_{1},\dots
,J_{n-3},J+1;m_{J_{n-2}}=J-1\rangle \nonumber \\ &&\times
|{\frac{1}{2}},{\frac{1}{2}}\rangle |{ \frac{1}{2}},{\frac{1}{2}}\rangle
\nonumber \\
 |{\rm BT}\rangle &\equiv& |J_{1},\dots ,J_{n-3},J,J+{\frac{1}{2}}
,J;m_{J}=J\rangle  \nonumber \\ &=&-\sqrt{\frac{2J+1}{2J+2}}|J_{1},\dots
,J_{n-3},J;m_{J_{n-2}}=J\rangle |{\frac{1}{2}},{\frac{1}{2}}\rangle |{\frac{1
}{2}},-{\frac{1}{2}}\rangle \nonumber \\
&&+{\frac{1}{\sqrt{(2J+2)(2J+1)}}}|J_{1},\dots ,J_{n-3},J;m_{J_{n-2}}=J\rangle
|{\frac{1}{2}},-{\frac{1}{2}}\rangle |{\frac{ 1}{2}},{\frac{1}{2}}\rangle
\nonumber \\ &&+\sqrt{\frac{2J}{(2J+1)(2J+2)}}|J_{1},\dots
,J_{n-3},J;m_{J_{n-2}}=J-1\rangle |{\frac{1}{2}},{\frac{1}{2}}\rangle |{
\frac{1}{2}},{\frac{1}{2}}\rangle \nonumber \\
 |{\rm TB}\rangle &\equiv& |J_{1},
\dots ,J_{n-3},J,J-{\frac{1}{2}} ,J;m_{J}=J\rangle \nonumber \\
&=&-\sqrt{\frac{2J}{2J+1}}|J_{1},\dots ,J_{n-3},J;m_{J_{n-2}}=J\rangle
|{\frac{1}{2}},-{\frac{1}{2}}\rangle |{\frac{ 1}{2}},{\frac{1}{2}}\rangle
\nonumber \\ &&+{\frac{1}{\sqrt{2J+1}}}|J_{1},\dots
,J_{n-3},J;m_{J_{n-2}}=J-1\rangle |{ \frac{1}{2}},{\frac{1}{2}}\rangle
|{\frac{1}{2}},{\frac{1}{2}}\rangle \nonumber
\\
 |{\rm BB}\rangle &\equiv& |J_{1},\dots ,J_{n-3},J-1,J-{\frac{1}{2}}
,J;m_{J}=J\rangle \nonumber \\ &=&|J_{1},\dots
,J_{n-3},J-1;m_{J_{n-2}}=J-1\rangle |{\frac{1 }{2}},{\frac{1}{2}}\rangle
|{\frac{1}{2}},{\frac{1}{2}}\rangle . \label{eq:twodeep}
\end{eqnarray}
Caution must be exercised in using these expressions near the boundary of
Table~(\ref{tab:strongdfs}), where some of the states may not exist.

\section{Enlarging Lemma}

Let ${\mathcal H}$ be a Hilbert space of dimension $d$ and let $|i\rangle \in
{\mathcal H}$. Assume we are given a set of Hamiltonians ${\sf H}_{1}$ that
generates $su(d-1)$ on the subspace of $ {\mathcal H}$ that does not contain
$|i\rangle $, and another set ${\sf H}_{2}$ that generates $su(2)$ on the
subspace of ${\mathcal H}$ spanned by $\left\{ |i\rangle ,|j\rangle \right\} $,
where $|j\rangle $ is another state in $ {\mathcal H}$. Then $[{\sf H}_{1},{\sf
H}_{2}]$ (all commutators) generates $su(d)$ on ${\mathcal H}$ under closure as
a Lie-algebra.

Proof: We explicitly construct the Lie-algebra $su(d)$ with the given
Hamiltonians. Let ${\tilde{{\mathcal H}}}\subset {\mathcal H}$ be the $d-1$
dimensional subspace ${\sf H}_{1}$ acts on. Let us show that we can generate
$su(2)$ between $|k\rangle \in {\tilde{{\mathcal H}}}$ and $|i\rangle $.

Let ${\bf X}_{ij}\equiv |i\rangle \langle j|+|j\rangle \langle i|\in {\sf H}
_{2}$ and ${\bf X}_{jk}\equiv |j\rangle \langle k|+|k\rangle \langle j|\in {\sf
H}_{1}$. Then
\begin{equation}
{\bf Y}_{ik}\equiv i[{\bf X}_{jk},{\bf X}_{ij}]=-i|i\rangle \langle
k|+i|k\rangle \langle i|
\end{equation}
acts as $\sigma _{y}$ on the states $|i\rangle ,|k\rangle $. Similarly
\begin{equation}
{\bf X}_{ik}\equiv i[{\bf Y}_{ij},{\bf X}_{jk}]=|i\rangle \langle k|+|k\rangle
\langle i|
\end{equation}
yields $\sigma _{x}$ on the space spanned by $|i\rangle ,|k\rangle $. These two
operations generate $su(2)$ on $|i\rangle ,|k\rangle $ for all $|k\rangle$ in
the subspace of ${\mathcal H}$ that does not contain $|i\rangle$. The Mixing
Lemma gives the desired result together with the observation that there we only
use elements in $[{\sf H}_{1},{\sf H}_{2}]$.

\section{Mixing Lemma}

Consider the division of an $n$ dimensional Hilbert space ${\mathcal H}$ into a
direct sum of two subspaces ${\mathcal H}_{1}\oplus {\mathcal H}_{2}$ of
dimensions $n_{1}$ and $n_{2}$ respectively. Suppose that $ |i_{n}\rangle $ is
an orthonormal basis for ${\mathcal H}_{n}$. Then the Lie algebras generated by
${\bf X}_{i_{1},i_{2}}=|i_{1}\rangle \langle i_{2}|+|i_{2}\rangle \langle
i_{1}|$, ${\bf Y}_{i_{1},i_{2}}=i|i_{1}\rangle \langle i_{2}|-i|i_{2}\rangle
\langle i_{1}|$, and ${\bf Z} _{i_{1},i_{2}}=|i_{1}\rangle \langle
i_{1}|-|i_{2}\rangle \langle i_{2}|$ generate $su(n)$.

Proof: We explicitly construct the elements of $su(n)$. Consider $i[ {\bf
X}_{i_{1},i_{2}},{\bf Y}_{j_{1},j_{2}}]$. Clearly, if $i_{1}\neq i_{2}\neq
j_{1}\neq j_{2}$ this equals zero and if $i_{1}=j_{1}$ and $ i_{2}=j_{2}$ then
this commutator is $-{\bf Z}_{i_{1},i_{2}}$. If, however, $ i_{1}=j_{1}$ and
$i_{2}\neq j_{2}$ this becomes
\begin{equation}
i[{\bf X}_{i_{1},i_{2}},{\bf Y}_{i_{1},j_{2}}]=-|i_{2}\rangle \langle
j_{2}|-|j_{2}\rangle \langle i_{2}|.
\end{equation}
Similarly:
\begin{equation}
i[{\bf X}_{i_{1},i_{2}},{\bf Y}_{j_{1},i_{2}}]=|i_{1}\rangle \langle
j_{1}|+|j_{1}\rangle \langle i_{1}|.
\end{equation}
Thus every $|i_{k}\rangle \langle j_{l}|+|j_{l}\rangle \langle i_{k}|$ is in
the Lie algebra. Similarly, $i[{\bf X}_{i_{1},i_{2}},{\bf X}_{j_{1},j_{2}}]$
yields
\begin{eqnarray}
i[{\bf X}_{i_{1},i_{2}},{\bf X}_{i_{1},j_{2}}] &=&i|i_{2}\rangle \langle
j_{2}|-i|j_{2}\rangle \langle i_{2}|  \nonumber \\ i[{\bf X}_{i_{1},i_{2}},{\bf
X}_{j_{1},i_{2}}] &=&i|i_{1}\rangle \langle j_{1}|-i|j_{1}\rangle \langle
i_{1}|.
\end{eqnarray}
Thus every $i|i_{k}\rangle \langle j_{l}|-i|j_{l}\rangle \langle i_{k}|$ is in
the Lie algebra. Taking the commutator of these with the $|i_{k}\rangle \langle
j_{l}|+|j_{l}\rangle \langle i_{k}|$ operators finally yields every $
|i_{k}\rangle \langle j_{l}|-|j_{l}\rangle \langle i_{k}|$. Since $su(n)$ can
be decomposed into a sum of overlapping $su(2)$'s\cite{Cahn:84a}, the Lie
algebra is the entire $su(n)$, as claimed.

\chapter{Controlled-phase on the four qubit strong collective decoherence DFS}
\label{ape}

In this Appendix we provide an analytically derived gate sequence for
performing a controlled-phase between the two four qubit strong collective
decoherence DFSs.  This sequence of operations was derived with the guidance of
two insights.  The first insight comes from Eq.~(\ref{eq:twodeep}).  This
equation describes how nearest neighbor exchange interactions act nontrivially
to mix only $BT$ and $TB$ pathways.  The second insight comes that executing
$\left( {\bf S}^{[k]} \right)^2$ can be executed as a sum of exchanges on the
first $k$ qubit and this can be used to subtract out the diagonal element of
Eq.~(\ref{eq:twodeep}) on the $TB$ and $BT$ pathways.  This allows us to
construct evolutions which flip between the different basis states of the
strong DFS basis (recall Section \ref{sec:strongbasis}).

In Figure \ref{fig:bumpy} we present the evolution of two conjoined four qubit
strong collective decoherence DFSs under that action of the following operators
\begin{eqnarray}
{\bf U}_1&=&\exp \left[ {i \pi \over \sqrt{3}} \left({\bf E}_{45} + {1 \over 2}
\left({\bf E}_{12} + {\bf E}_{13} + {\bf E}_{14} + {\bf E}_{23} + {\bf E}_{24}
+ {\bf E}_{34} \right) \right) \right] \nonumber \\
 {\bf U}_2&=&\exp \left[ {i \pi \over 4 \sqrt{2}} \left(-3 {\bf E}_{56}-{2
 \over 3} ({\bf E}_{68}+{\bf E}_{68}+{\bf E}_{78} ) \right) \right] \nonumber
 \\
 {\bf U}_3&=&\exp \left[ {i \pi \over 4 \sqrt{2}} \left(-3 {\bf E}_{34}-{2
 \over 3} ({\bf E}_{12}+{\bf E}_{13}+{\bf E}_{32} ) \right) \right] \nonumber
 \\
 {\bf U}_4&=&\exp \left[ {i \pi \over \sqrt{3}}\left({\bf E}_{23}+ {1 \over 2} {\bf E}_{12} \right)
 \right]\nonumber \\
 {\bf U}_5&=&\exp \left[ {i \pi \over \sqrt{3}} \left({\bf E}_{67}+{1 \over 2}
 {\bf E}_{78} \right) \right], \nonumber
\end{eqnarray}
and also the slightly more mysterious
\begin{eqnarray}
 {\bf U}_A&=& \exp \left[ -{i \over 2}\cos^{-1}\left(-{1 \over 3}\right) {\bf
E}_{45} \right] \nonumber \\
 {\bf U}_B&=& \exp \left[ -{i \pi \over 2}\left( {\bf E}_{12}+{\bf E}_{13}+{\bf E}_{14} + {\bf E}_{23}+{\bf E}_{24}+{\bf E}_{34} \right)
 \right]\nonumber \\
 {\bf U}_6&=& {\bf U}_A {\bf U}_B {\bf U}_A^\dagger {\bf U}_B^\dagger  {\bf U}_A {\bf U}_B {\bf U}_A^\dagger {\bf
 U}_B^\dagger.
\end{eqnarray}
In Figure~\ref{fig:bumpy} we demonstrate the result of these gates on the
original DFS states.  This figure is accurate up to single qubit phases (i.e.
phases which can be generated by single qubit rotations).  The first five ${\bf
U}_i$ acts imply to manipulate the encoded basis states to new strong DFS basis
states.  The final ${\bf U}_6$ was discovered by noting that after the basis
states have been transferred to the indicated strong DFS basis states, ${\bf
E}_{45}$ acts to mix two of these basis states with other strong DFS basis
states and ${\bf E}_{12}+{\bf E}_{13}+{\bf E}_{14}+{\bf E}_{23}+{\bf
E}_{24}+{\bf E}_{34}$ acts as a phase on these strong basis states.

In order to limit the total number of gates, one can further note that ${\bf
U}_2$ and ${\bf U}_3$ can be executed at the same time as can ${\bf U}_4$ and
${\bf U}_5$ because these gates operate on completely separate qubits.  Thus in
Figure~\ref{fig:bumpycnot} we present the layout for this quantum circuit and
from this diagram it becomes obvious that ${\bf U}_4$ is not needed in the gate
array.

\begin{figure}[h]
\psfig{figure=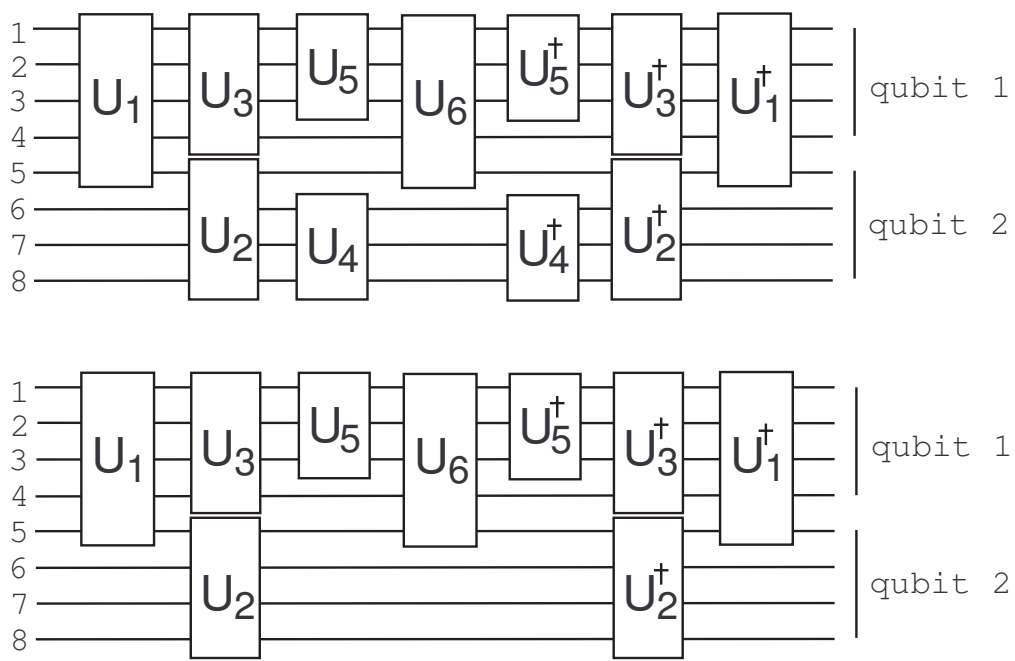,width=6in} \label{fig:bumpycnot} \caption{\em
Encoded controlled-phase gate sequence}
\end{figure}

\begin{figure}[h]
\psfig{figure=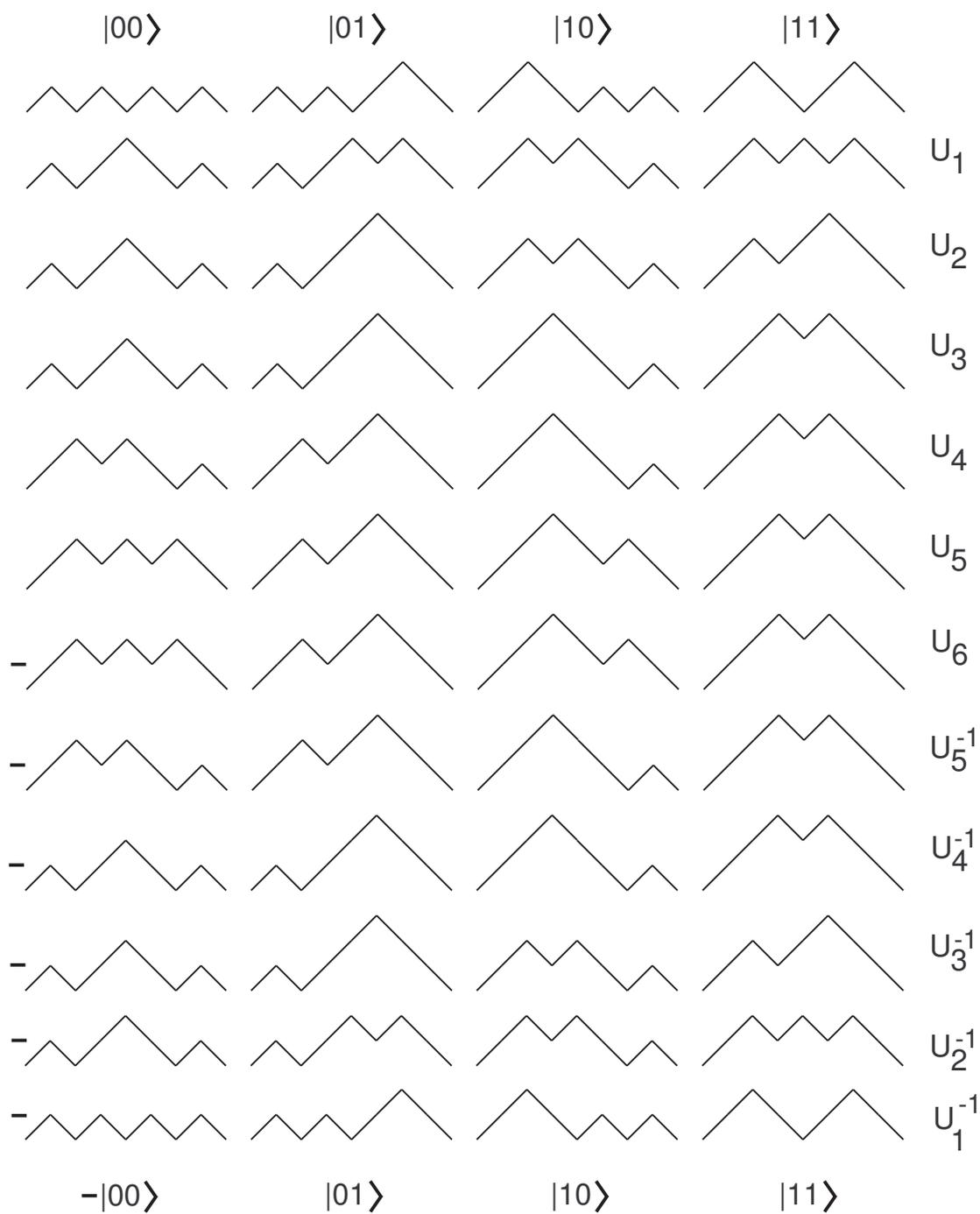,width=6in} \label{fig:bumpy} \caption{\em
Controlled-phase diagram demonstrating the usefulness of the strong collective
decoherence basis}
\end{figure}

\end{document}